\documentclass[a4paper,11pt]{article}
\pdfoutput=1 

\usepackage{jcappub} 

\usepackage[T1]{fontenc} 
\usepackage{mathrsfs}
\usepackage{amsmath}
\usepackage{upgreek}

\title{\boldmath Influences of tilted thin accretion disks on the observational appearance of hairy black holes in Horndeski gravity}

\author[a,1]{Shiyang Hu,\note{Corresponding author.}}
\author[b]{Dan Li,}
\author[c]{Chen Deng,}
\author[a,d]{Xin Wu,}
\author[a]{and Enwei Liang}

\affiliation[a]{School of Physical Science and Technology $\&$ Guangxi Key Laboratory for Relativistic Astrophysics, Guangxi University,\\Nanning, 530004 People's Republic of China}
\affiliation[b]{School of Mathematics and Physics, University of South China,\\Hengyang, 421001 People's Republic of China}
\affiliation[c]{School of Astronomy and Space Science, Nanjing University,\\Nanjing, 210023 People's Republic of China}
\affiliation[d]{School of Mathematics, Physics and Statistics $\&$ Center of Application and Research of Computational Physics, Shanghai University of Engineering Science,\\Shanghai, 201620 People's Republic of China}

\emailAdd{husy@st.gxu.edu.cn}
\emailAdd{ld$\_$physics@163.com}
\emailAdd{dengchen@smail.nju.edu.cn}
\emailAdd{wuxin$\_$1134@sina.com}
\emailAdd{lew@gxu.edu.cn}

\abstract{Research on the observational appearance of black holes, both in general relativity and modified gravity, has been in full swing since the Event Horizon Telescope Collaboration announced photos of M87$^{*}$ and Sagittarius A$^{*}$. Nevertheless, limited attention has been given to the impact of tilted accretion disks on black hole images. This paper investigates the $230$ GHz images of non-rotating hairy black holes illuminated by tilted, thin accretion disks in Horndeski gravity with the aid of a ray tracing method. The results indicate that reducing the scalar hair parameter effectively diminishes image luminosity and extends both the critical curve and the inner shadow. This trend facilitates the differentiation between hairy black holes and Schwarzschild black holes, especially in certain parameter spaces where the current Event Horizon Telescope array is capable of capturing such variations. Furthermore, we observe that the inclination of the tilted accretion disk can mimic the observation angle, consequently affecting image brightness and the morphology of the inner shadow. In specific parameter spaces, alterations in the tilt or position of the accretion disk can lead to a drift in the light spot within the images of hairy black holes. This finding may establish a potential correlation between the precession of the tilted accretion disk and image features. Additionally, through an examination of images depicting hairy black holes surrounded by two thin accretion disks, we report the obscuring effect of the accretion environment on the inner shadow of the black hole.}

\begin{document}
\maketitle
\flushbottom

\section{Introduction}
Black holes (BHs), the extremely compact objects predicted in the theory of general relativity, existed only as mathematical formulas widely disseminated in literature until several scholars independently recognized that Cygnus X-1 could be a BH \cite{Braes $&$ Miley (1972),Webster $&$ Murdin (1972)}. For nearly half a century thereafter, many pioneering explorations, both in astronomical observations and theoretical analysis, have been devoted to confirming the presence of BHs in the universe. Gravitational wave bursts from binary BHs mergers captured by the Laser-Interferometer Gravitational Wave Observatory (LIGO) \cite{Abbott et al. (2016)}, as well as the stochastic gravitational wave background detected by the Pulsar Timing Array \cite{Xu et al. (2023),Agazie et al. (2023),Antoniadis et al. (2023)}, provide indirect support for the existence of BHs. An even more inspiring achievement was the release of photos of M87$^{*}$ \cite{Akiyama et al. (2019a)} and Sagittarius A$^{*}$ \cite{Akiyama et al. (2022a)} by the Event Horizon Telescope Collaboration (EHT). It is believed that the bright rings appearing in these photos originate from the accretion flow around the BHs that are hidden in the shadows inside the rings. As a milestone on the road to unveiling the mysteries of BHs, the images of these celestial objects not only provide cast-iron and direct evidence for their existence but also present a completely new opportunity for testing the theories of general relativity \cite{Psaltis (2019),Gralla et al. (2020),Akiyama et al. (2022b),Younsi et al. (2023)} and exploring the accretion mechanism in the vicinity of BHs \cite{Akiyama et al. (2019b),Akiyama et al. (2022c)}.

The features of the BH images are significantly influenced by the accretion flow. Several authors have numerically simulated images of spherically accreting BHs and suggested that the boundaries of the BH shadow overlap with the photon ring and are hardly affected by the accretion details \cite{Narayan et al. (2019),Zeng et al. (2020),Gan et al. (2021a),Guo et al. (2021),Saurabh $&$ Jusufi (2021),He et al. (2022)}, including the dynamic characteristics, emission region, and specific emissivity profile of accreting gas. However, this dependency was found to fail in the case of thin disk accretion. Gralla et al. classified the disk emissions by scrutinizing the photon deflections in the vicinity of a Schwarzschild BH endowed with a geometrically and optically thin accretion disk into direct emission, lensing ring emission, and photon ring emission. Then, they pointed out that the size of the BH shadow is closely related to the radiation spectrum of the accretion disk \cite{Gralla et al. (2019)}. In particular, when the emission region of the accretion disk extends down to the event horizon, the BH shadow is confined to a small region, with a clear separation between its boundary and the critical curve, which represents the asymptotic image of the photon ring. This investigation provided new insights into the interpretation of BH photos and triggered a flood of work regarding the influences of thin accretion disks on BH images \cite{Peng et al. (2021),Gan et al. (2021b),Chael et al. (2021),Guo et al. (2022),Chakhchi et al. (2022),Hu et al. (2022),Wang et al. (2023),Meng et al. (2023),Silva et al. (2023),Yang et al. (2023a)}. However, it is worth noting that all of these works restricted the accretion disk to the equatorial plane. Consequently, the above investigations do not tell the whole story and it is natural to wonder what new signatures a tilted thin accretion disk would introduce to the BH image. Unfortunately, research in this domain has received little attention compared to the pervasive interest of the equatorial accretion disk scenario.

Indeed, a substantial body of observational evidence has confirmed the prevalence of tilted accretion disks around astrophysical BHs due to the widespread misalignment in angular momentum between the accreting matter and the BH \cite{Maccarone (2002),Caproni et al. (2006),van den Eijnden et al. (2017),Cui et al. (2023)}. Therefore, studies on morphological transitions in tilted accretion disks are crucial for a comprehensive understanding of BH spin evolution \cite{White et al. (2019)}, the Lense-Thirring effect \cite{Mashhoon et al. (1984)}, the formation of relativistic jets \cite{Mckinney et al. (2013),Liska et al. (2018),Liska et al. (2019)}, and accretion physics. Several authors found that in the tilted accretion disk scenario, accretion onto the BH predominantly occurs through plunging streams near high latitude regions, which is distinct from the case of an aligned accretion disk \cite{Fragile $&$ Anninos (2005),Fragile et al. (2007)}. They also pointed out that the tilted disk undergoes a periodic precession of $3$ ($M_{\odot}/M$) Hz in response to the Lense-Thirring effect, causing the inner disk to warp as well. Liska et al. developed a fully $3$D GPU-accelerated general relativistic magnetohydrodynamic (GRMHD) code, \texttt{H-AMR} \cite{Liska et al. (2022)}, and simulated the evolution of a moderately thin accretion disk (dimensionless scaleheight $h/r = 0.1$) tilted at $60^{\circ}$ around a rapidly rotating BH \cite{Liska et al. (2023a)}. The results indicate that the precession of the tilted accretion disk is synchronized with that of the relativistic jet. However, there is a phase lag of $10-40^{\circ}$ between the disk and the corona. More interestingly, with the help of GRMHD, Liska et al. investigated the structural evolution of very thin (dimensionless scaleheight $h/r \leq 0.03$), highly inclined accretion disks and observed that the disk can be torn not only into two independently precessing sub-disks, but even into three sub-disks within certain timescales \cite{Liska et al. (2021),Liska et al. (2023b),Kaaz et al. (2023),Musoke et al. (2023)}. This occurs because the differential Lense-Thirring torques overwhelm the viscous torques that maintain the integrity of the accretion disk. It is worth noting that the angular momentum of the inner disk can align with the BH rotation within a short timescale due to the Bardeen-Petterson effect \cite{Liska et al. (2019),Liska et al. (2021),Bardeen $&$ Petterson (1975),Nealon et al. (2015)}, while the outer accretion disk slowly aligns through the combination of the precessing jet torque and the frame dragging effect. As a result, these sub-disks exhibit a discontinuity in tilt angle and a sharp decrease in density. In brief, the morphology of the tilted accretion disk varies depending on the disk inclination, viscosity coefficient, dimensionless scaleheight, ion-electron temperature ratio, and BH parameters. This evolution, apparently, gives peculiar observational features to astrophysical BHs.

Chatterjee et al. conducted numerical simulations of a rapidly spinning Kerr BH with a tilted accretion disk at various observational bands \cite{Chatterjee et al. (2020)}, using the \texttt{H-AMR} and general relativistic radiative transfer (GRRT) code, \texttt{BHOSS} \cite{Younsi et al. (2016)}. The results indicate that altering the inclination of the accretion disk can impact the characteristics of the resulting images. Specifically, the north-south asymmetry of the emission ring amplifies with increasing disk tilt. Through a detailed examination of BH images across diverse parameter spaces, the authors suggested the possibility of M$87^{*}$ featuring a tilted disk/jet system. This proposition has been substantiated by recent investigations into the periodic precession of the M$87^{*}$ jet \cite{Cui et al. (2023)}. In a parallel study \cite{White et al. (2020)}, it was observed that a tilted disk introduces a non-axisymmetric shock in the foreground of the emission region. Consequently, the shadows cast by the BH exhibit reduced circularity compared to the aligned accretion disk scenario. While the utilization of the GRMHD and GRRT methodologies for studying BH images is deemed entirely reliable, it remains a complex and time-consuming process. Hence, it is imperative to generalize the static, tilted accretion disk model within the framework proposed by the authors in \cite{Gralla et al. (2019)}. This extension aims to facilitate a more streamlined exploration of novel, qualitative features contributed by tilted disks to BH images, addressing a current research gap. Furthermore, and most importantly, BH images are affected by the BH mass, spin, charge, and parameters in various theories of modified gravity. Since the current resolution of images is not sufficient to rule out the possibility of modified gravity \cite{Mizuno et al. (2018),Gralla (2021),Vagnozzi et al. (2023)}, studying the observational appearance of BHs in modified gravity is necessary.

Horndeski gravity, proposed in $1974$ \cite{Horndeski (1974)}, is an inspiring case of the four-dimensional scalar-tensor theory that includes a scalar field as well as an energy-momentum tensor field. Similar to general relativity, Horndeski's theory possesses the same diffeomorphism invariance and second-order field equations. Given the superiority of the Horndeski theory in explaining the accelerated expansion of the universe \cite{Kobayashi (2011),Kobayashi (2019)}, this theory and its BH solutions have attracted a great deal of attention in the scientific community \cite{Anabalon et al. (2014),Maselli et al. (2015),Bhattacharya $&$ Chakraborty (2017),Tattersall $&$ Ferreira (2018),Kreisch $&$ Komatsu (2018),Afrin $&$ Ghosh (2022),Jha et al. (2023)}. In particular, the authors of \cite{Bergliaffa et al. (2021)} presented a static, spherically symmetric BH solution in Horndeski gravity. The solution inevitably exhibits hairiness due to the presence of an additional matter field and is crucial for testing the no-hair theorem. Consequently, the properties of the Horndeski hairy BH have been studied extensively.

Kumar et al. investigated the strong gravitational lensing effects, such as the deflection angles, the relative magnification of lensing images, and the time delay of light rays, caused by supermassive non-rotating hairy BHs in Horndeski gravity, a potential method of distinguishing Horndeski BHs from their counterparts in general relativity using relativistic lensing images was discussed in detail \cite{Kumar et al. (2022)}. They also pointed out that the theoretical shadow size of the hairy BH is consistent with EHT observations of M87$^{*}$ and reported constraints on the scalar hair parameter. A study with respect to the inference of the hair parameter using the shadow image of Sagittarius A$^{*}$ has also been completed \cite{Vagnozzi et al. (2023)}. In order to explore the influences of plasma on the gravitational lensing effects of the hairy BH in Horndeski gravity, the authors in \cite{Atamurotov et al. (2022)} studied the deflection angles of light rays propagating in this spacetime under the framework of the weak field approach. They found that the light deflection angles depend not only on the scalar hair parameter but also on the distribution model of the plasma (uniform distribution or singular isothermal sphere distribution). Note that the features of the strong gravitational field of Horndeski BHs can be revealed by examining the motions of massive particles in the vicinity of the BH. Following this line of thought, Lin and Deng simulated a series of periodic and quasiperiodic orbits with different leaves, whirl numbers, and precession around the hairy BH by the elaborate selection of the particles' energy and angular momentum \cite{Lin $&$ Deng (2023)}. Their study obtained a preliminary bound on the hairy BH based on matching particles' relativistic periastron precession with that of the S2 star reported by the GRAVITY Collaboration, which is about $3-4$ orders of magnitude tighter than the constraint obtained by the shadow observations of M87$^{*}$. Quasiperiodic oscillation in the observed electromagnetic spectrum is a well-known astrophysical phenomenon that is expected to be explained by the quasiperiodic motion of massive particles near the BH. The author of \cite{Rayimbaev et al. (2023)} investigated the possible values of the upper and lower frequencies for quasiperiodic oscillations around non-rotating hairy BHs in Horndeski's theory and discovered a relationship between quasiperiodic oscillations and the radius of the innermost stable circular orbit of the massive particle. Using observational data of quasiperiodic oscillations from the microquasars GRS $1915$+$105$ and XTE $1550$-$564$, they also constrained the scalar hair parameter and the mass of hairy BHs. The propagations of external fields, such as scalar field, electromagnetic field, and Dirac field in Horndeski gravity were studied in \cite{Yang et al. (2023b)}. It was shown that the hairy BH is stable under the perturbations of those external fields. Overall, research on spherically symmetric hairy BHs in Horndeski gravity has attracted many scholars and covers various aspects of astrophysics. Meanwhile, further revealing the observational appearance of these BHs is of great significance for future observations.

Recently, Wang et al. studied the variations of the photon sphere and the BH shadow with the Horndeski hairy parameter \cite{Wang et al. (2023)}. The results showed that the photon sphere and shadow radii are monotonically decreasing functions of the hair parameter. In addition, they exhibited the observational appearance of hairy BHs surrounded by equatorial thin accretion disks, and well demonstrated the contributions of direct emissions, lensing ring emissions, and photon ring emissions to the luminosities of BH images, respectively. The effects of the hair parameter on the qualitative observational signatures of the BH are displayed intuitively. However, all these findings are based on the assumption that the remote observer is fixed in the north pole direction of the hairy BH, which is rarely encountered in natural astrophysical environments. Therefore, exploring the features of hairy BH images under different observation positions in Horndeski gravity is necessary, which is one motivation for the present paper. Another motivation stems from the intriguing open problem of what kind of new physical features a static, tilted accretion disk would introduce to the BH images. Despite its importance, this topic has received little attention in the literature.

The remainder of this paper is organized as follows. In section 2, we briefly review the metric of the hairy BH in Horndeski gravity and explore the inclined stable circular orbits of massive particles. These orbits provide theoretical evidence for the existence of a tilted accretion disk in the vicinity of the BH. In section 3, we simulate the observational images at $230$ GHz of the target BH surrounded by a tilted, thin accretion disk with the help of a ray tracing algorithm. The effects of disk inclination on BH images and redshift distributions are elucidated in detail. Additionally, we are intrigued by images of the hairy BH with multiple disk structures, which are further investigated and exhibited in section 4. The results and discussions are concluded in section 5. Throughout this paper, the dimensionless operations and geometric units are taken, consequently the speed of light $c$, the gravitational constant $G$, and the BH mass $M$ are taken as $1$.
\section{Tilted stable circular orbits of massive particles in Horndeski gravity}
In Boyer-Lindquist coordinates $x^{\mu}=(t,r,\theta,\varphi)$, the static spherically symmetric BH with scalar hair in Horndeski gravity can be found in \cite{Afrin $&$ Ghosh (2022),Atamurotov et al. (2022),Lin $&$ Deng (2023),Kumar et al. (2022),Wang et al. (2023)} as
\begin{eqnarray}\label{1}
\textrm{d}s^{2}=g_{tt}\textrm{d}t^{2}+g_{rr}\textrm{d}r^{2}+g_{\theta \theta}\textrm{d}\theta^{2}+g_{\varphi \varphi}\textrm{d}\varphi^{2},
\end{eqnarray}
where $g_{\mu \nu}$ is the covariant metric tensor, which contains only diagonal elements that read as
\begin{eqnarray}\label{2}
-g_{tt}=\frac{1}{g_{rr}}=f(r),\quad g_{\theta \theta}=r^{2},\quad g_{\varphi \varphi}=r^{2}\sin^{2}\theta,
\end{eqnarray}
with dimensionless lapse function
\begin{equation}\label{3}
f(r)=1-\frac{2}{r}+\frac{h}{r}\ln\left(\frac{r}{2}\right).
\end{equation}
Here, $h$ is the so-called scalar hair parameter with the unit of BH mass in the geometric units system. The spacetime described by the spherically symmetric Horndeski metric \eqref{1} is asymptotically flat and degenerates to the classical Schwarzschild case when the scalar hair is absent ($h \rightarrow 0$). Interestingly, by examining the variation of the lapse function \eqref{3} with radius $r$, previous studies have confirmed that the event horizon $r_{+}$ of the hairy BH is located at $2M$ when the scalar hair parameter satisfies the condition $h \geq -2$ \cite{Wang et al. (2023),Afrin $&$ Ghosh (2022),Lin $&$ Deng (2023),Rayimbaev et al. (2023),Yang et al. (2023b)}.

The motion of particles in Horndeski spacetime is governed by the Lagrangian formulism
\begin{equation}\label{4}
\mathscr{L} = \frac{1}{2} g_{\mu \nu} \dot{x}^{\mu} \dot{x}^{\nu},
\end{equation}
where $\dot{x}^{\mu}$ is the derivative of the generalized coordinates with respect to the affine parameter $\lambda$, which represents the four-velocity of the particle. Note that the Lagrangian \eqref{4} corresponds to massive particles when $\mathscr{L} = -1/2$, while it corresponds to photons when $\mathscr{L}$ takes the value of $0$. The existence of Killing vectors due to time-translational and rotational invariance implies that two conserved quantities appear in the dynamic system \eqref{4} and, respectively, correspond to the specific energy $E$ and angular momentum $L$ of the particle,
\begin{equation}\label{5}
E = -g_{tt}\dot{t},
\end{equation}
\begin{equation}\label{6}
L = g_{\varphi \varphi} \dot{\varphi}.
\end{equation}
Hence, the Lagrangian $\mathscr{L}$ can be rewritten as
\begin{equation}\label{7}
\mathscr{L} = \frac{1}{2} \left( \frac{E^{2}}{g_{tt}}+\frac{L^{2}}{g_{\varphi \varphi}}+g_{rr} \dot{r}^{2}+g_{\theta \theta} \dot{\theta}^{2}\right).
\end{equation}

The present paper does not confine the accretion disk to the equatorial plane of the BH, as has been extensively adopted in the existing literature. Instead, we mainly focus on the effects of a tilted, geometrically thin accretion disk on the BH images. The main body of this disk is composed of a series of inclined, stable circular orbits with different radii. For this purpose, we need to study the circular motion of the massive particle in an arbitrary plane with the aid of the effective potential. According to Eqs. \eqref{2} and \eqref{7} with $\mathscr{L} = -1/2$, the radial effective potential of the massive particle in the plane of tilt $\sigma$ can be deduced as
\begin{equation}\label{8}
V_{\textrm{eff}} = E = \sqrt{f(r)\left[1+\frac{L^{2}}{r^{2}\sin^{2}(\pi/2-\sigma)}\right]}.
\end{equation}
Figure 1 illustrates the behavior of the effective potential for test particles with various specific angular momentum $L$ at a fixed orbital plane inclination of $\sigma=60^{\circ}$. For each curve, the unstable circular orbit corresponds to the local maximum of the potential, while the position of the stable circular orbit is located at the local minimum, marked by a pentagram. Then, the radius of the stable circular orbit $r_{\textrm{SCO}}$ can be obtained based on conditions
\begin{equation}\label{9}
\frac{\partial V_{\textrm{eff}}}{\partial r} =0, \quad \frac{\partial^{2} V_{\textrm{eff}}}{\partial r^{2}} > 0.
\end{equation}
When replacing the latter criteria in Eq. \eqref{9} with $\partial^{2} V_{\textrm{eff}}/\partial r^{2} = 0$, the radius of the innermost stable circular orbit $r_{\textrm{ISCO}}$ of the massive particle can be calculated. The results in figure 1 have confirmed that in Horndeski spacetime \eqref{1}, circular orbits of massive particles exist not only in the equatorial plane of the BH but are also stable in the non-equatorial plane. This phenomenon is expected due to the spherical symmetry of spacetime.
\begin{figure*}
\center{
\includegraphics[width=7cm]{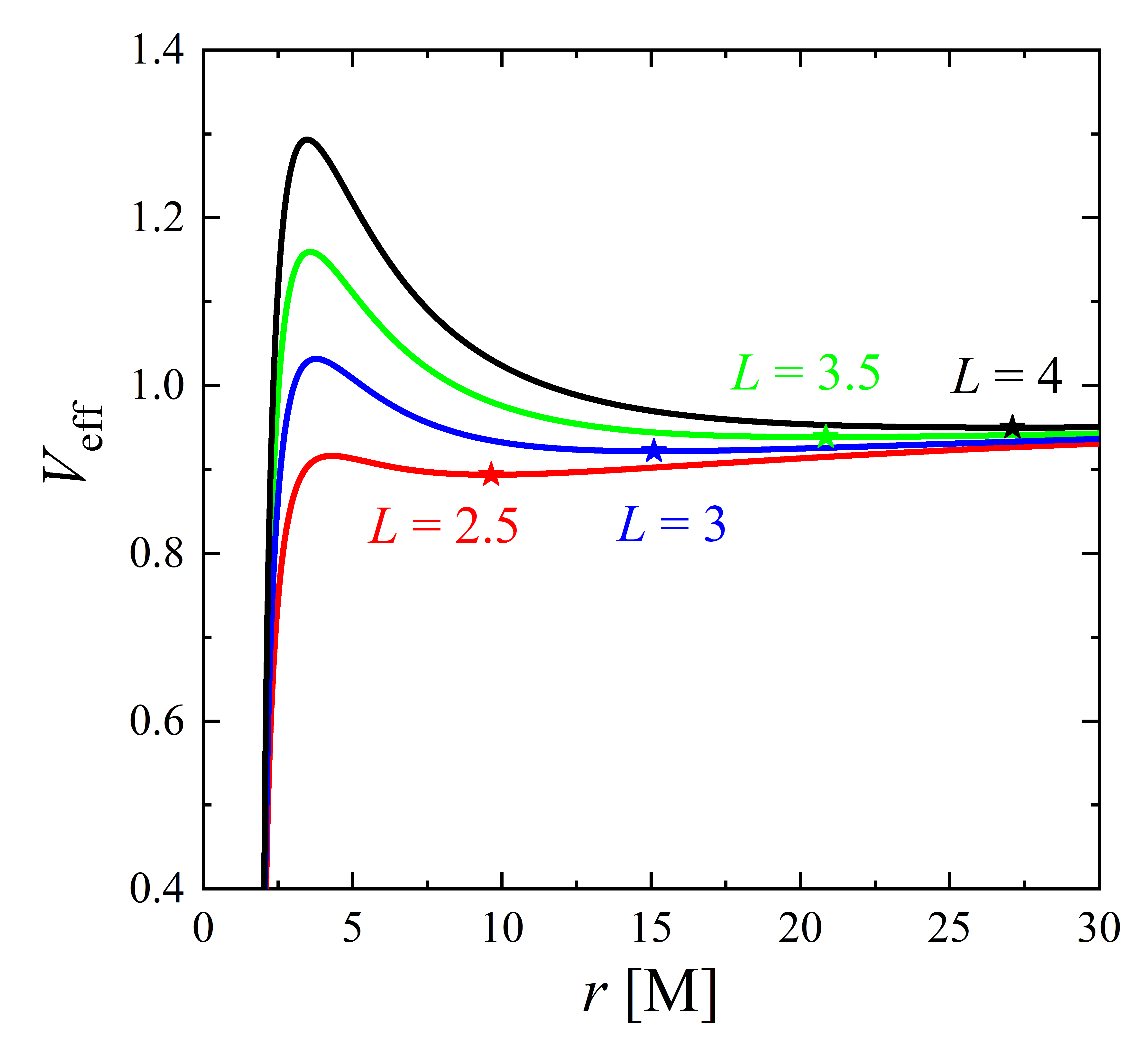}
\caption{Dependence of the effective potential $V_{\textrm{eff}}$ on the radius $r$ for several values of the particle's angular momentum. Here, the orbital plane inclination and scalar hair parameter are fixed at $\sigma=60^{\circ}$ and $h=-1$, respectively. The presence of a local minimum in each potential suggests the existence of tilted stable circular orbits of massive particles around the hairy BH in Horndeski gravity.}}\label{fig1}
\end{figure*}
\begin{figure*}
\center{
\includegraphics[width=6cm]{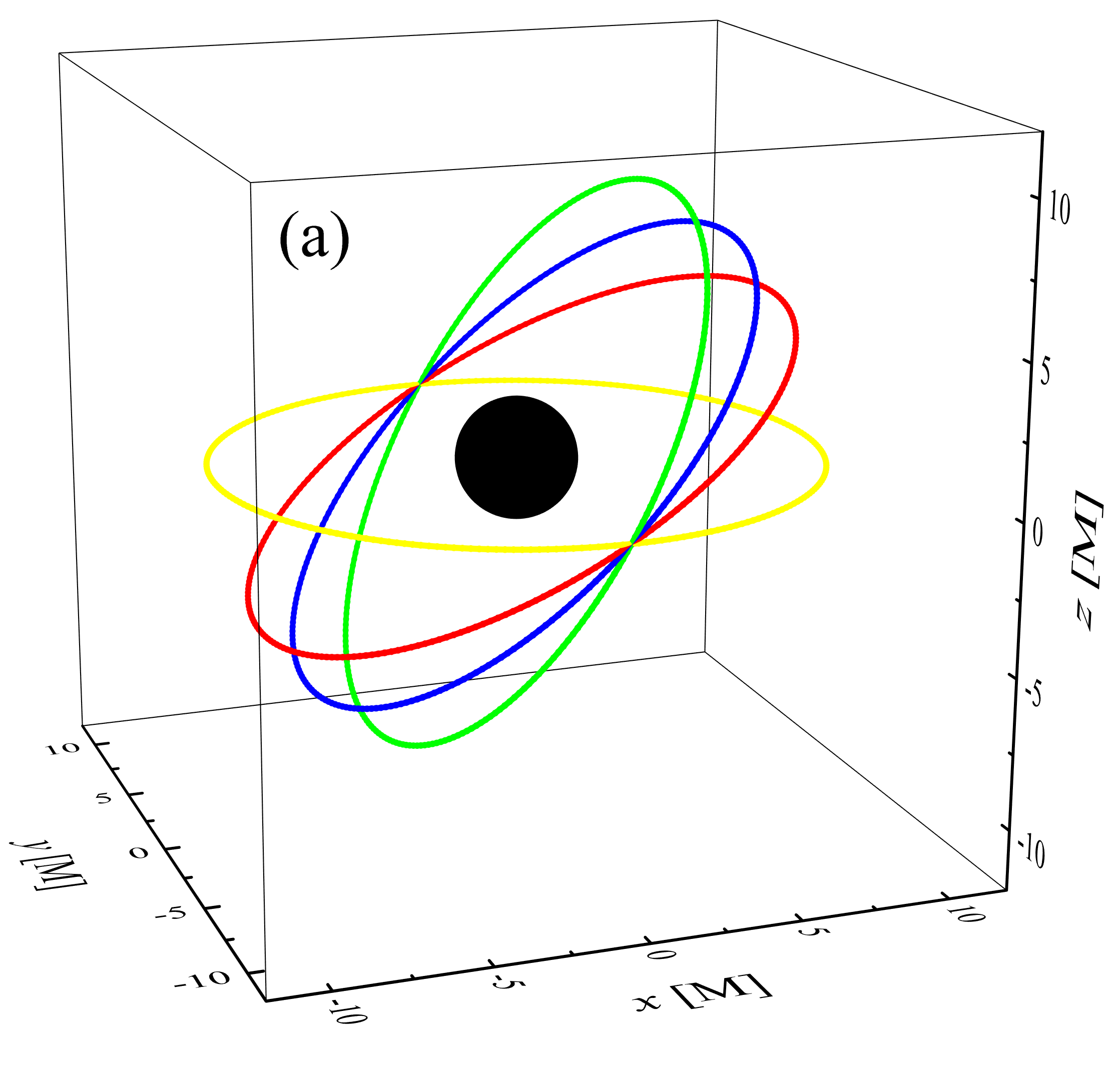}
\includegraphics[width=6cm]{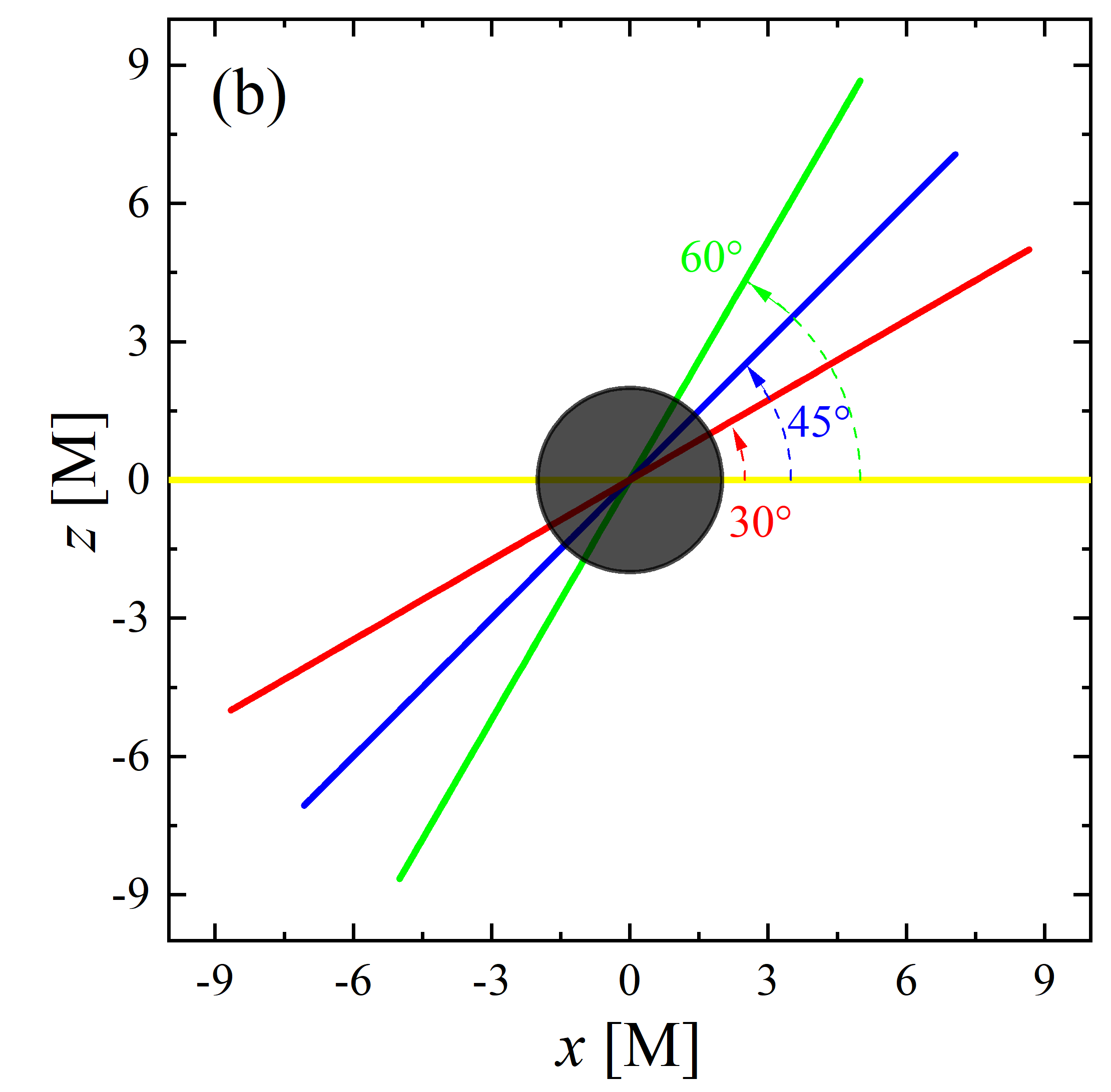}
\caption{Several tilted stable circular orbits of timelike particles in pseudo-Cartesian coordinates (a) and their projections in the $xoz$ plane (b). The green, blue, red, and yellow lines correspond to the stable circular orbits with orbital inclinations $\sigma$ of $60^{\circ}$, $45^{\circ}$, $30^{\circ}$, and $0^{\circ}$, respectively. The black sphere (disk) represents the event horizon of the Horndeski BH with the scalar hair parameter of $h=-1$. These orbits are simulated with the help of a fifth- and sixth-order Runge-Kutta-Fehlberg integrator with adaptive step sizes.}}\label{fig2}
\end{figure*}

Further, we simulate four representative stable circular orbits with different orbital inclinations and the same radius $r = 10M$, as displayed in figure 2. It is noteworthy that the values of specific angular momentum $L$ can be calculated through Eq. \eqref{9} as $5.0650$, $4.3864$, $3.5815$, and $2.5325$ for the yellow (equatorial), red, blue, and green orbit, respectively. Concurrently, these orbits share a common specific energy of $0.8961$. Therefore, we realize that the specific angular momentum of the massive particle moving in the tilted stable circular orbit varies with different orbital inclinations. In contrast, the specific energy of the particle remains independent of the orbital inclination as long as the orbital radius remains constant. These findings can be strongly supported by figure 3, which illustrates the variations of specific angular momentum and energy pertaining to stable circular orbits of massive particles across varying radii and orbital inclinations. One can observe that when the radius $r$ is increased at a fixed $\sigma$, both the specific angular momentum and energy increase. However, for a given value of $r$, only the specific angular momentum is affected by the orbital inclination. In addition, it is unveiled that the scalar hair parameter exerts a positive effect on the specific energy of the massive particle but a negative effect on the particle's specific angular momentum.

So far, we have theoretically and numerically confirmed the existence of tilted stable circular orbits around the hairy BH in Horndeski gravity. These orbits can, in principle, give rise to tilted accretion disks that illuminate BHs. More importantly, we have obtained the specific energy and angular momentum of stable circular orbits with arbitrary orbital inclination. These motion constants play a crucial role in calculating the redshift factors of light rays in the following section.
\begin{figure*}
\center{
\includegraphics[width=4.5cm]{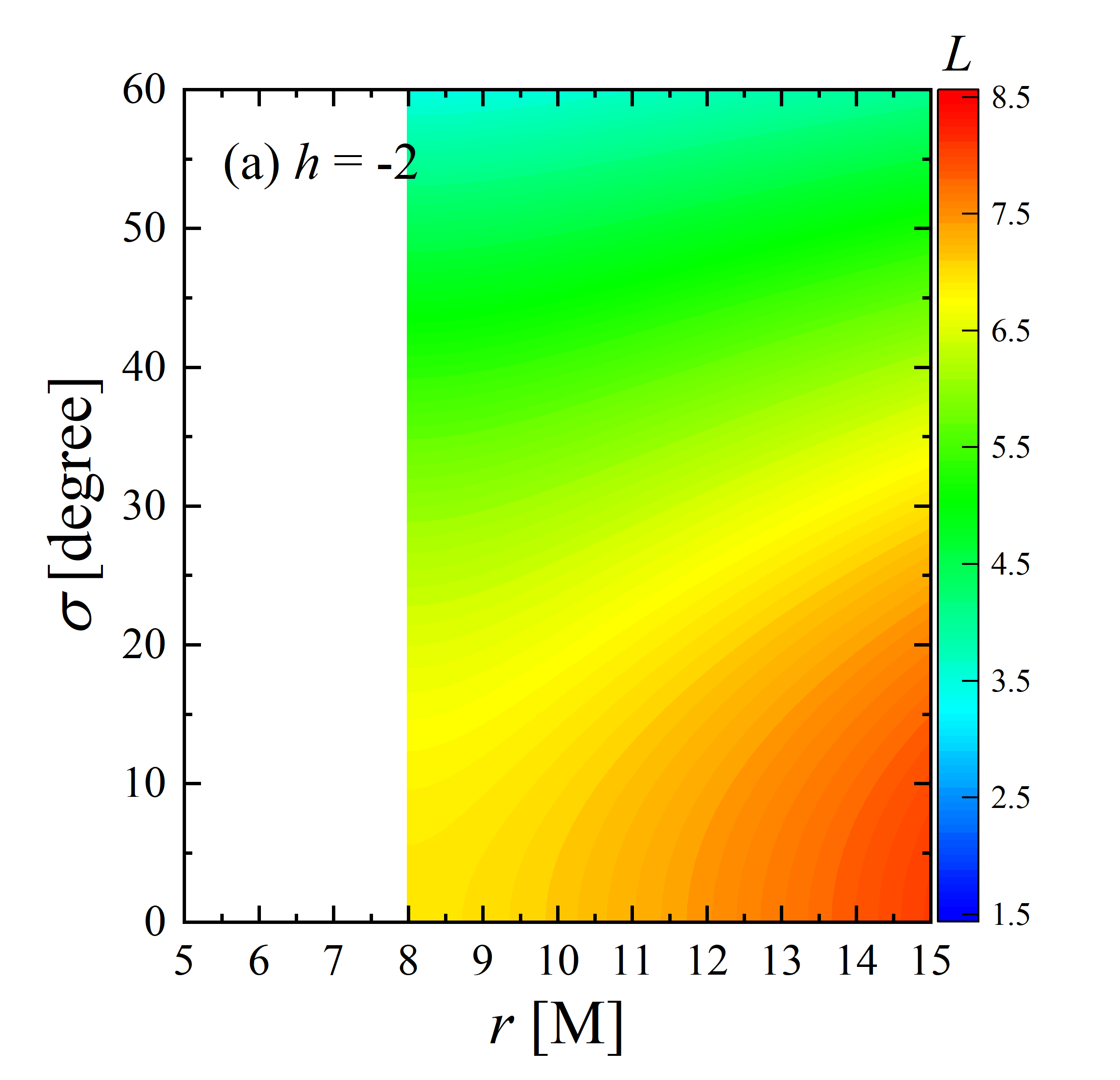}
\includegraphics[width=4.5cm]{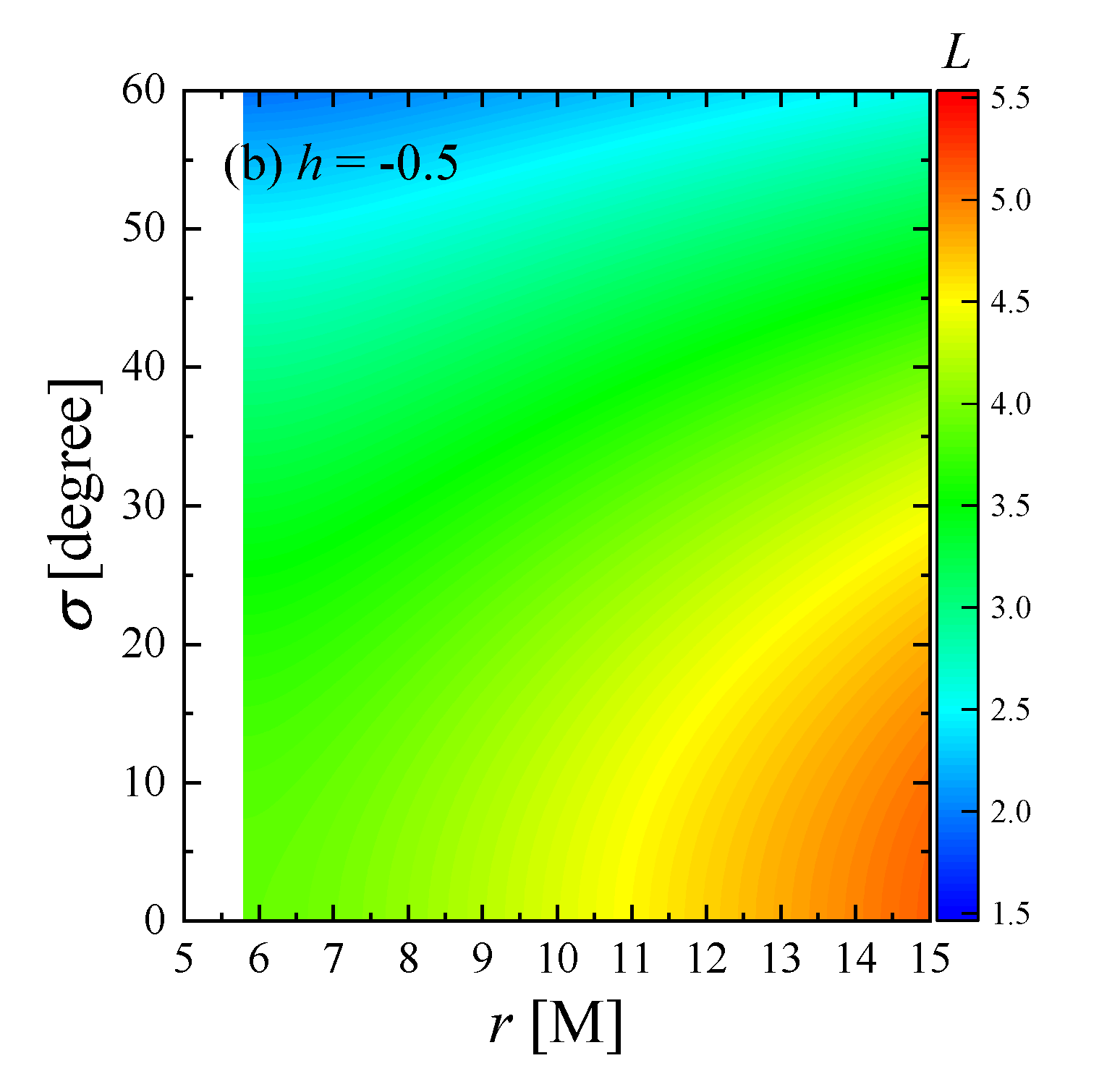}
\includegraphics[width=4.5cm]{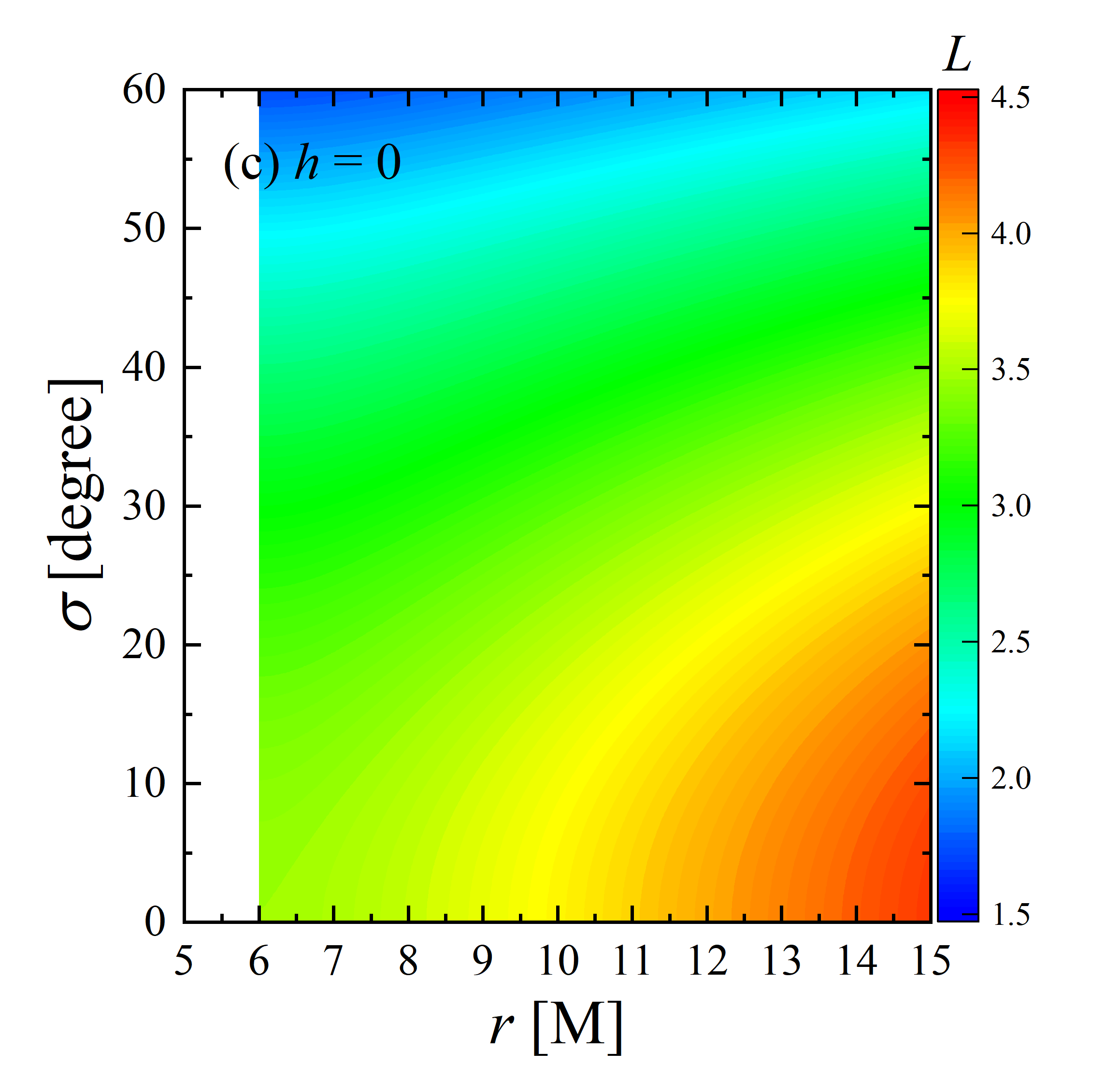}
\includegraphics[width=4.5cm]{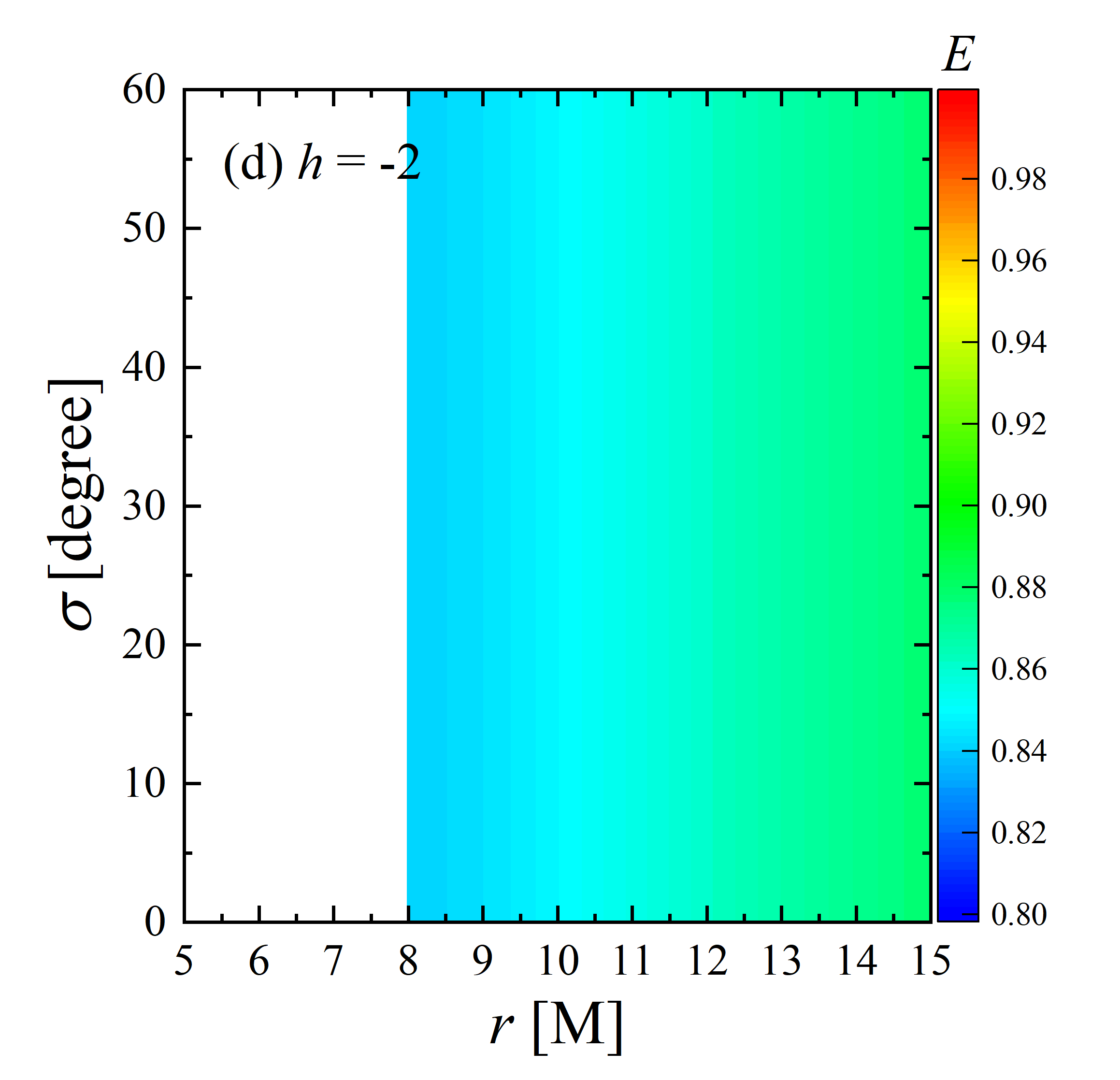}
\includegraphics[width=4.5cm]{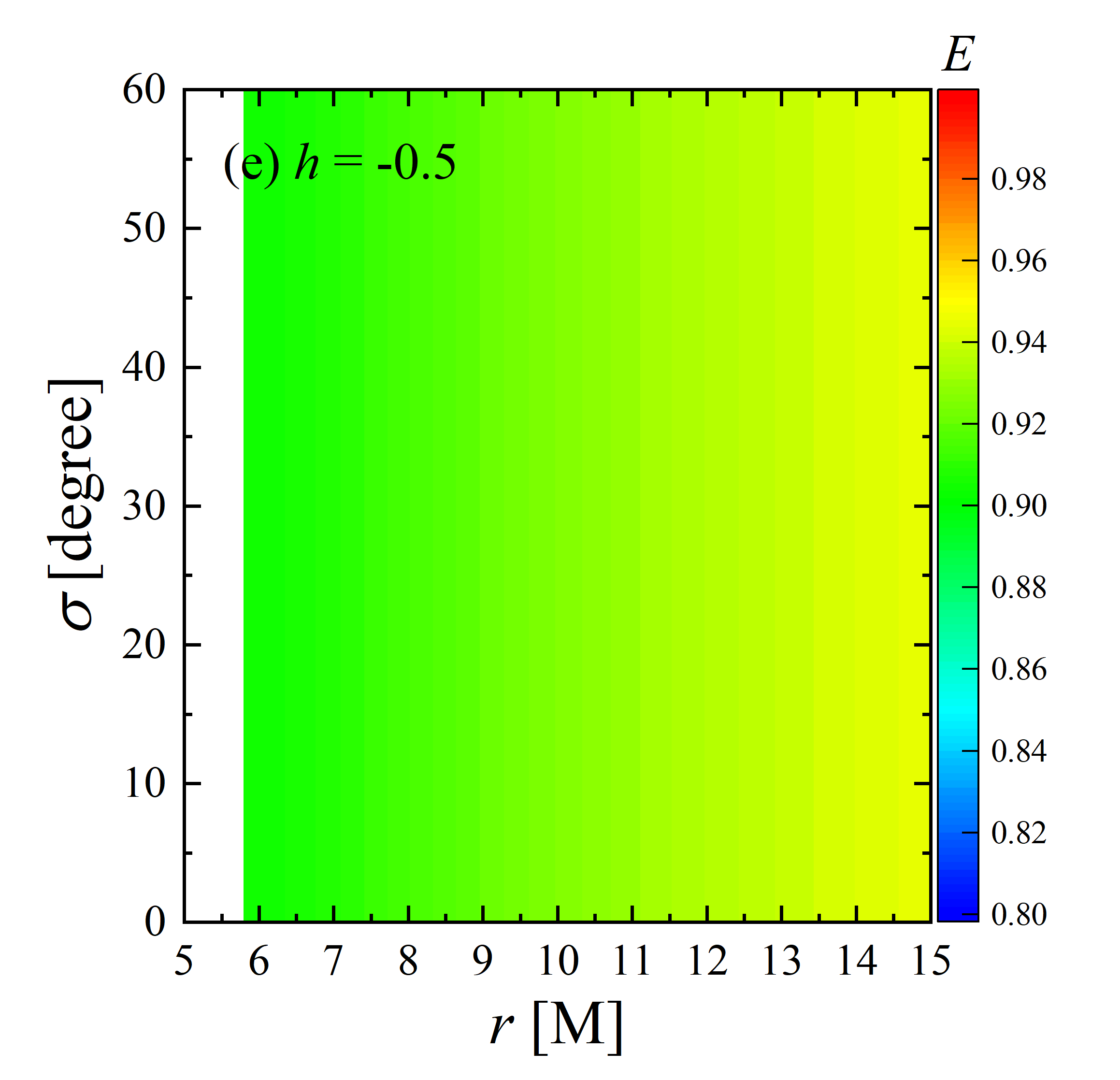}
\includegraphics[width=4.5cm]{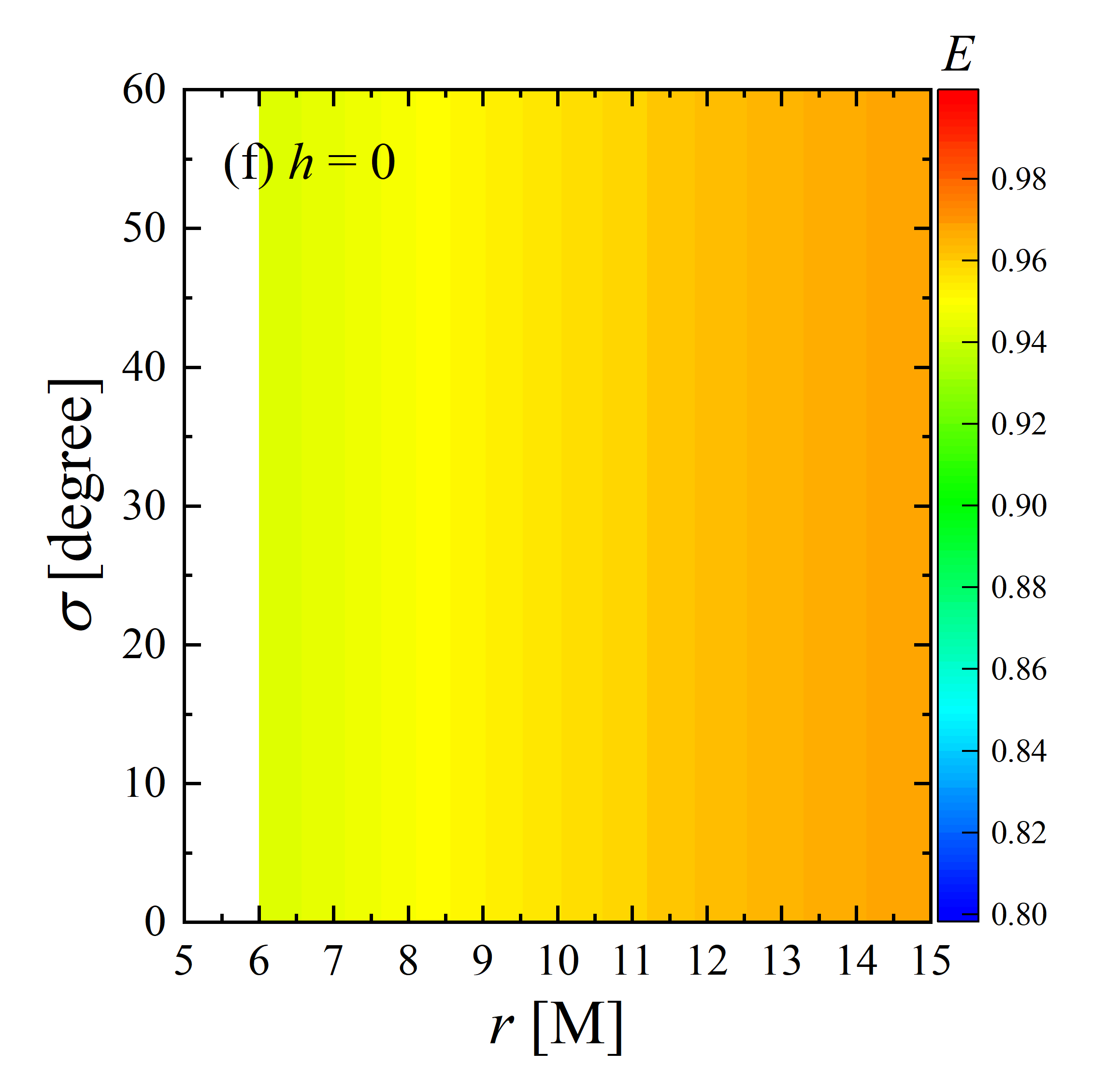}
\caption{The distributions of the specific angular momentum $L$ (first row) and the energy $E$ (second row) in the two-dimensional space of radius $r$ and orbital inclination $\sigma$ for the stable circular orbits under different values of the scalar hair parameter $h$. In each panel, the starting point of $r$ corresponds to the radius of the innermost stable circular orbit. The radius $r$ positively influences the specific angular momentum and energy, while the variation of $\sigma$ only changes the values of specific angular momentum. More specifically, $L$ decreases with increasing $\sigma$ for a fixed $r$.}}\label{fig3}
\end{figure*}
\section{Images of hairy black holes surrounded by a tilted thin accretion disk}
In this section, we present novel observational signatures of hairy BHs in Horndeski's theory that are caused by the tilt of the thin accretion disk. To this end, we commence with an introduction to \texttt{Odyssey}, a widely used backward ray tracing algorithm developed by the authors in \cite{Pu et al. (2016)} based on the photon initial conditions proposed by \cite{Younsi et al. (2016)}. Subsequently, we calculate the redshift factors of light rays radiated by the equatorial and non-equatorial accretion flow moving in stable circular orbits with radii greater than $r_{\textrm{ISCO}}$ and in plunging orbits with radii of $r_{+} < r < r_{\textrm{ISCO}}$, respectively. Finally, the $230$ GHz images of hairy BHs are simulated.
\begin{figure*}
\center{
\includegraphics[width=8cm]{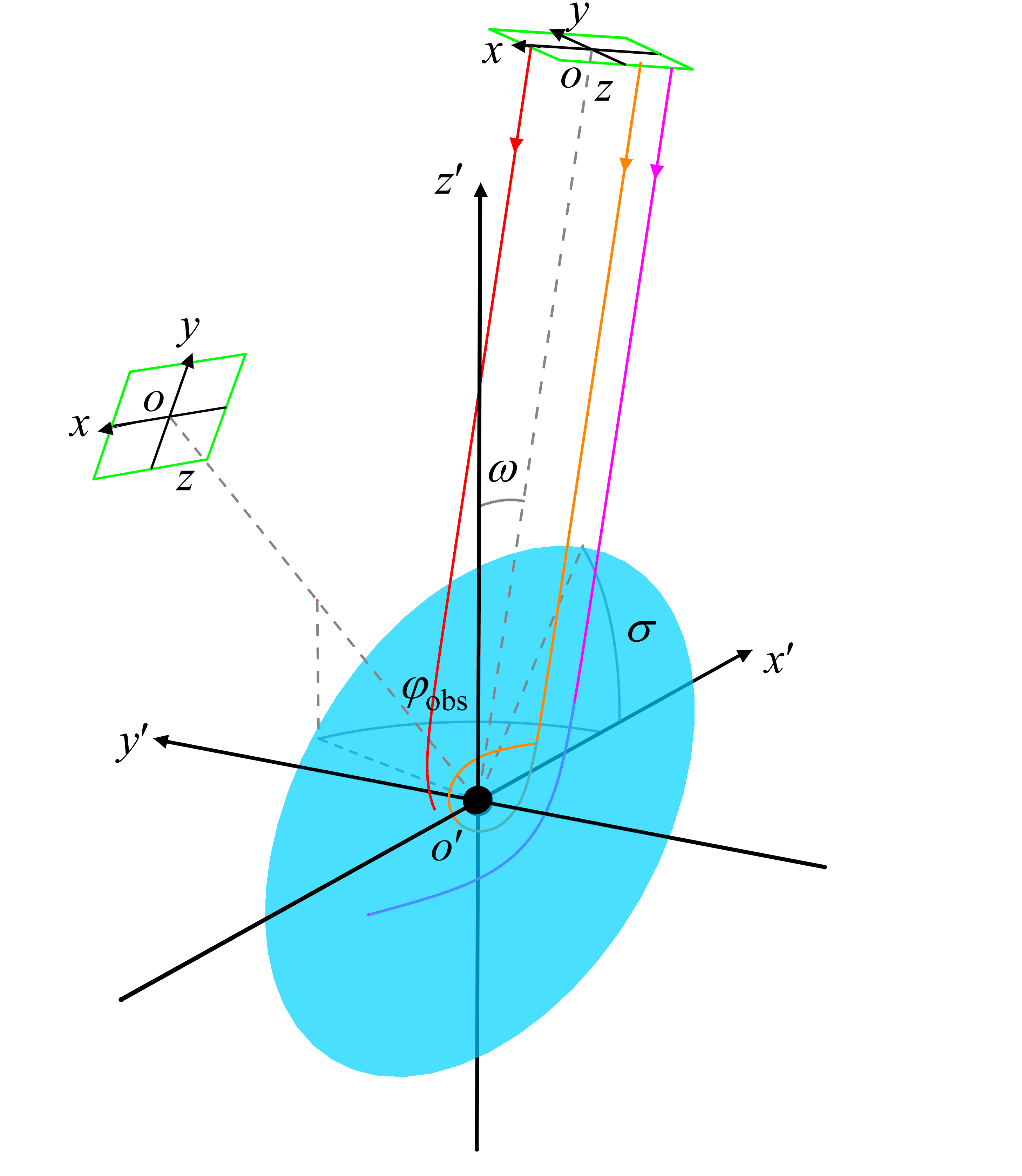}
\caption{The schematic map of the backward ray tracing method. Here, the observation plane and the space around the hairy BH are inscribed in the local coordinate systems $(x,y,z)$ and $(x^{\prime},y^{\prime},z^{\prime})$, respectively. The solid blue disk represents the tilted accretion disk, where the angle of inclination is denoted by $\sigma$, which takes values in the range of $[0,\pi/2]$. The observation angle and azimuth of the distant observer are denoted by $\omega$ and $\varphi_{\textrm{obs}}$, respectively, where $\omega$ falls within the range of $[0,\pi/2]$ and $\varphi_{\textrm{obs}}$ is within the interval of $[0,2\pi)$. In the plot, we also show three light rays represented by red, purple, and orange lines. They are ejected from different pixels on the observation screen and eventually intersect the accretion disk once, twice, and three times, respectively.}}\label{fig4}
\end{figure*}
\subsection{Ray tracing method}
Light rays radiated by an accretion disk can be (a) absorbed by the BH; (b) scattered to infinity by the BH; (c) received by an observer and form the BH images with different orders. These images can be simulated by tracing the light rays backward in time from the observer's image screen to the light source near the BH. To date, a number of ray tracing algorithms have been developed and efficiently applied to various astrophysical problems \cite{Johannsen $&$ Psaltis (2010),Vincent et al. (2011),Psaltis $&$ Johannsen (2012),Bambi (2012),Younsi et al. (2012),Chen et al. (2013),Yang $&$ Wang (2013),James et al. (2015),Cunha et al. (2016),Kimpson et al. (2019),Gralla $&$ Lupsasca (2020),Hu et al. (2021),Zhong et al. (2021),Lin et al. (2021),Cadavid et al. (2022),Garnier (2023)}. Here, we have chosen \texttt{Odyssey} \cite{Pu et al. (2016)} as a robust tool for tracking the evolution of light rays. As shown in figure 4, the Horndeski BH is located at point $o^{\prime}$, the origin of the local frame $(x^{\prime},y^{\prime},z^{\prime})$, in which $\overline{o^{\prime}z^{\prime}}$ aligns with the north pole of the BH, and $\overline{x^{\prime}o^{\prime}y^{\prime}}$ represents the equatorial plane. The optically and geometrically thin accretion disk (colored blue), with an angle $\sigma$ to the equatorial plane and an inner boundary at the event horizon of the BH, serves as the light source that illuminates the BH. It is essential to emphasize that the motion of the accretion flow is directed along the $\varphi$ axis (prograde accretion flow). This implies that, for an observer on the $x^{\prime}$-axis, the left and right sides correspond to the approaching and receding portions of the accretion disk, respectively. The observation screen (green blocks), with an observation angle $\omega$ and azimuth $\varphi_{\textrm{obs}}$, is depicted by a Cartesian coordinate system $(x,y,z)$ with the $z$-axis oriented towards the BH singularity. Note that in the tilted accretion disk scenario, the BH images are sensitive to the azimuthal position of the observer, differing from the equatorial disk case. The propagation of a photon is determined by its initial conditions. As displayed in figure 4, the red, orange, and purple solid lines represent three light rays with varying initial conditions, each with a unique fate. Light rays that pass through the accretion disk once (red line) and twice (purple line) are associated with the direct and lensing emissions, which form the direct and lensed images of the BH, respectively. Light rays that cross through the accretion disk three (orange line) or more times correspond to the photon ring emissions, which are closely related to the critical curve \cite{Gralla et al. (2019)}. Next, we need to determine the initial position and momentum of the photon corresponding to each pixel in the observation plane in Boyer-Lindquist coordinates.

Assuming that in the observer's coordinate system $(x,y,z)$, the photon's initial position in the observation plane is $(x,y,0)$. As outlined in \cite{Younsi et al. (2016)} and \cite{Pu et al. (2016)}, this coordinate can be transformed and denoted in the BH's coordinate system $(x^{\prime},y^{\prime},z^{\prime})$ as
\begin{eqnarray}\label{10}
x^{\prime} &=& \left(r_{\textrm{obs}}\sin\omega-y\cos\omega\right)\cos\varphi_{\textrm{obs}}-x\sin\varphi_{\textrm{obs}}, \nonumber \\
y^{\prime} &=& \left(r_{\textrm{obs}}\sin\omega-y\cos\omega\right)\sin\varphi_{\textrm{obs}}+x\cos\varphi_{\textrm{obs}}, \nonumber \\
z^{\prime} &=& r_{\textrm{obs}}\cos\omega+y\sin\omega,
\end{eqnarray}
where $r_{\textrm{obs}}$ denotes the distance between observer and the BH. The photon position \eqref{10} can be further expressed in Boyer-Lindquist coordinate system as
\begin{eqnarray}\label{11}
r &=& \sqrt{x^{\prime2}+y^{\prime2}+z^{\prime2}}, \nonumber \\
\theta &=& \arccos\left(\frac{z^{\prime}}{r}\right), \nonumber \\
\varphi &=& \textrm{atan2}\left(y^{\prime},x^{\prime}\right).
\end{eqnarray}
In order to track the light ray, one also needs to know its initial velocity. It is reasonable to expect that in an asymptotically flat spacetime, the photon arrives at a distant observation screen with a velocity almost parallel to the $z$-axis. Consequently, the velocity of the photon can be safely set as $(\dot{x},\dot{y},\dot{z})=(0,0,1)$. Substituting this condition into the differential of Eq. \eqref{10}, one obtains the photon's velocity in the BH coordinates system $(x^{\prime},y^{\prime},z^{\prime})$ as
\begin{eqnarray}\label{12}
\dot{x}^{\prime} &=& -\sin\omega\cos\varphi_{\textrm{obs}}, \nonumber \\
\dot{y}^{\prime} &=& -\sin\omega\sin\varphi_{\textrm{obs}}, \nonumber \\
\dot{z}^{\prime} &=& -\cos\omega.
\end{eqnarray}
Substituting the above expressions into the differential of Eq. \eqref{11}, the light ray's velocity components in Boyer-Lindquist coordinates are solved as
\begin{eqnarray}\label{13}
\dot{r} &=& -\sin\theta\sin\omega\cos\left(\varphi-\varphi_{\textrm{obs}}\right)-\cos\theta\cos\omega, \nonumber \\
\dot{\theta} &=& \frac{\sin\theta\cos\omega-\cos\theta\sin\omega\cos\left(\varphi-\varphi_{\textrm{obs}}\right)}{r}, \nonumber \\
\dot{\varphi} &=& \frac{\sin\omega\sin\left(\varphi-\varphi_{\textrm{obs}}\right)}{r\sin\theta}.
\end{eqnarray}
With Eqs.\eqref{10}, \eqref{11}, and \eqref{13} in hand, the initial condition of the photon corresponding to each pixel in the observation plane $\overline{xoy}$ is easy to calculate.

Next, the null geodesic equations for photons moving in Horndeski spacetime \eqref{1} must be determined. According to Legendre transformation, the Hamiltonian $\mathscr{H}$ for the photon is given by
\begin{eqnarray}\label{14}
\mathscr{H} &=& p_{\mu}\dot{x}^{\mu}-\mathscr{L} = \frac{1}{2}g^{\mu\nu}p_{\mu}p_{\nu}.
\end{eqnarray}
Here, $g^{\mu\nu}$ is the contravariant metric tensor, which satisfies the condition $g^{\mu\nu}g_{\mu\nu}=1$. The conjugate momentum of the photon, denoted as $p_{\mu}$, is governed by
\begin{eqnarray}\label{15}
p_{t} = g_{tt}\dot{t} = -\mathscr{E}, \quad p_{\varphi} = g_{\varphi\varphi}\dot{\varphi} = \mathscr{J}, \quad p_{r} = g_{rr}\dot{r}, \quad p_{\theta} = g_{\theta\theta}\dot{\theta},
\end{eqnarray}
where $\mathscr{E}$ and $\mathscr{J}$ correspond to the specific energy and angular momentum, which are motion constants along the photon path. Substituting Eq. \eqref{13} into Eq. \eqref{15}, we have the initial value of the photon's conjugate momentum in the observer screen. It is worth mentioning that the value of $\mathscr{E}$ (or $p_{t}$) is not directly obtained through the above equations since $\dot{t}$ does not appear in Eq. \eqref{13}. Nevertheless, this motion constant can be determined by the constraint of $\mathscr{H}=0$. With the help of a fifth- and sixth-order Runge-Kutta-Fehlberg integrator with variable step sizes, the photon motion can be simulated by the canonical equations,
\begin{eqnarray}\label{16}
\dot{x}^{\mu} = \frac{\partial\mathscr{H}}{\partial p_{\mu}}, \quad \dot{p}_{\mu} = -\frac{\partial\mathscr{H}}{\partial x^{\mu}}.
\end{eqnarray}

Now, we have the ability to trace the evolution of the light ray originating from each pixel of the observation plane and determine whether the ray intersects the accretion disk or the BH event horizon. The light rays in the former case contribute to the brightness of the BH image, while the BH shadow is associated with the latter scenario.
\subsection{Redshift factors}
Before simulating a BH image, one critical aspect needs to be elaborately addressed: determining the redshift factor of the light ray. Here, we consider that the emission region of the thin accretion disk extends all the way to the event horizon of the hairy BH. The accreted massive particles either move in stable circular orbits with radii $r \geq r_{\textrm{ISCO}}$ or accelerate along the plunging orbits in the region of $r_{+} < r < r_{\textrm{ISCO}}$. In either case, the relative motion between the massive particles (light source) and the observer causes a Doppler effect in the observed light rays, which remarkably modifies the BH images. The generic expression for the redshift factor can be read as
\begin{equation}\label{17}
g = \frac{p_{\mu}u^{\mu}_{\textrm{o}}}{p_{\nu}u^{\nu}_{\textrm{e}}},
\end{equation}
where $p_{\mu}$ is the conjugate momentum of the photon, $u^{\mu}_{\textrm{o}}=(1,0,0,0)$ is the four-velocity of the static remote observer, and $u^{\nu}_{\textrm{e}}=(\dot{t_{\textrm{e}}},\dot{r_{\textrm{e}}},\dot{\theta_{\textrm{e}}},\dot{\varphi_{\textrm{e}}})$ denotes the four-velocity of the emitting source. Note that $p_{\mu}$ can be obtained using the ray tracing method; therefore, the key to calculating the redshift factors is determining the four-velocity of the emitting source as it moves in different orbits.
\begin{figure*}
\centering
\includegraphics[width=3.7cm]{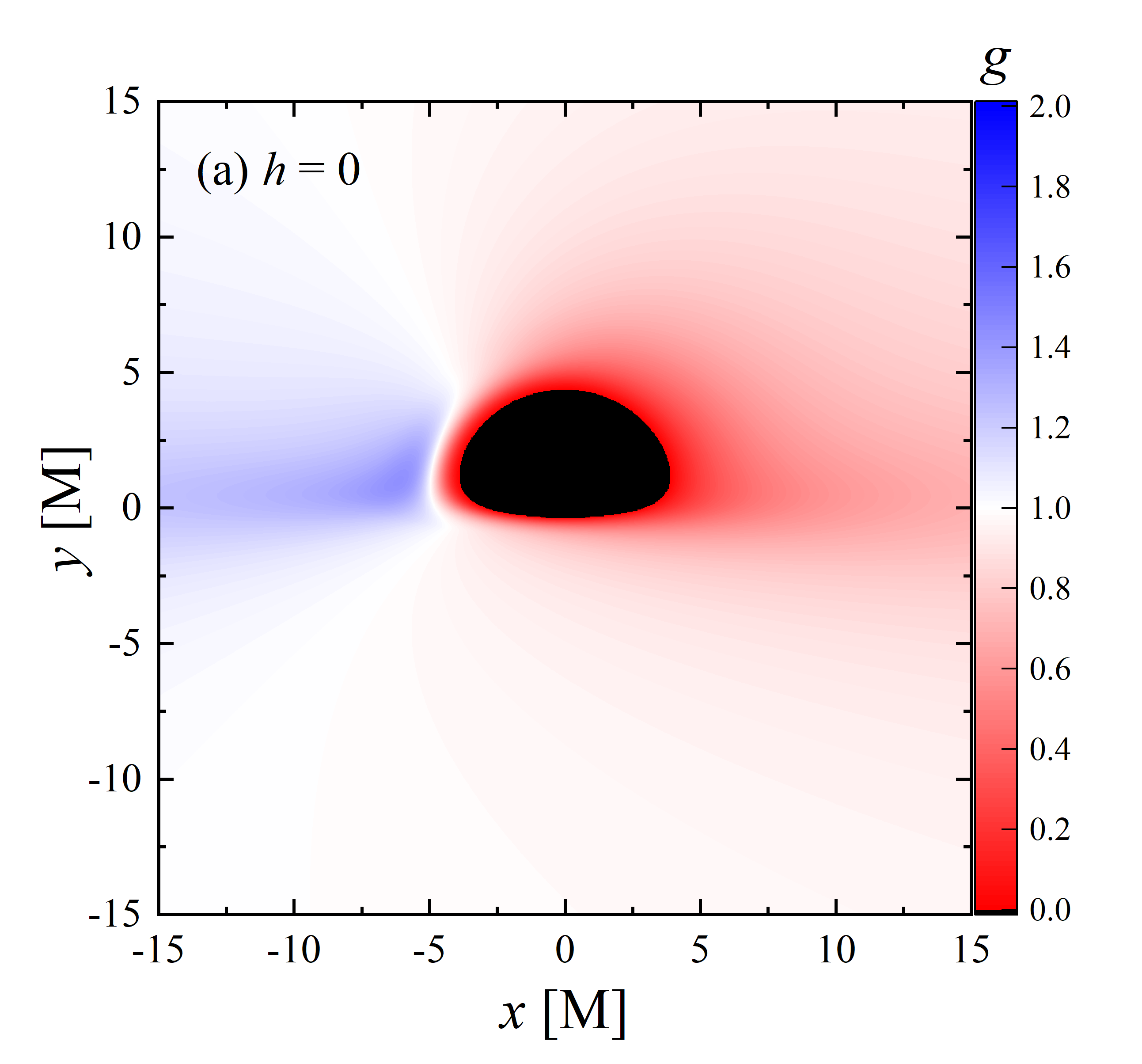}
\includegraphics[width=3.7cm]{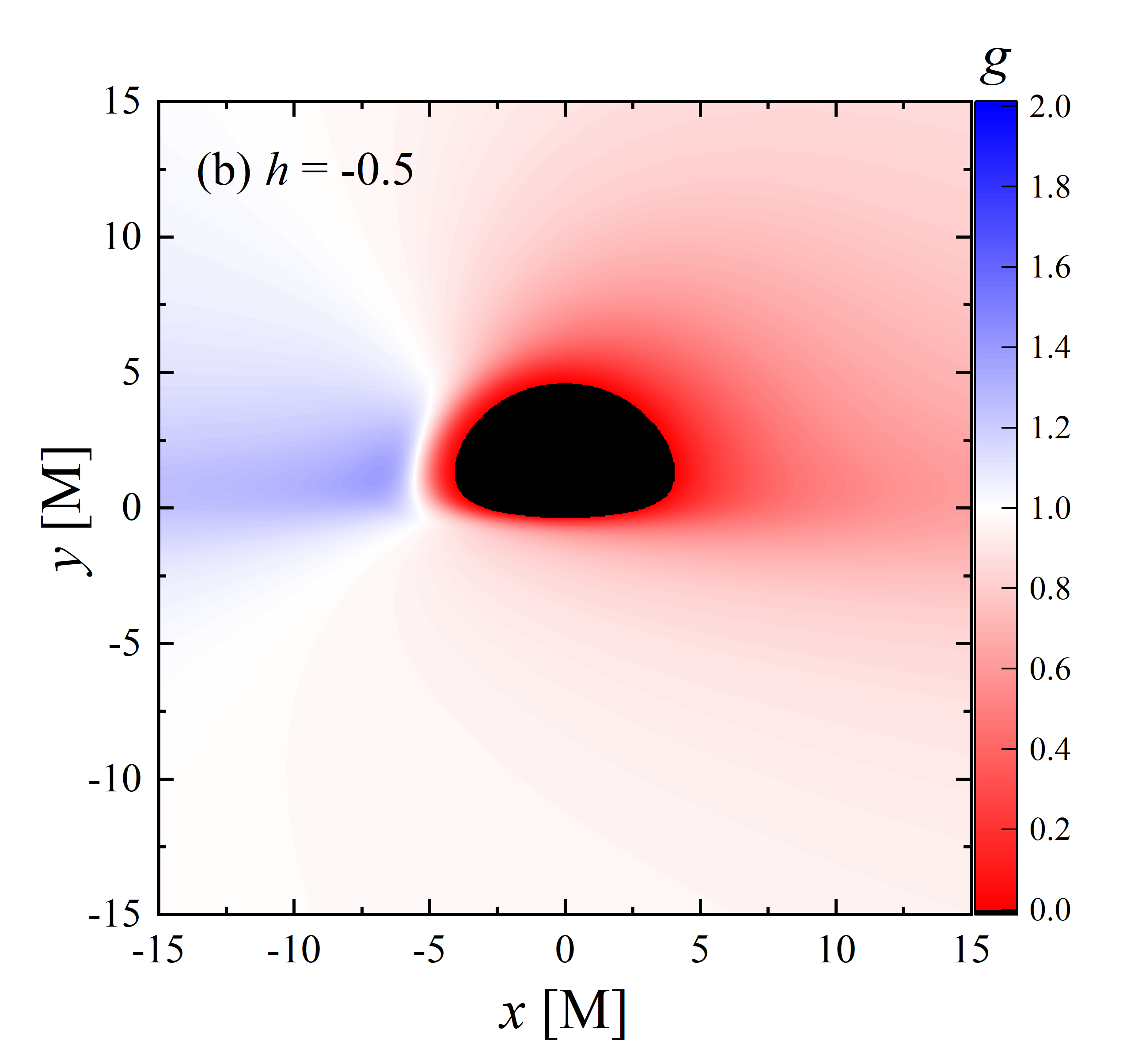}
\includegraphics[width=3.7cm]{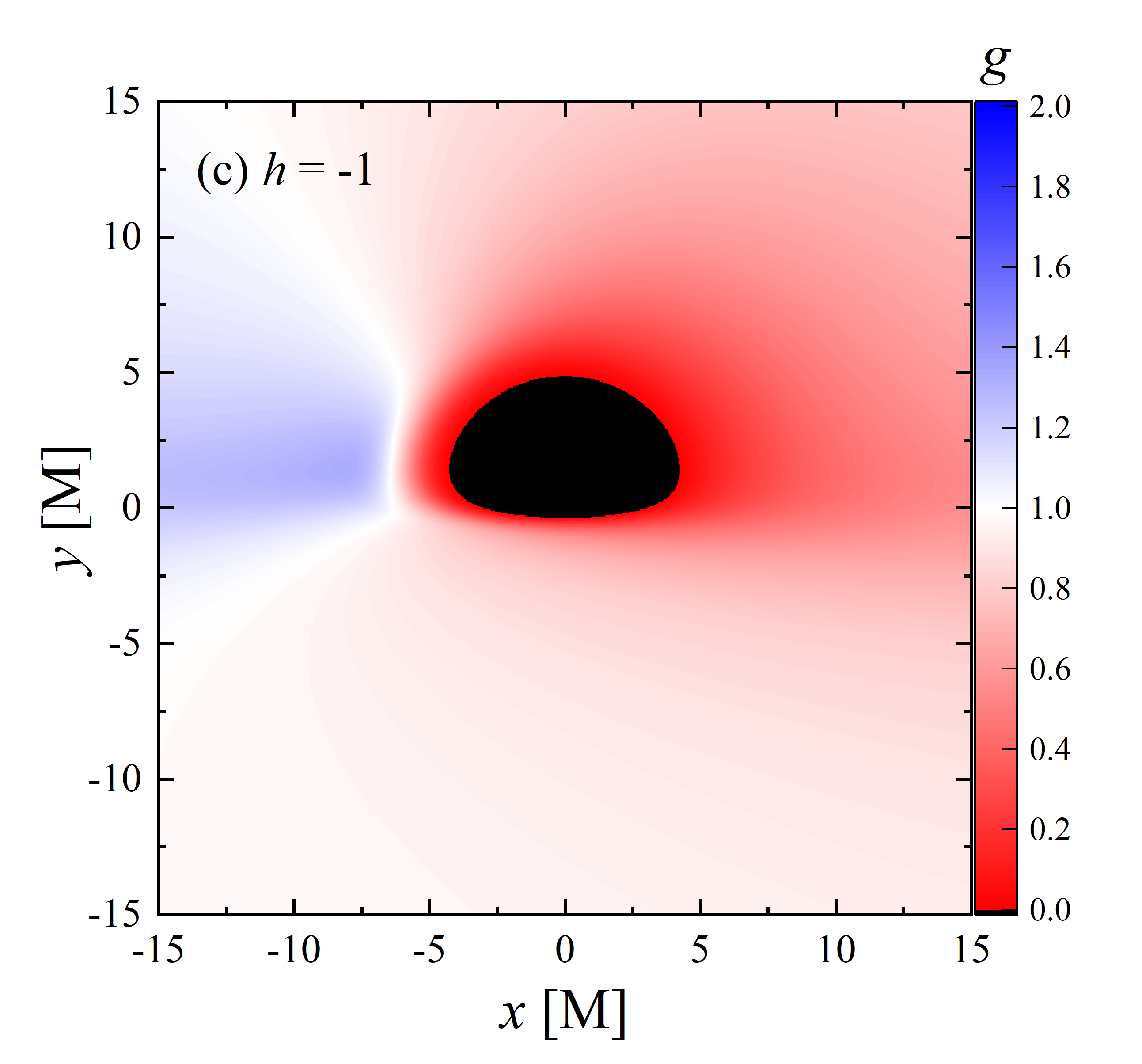}
\includegraphics[width=3.7cm]{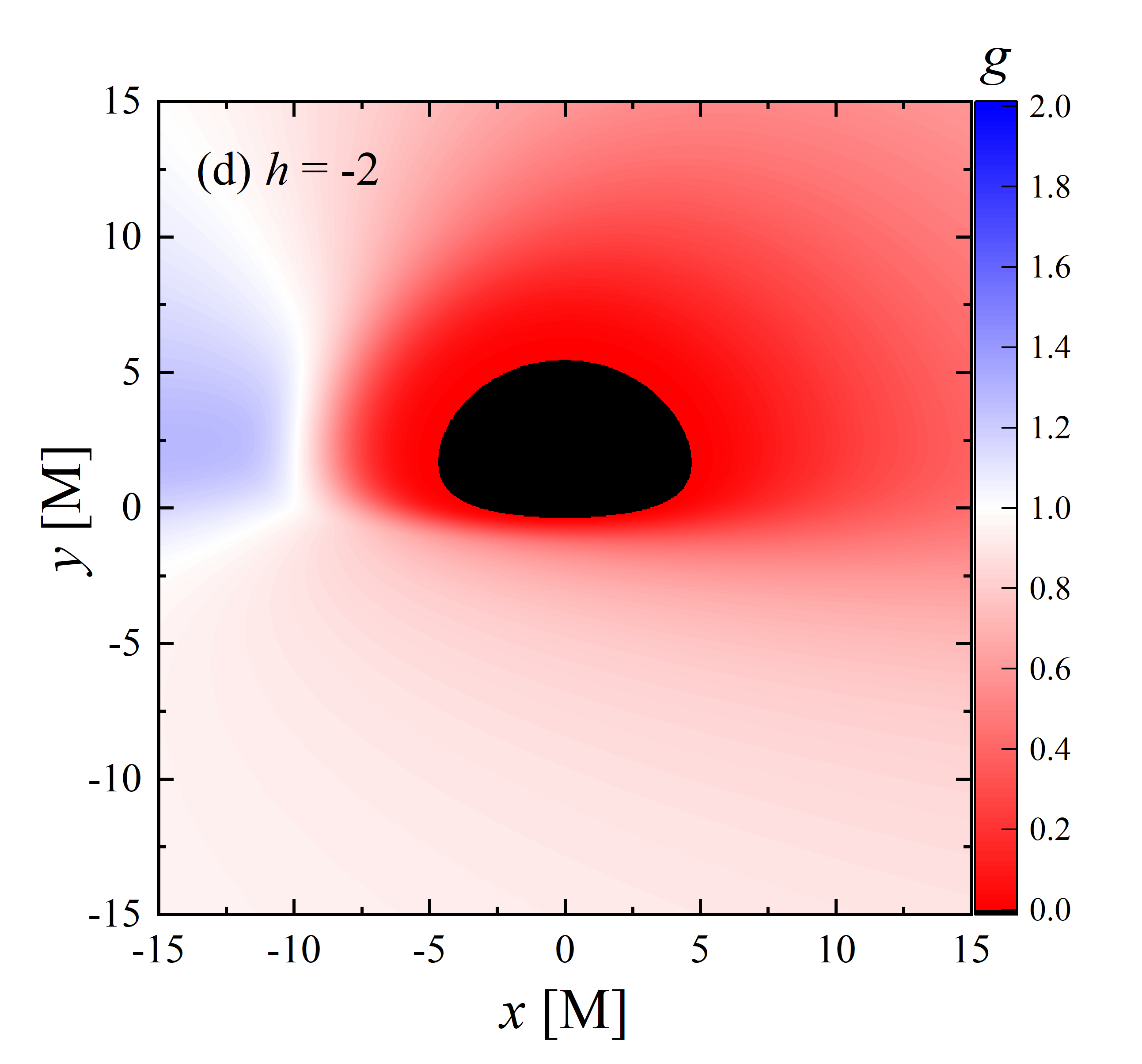}
\caption{The distributions of redshift factors in direct images of hairy BHs surrounded by an equatorial accretion disk under different values of the scalar hair parameter. Redshifts and blueshifts correspond to $g < 1$ and $g > 1$, respectively. Here, we have the observation angle of $\omega = 80^{\circ}$ and the observation azimuth of $\varphi_{\textrm{obs}}=0^{\circ}$.}\label{fig5}
\end{figure*}
\begin{figure*}
\center{
\includegraphics[width=3.7cm]{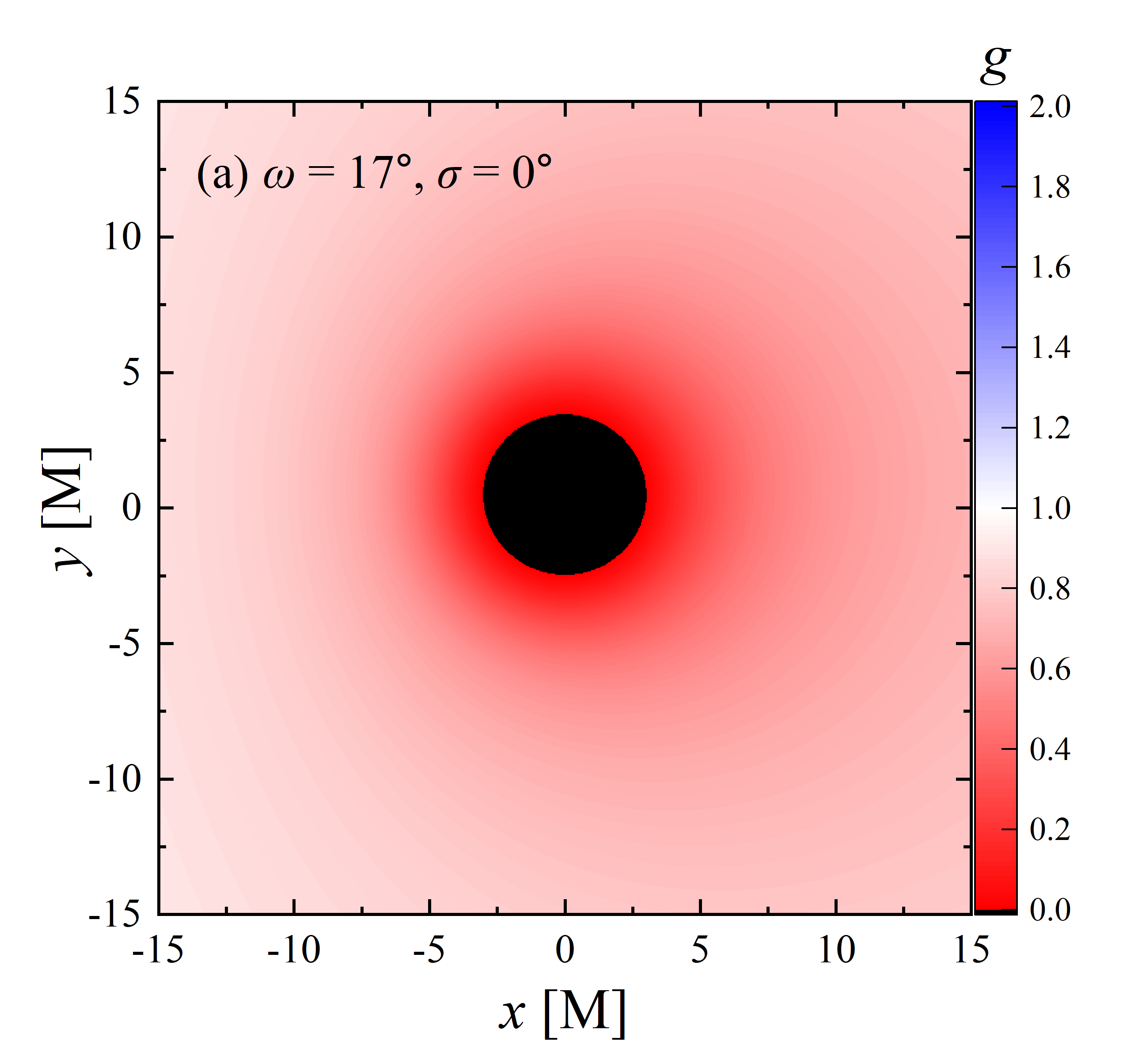}
\includegraphics[width=3.7cm]{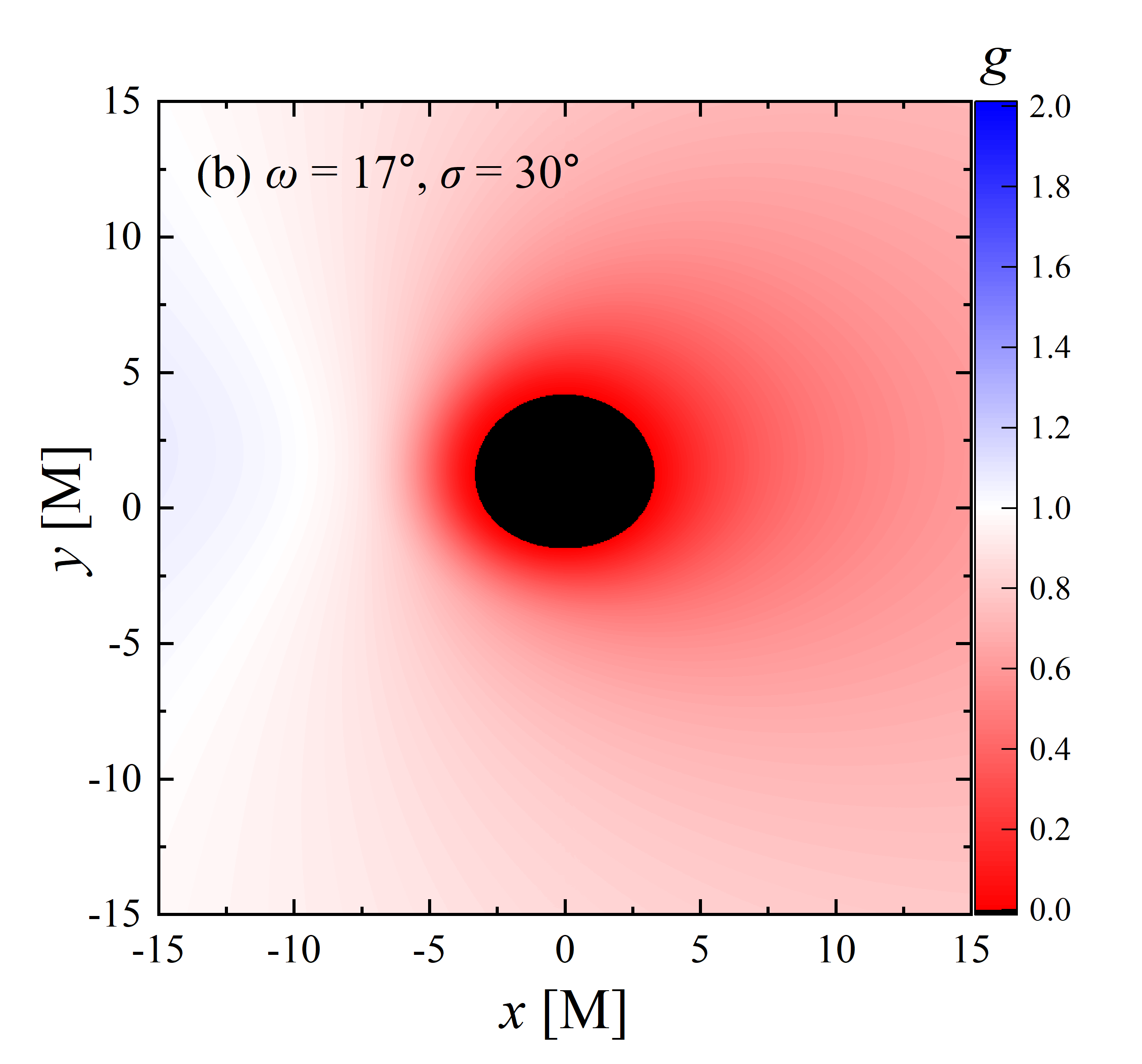}
\includegraphics[width=3.7cm]{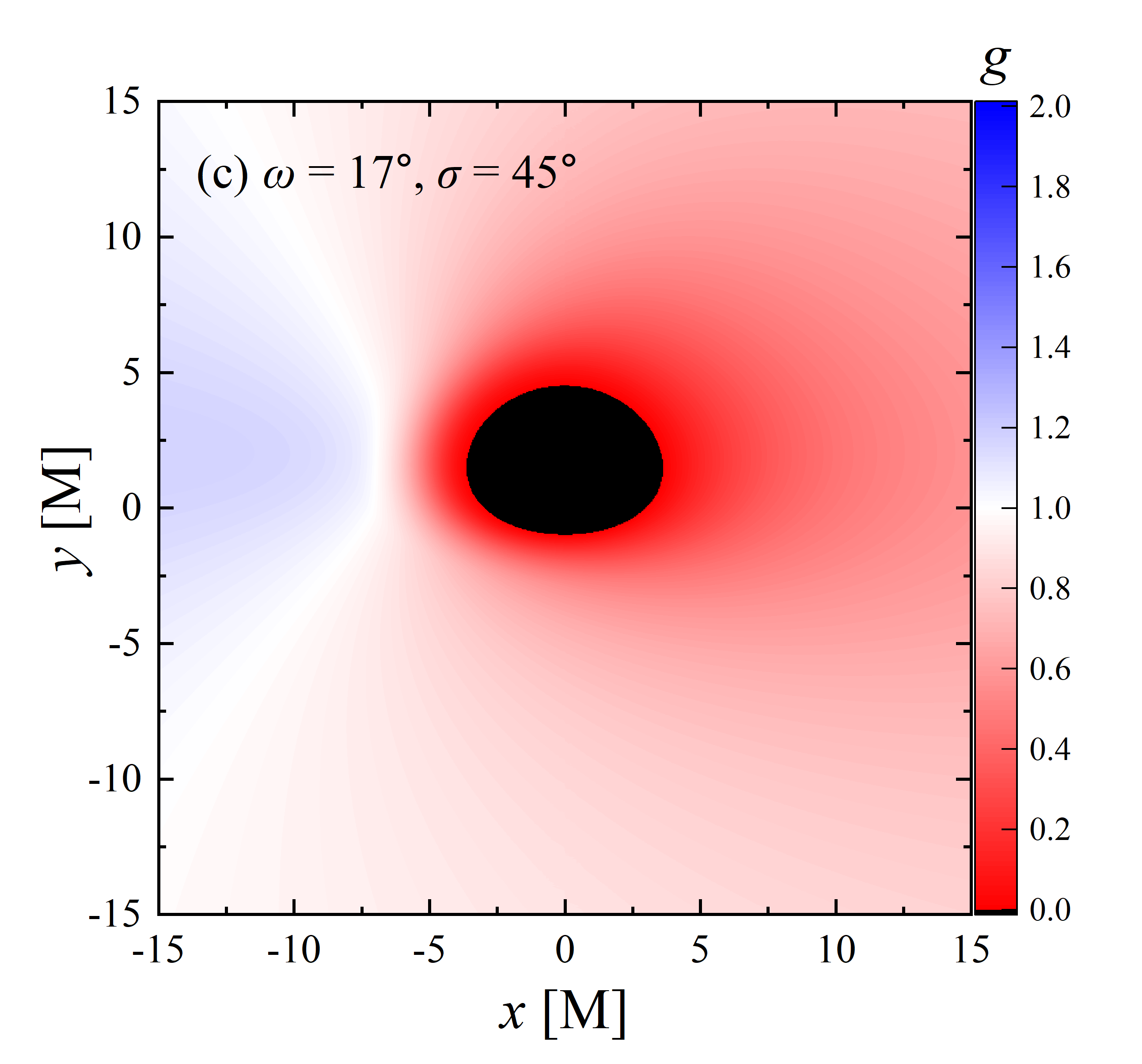}
\includegraphics[width=3.7cm]{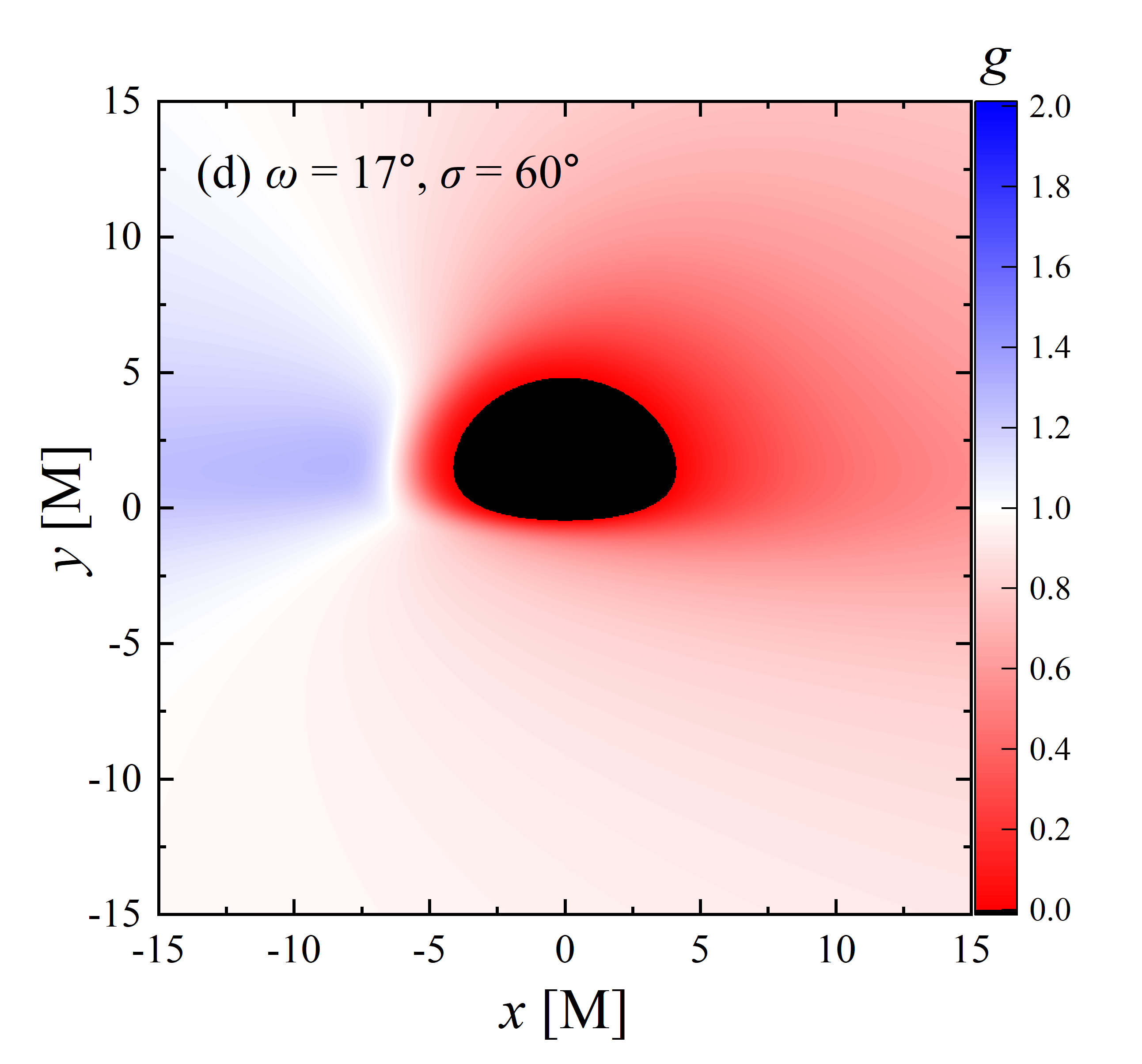}
\includegraphics[width=3.7cm]{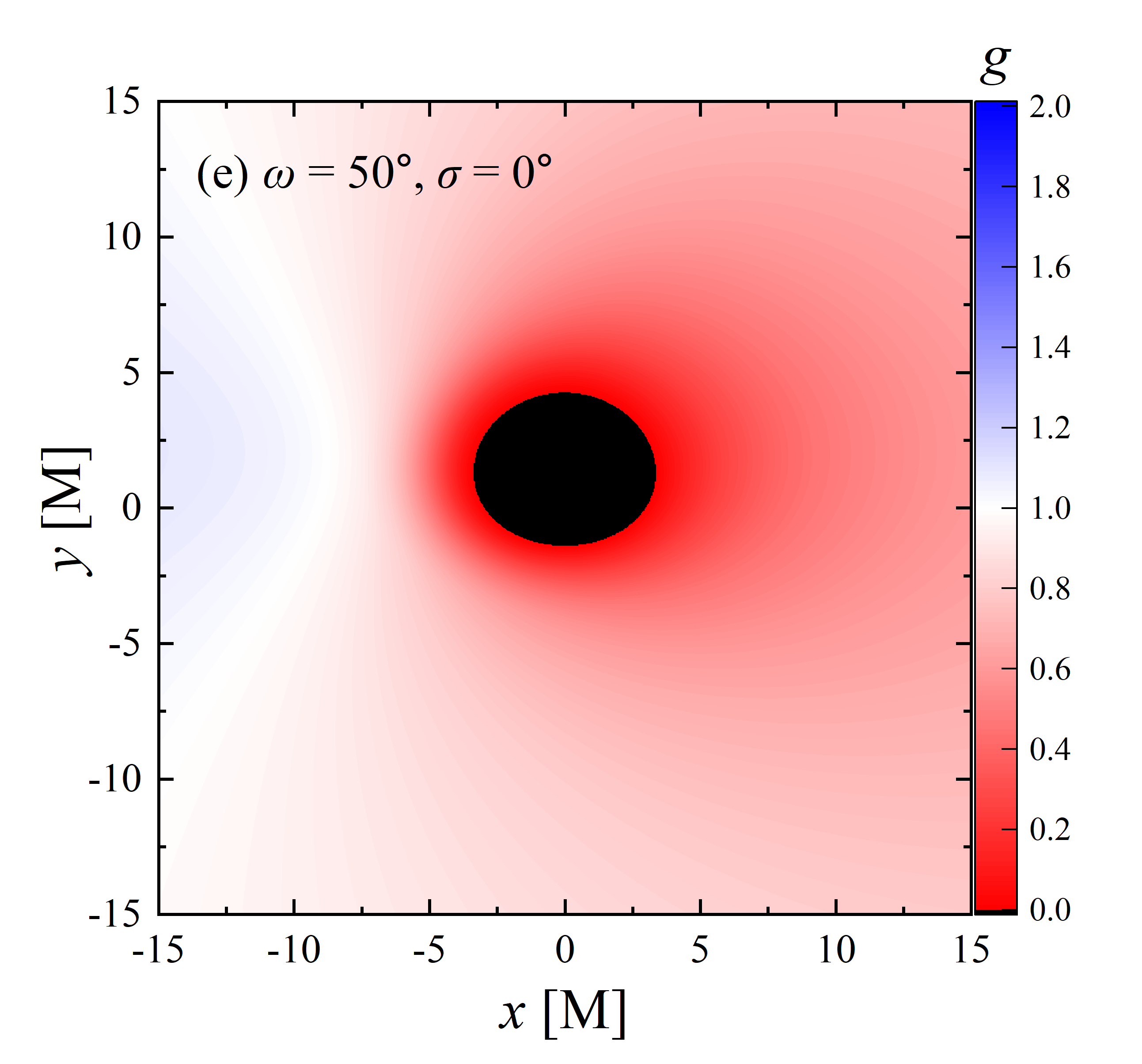}
\includegraphics[width=3.7cm]{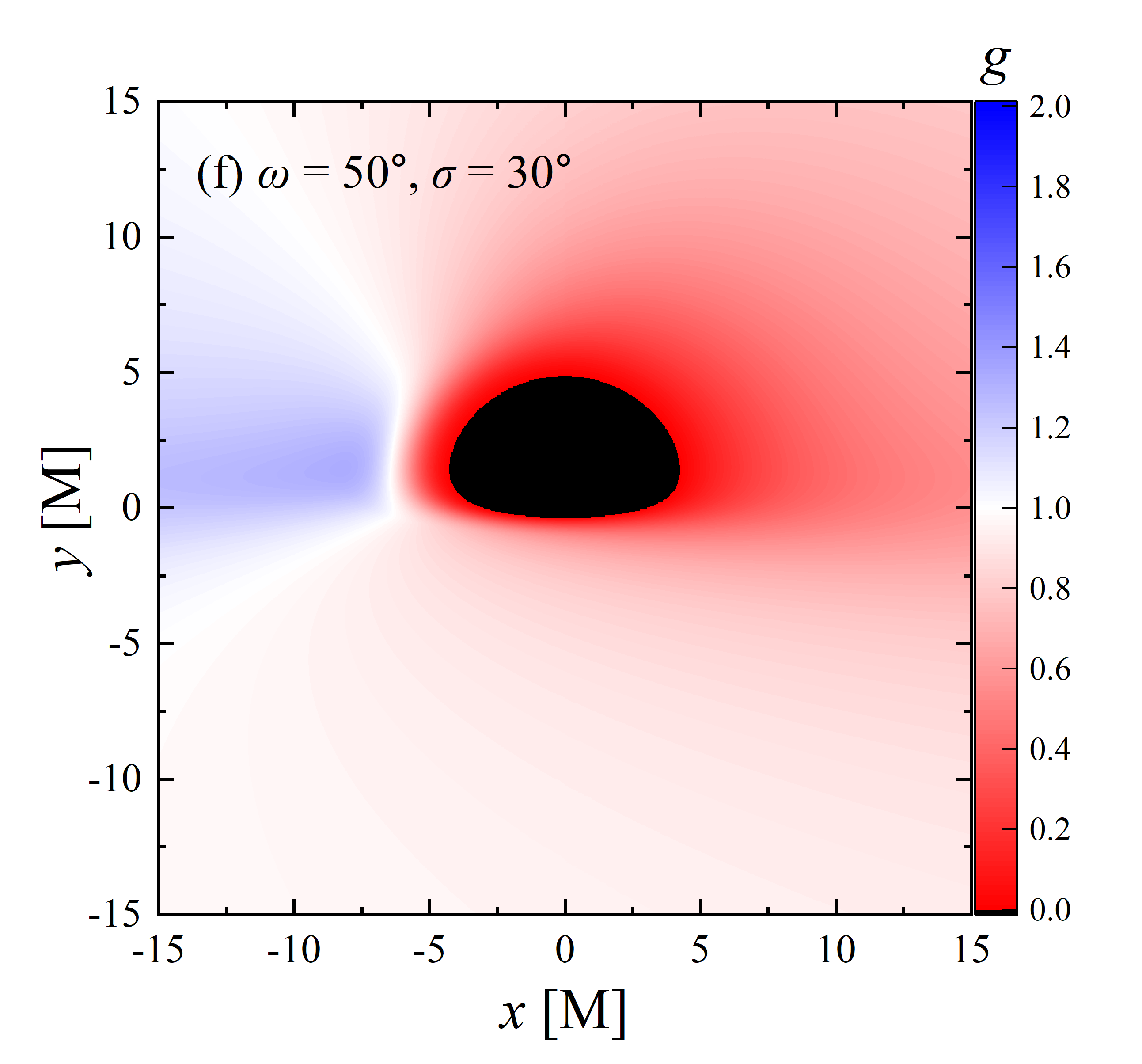}
\includegraphics[width=3.7cm]{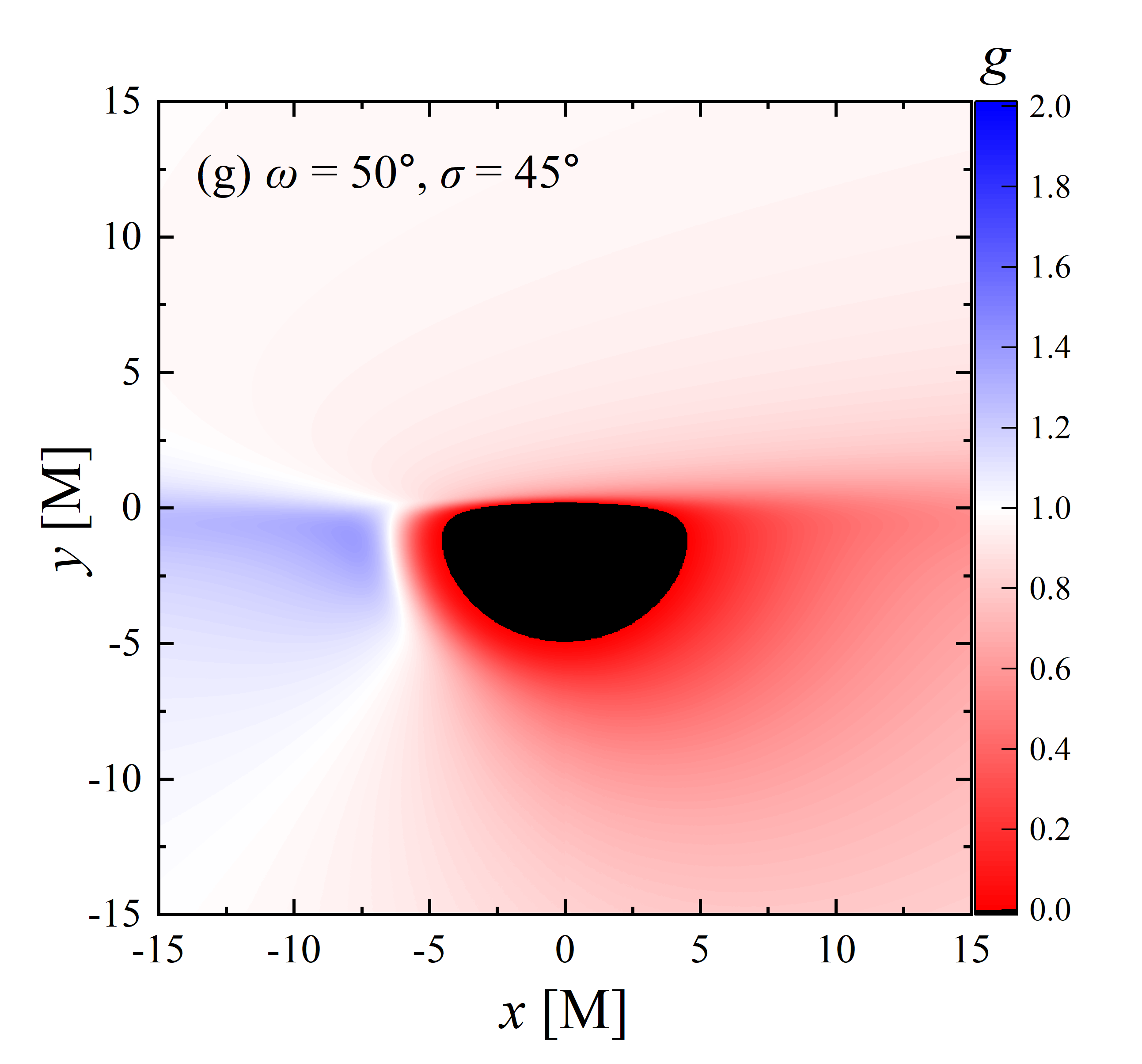}
\includegraphics[width=3.7cm]{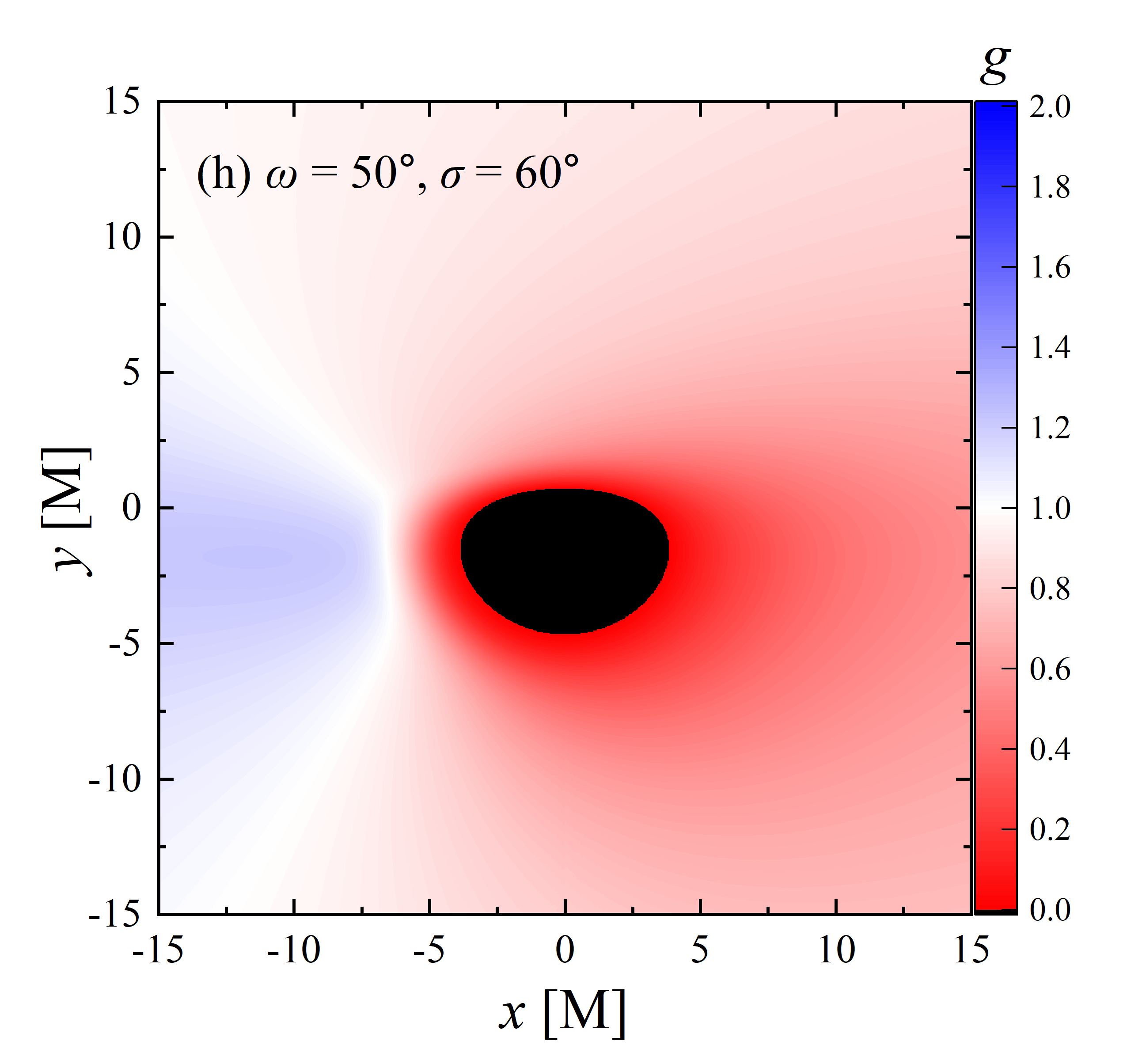}
\includegraphics[width=3.7cm]{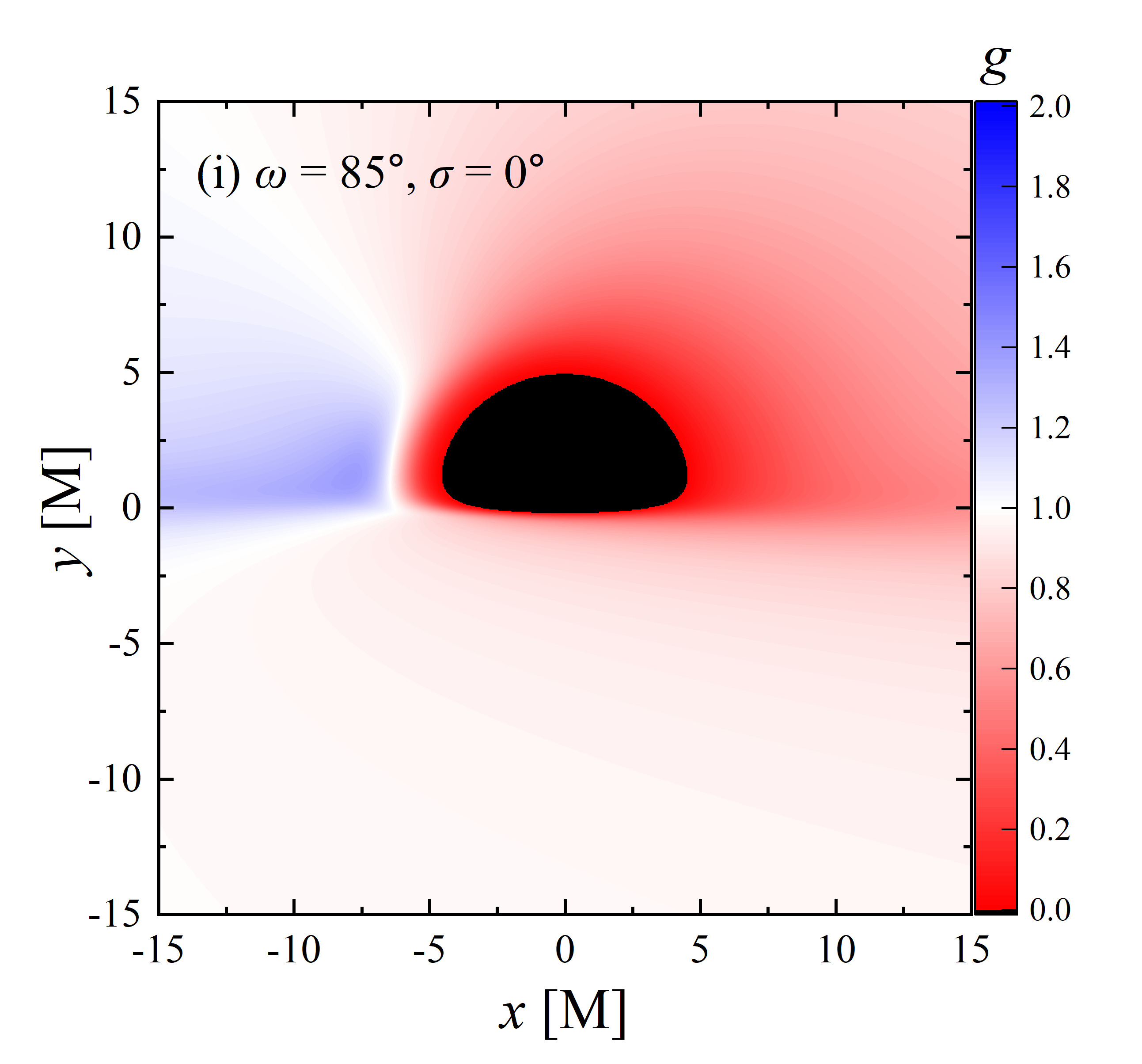}
\includegraphics[width=3.7cm]{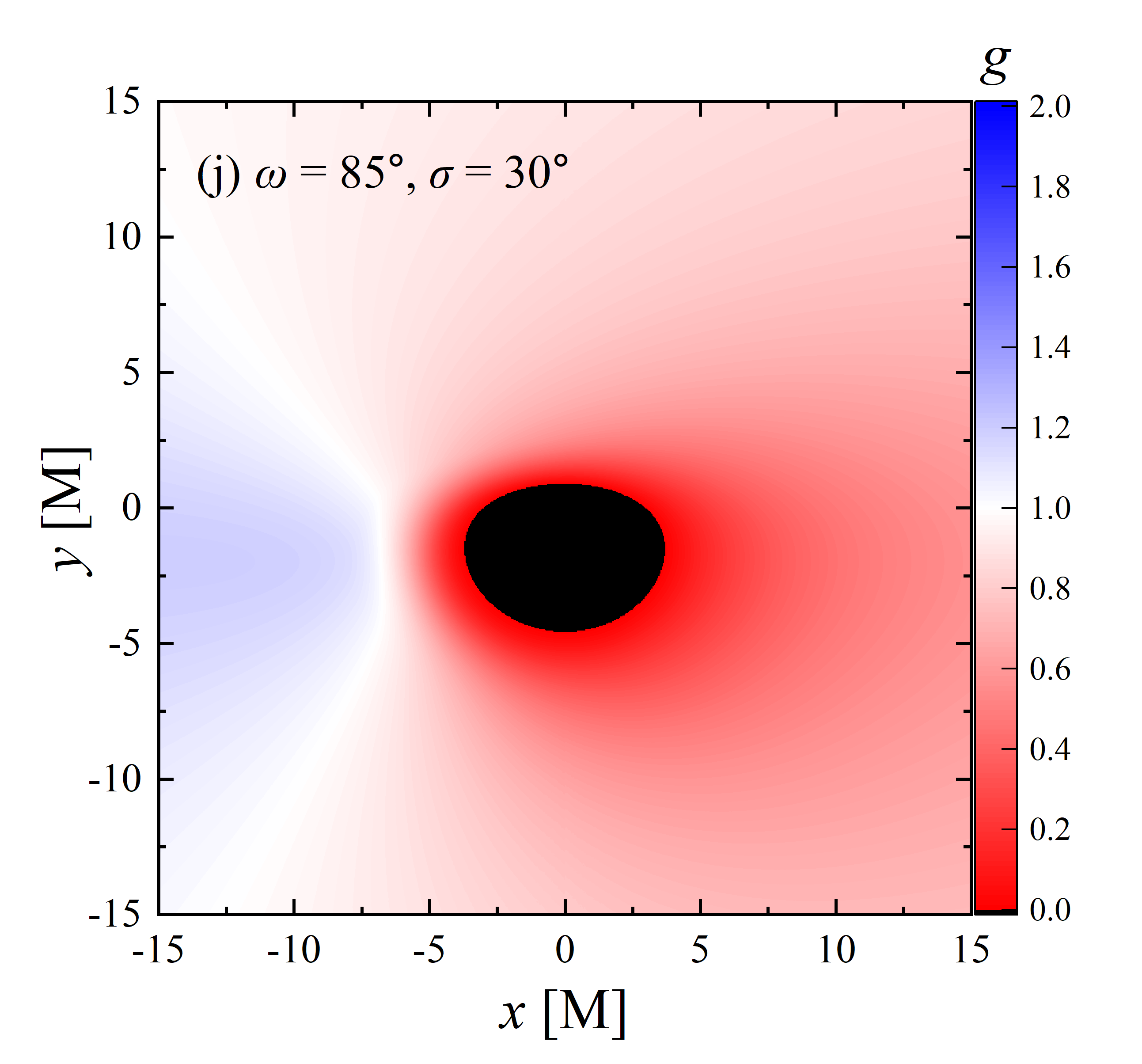}
\includegraphics[width=3.7cm]{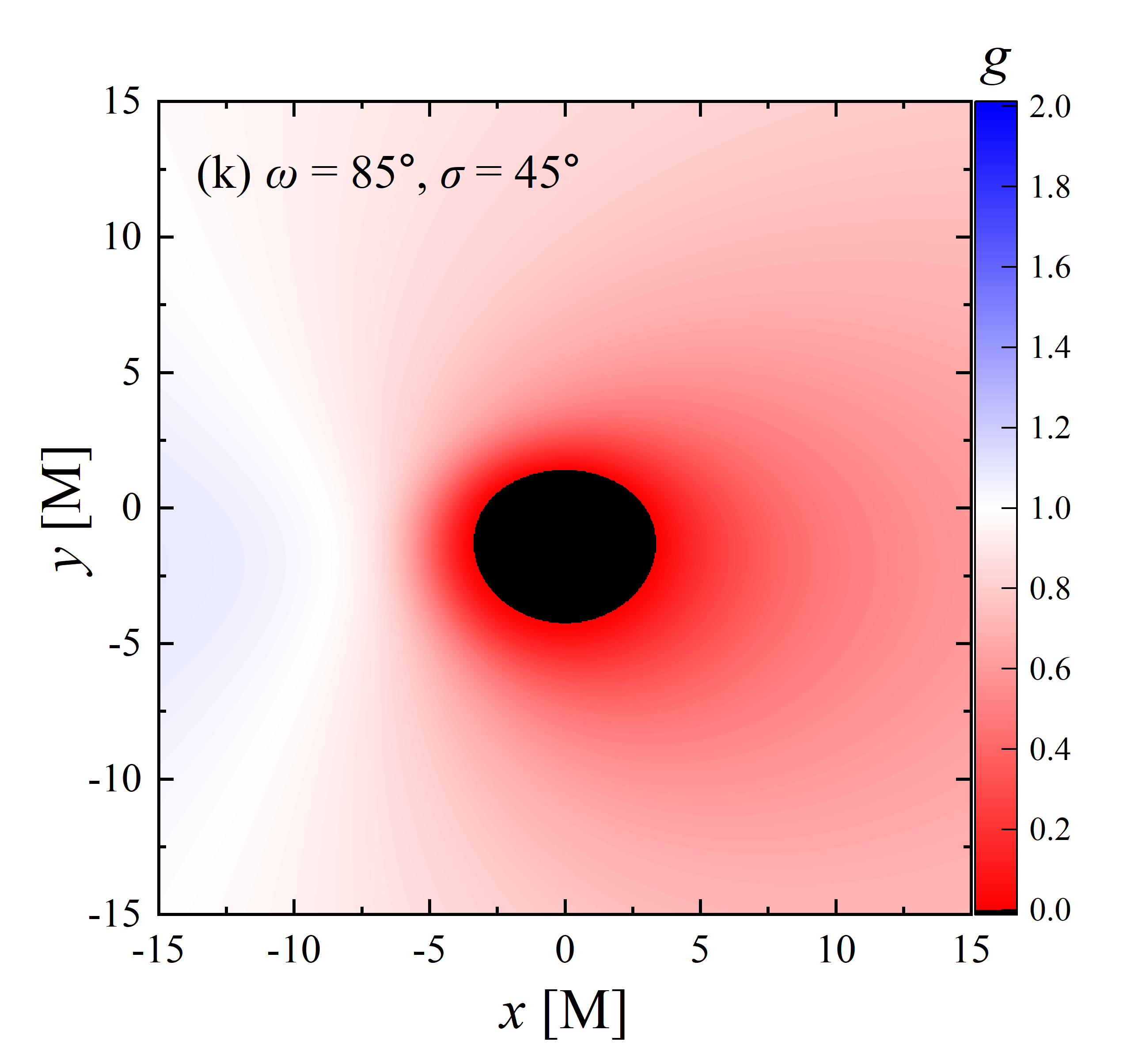}
\includegraphics[width=3.7cm]{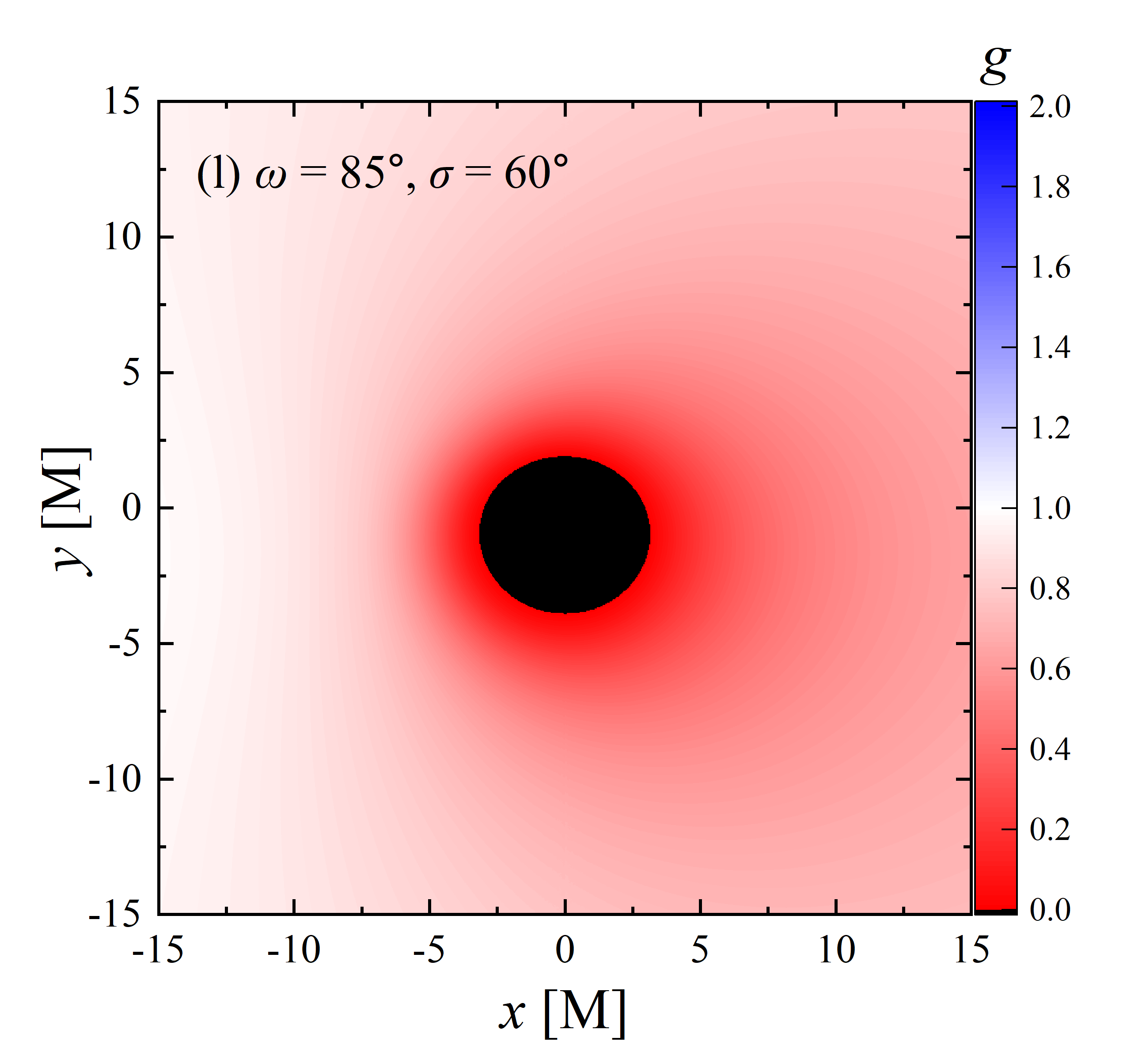}
\caption{The distributions of redshift factors in direct images of the hairy BH surrounded by a tilted accretion disk with varying observation angle $\omega$ and disk inclination $\sigma$. Here, we have the scalar hair parameter of $h = -1$.}}\label{fig6}
\end{figure*}
\subsubsection{Tilted stable circular orbit}
When a massive particle with coordinates $(r_{\textrm{e}},\theta_{\textrm{e}},\varphi_{\textrm{e}})$ is moving in a tilted stable circular orbit with an orbital inclination of $\sigma$, it is reasonable to expect that this particle has a radial velocity of zero and undergoes latitudinal oscillations within the range of $[\pi/2-\sigma,\pi/2+\sigma]$. This implies that the value of velocity in $\theta$-component is zero when $\theta_{\textrm{e}}=\pi/2-\sigma$ or $\theta_{\textrm{e}}=\pi/2+\sigma$. Consequently, the specific energy $E$ and angular momentum $L$ of the particle can be inferred using Eqs \eqref{8} and \eqref{9}, and the velocity components $\dot{t_{\textrm{e}}}$ and $\dot{\varphi_{\textrm{e}}}$ can be further derived from relations \eqref{5} and \eqref{6}. The last velocity component, $\dot{\theta_{\textrm{e}}}$, can be solved from Lagrangian \eqref{7} as
\begin{eqnarray}\label{18}
\dot{\theta_{\textrm{e}}} = \pm \sqrt{\frac{1}{g_{\theta\theta}}\left(-1-\frac{E^{2}}{g_{tt}}-\frac{L^{2}}{g_{\varphi\varphi}}\right)},
\end{eqnarray}
where the sign ``$\pm$'' is determined by the direction of motion of the particle. Specifically, $\dot{\theta_{\textrm{e}}} > 0$ for $\varphi_{\textrm{e}}$ within the range $(0^{\circ}, 180^{\circ})$, while $\dot{\theta_{\textrm{e}}}$ takes negative values when $\varphi_{\textrm{e}}$ is greater than $180^{\circ}$. The values of $g_{tt}$, $g_{\theta\theta}$, and $g_{\varphi\varphi}$ are evaluated at $r=r_{\textrm{e}}$, $\theta=\theta_{\textrm{e}}$. Hence, we have the formulation of redshift factors for light rays radiated by a light source that moves in a tilted stable circular orbit, expressed as
\begin{eqnarray}\label{19}
g_{\textrm{SCO}} = \frac{1}{\dot{t_{\textrm{e}}}+\dot{\varphi_{\textrm{e}}}p_{\varphi}/p_{t}\pm\dot{\theta_{\textrm{e}}}p_{\theta}/p_{t}}.
\end{eqnarray}
The above equation reduces to a form that is applicable to the equatorial case when the conditions $\sigma = \dot{\theta_{\textrm{e}}} = 0$ and $\theta_{\textrm{e}}=\pi/2$ are satisfied.
\begin{figure*}
\center{
\includegraphics[width=3.7cm]{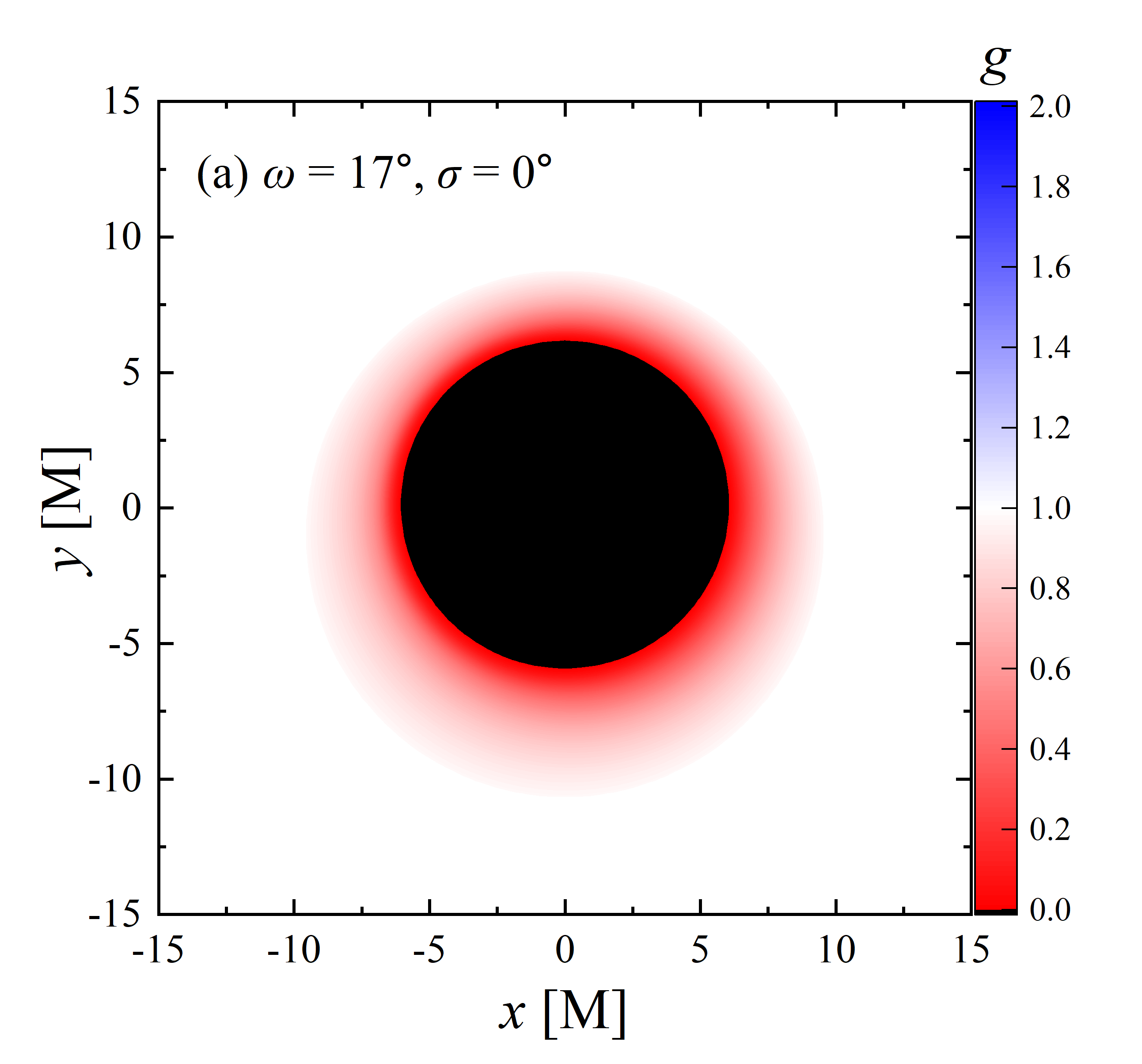}
\includegraphics[width=3.7cm]{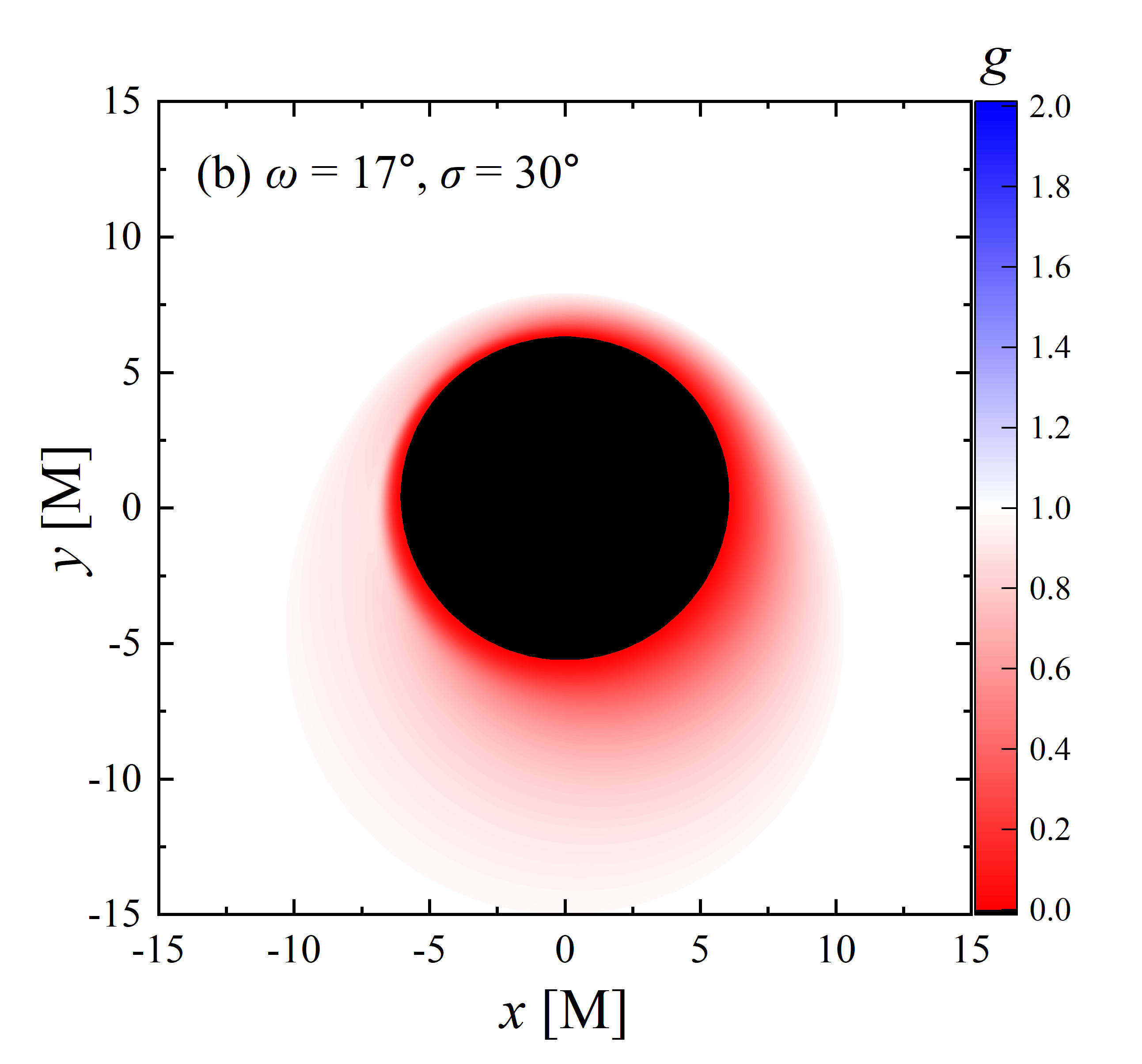}
\includegraphics[width=3.7cm]{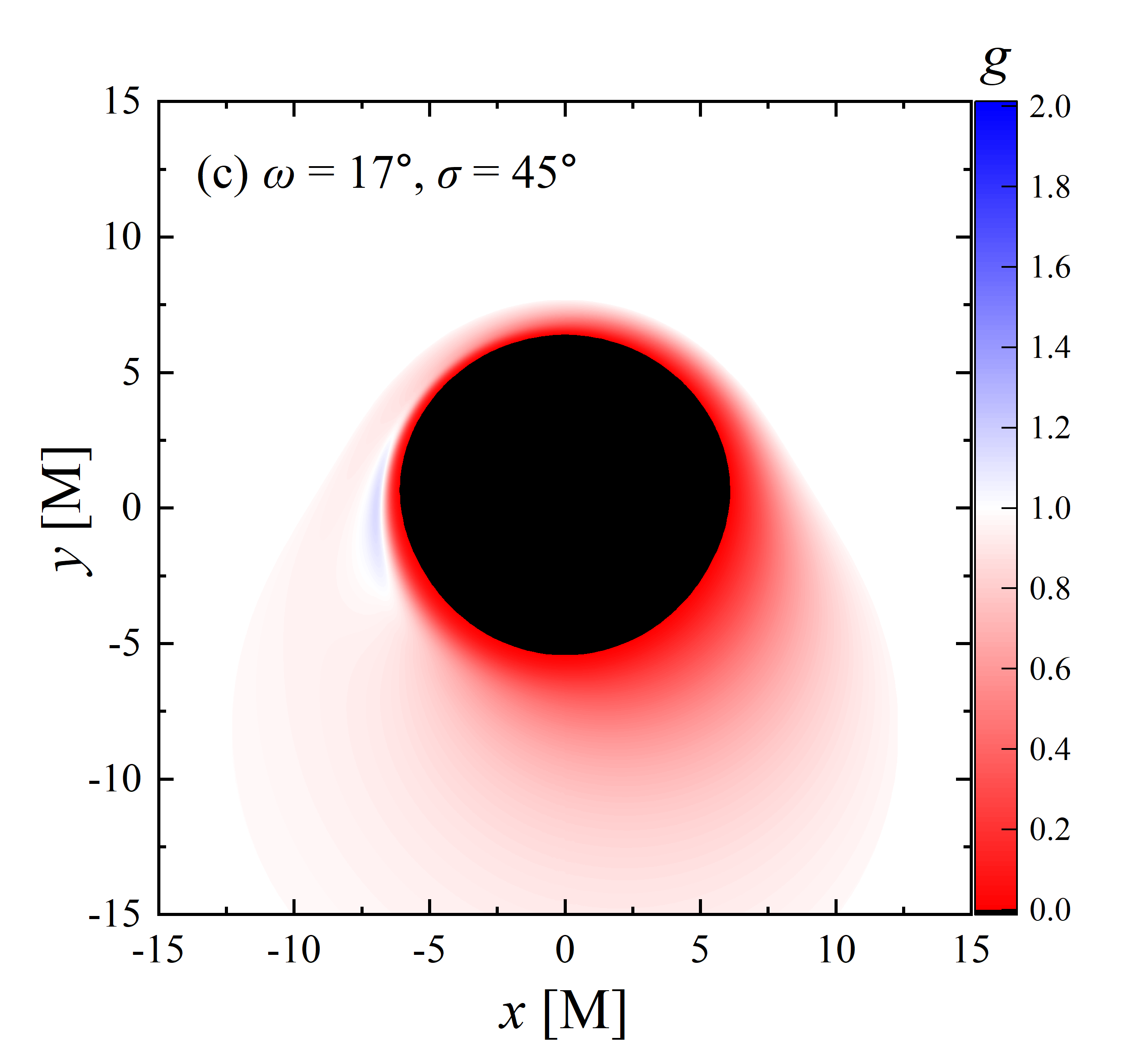}
\includegraphics[width=3.7cm]{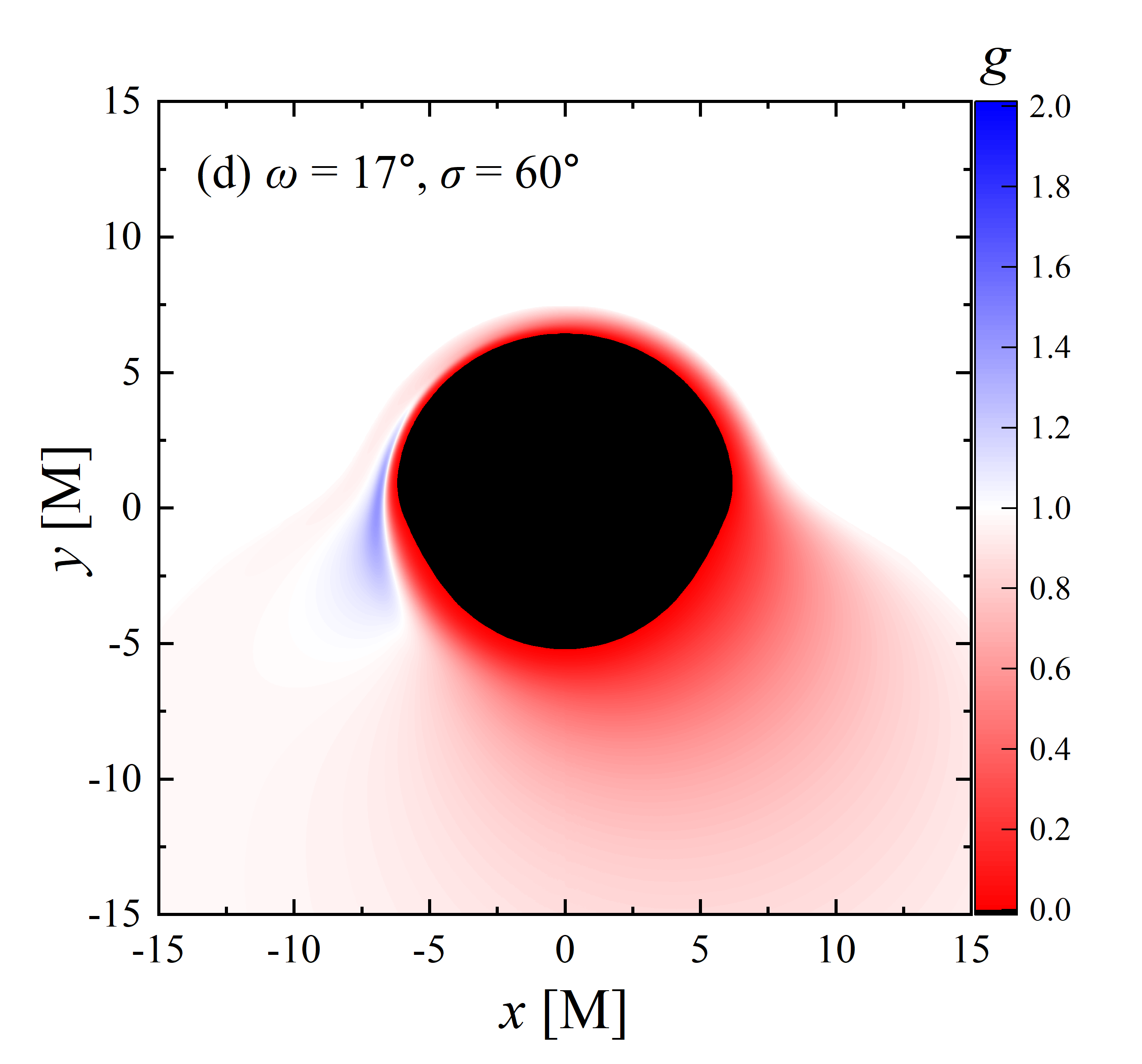}
\includegraphics[width=3.7cm]{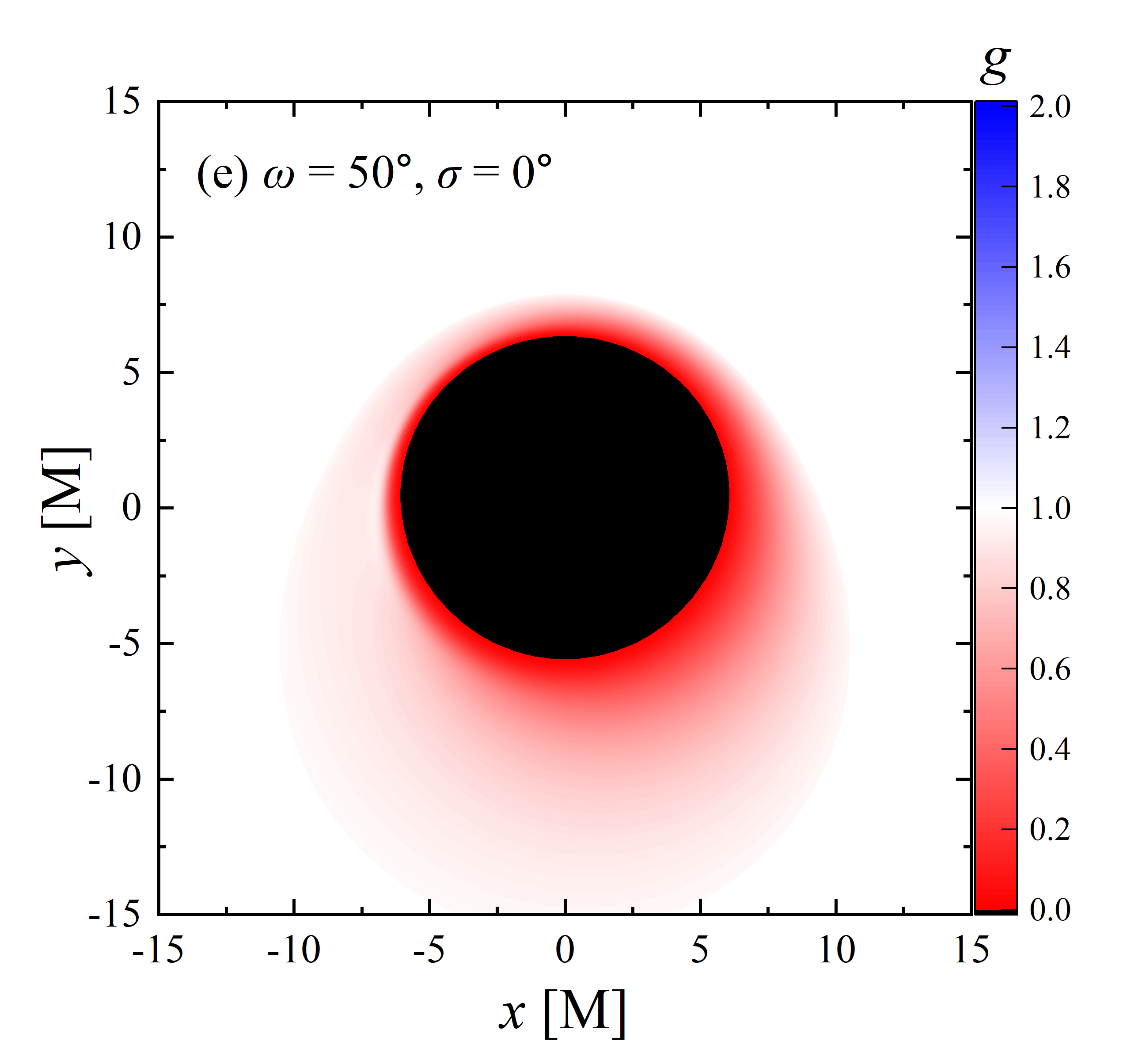}
\includegraphics[width=3.7cm]{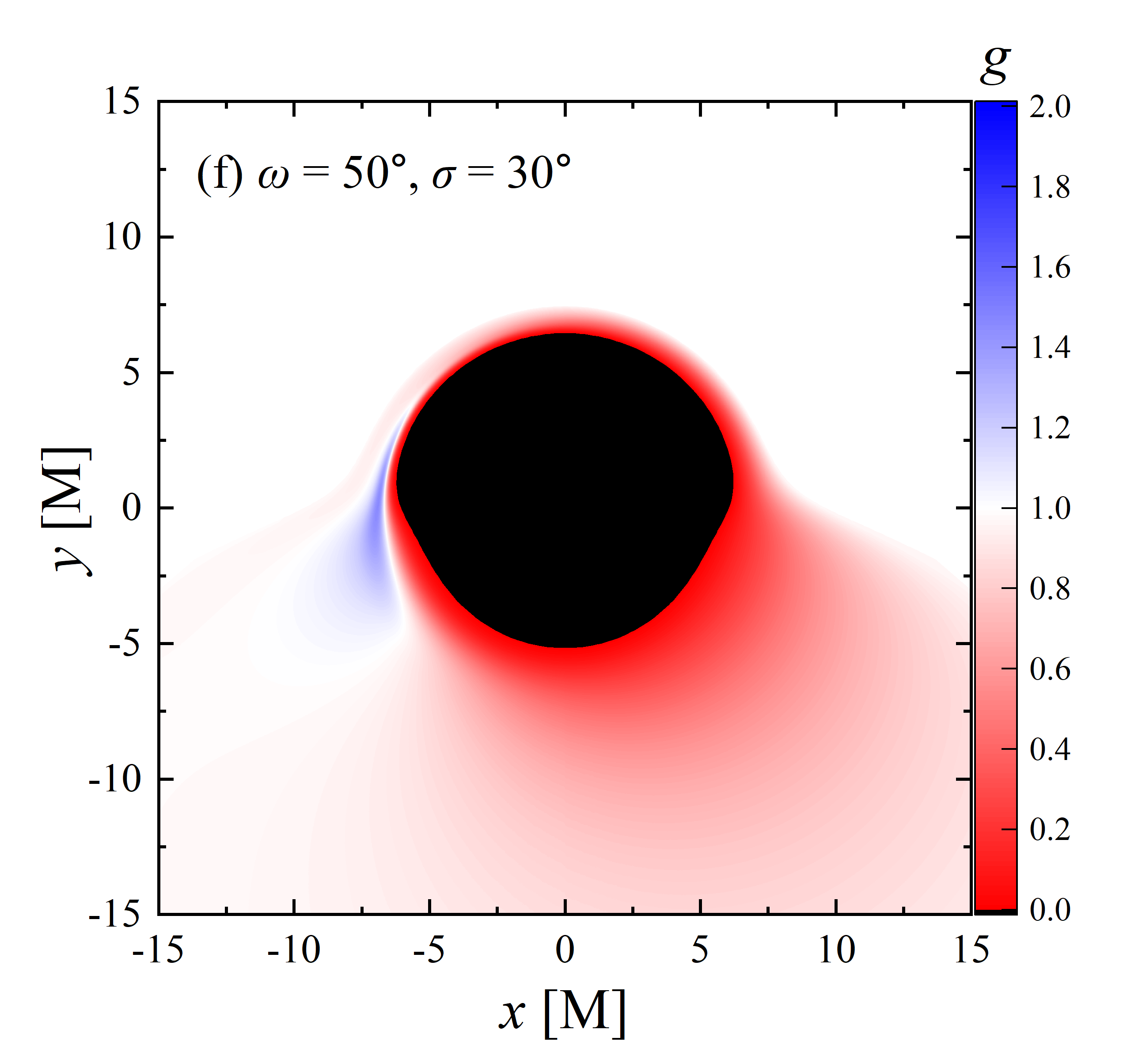}
\includegraphics[width=3.7cm]{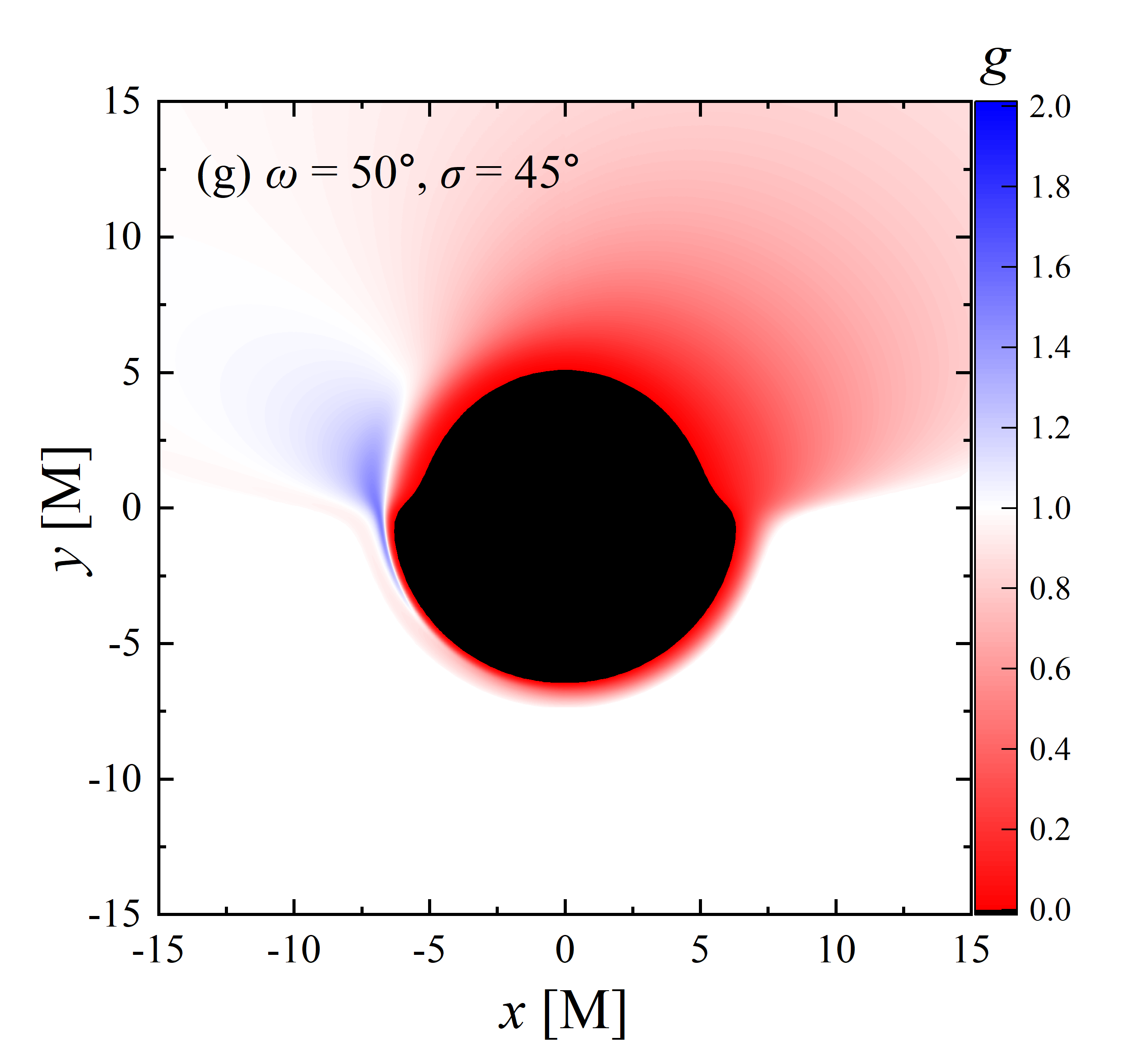}
\includegraphics[width=3.7cm]{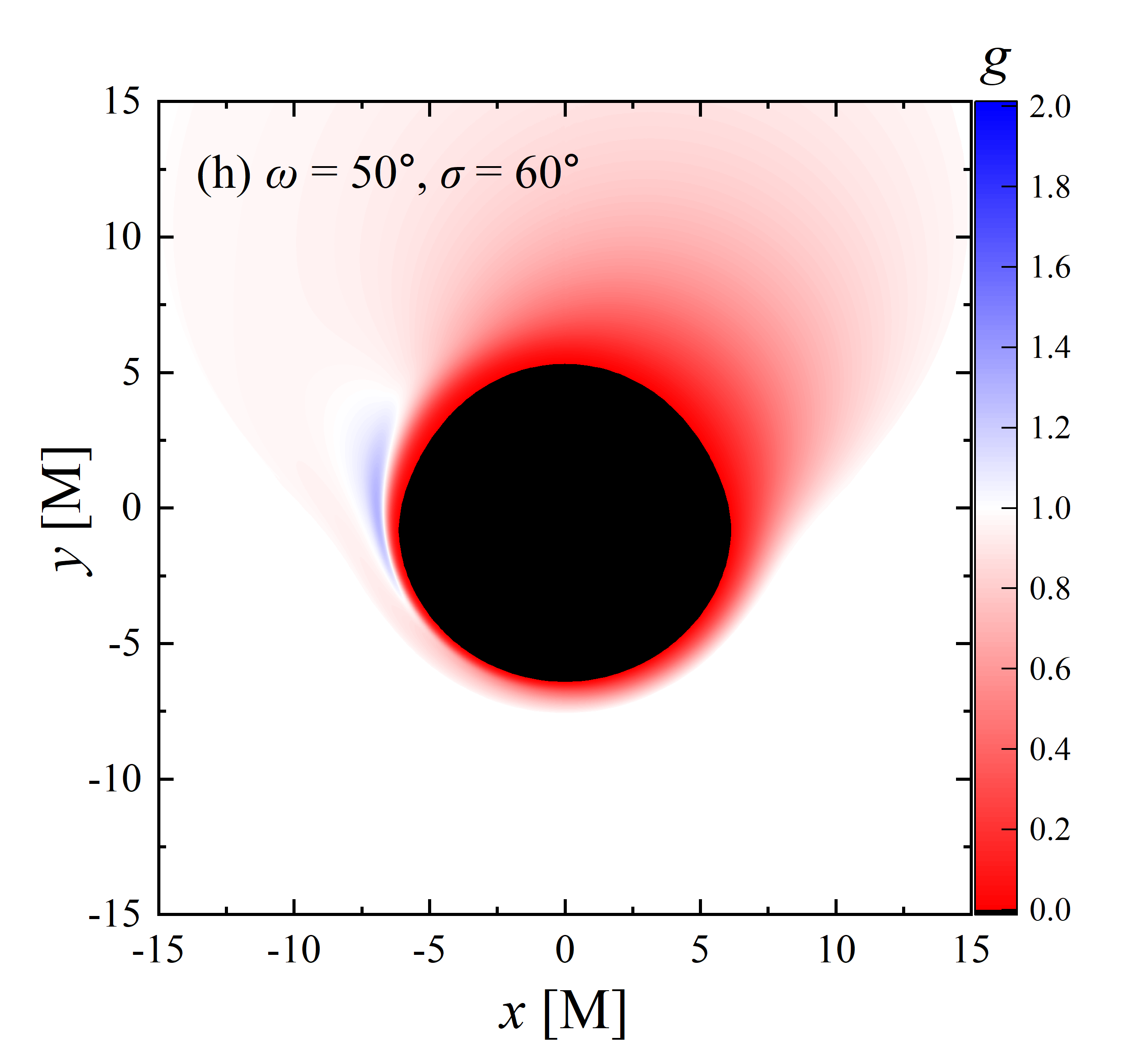}
\includegraphics[width=3.7cm]{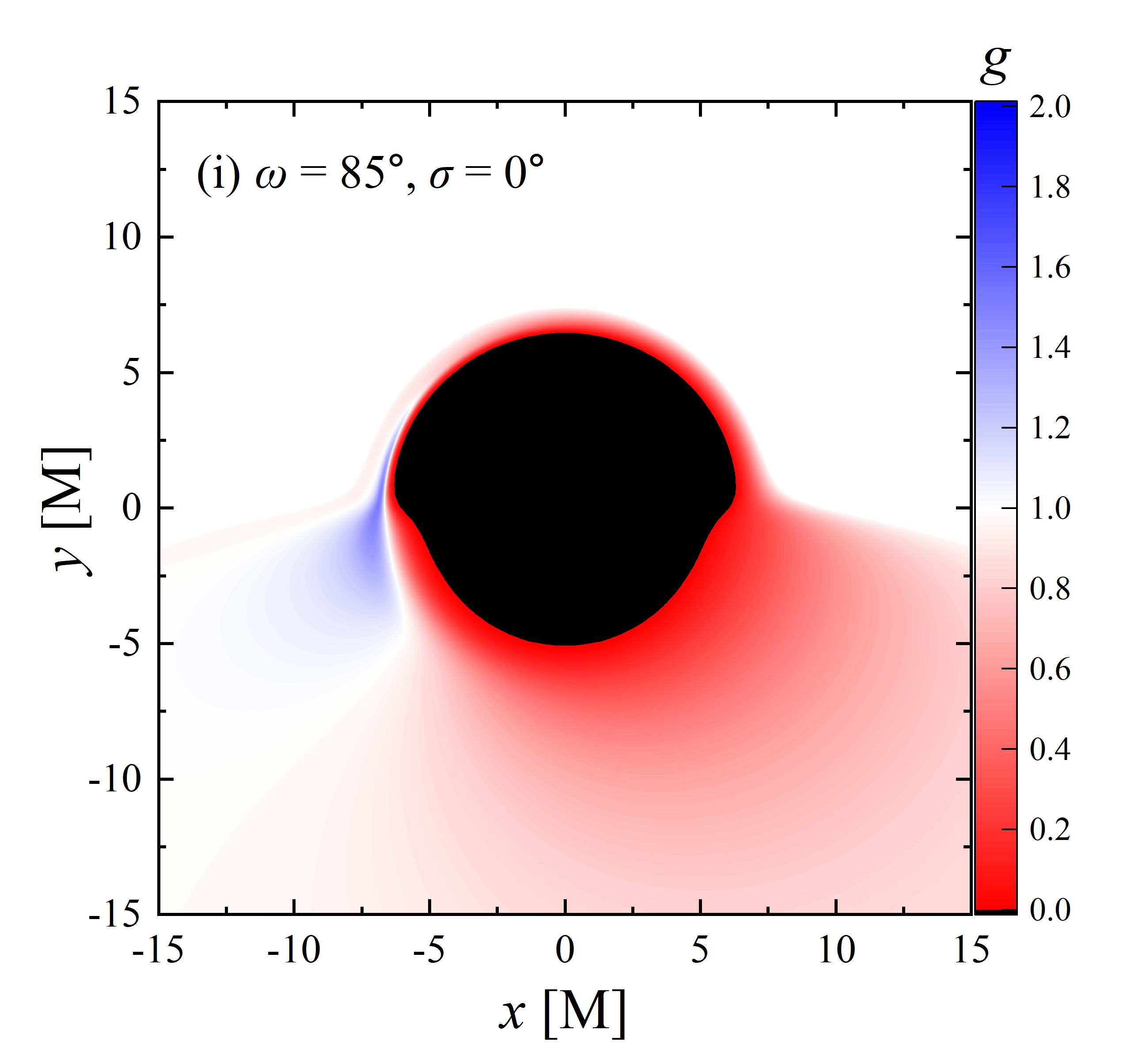}
\includegraphics[width=3.7cm]{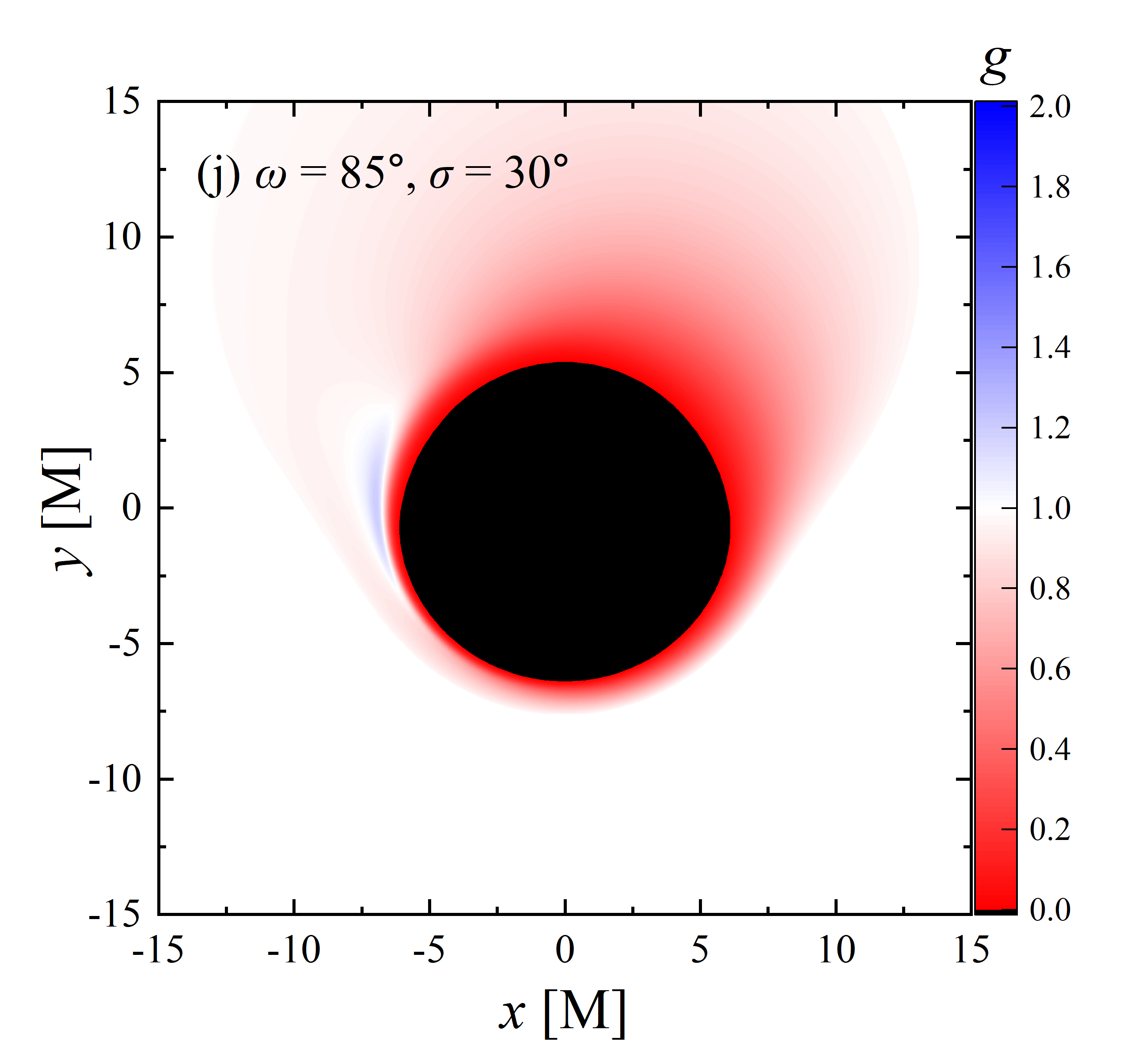}
\includegraphics[width=3.7cm]{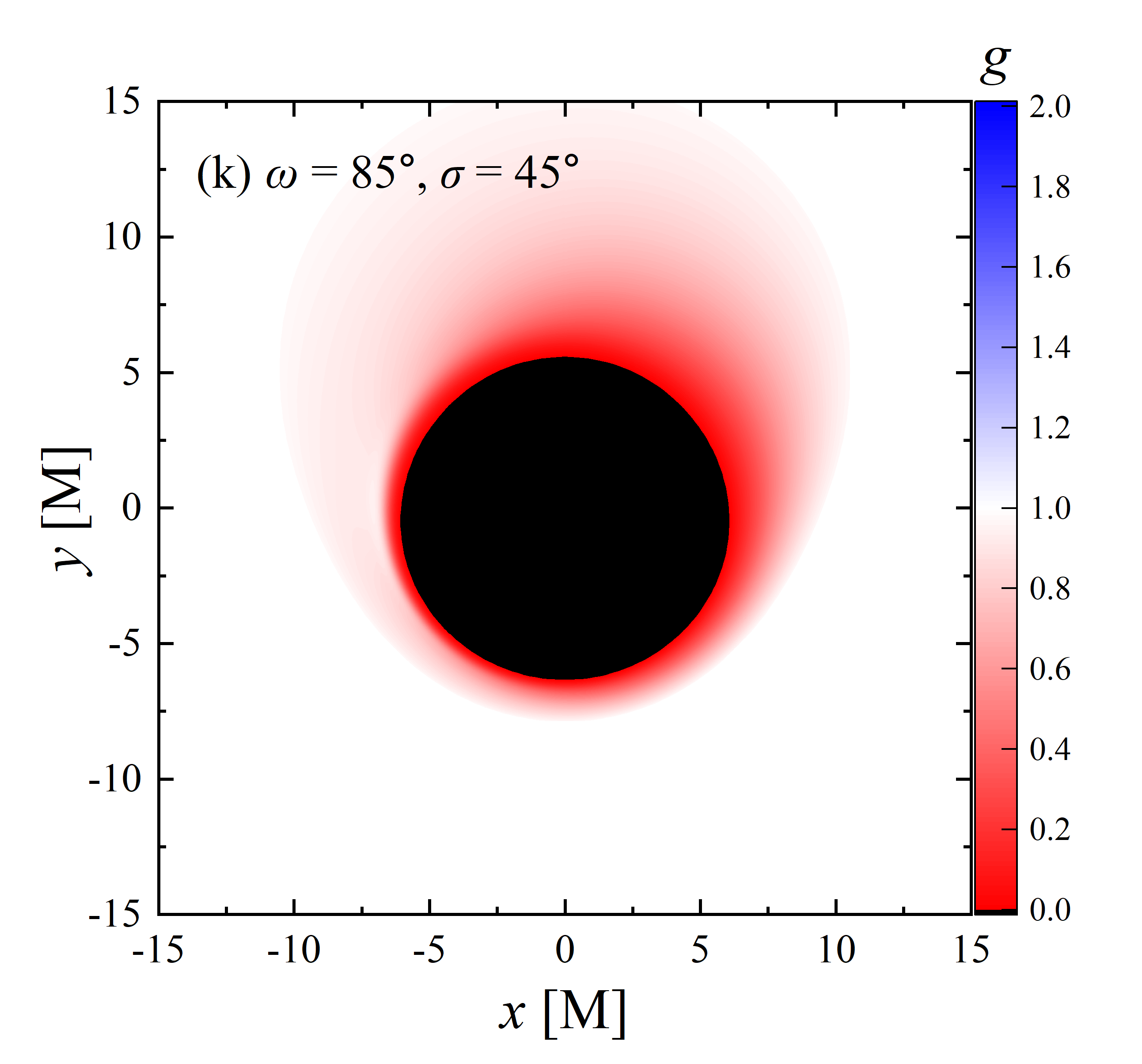}
\includegraphics[width=3.7cm]{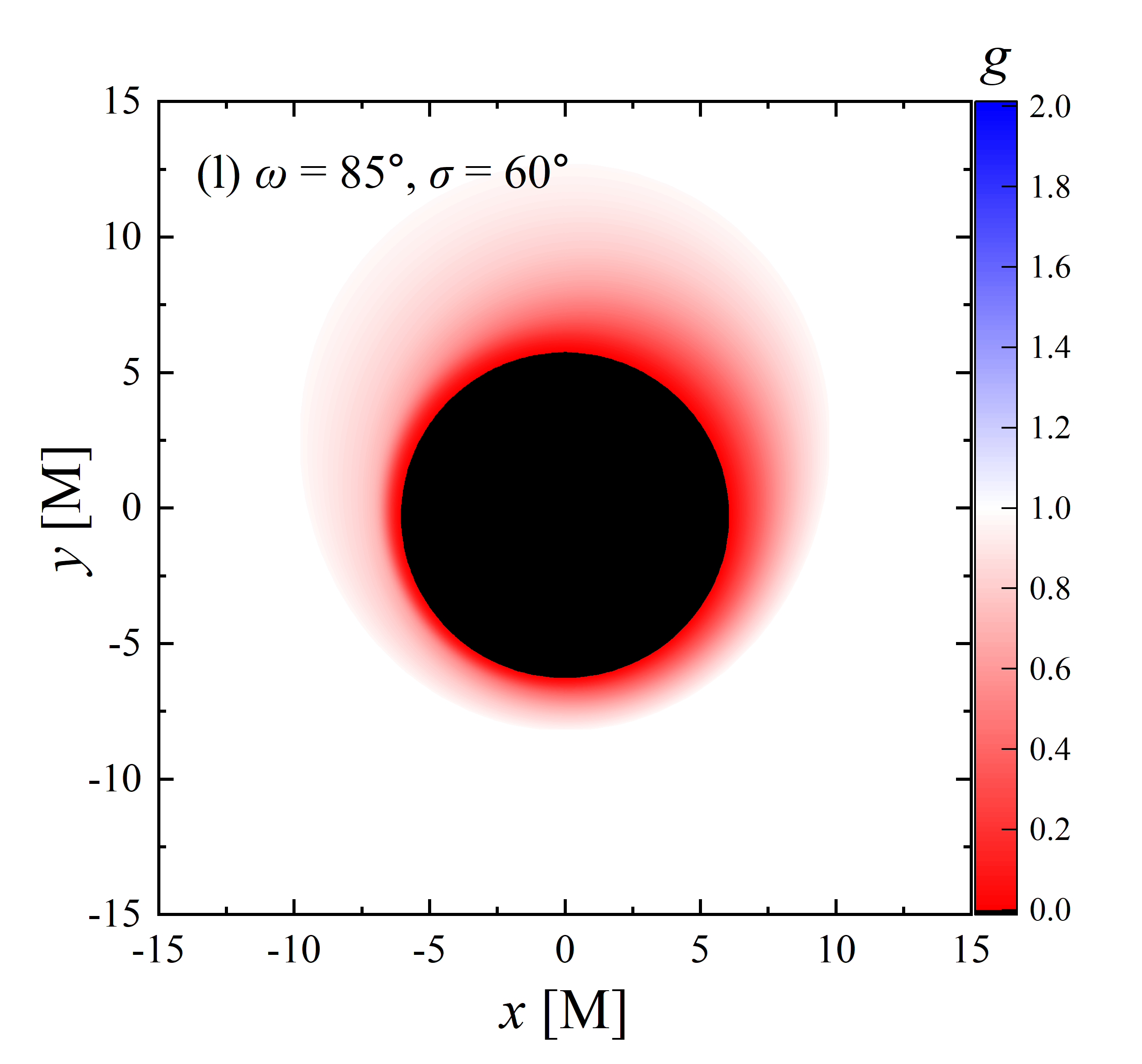}
\caption{Similar to figure 6, but for the lensed images.}}\label{fig7}
\end{figure*}
\subsubsection{Tilted plunging orbit}
The massive particle will be accelerated along the plunging orbit once it crosses the innermost stable circular orbit. The authors of \cite{Chael et al. (2021)} pointed out that the specific angular momentum of the plunging orbit follows a power-law fitting function in relation to the coordinate $r_{\textrm{e}}$. For simplicity, we assume that the particle's specific energy and angular momentum are the same as those of the innermost stable circular orbit (marked by $E_{\textrm{ISCO}}$ and $L_{\textrm{ISCO}}$), as also stated in \cite{Gralla et al. (2020),Hou et al. (2022),Cunningham (1975)}. These two motion constants can be calculated according to the condition of $\partial V_{\textrm{eff}}/\partial r = \partial^{2} V_{\textrm{eff}}/\partial r^{2} = 0$ where $V_{\textrm{eff}}$ can be found in Eq \eqref{8}.

Due to the integrability of the hairy BH, there is also a motion constant governing the latitudinal oscillations of the massive particle, known as the Carter constant \cite{Carter (1968)}. Hence, we have
\begin{eqnarray}\label{20}
\dot{\theta_{\textrm{e}}} = \pm \frac{1}{r^{2}_{\textrm{e}}} \sqrt{Q_{\textrm{ISCO}}-\frac{L_{\textrm{ISCO}}^{2}}{\sin^{2}\theta_{\textrm{e}}}},
\end{eqnarray}
where $Q_{\textrm{ISCO}}$ is the Carter constant for a massive particle moving in the innermost stable circular orbit with orbital inclination $\sigma$. In particular, $Q_{\textrm{ISCO}}=L^{2}_{\textrm{ISCO}}/\sin^{2}(\pi/2-\sigma)$. Substituting $\dot{\theta_{\textrm{e}}}$, $E_{\textrm{ISCO}}$, and $L_{\textrm{ISCO}}$ into Lagrangian \eqref{7}, we obtain the expression for the radial velocity of a particle moving in a plunging orbit as
\begin{eqnarray}\label{21}
\dot{r_{\textrm{e}}} = - \sqrt{\frac{1}{g_{rr}}\left(-1-\frac{E^{2}_{\textrm{ISCO}}}{g_{tt}}-\frac{L^{2}_{\textrm{ISCO}}}{g_{\varphi\varphi}}-g_{\theta\theta}\dot{\theta^{2}_{\textrm{e}}}\right)},
\end{eqnarray}
where the negative sign indicates that the particle is spiralling towards the BH. The value of covariant metric tensor $g_{\mu\nu}$ is evaluated at $r=r_{\textrm{e}}$ and $\theta=\theta_{\textrm{e}}$. Finally, we rearrange Eq. \eqref{17} as
\begin{eqnarray}\label{22}
g_{\textrm{PO}} = \frac{1}{\dot{t_{\textrm{e}}}+\dot{\varphi_{\textrm{e}}}p_{\varphi}/p_{t}\pm\dot{\theta_{\textrm{e}}}p_{\theta}/p_{t}+\dot{r_{\textrm{e}}}p_{r}/p_{t}}.
\end{eqnarray}
\begin{figure*}
\center{
\includegraphics[width=3.7cm]{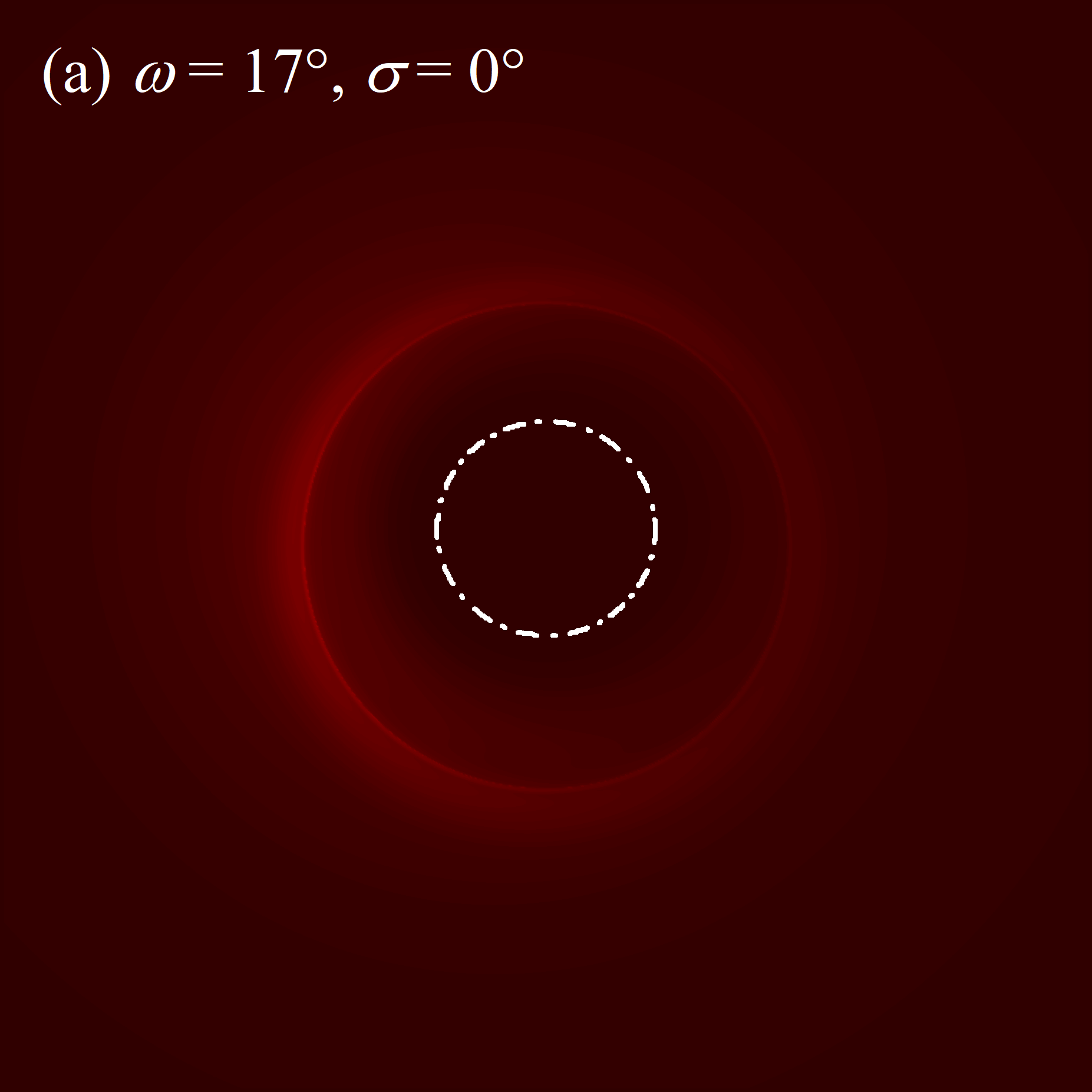}
\includegraphics[width=3.7cm]{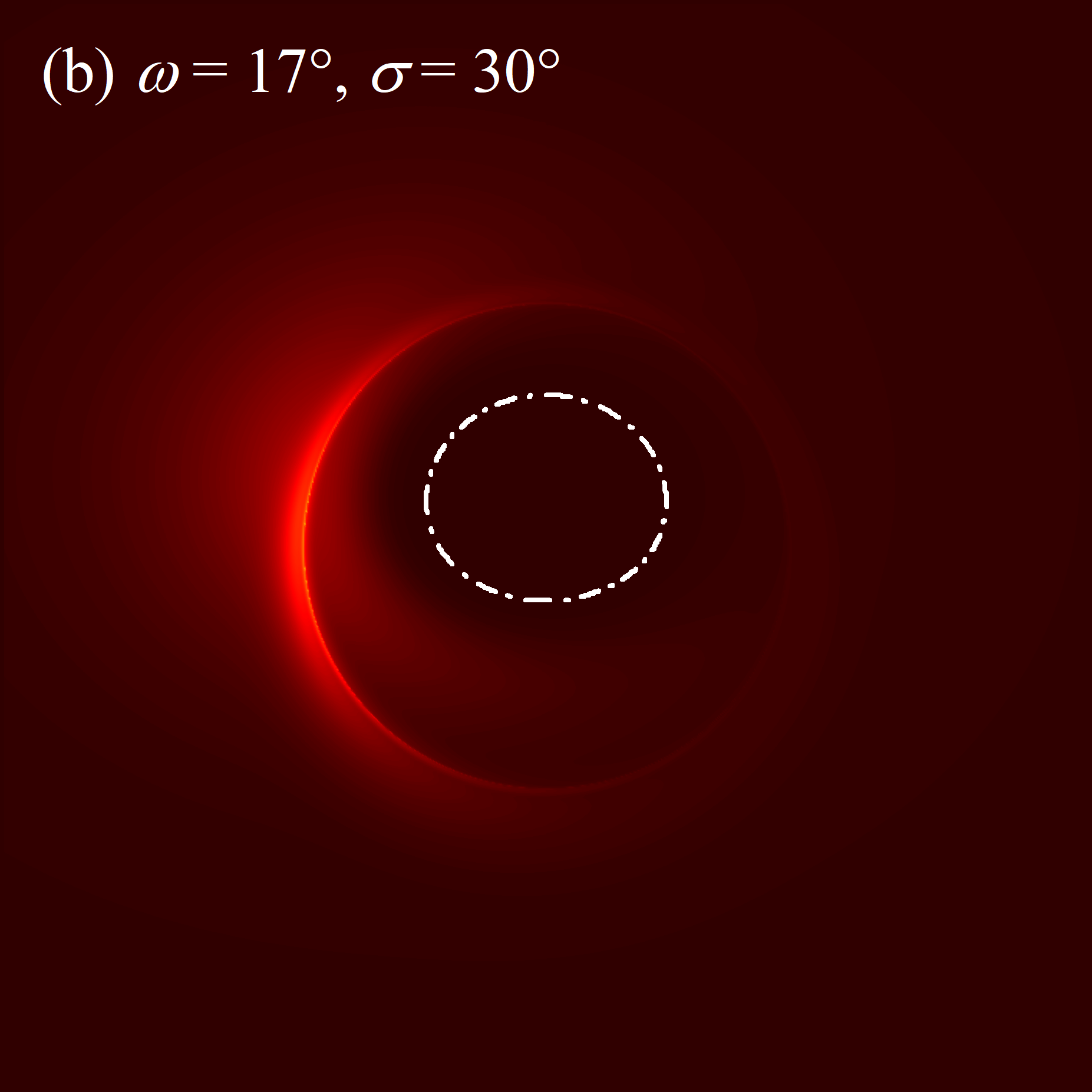}
\includegraphics[width=3.7cm]{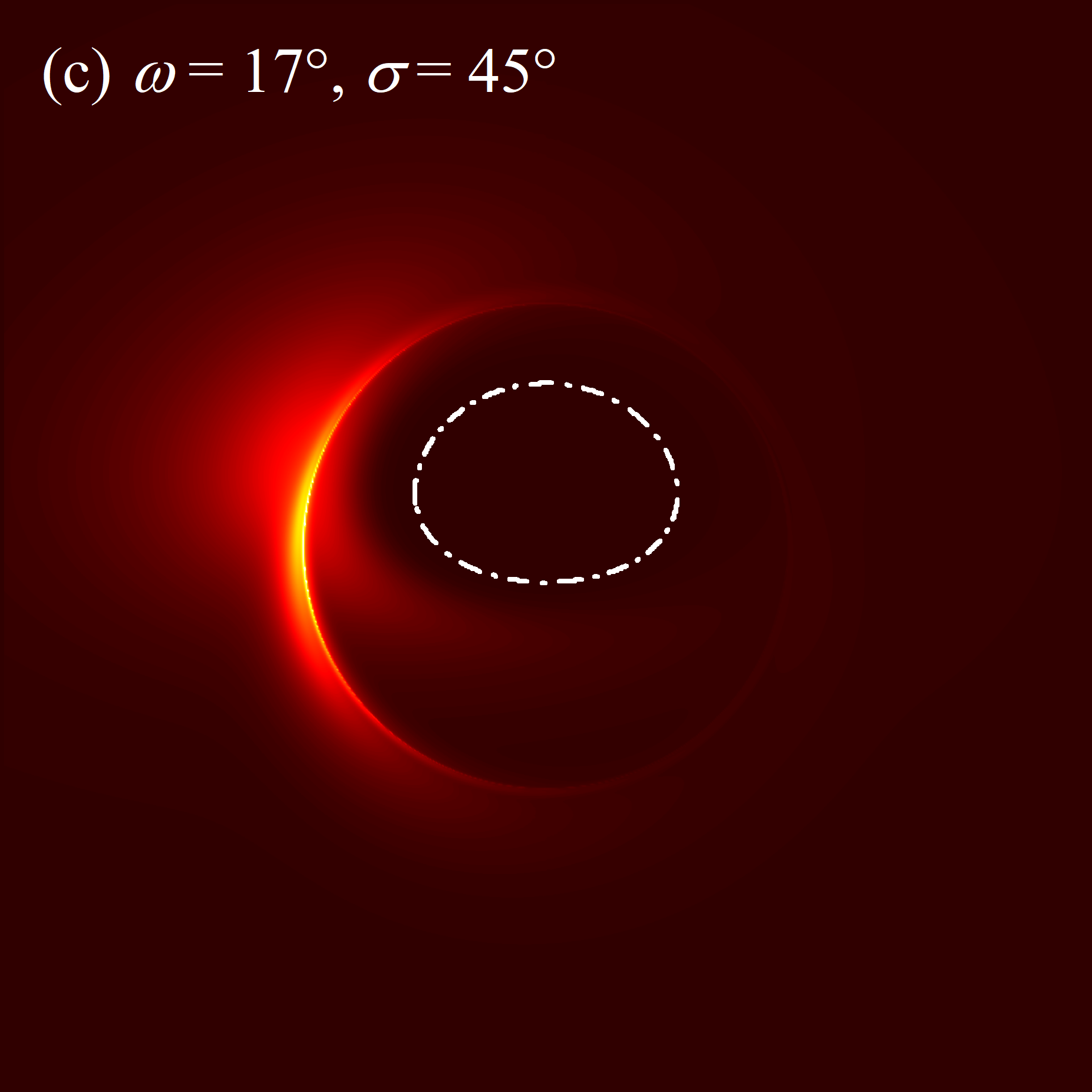}
\includegraphics[width=3.7cm]{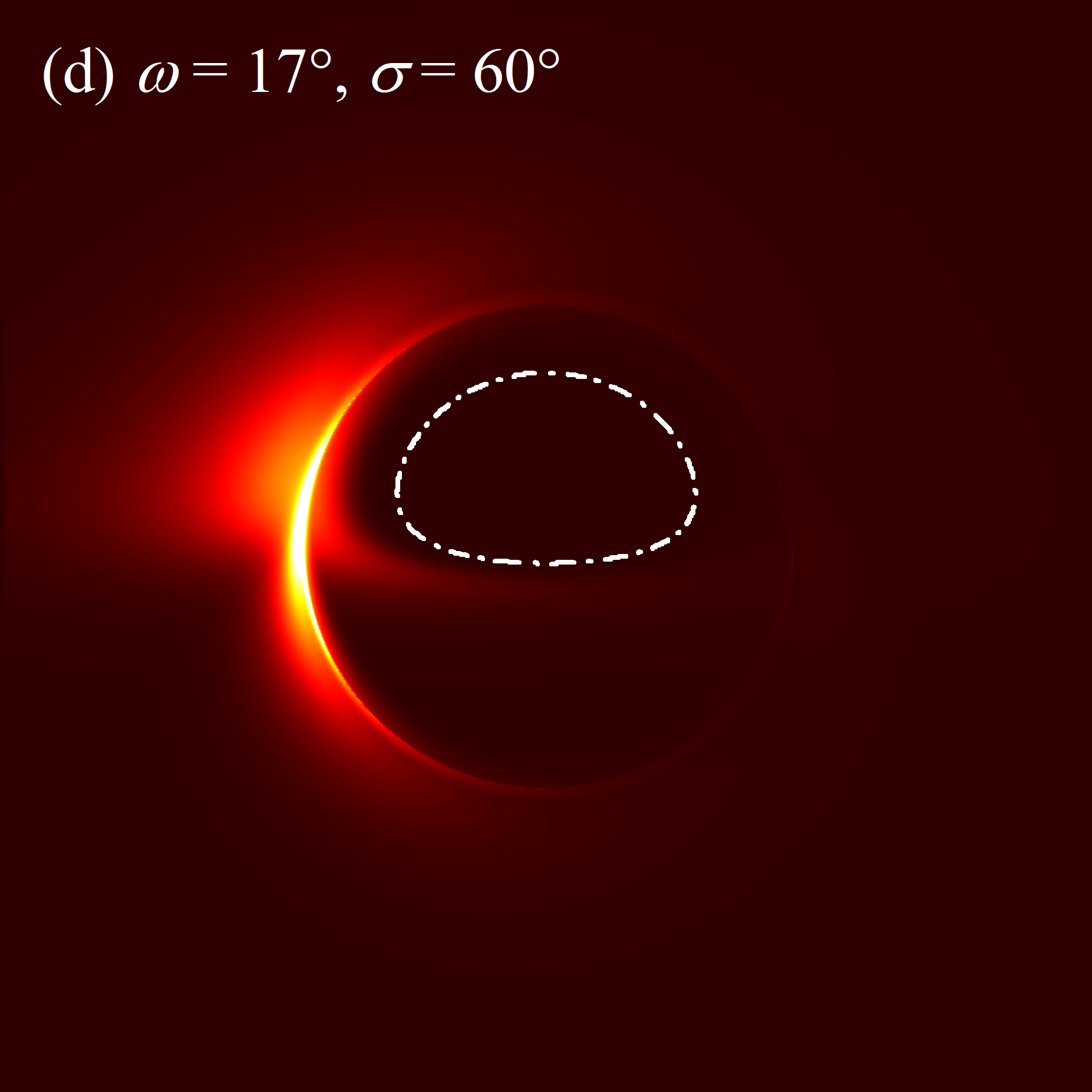}
\includegraphics[width=3.7cm]{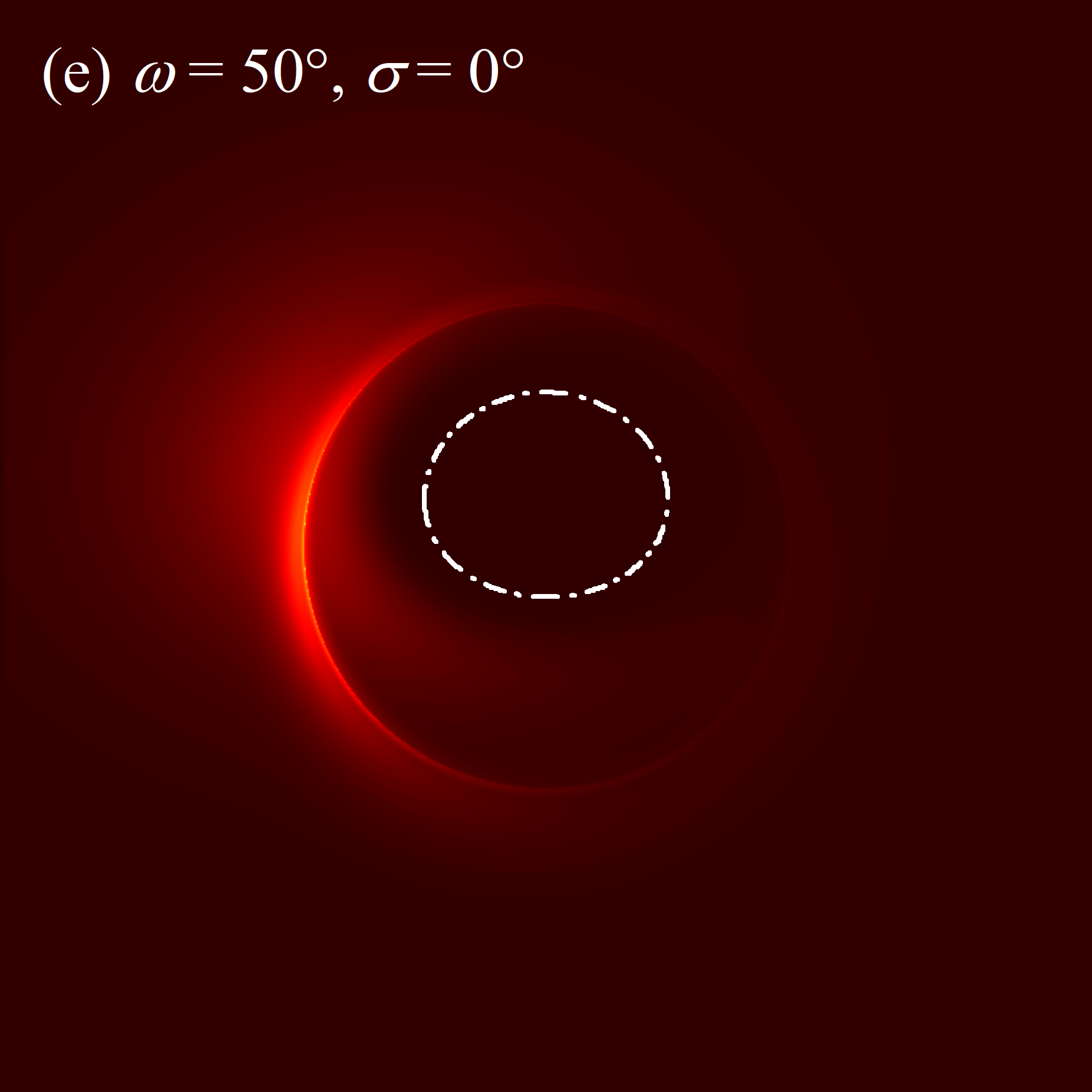}
\includegraphics[width=3.7cm]{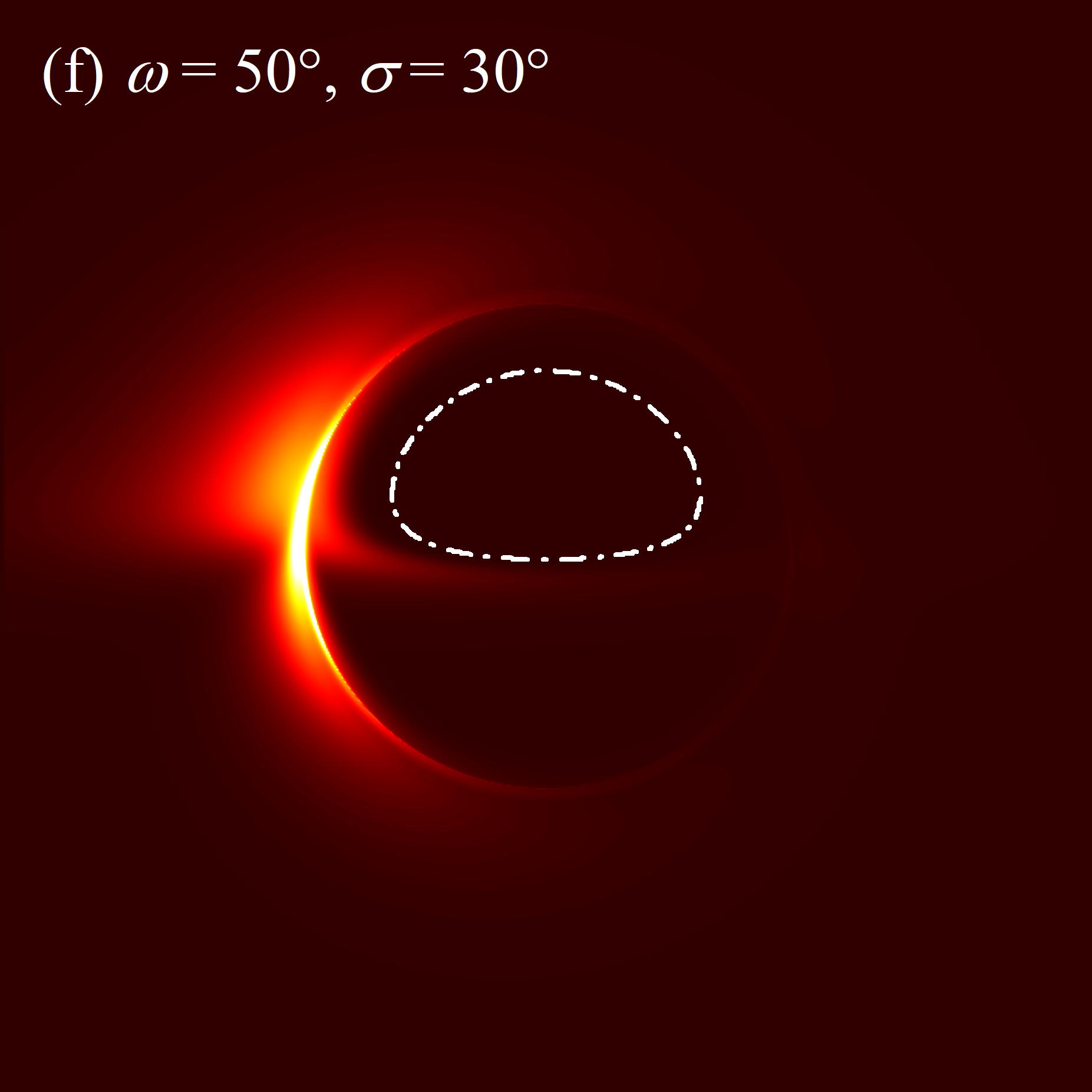}
\includegraphics[width=3.7cm]{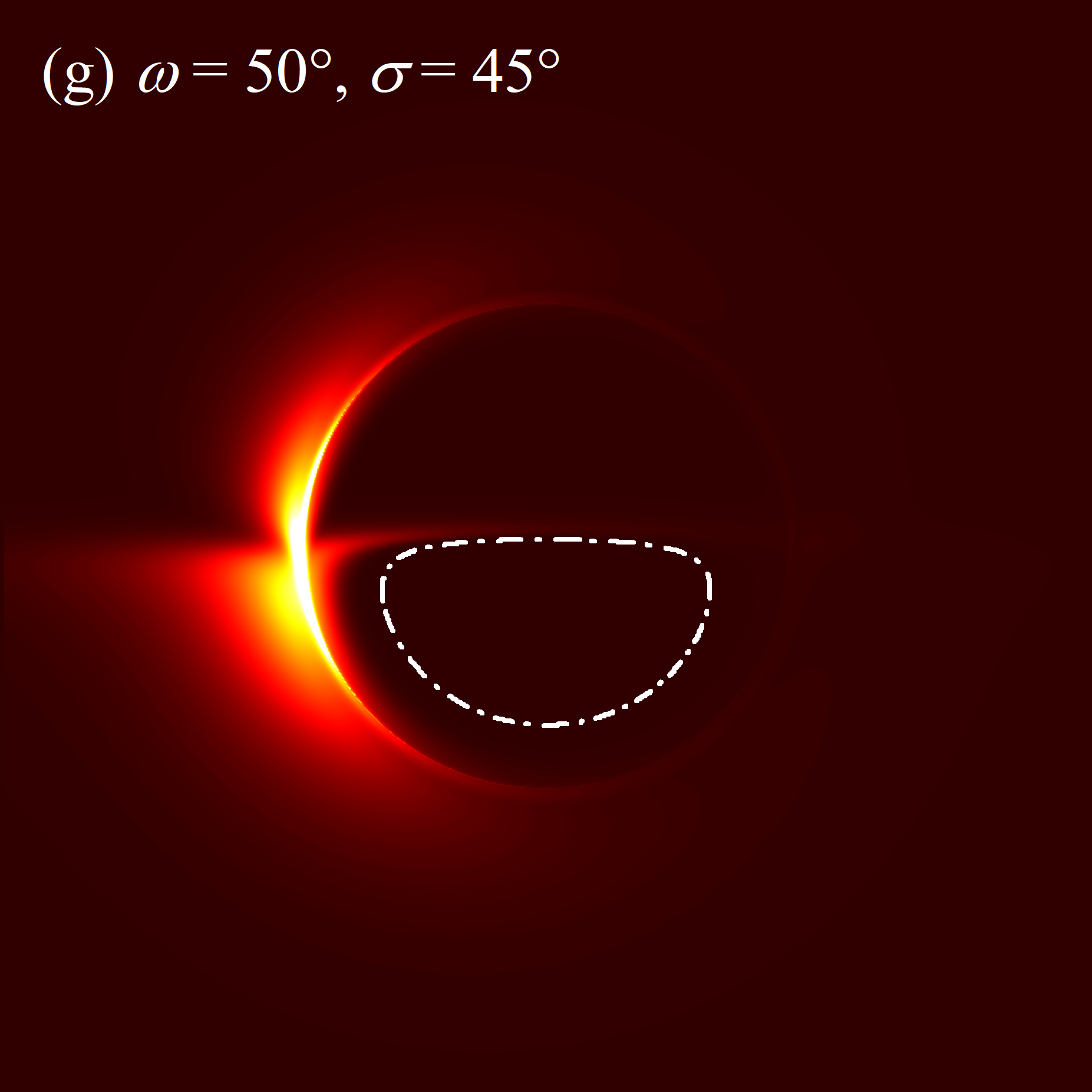}
\includegraphics[width=3.7cm]{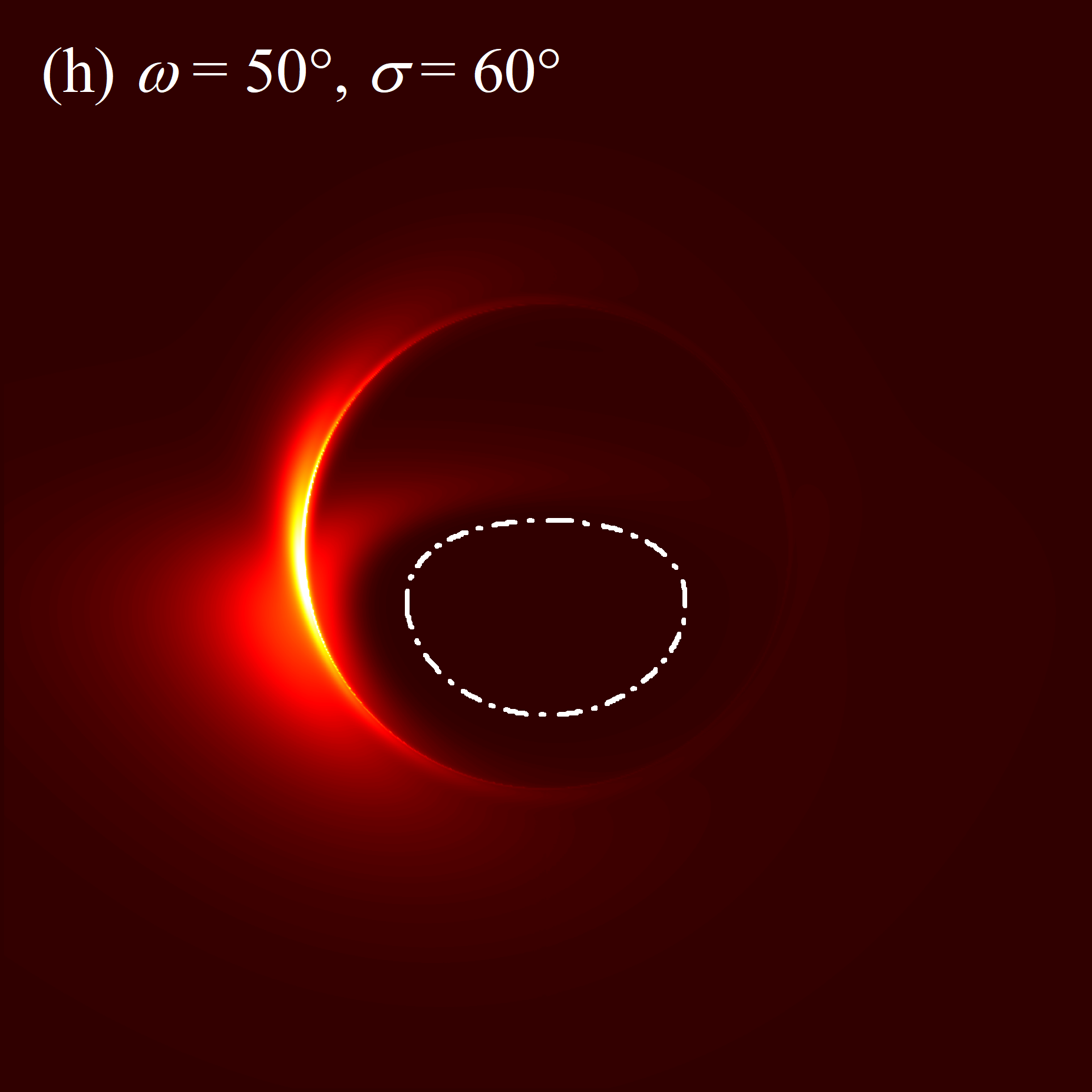}
\includegraphics[width=3.7cm]{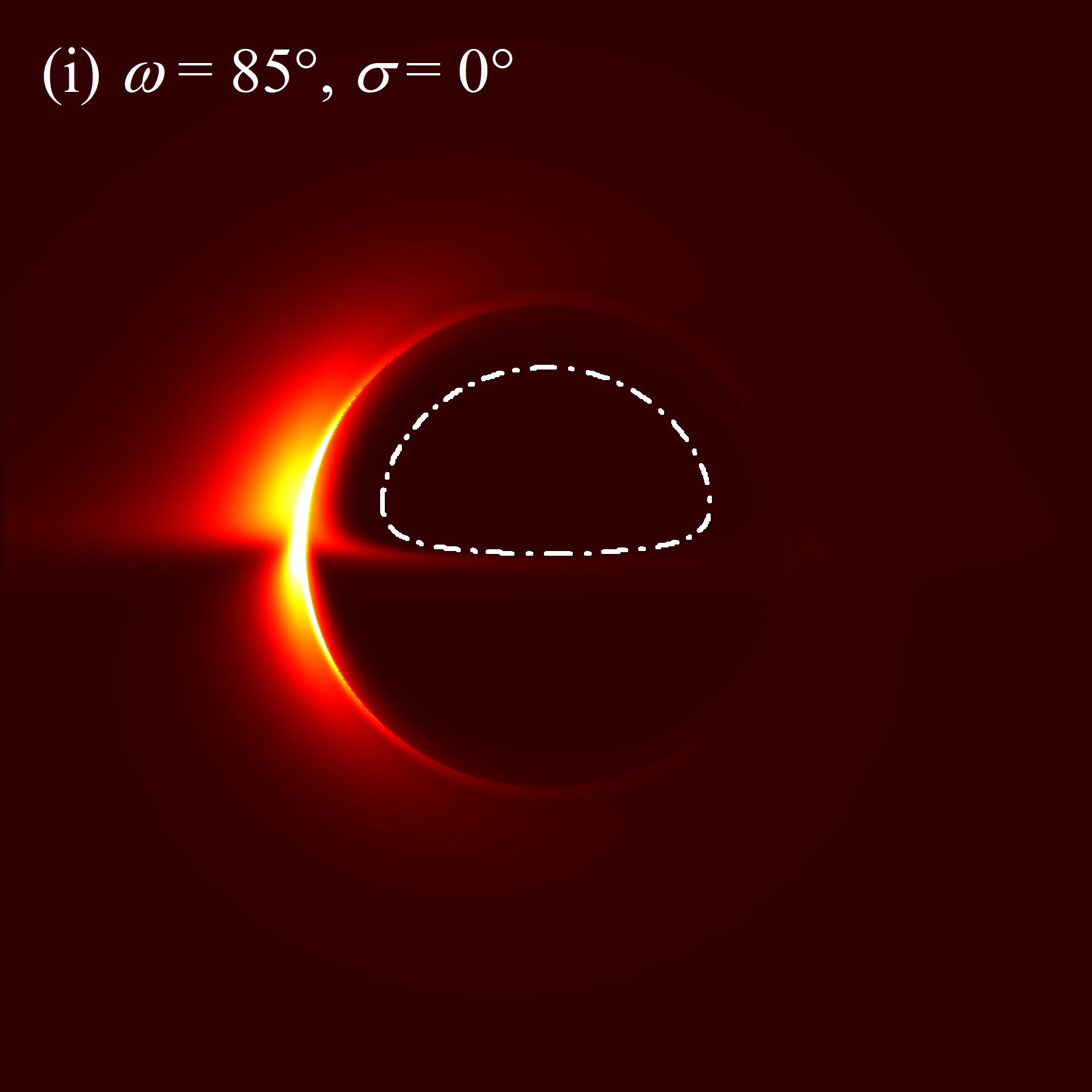}
\includegraphics[width=3.7cm]{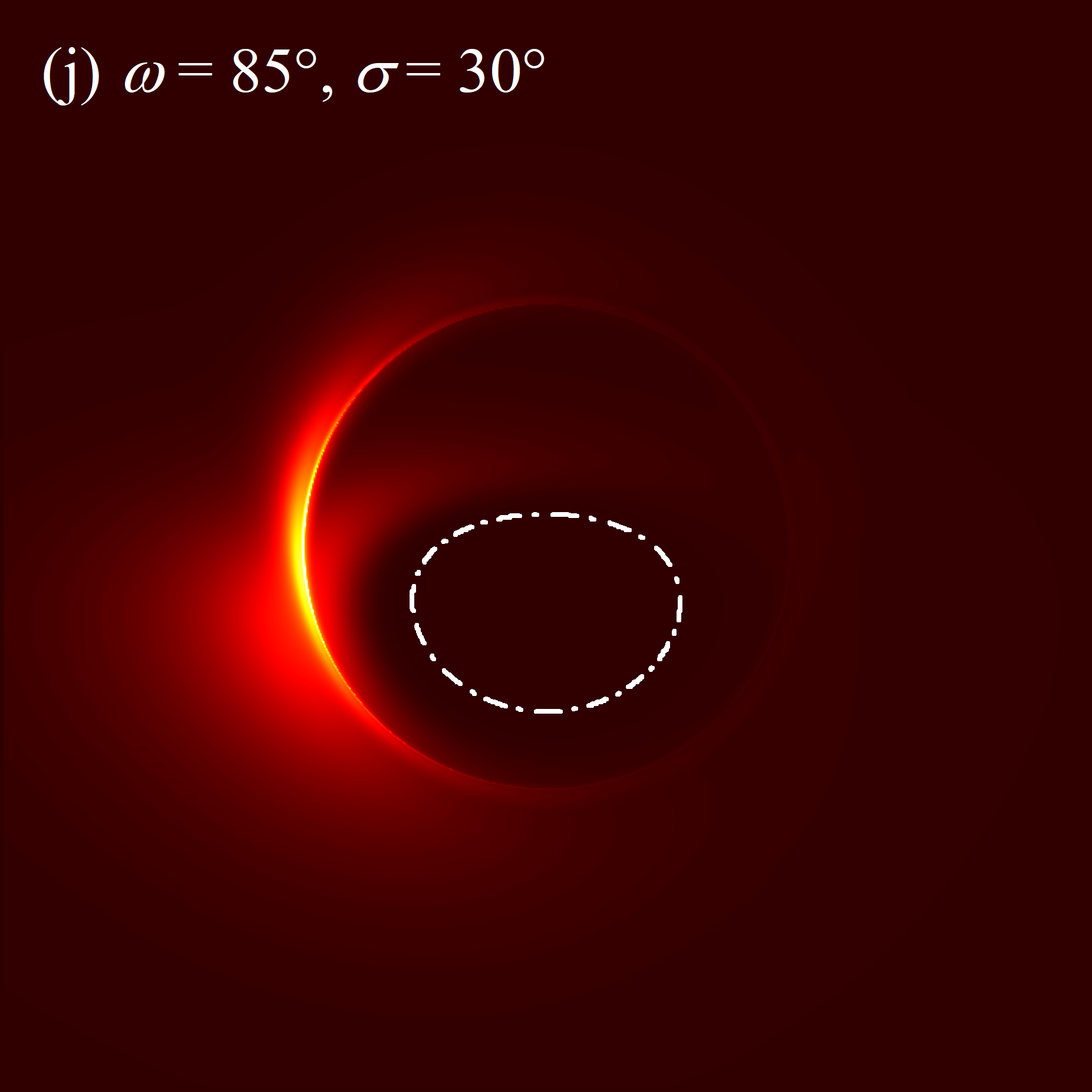}
\includegraphics[width=3.7cm]{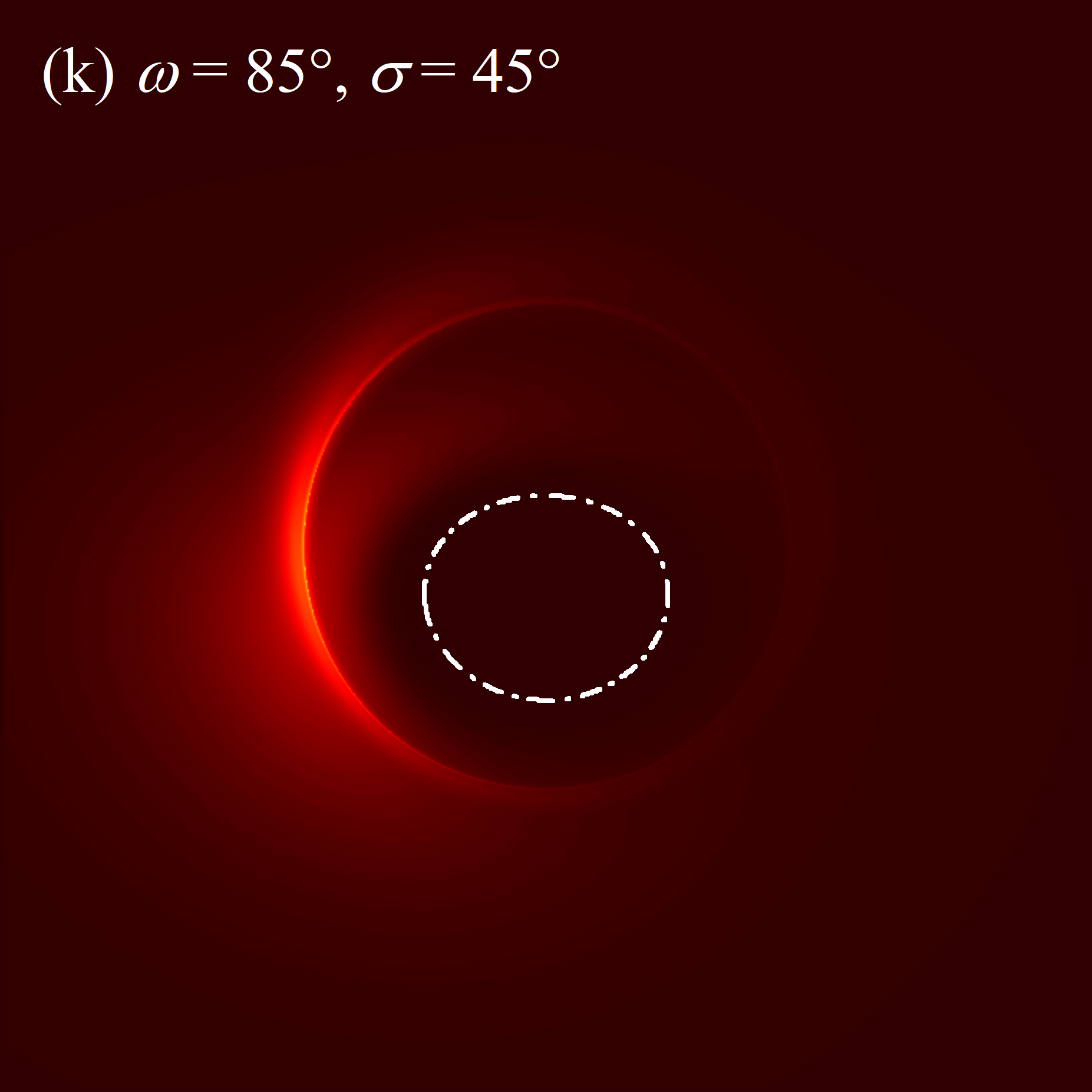}
\includegraphics[width=3.7cm]{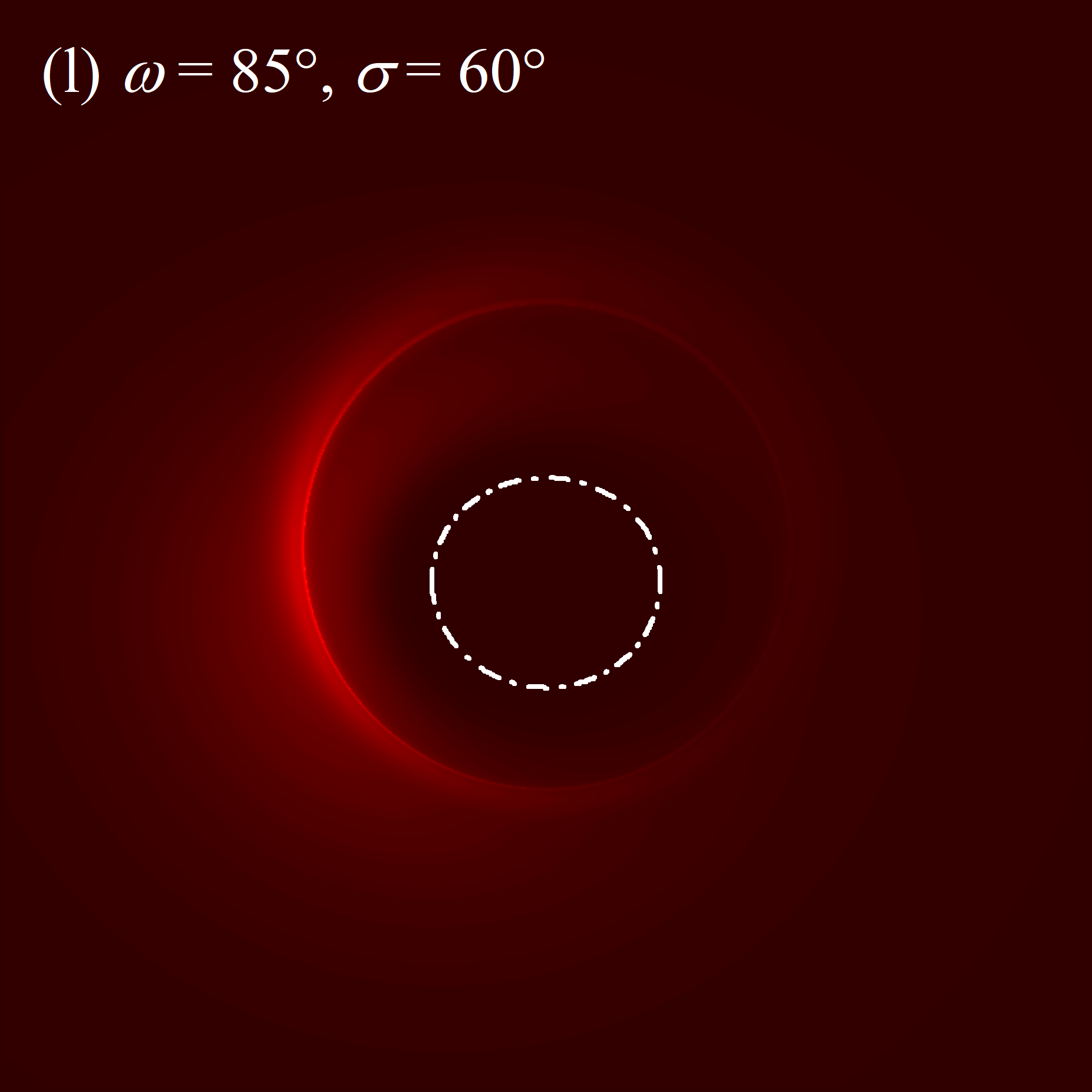}
\caption{Images of the hairy BH illuminated by a tilted thin accretion disk with various inclinations. Each row has the same observation angle, and the disk inclinations in each column are fixed. Two features are observed in each panel: one is the inner shadow surrounded by a white dashed line, and the other is a bright critical curve with a light spot appearing on its left side. There is also a degeneracy found between the observation angle $\omega$ and the inclination of the accretion disk $\sigma$. More precisely, the images corresponding to different observation angles can be mimicked by carefully adjusting the inclination of the accretion disk at a fixed $\omega$. Here, we have the scalar hair parameter of $h = -1$ and the observation azimuth of $\varphi_{\textrm{obs}}=0^{\circ}$. The physical size of imaging screen is $15 \times 15 M$, which is applicable to all of the subsequent images.}}\label{fig8}
\end{figure*}
\subsubsection{Distributions of redshift factors}
We fix the observation distance to $1000M$ and consider an imaging screen of a $15 \times 15 M$ view field. The image resolution is $700 \times 700$ pixels. Figure 5 illustrates the distributions of the redshift factor $g$ in the direct images cast by the hairy BH surrounded by an equatorial accretion disk ($\sigma=0$) under different values of the scalar hair parameter $h$, as viewed from an inclination of $80^{\circ}$. Each redshift factor is visualized using a continuous color map in which red is associated with the redshift ($g < 1$) and blue corresponds to the blueshift ($g > 1$). The dark region at the center in each panel represents the inner shadow of the BH \cite{Chael et al. (2021)}, which is closely associated with the event horizon since the light rays corresponding to this area do not intersect with the accretion disk. Instead, they penetrate the BH's event horizon directly. Although the event horizon of the hairy BH is independent of the scalar hair parameter, $h$ can affect the gravitational field of the BH. Hence, we observe a gradual expansion of the inner shadow as $h$ decreases from $0$. Moreover, we find that a red ring always encircles the inner shadow. This ring is contributed by the light rays with significant redshifts radiated by the emitters as they accelerate towards the BH along the plunging orbits. It is noteworthy that as the parameter $h$ decreases, the characteristic red ring becomes thicker, and the blueshift on the left side of the image is suppressed.

In figure 6, we systematically explore the influences of the observation inclination $\omega$ as well as the disk inclination $\sigma$ on the distributions of redshift factors in direct images of a hairy BH characterized by a scalar hair parameter of $h = -1$. The observation inclinations of panels in the same row are consistent, and all plots in the same column correspond to the same tilt of the accretion disk. It is found that in the equatorial accretion scenario (first column), an increase in the observation angle induces a blueshift region in the image and changes the silhouette of the inner shadow such that the inner shadow extends horizontally and becomes flatter in the vertical direction. In some specific cases (e.g., the first row), similar variations can also be observed with an increase in the inclination of the accretion disk. This arises from the fact that, for a static observer in a spherically symmetric spacetime, altering the inclination of the accretion disk is equivalent to modifying the observer's observation angle. Therefore, we point out that there is a degeneracy between the observation angle and the disk inclination, which makes discerning the observation angle from a BH image a Herculean task (Panels (d) and (f) provide an illustrative example). In addition, we identify novel inverted images in specific parameter spaces, including panels (g), (h), and (j)-(l). This phenomenon occurs when the observer is positioned below the tilted accretion disk. We also examine the distributions of redshift factors in the lensed images of the hairy BH, as arranged in figure 7. It is shown that the redshift factors are dramatically affected by the observation angle and the disk inclination. Specifically, the blueshift is enhanced as the observer approaches the accretion disk.

In fact, we can introduce a new parameter to discuss the synergistic effect of the observation angle and the disk inclination on the redshift factors. That is the ``effective observation angle'', $\Theta$, an angle between the observer's line of sight and the accretion disk, which is mathematically defined as $\Theta=\pi/2-(\omega+\sigma)$. The blueshift is strengthened as the absolute value of $\Theta$ decreases. A negative $\Theta$ indicates that the observer is below the accretion disk, resulting in a inverted image.
\subsection{Observational appearance of hairy black holes}
To obtain the observational appearance of hairy BHs, it is essential to calculate the brightness of the light ray corresponding to each pixel in the imaging screen. Typically, however, such calculation requires careful handling of GRMHD simulations and GRRT equations, making the process complex and time-consuming. Fortunately, based on the results of \cite{Gralla et al. (2020)}, the authors in \cite{Chael et al. (2021)} introduced a concise, analytic formula for computing observed intensities as
\begin{eqnarray}\label{23}
I_{\textrm{obs}} = \sum^{N_{\textrm{MAX}}}_{n = 1} \kappa_{n}\Upsilon(r_{\textrm{e}})g^{3},
\end{eqnarray}
where $N_{\textrm{MAX}}$ denotes the maximum number of times the light ray crosses the accretion disk, $\kappa_{n}$ is the fudge factor, which takes the value of $1$ for $n = 1$ and $2/3$ for $n > 1$. $g$ is the redshift factor, which has been properly addressed in the preceding section. $\Upsilon(r_{\textrm{e}})$ is the emissivity of the accretion disk, which can be expressed by a second-order polynomial in log-space as
\begin{eqnarray}\label{24}
\log\left[\Upsilon(r_{\textrm{e}})\right] = p_{1}\log\left(\frac{r_{\textrm{e}}}{r_{+}}\right)+p_{2}\left[\log\left(\frac{r_{\textrm{e}}}{r_{+}}\right)\right]^{2},
\end{eqnarray}
where $r_{\textrm{e}}$ is the radial coordinate of the light ray as it intersects the accretion disk, which can be obtained by ray tracing method. The parameters $p_{1}$ and $p_{2}$ are matching parameters with respect to the observation frequency. Here, similar to \cite{Chael et al. (2021),Hou et al. (2022)}, we set $p_{1} = -2$ and $p_{2} = -1/2$ to match the $230$ GHz time-averaged images from the radiative GRMHD simulation.
\begin{figure*}
\center{
\includegraphics[width=3.7cm]{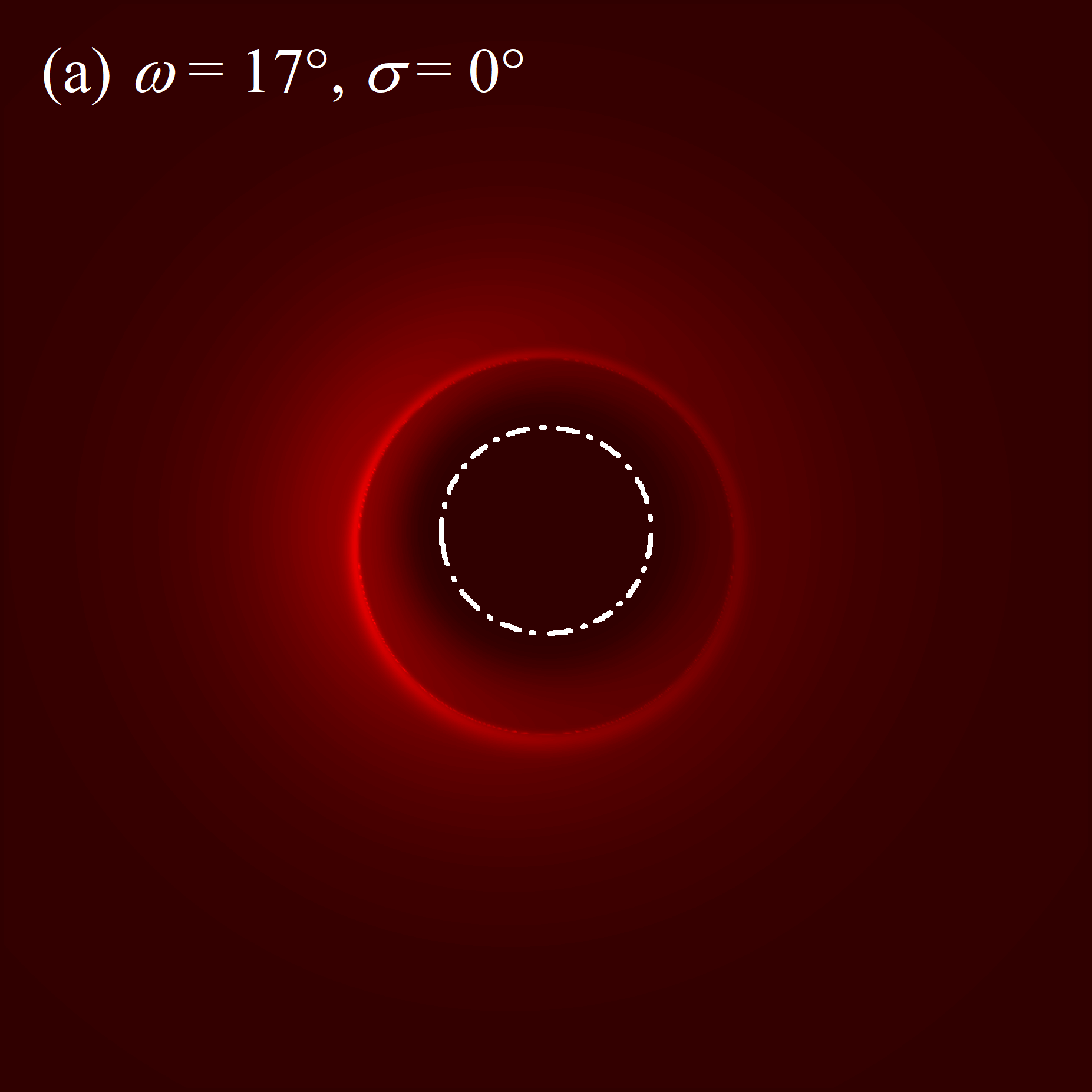}
\includegraphics[width=3.7cm]{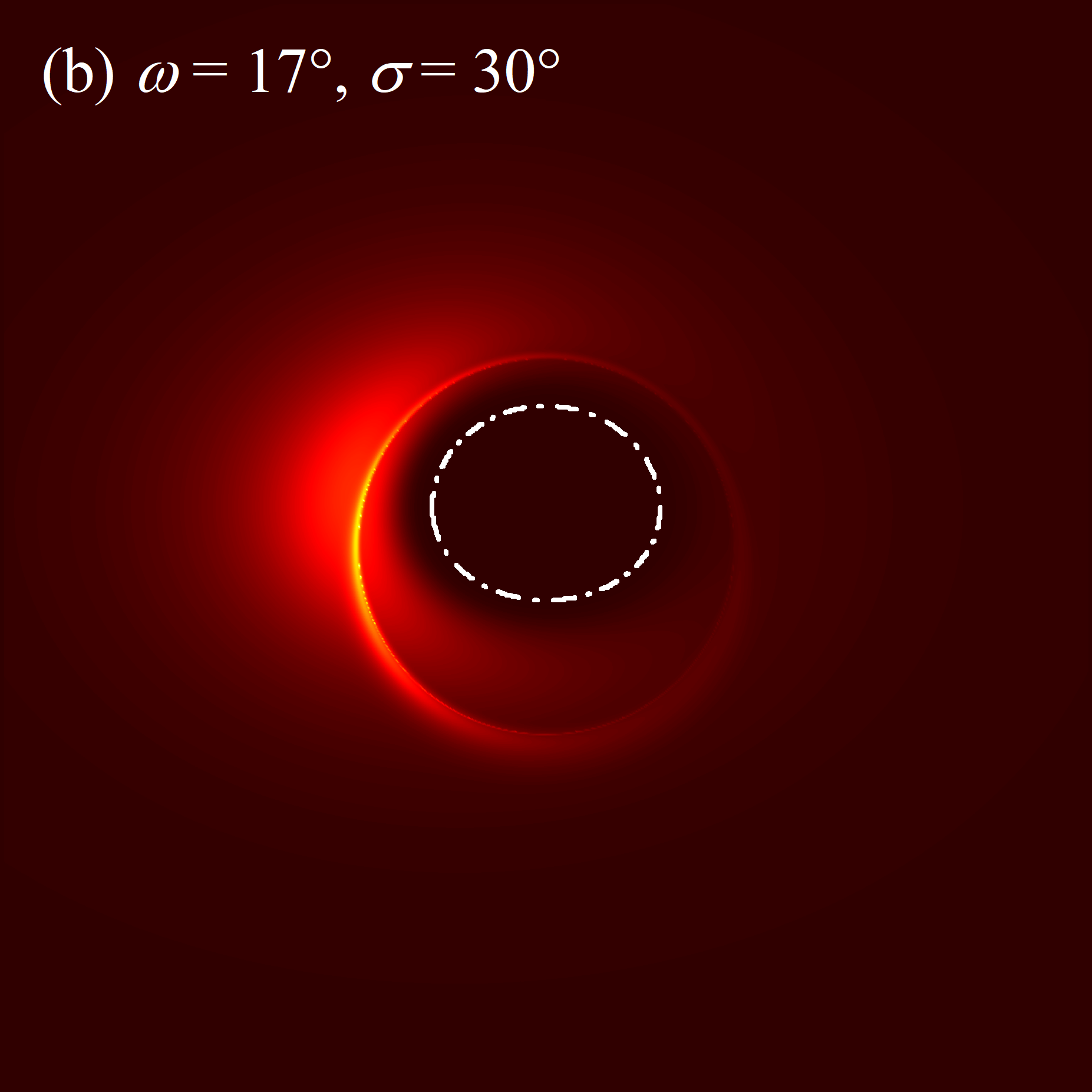}
\includegraphics[width=3.7cm]{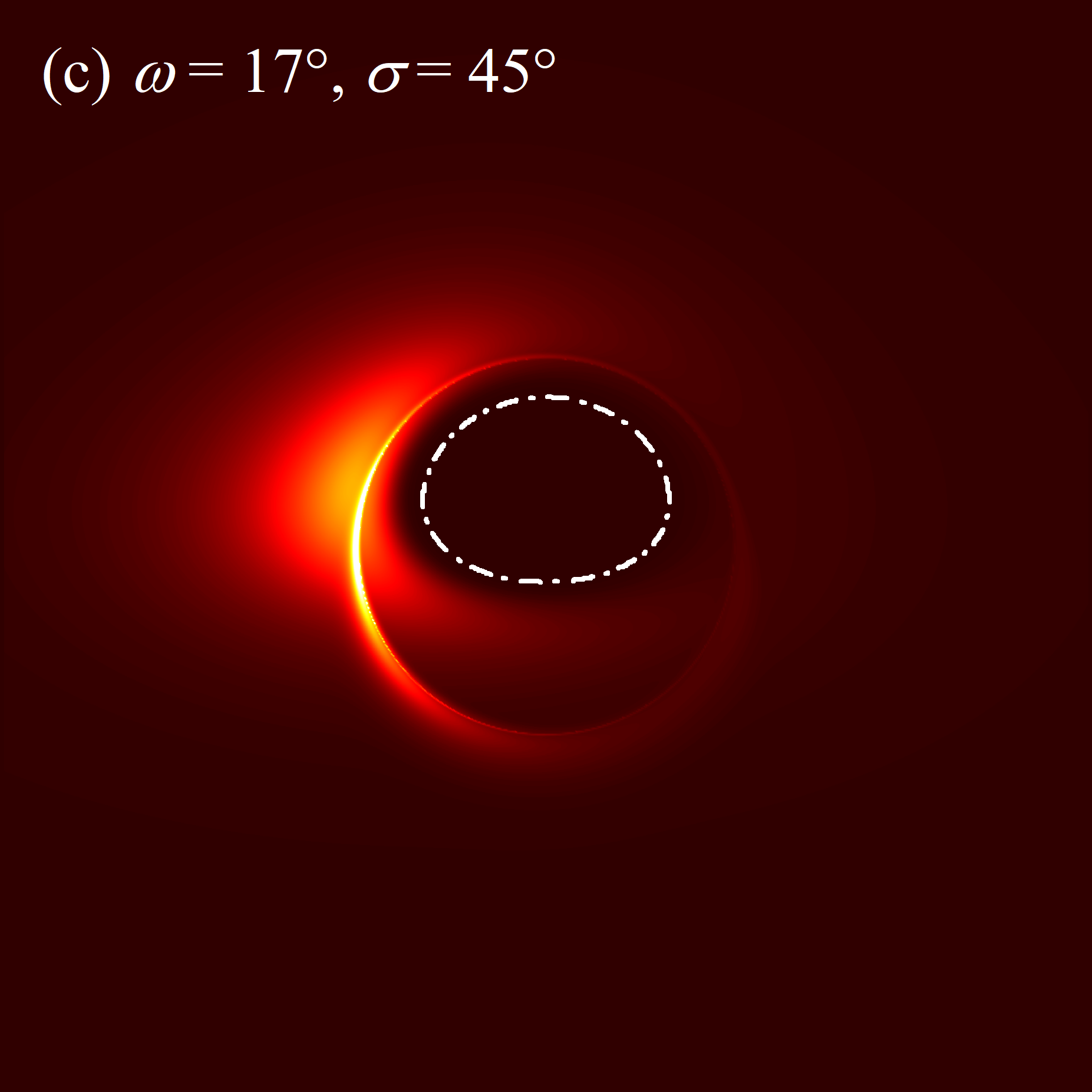}
\includegraphics[width=3.7cm]{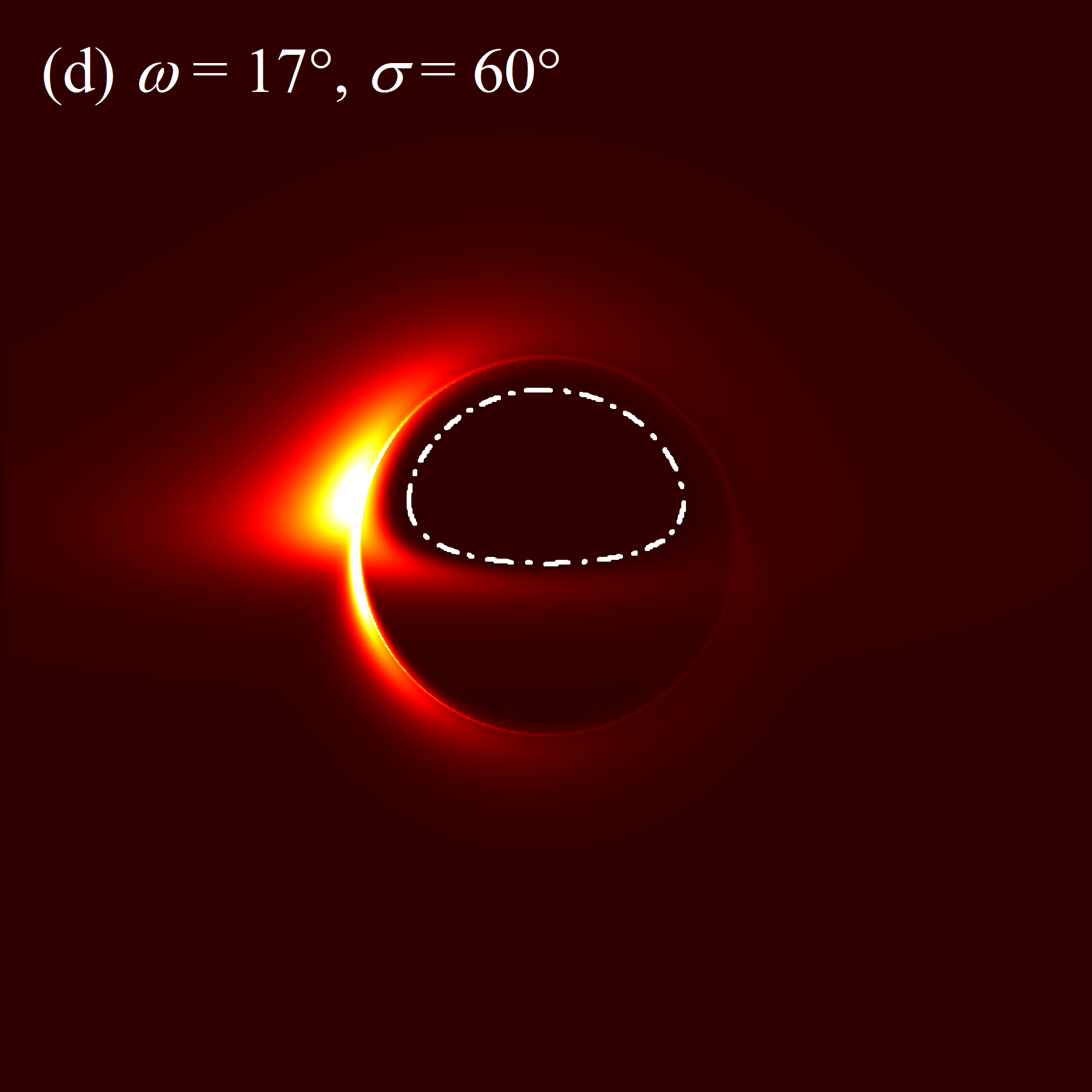}
\includegraphics[width=3.7cm]{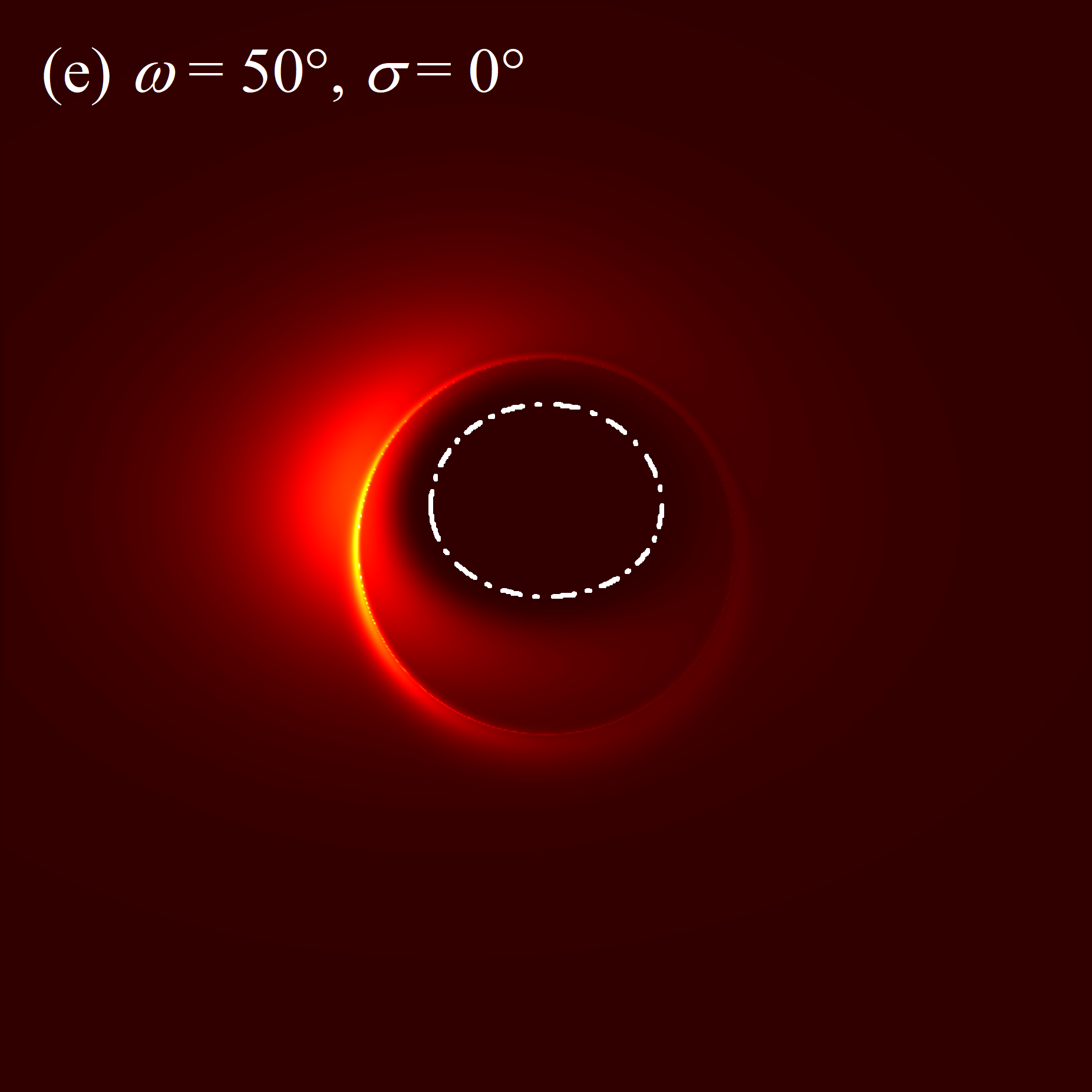}
\includegraphics[width=3.7cm]{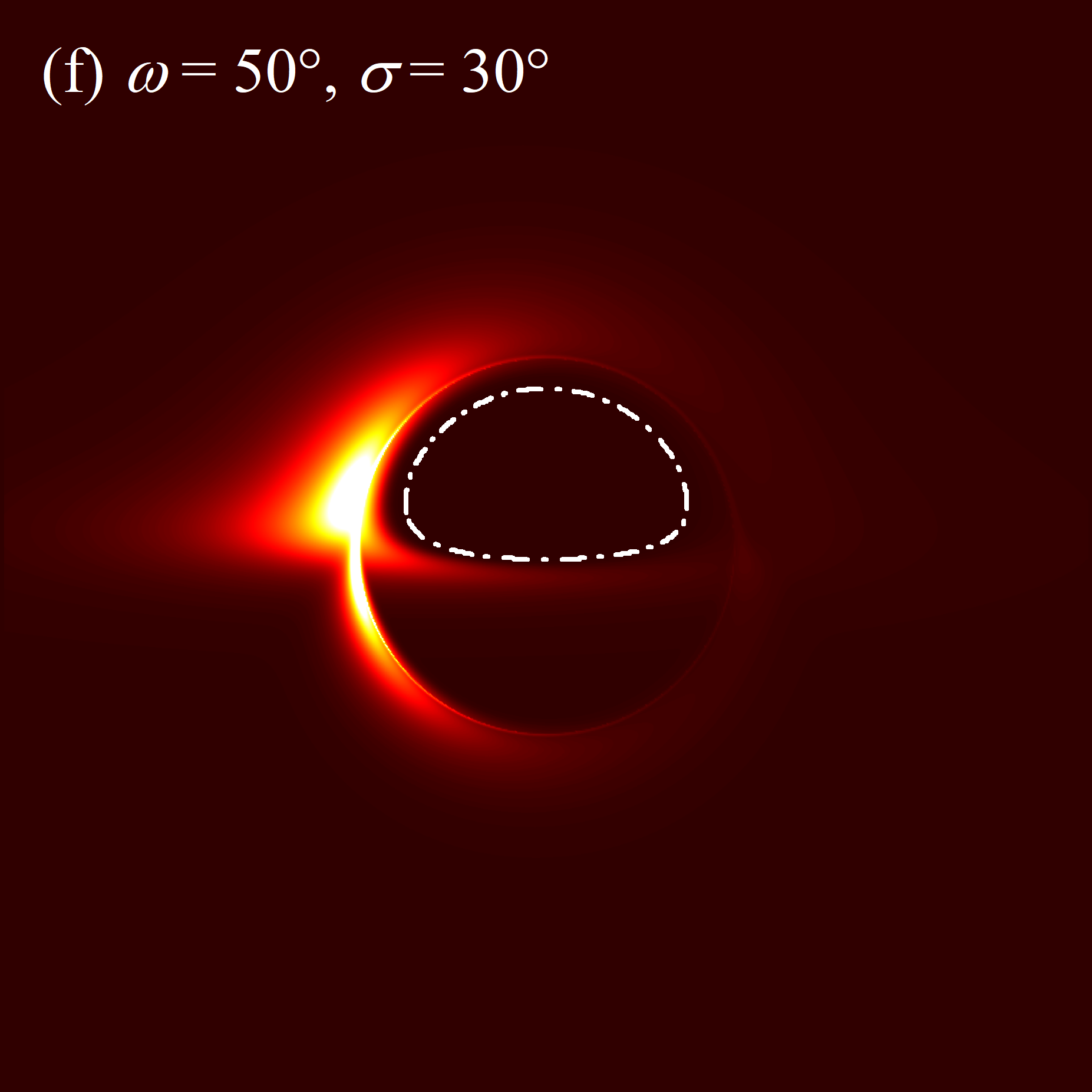}
\includegraphics[width=3.7cm]{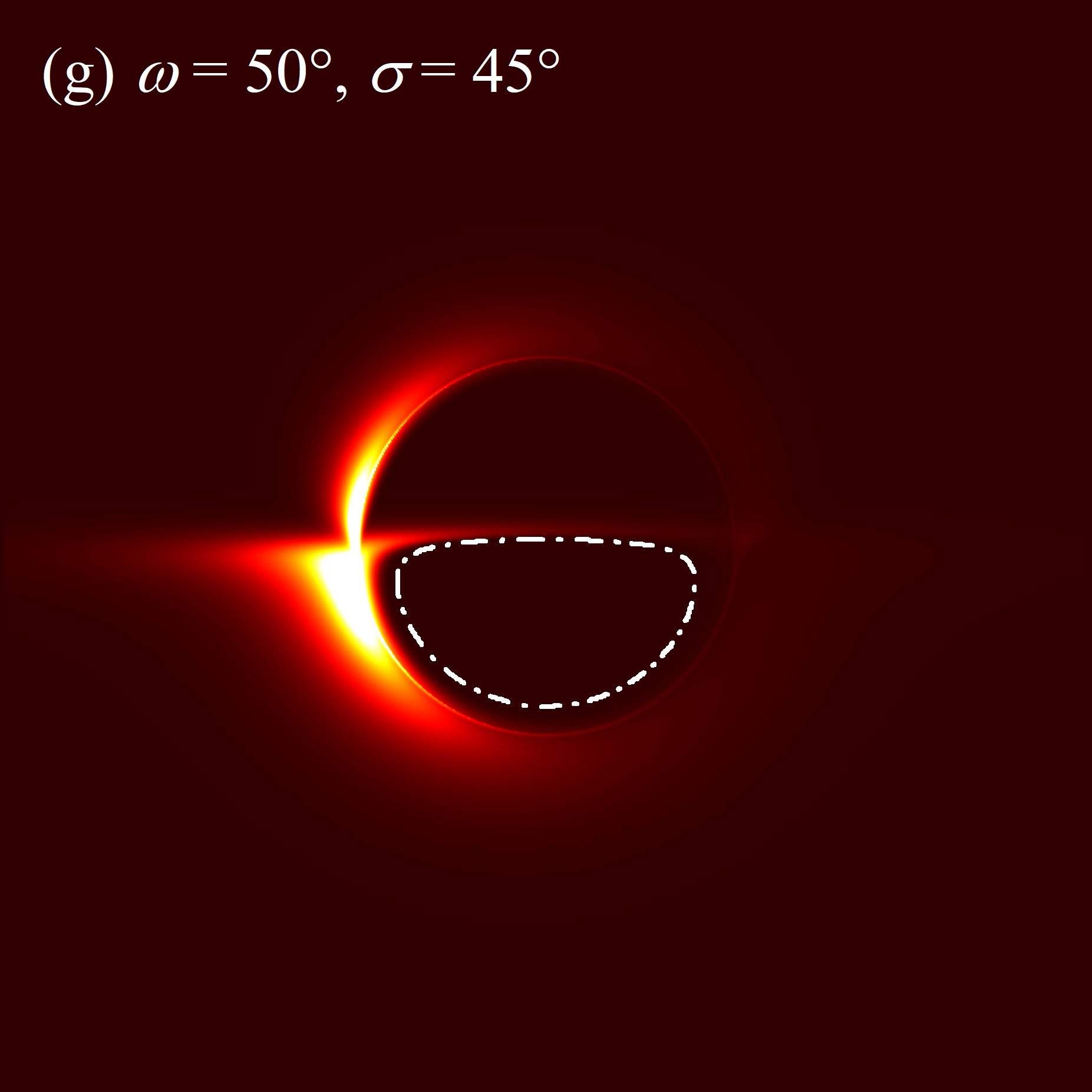}
\includegraphics[width=3.7cm]{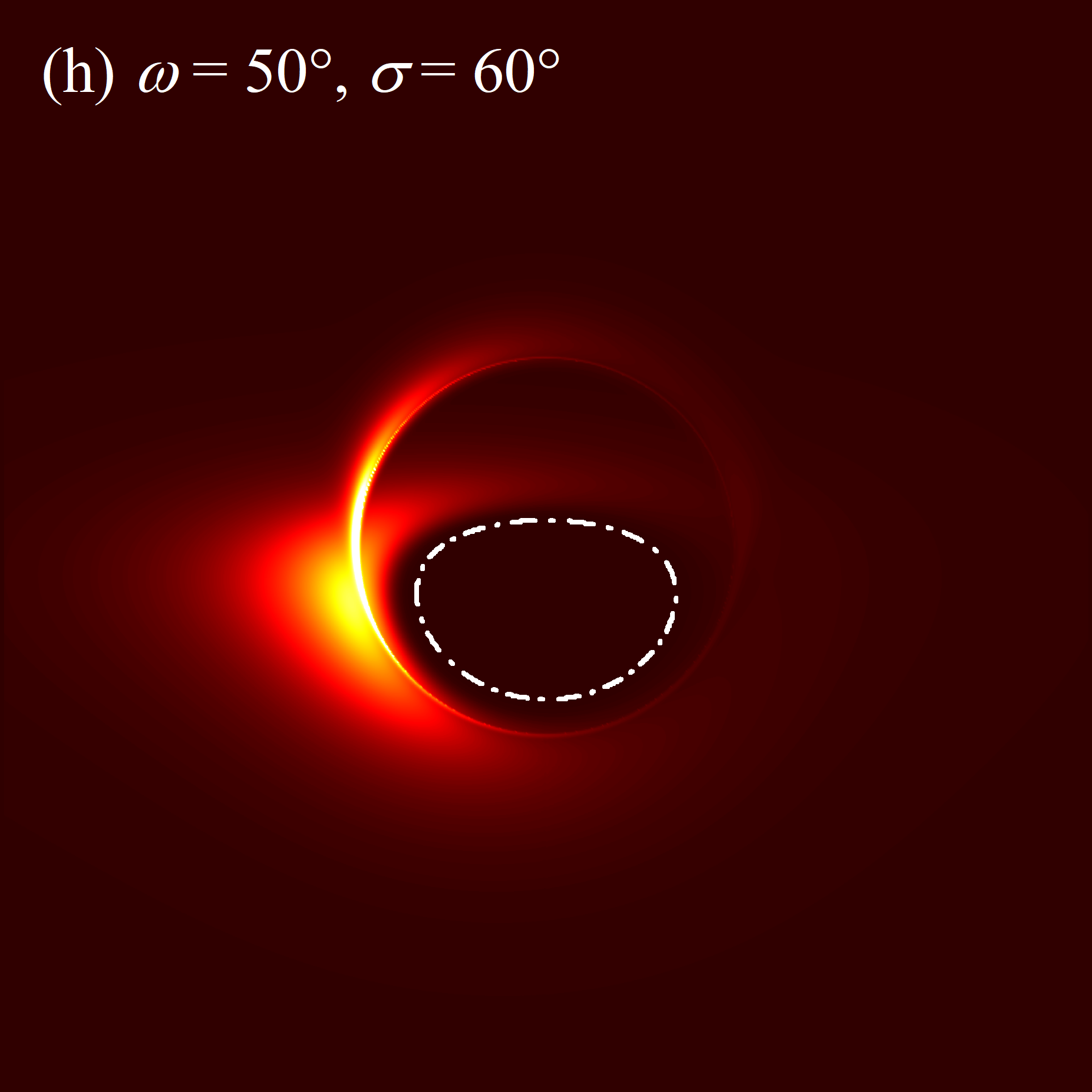}
\includegraphics[width=3.7cm]{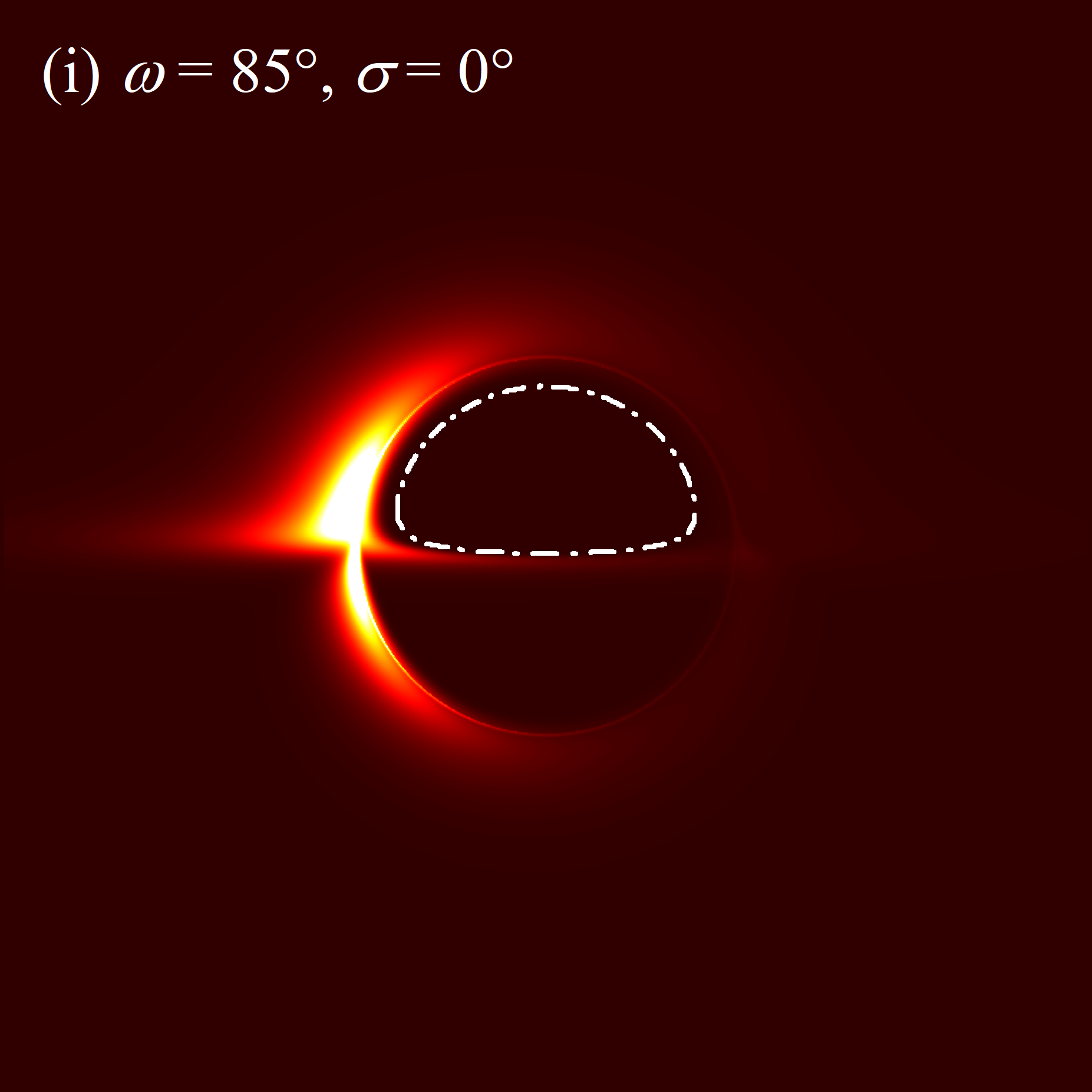}
\includegraphics[width=3.7cm]{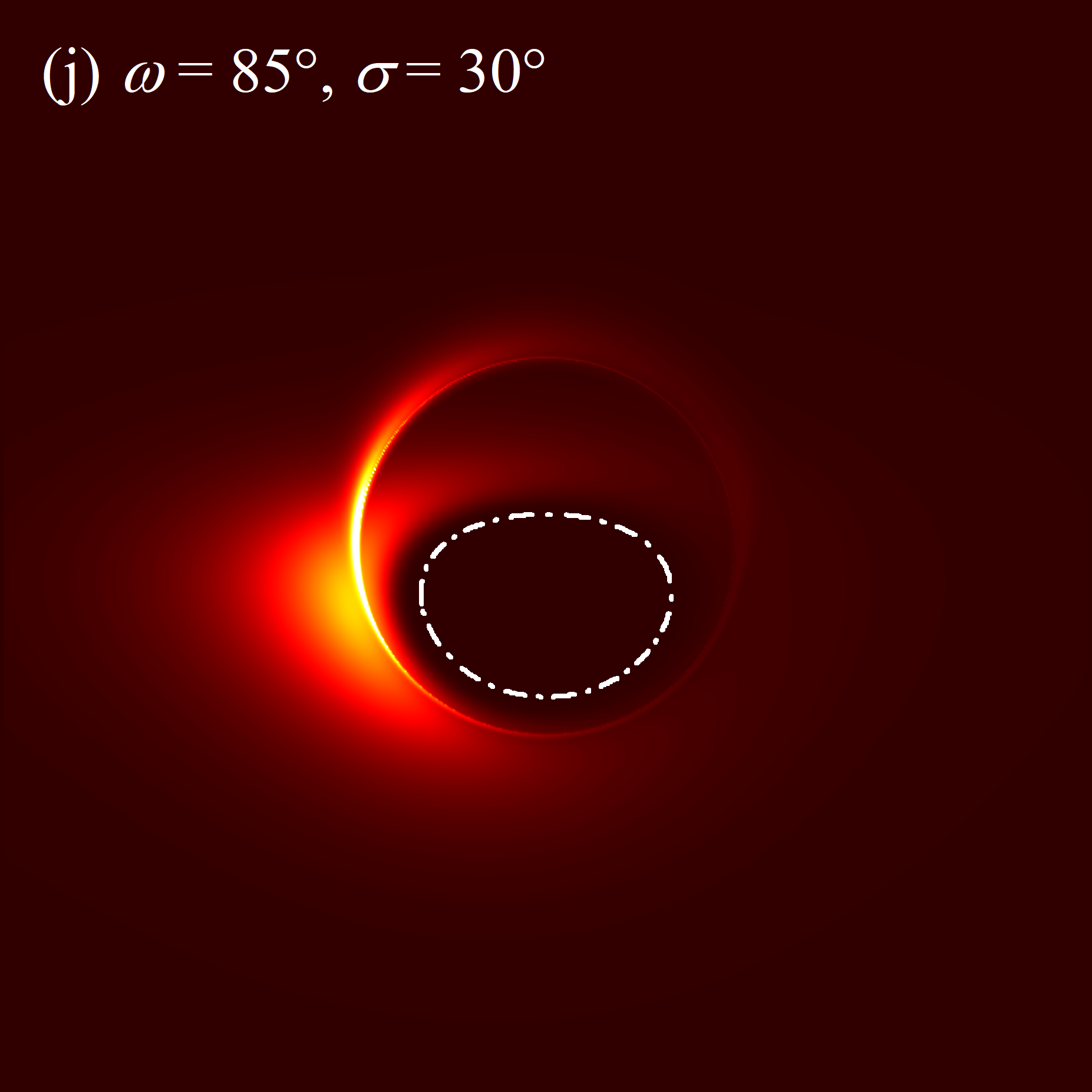}
\includegraphics[width=3.7cm]{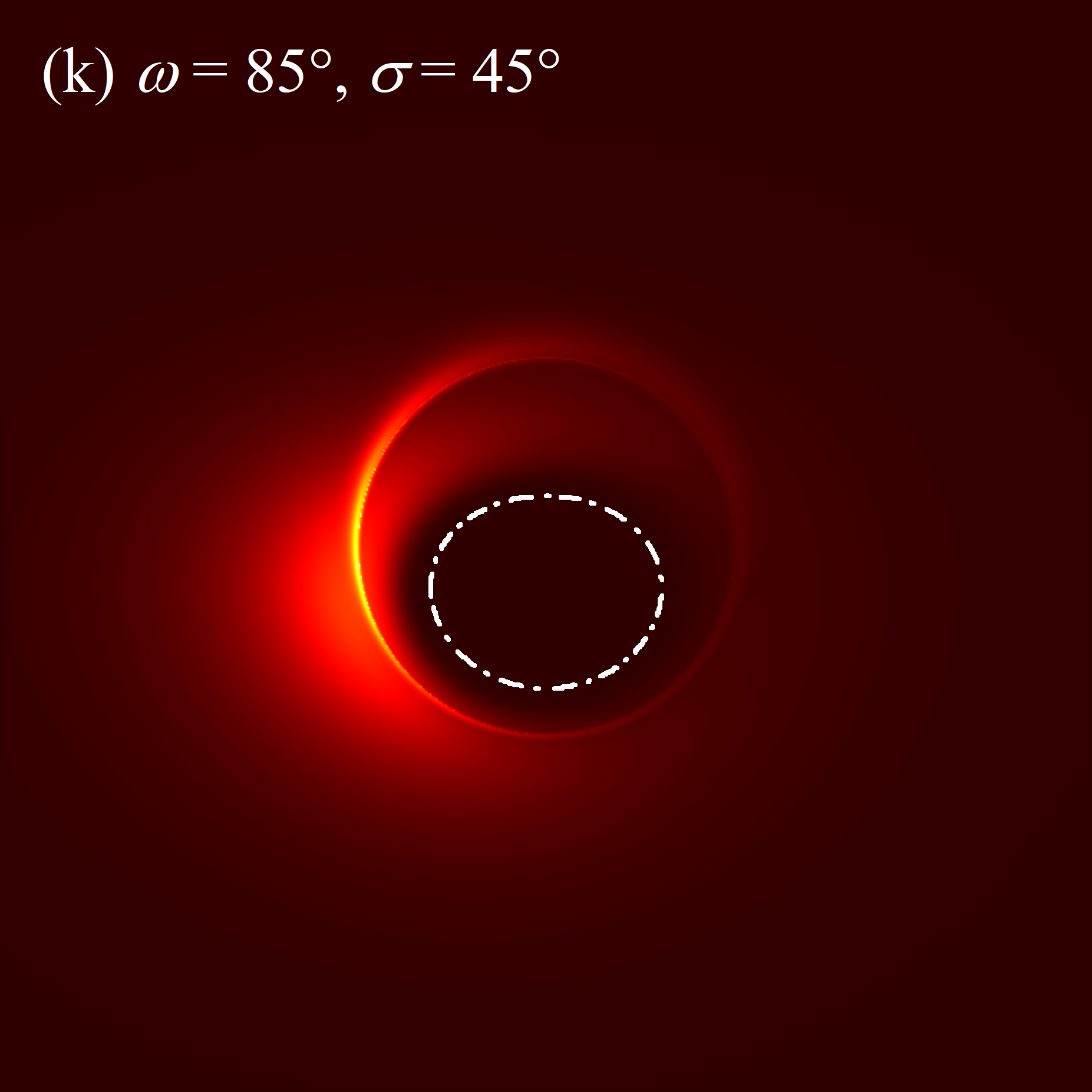}
\includegraphics[width=3.7cm]{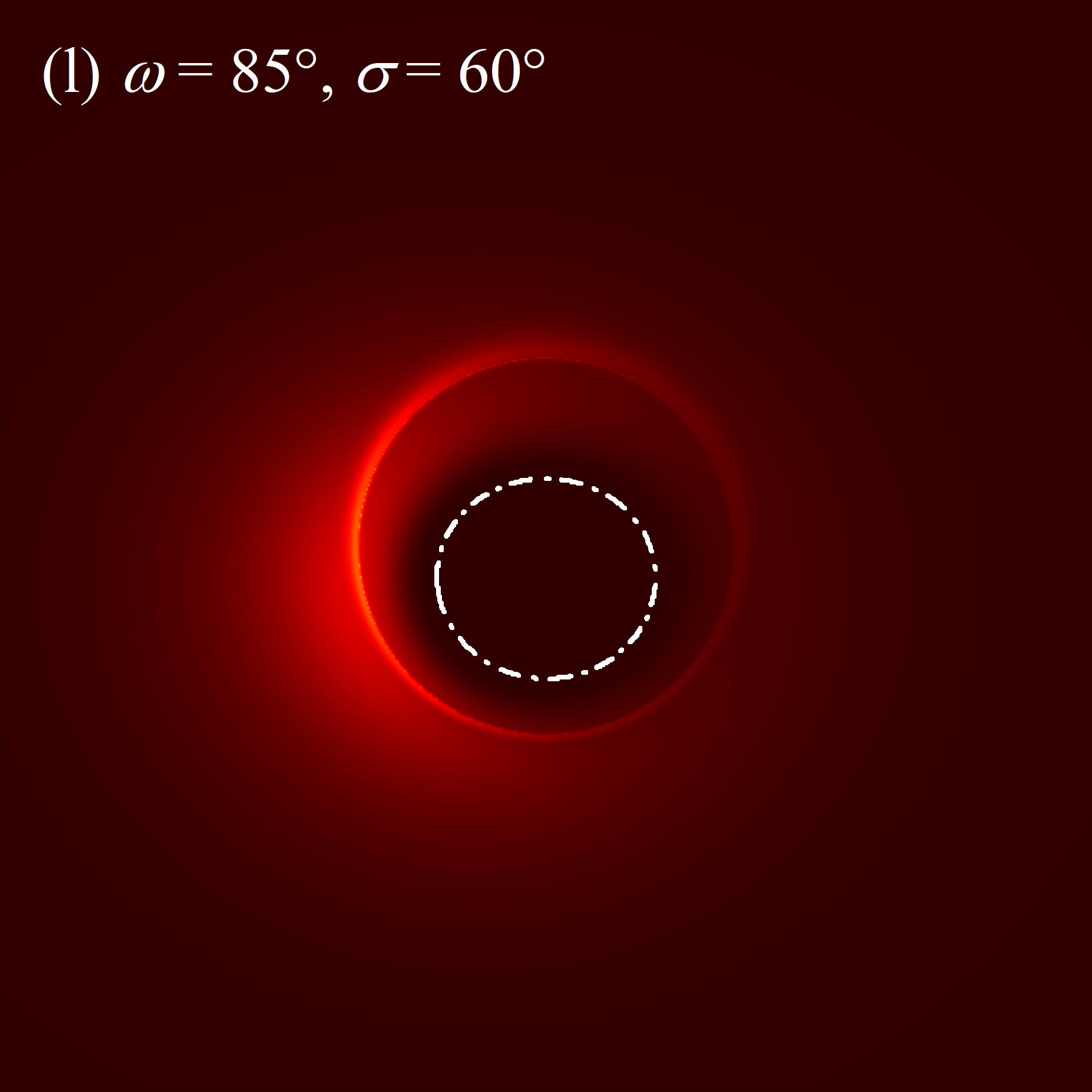}
\caption{Images of the Schwarzschild BH ($h = 0$) surrounded by a tilted thin accretion disk with various inclinations. It is found that the image of the Schwarzschild BH is brighter than that of the hairy BH, but the critical curve of the former is smaller than that of the latter.}}\label{fig9}
\end{figure*}

In figure 8, we present $230$ GHz images of the hairy BH surrounded by a tilted thin accretion disk with various inclinations $\sigma$, as viewed from different observation angles $\omega$. The intensity of each pixel in images is visualized by means of a continuous, linear color spectrum, as dull red corresponds to the minimum value, while white is associated with the maximal luminosity. In each plot, we observe an impressive bright thin ring with a right-left brightness asymmetry appears near the critical curve \cite{Gralla et al. (2019)}. This luminous ring arises from the multiple crossings of light rays through the accretion disk, elucidating its heightened brightness relative to the rest of the image. We find that the radius of the critical curve is approximately $6.7M$ in each panel, which suggests that the size of the characteristic ring does not vary with the observation angle or the inclination of the accretion disk. The region of brightness depression inside the critical curve, depicted by the white dashed line, represents the inner shadow of the hairy BH, as well as the observed profile cast by the event horizon. Clearly, at an observation angle of $17^{\circ}$, the tilted accretion disk introduces a moderately bright streak to the foreground of the image, subsequently obscuring and compressing the inner shadows. Given the degeneracy between $\omega$ and $\sigma$, we use the effective observation angle, $\Theta=\pi/2-(\omega+\sigma)$, to delineate the variation of the inner shadow. That is, as the absolute value of $\Theta$ decreases (the observer moves closer to the accretion disk), the inner shadow extends horizontally and compresses vertically; when $\Theta$ is negative, there is an inverted inner shadow, meaning that the main body of the inner shadow is situated below the line of sight. This behavior of the inner shadow helps us to infer the position of the accretion disk from the hairy BH images when the observation angle is given. Around the silhouette of the inner shadow, a dark ring with extremely low luminosity is observed. This is caused by the significant redshifts of the light rays contributing to the formation of this dark ring. In addition, we note that there is an accumulation of brightness on the left side of the image, resulting in a light spot along the left segment of the critical curve. At the same time, the size and brightness of this flare increases as the absolute value of $\Theta$ decreases. This is because the light rays responsible for the flare originate from the approaching side of the accretion flow, and the blueshift is enhanced as $\Theta$ approaches zero.
\begin{figure*}
\center{
\includegraphics[width=5cm]{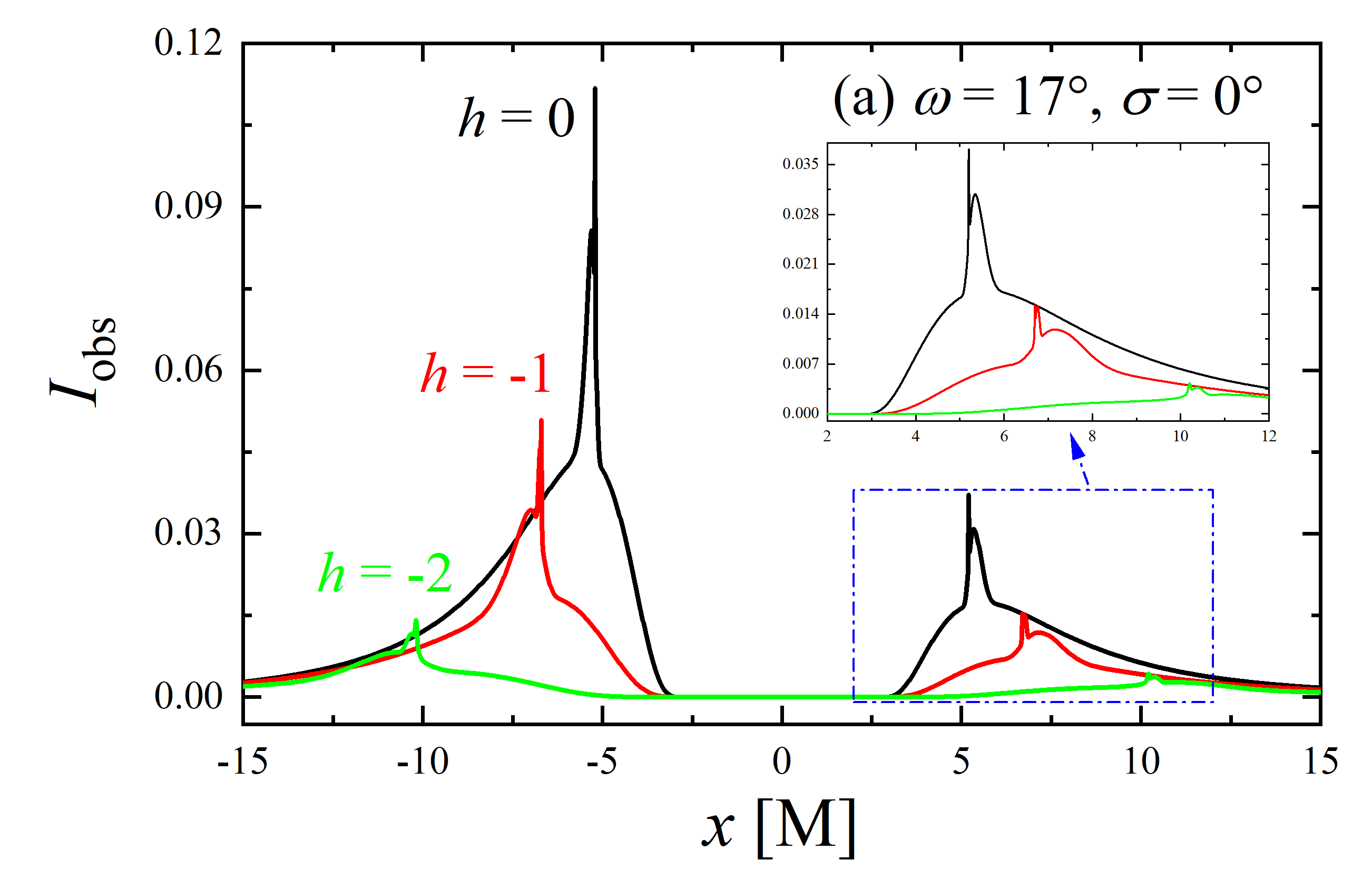}
\includegraphics[width=5cm]{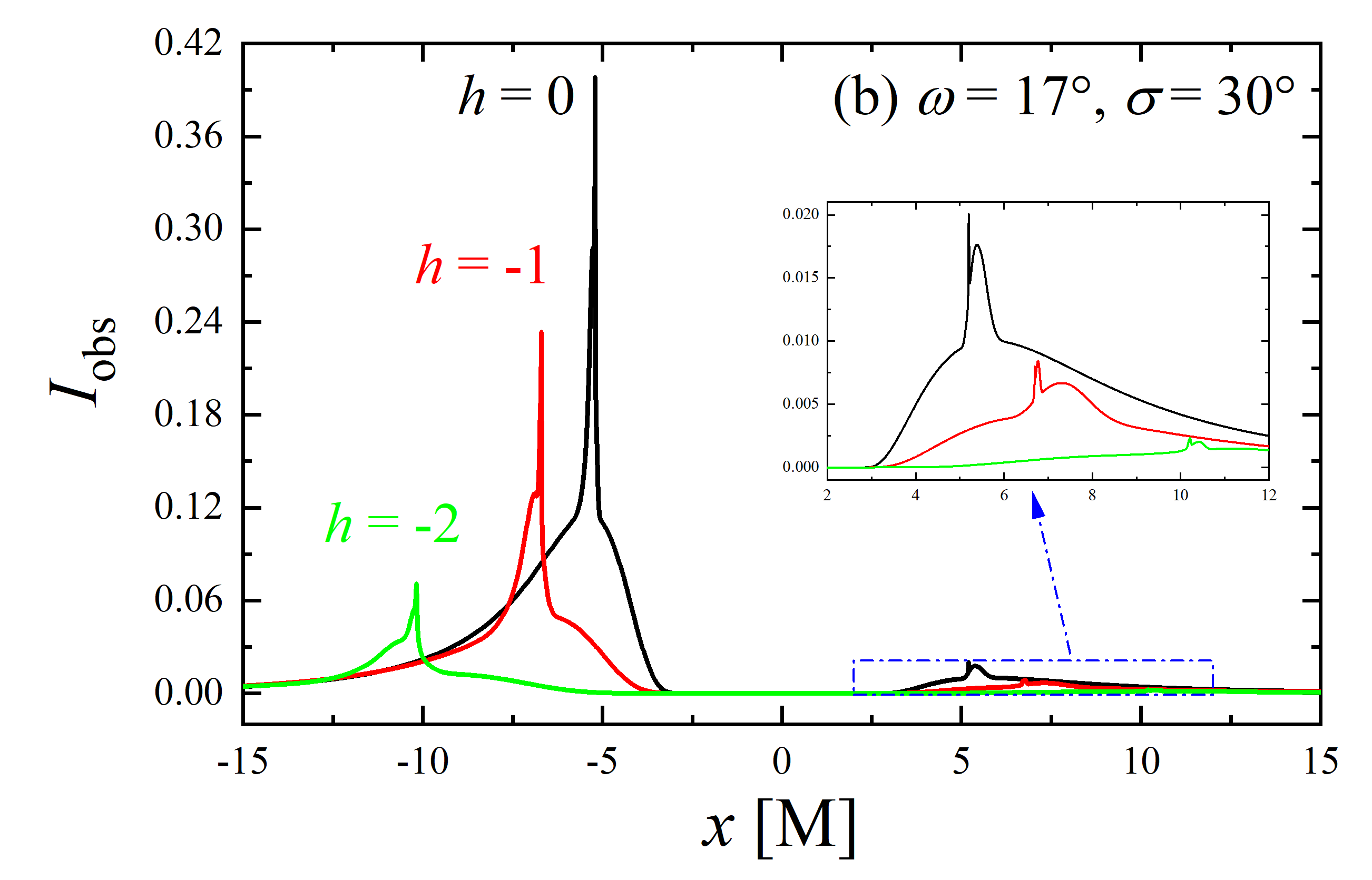}
\includegraphics[width=5cm]{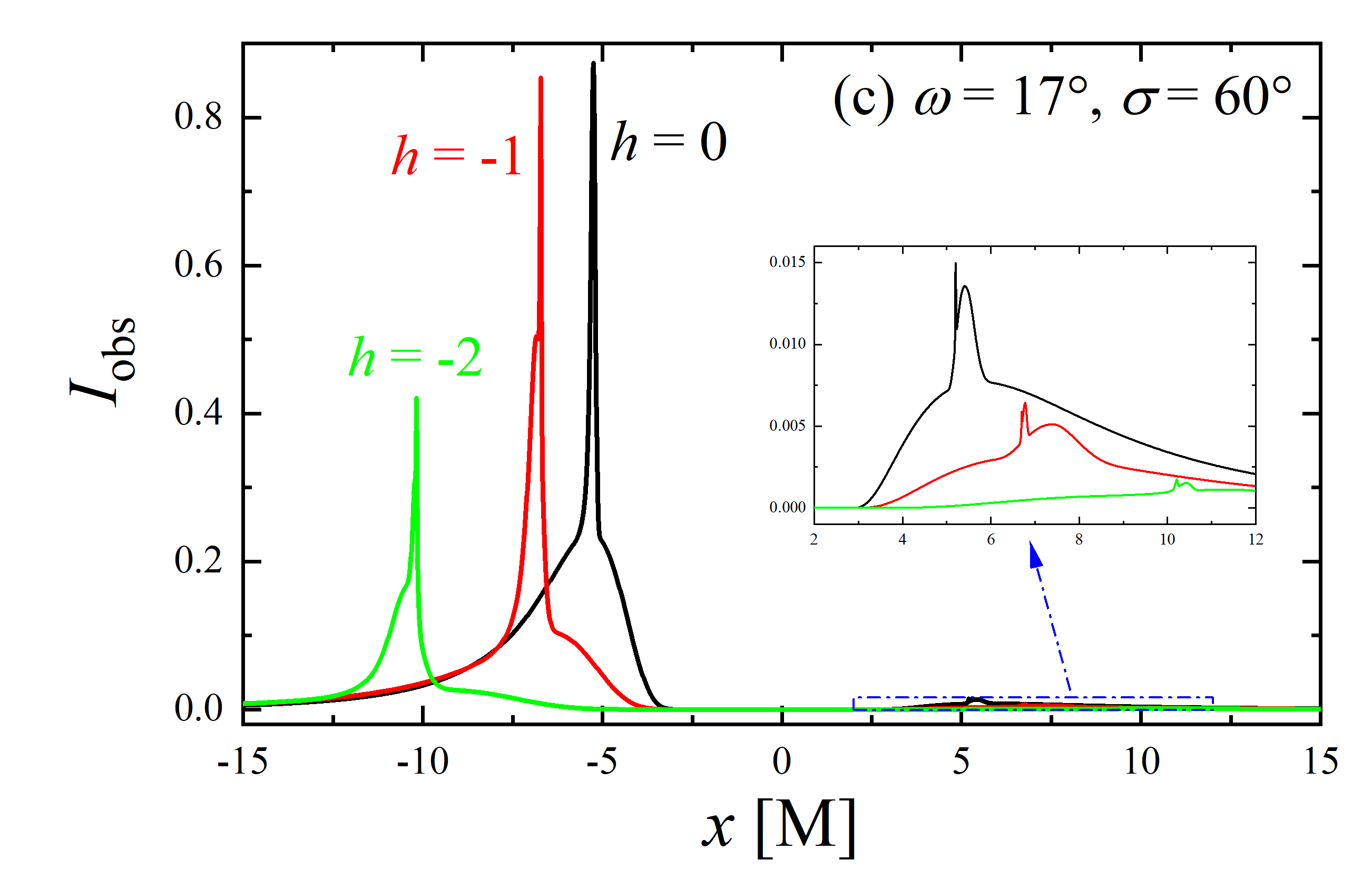}
\includegraphics[width=5cm]{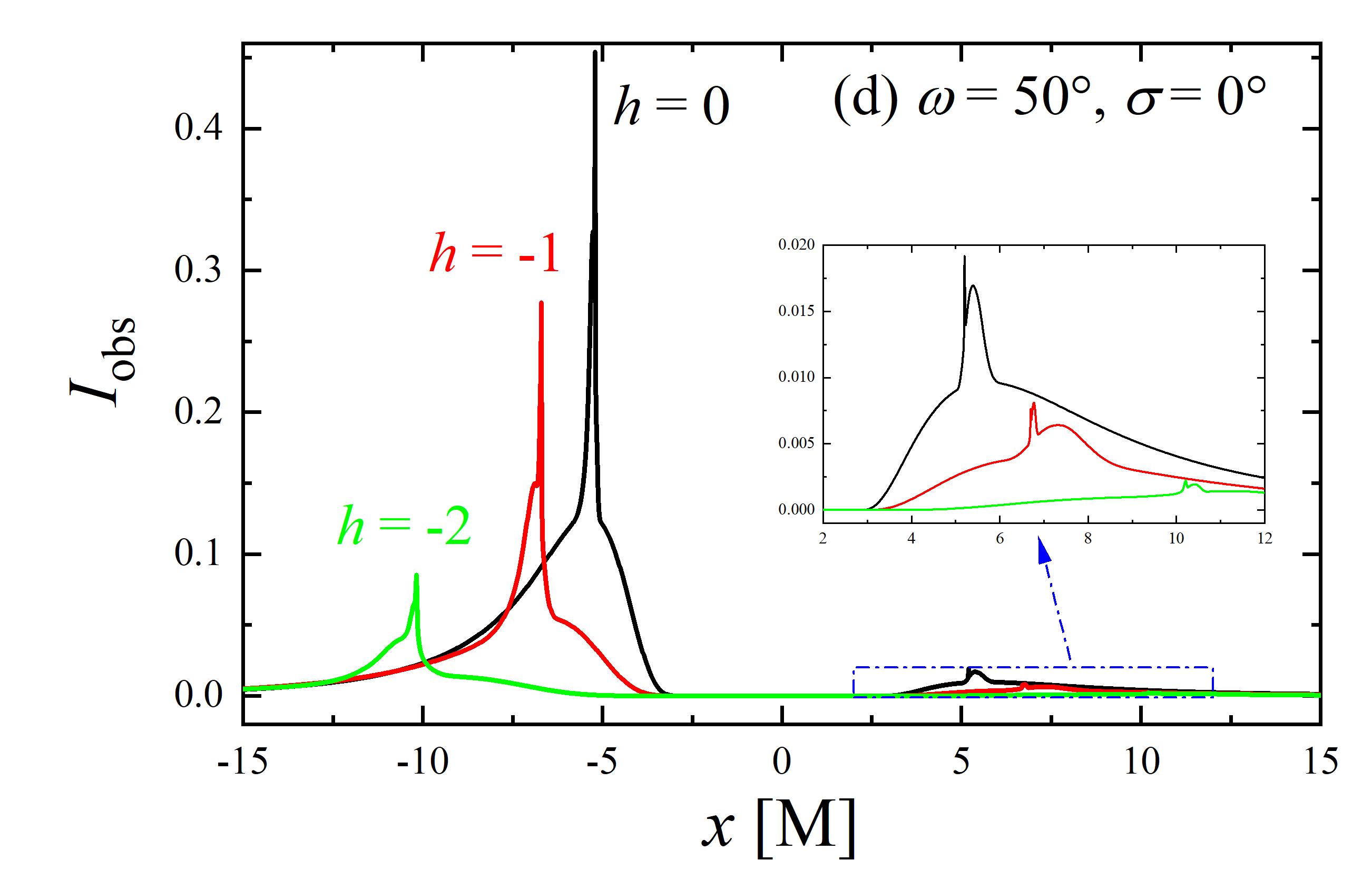}
\includegraphics[width=5cm]{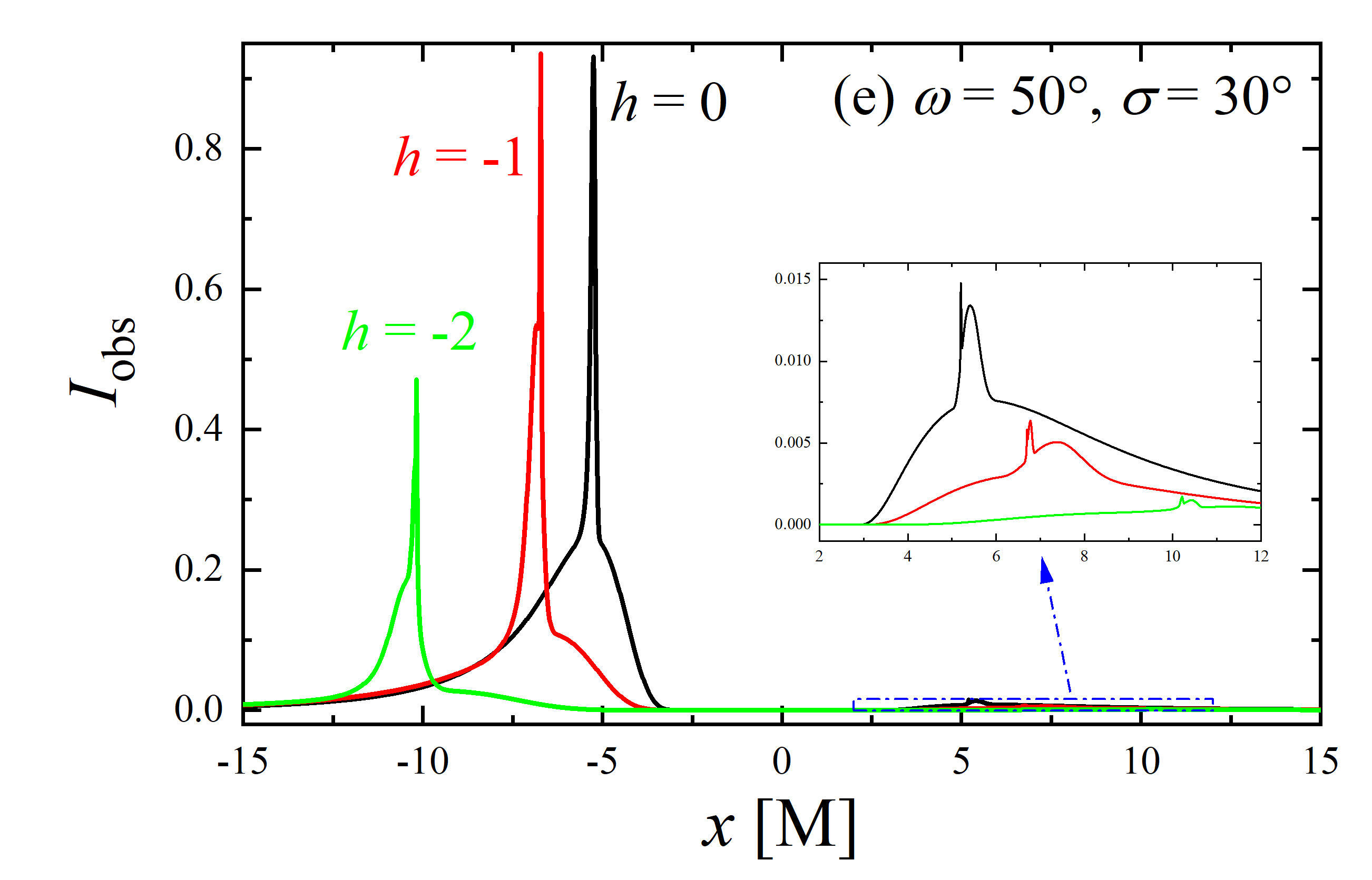}
\includegraphics[width=5cm]{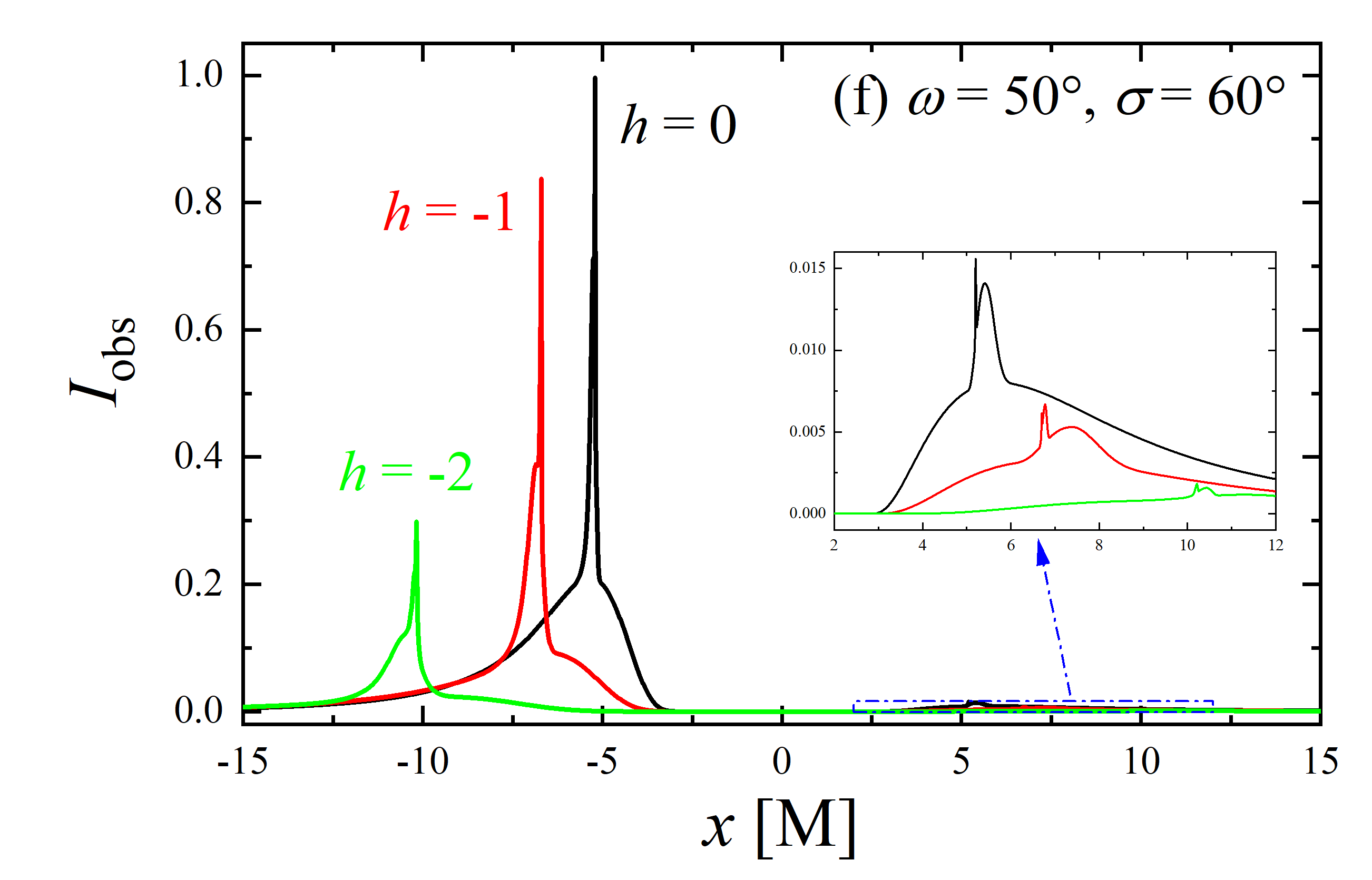}
\includegraphics[width=5cm]{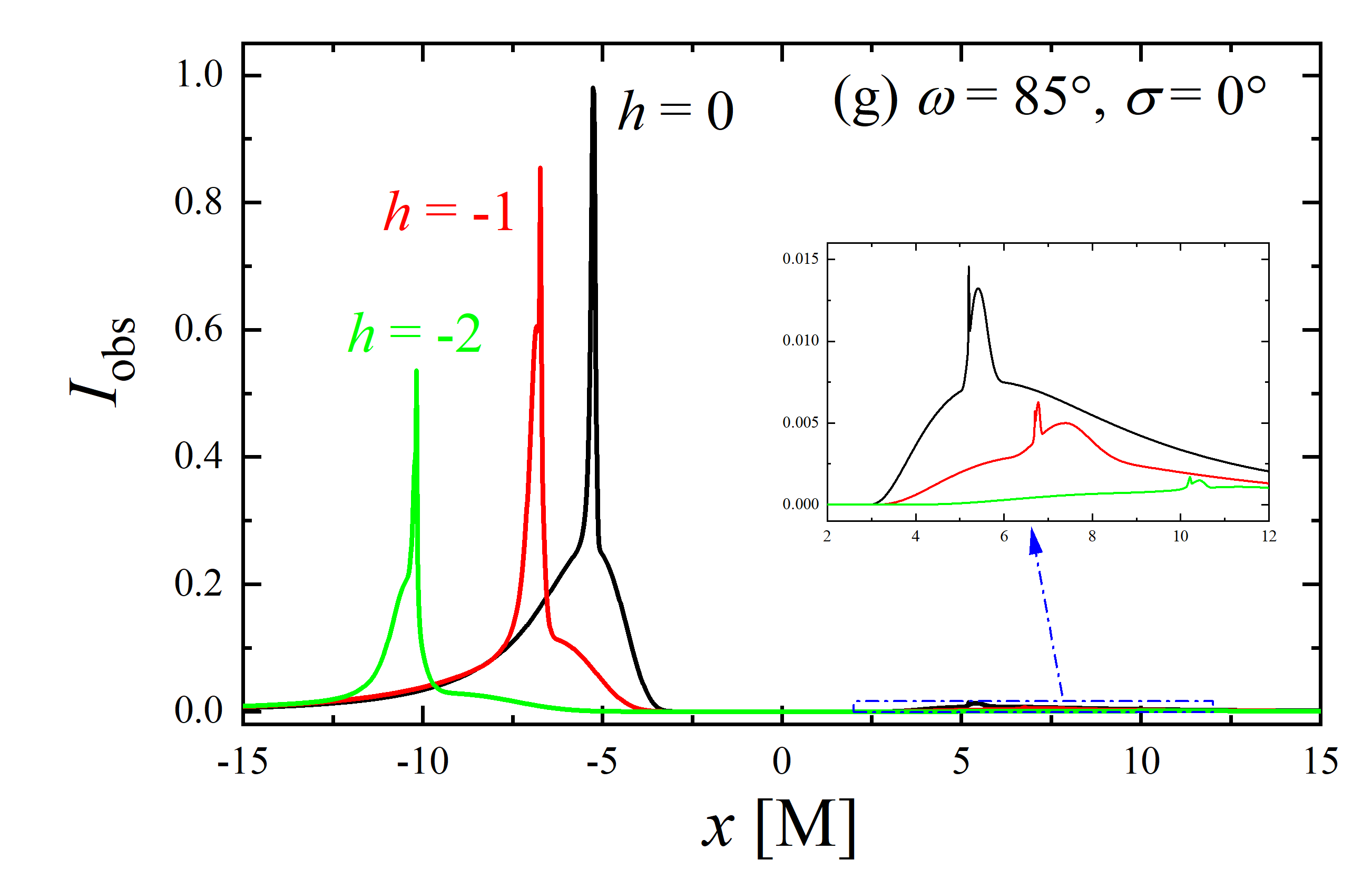}
\includegraphics[width=5cm]{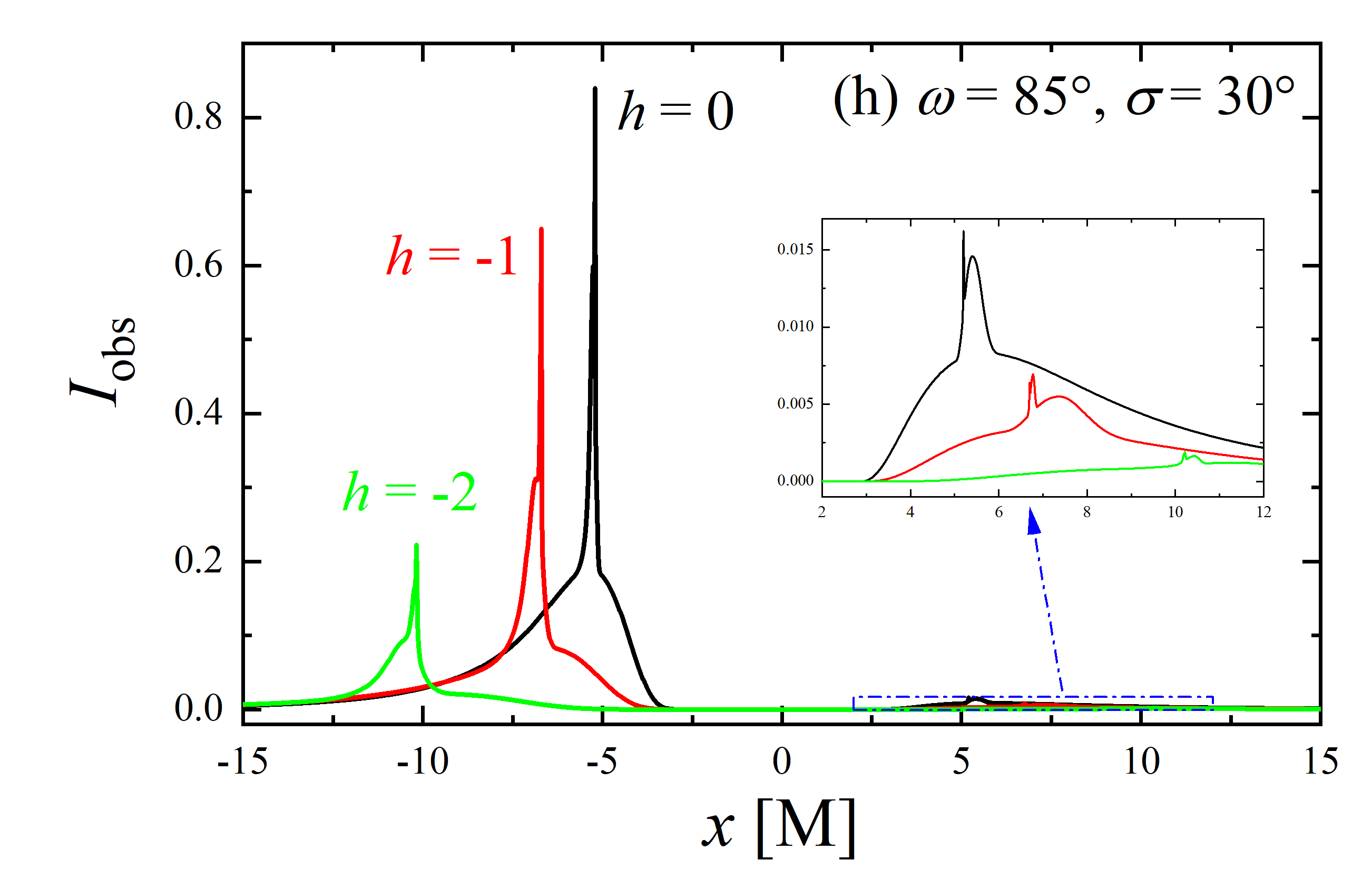}
\includegraphics[width=5cm]{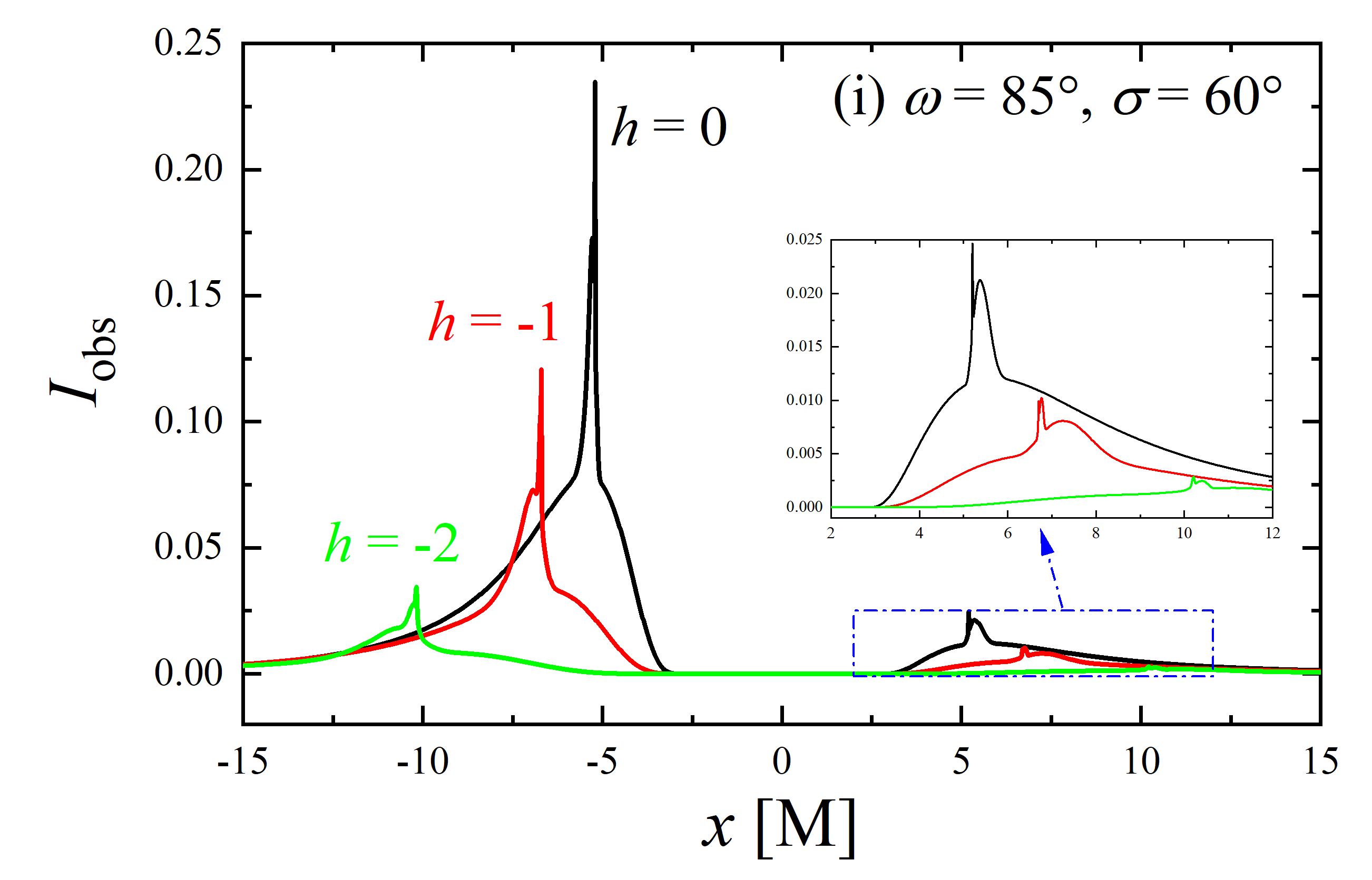}
\caption{Distributions of the observed flux along the $x$-axis of the observation screen for hairy BHs with different values of $h$. It is shown that the flux curve is sensitive to the observation angle, disk inclination, and the scalar hair parameter. An increase in $h$ can effectively enhance the intensity and shift the peak position towards the center.}}\label{fig10}
\end{figure*}

We now shift our focus to explore the influences of the scalar hair parameter $h$ on the images of the hairy BH and highlight the disparities in the observational signatures between hairy BHs and Schwarzschild BHs. To this end, we simulate the observational appearance of a Schwarzschild BH endowed with a tilted thin accretion disk of various inclinations under different observation angles, as demonstrated in figure 9. Our findings show that the image of the Schwarzschild BH at $230$ GHz is considerably brighter compared to that of the hairy BH, regardless of the observation angle and the inclination of the accretion disk. This indicates that increasing $h$ decreases the observed intensity of the hairy BH. Meanwhile, the bright ring of the Schwarzschild BH is smaller than that of the hairy BH because a shift of $h$ from $0$ to $-1$ expands the photon ring of the hairy BH. To gain further insights, we extracted the intensity distributions along the $x$-axis of the BH images across different parameter spaces, as illustrated in figure 10, where $h = 0$ corresponds to the Schwarzschild case. We find that each intensity curve manifest two peaks: the prominent left peak corresponds to the approaching side of the accretion flow, while the right peak, with lower intensity, corresponds to the receding side of the accretion disk. The positions of these peaks are solely determined by the scalar hair parameter $h$. In particular, as $h$ weakens, the positions of the zeniths shift to either side along the $x$-axis, which unambiguously suggests that increasing $h$ has a negative impact on the dimensions of the critical curve. On the other hand, the peak value is sensitive to the effective observation angle $\Theta$ and the parameter $h$. It is confirmed that a reduction in $h$ or an increase in the absolute value of $\Theta$ can effectively suppress the observed flux of the hairy BH. So, what effect does the parameter $h$ have on the inner shadow of hairy BHs observed at different inclinations? We address this inquiry by simulating the inner shadow of the hairy BH illuminated by an equatorial accretion disk under different values of $h$ and $\omega$. As displayed in figure 11, we find that reducing $h$ amplifies the inner shadow, and the degree of this magnification is determined by the viewing angle. Specifically, when the observer's line of sight is perpendicular to the accretion disk (first column), the impact of $h$ on the inner shadow is negligible. It is also found that the shape of the inner shadow remains independent of the scalar hair parameter, instead, being linked to the relative positions of the observer and the accretion disk. The findings from figures $8$-$11$ can serve as a potential, efficient tool for identifying hairy BHs from Schwarzschild BHs.

In the case of equatorial accretion, the observed images of hairy BHs remain invariant with respect to the observer's azimuth $\varphi_{\textrm{obs}}$. Conversely, in the tilted accretion disk scenario, it is reasonable to believe that the BH images are notably affected by the azimuth of the observer. Hence, we investigate the observational images of a hairy BH surrounded by a tilted accretion disk as viewed by observers with different values of $\varphi_{\textrm{obs}}$, as presented in figure 12 for $\omega = 17^{\circ}$ and in figure 13 for $\omega = 85^{\circ}$. Our findings reveal that, as $\varphi_{\textrm{obs}}$ varies, the symmetry of the inner shadow with respect to the $y$-axis is disrupted, and the position of the light spot shifts. More precisely, the inner shadow and light spot revolve around the center of the image with the variation of $\varphi_{\textrm{obs}}$, and the angle of rotation depends on the inclination of the accretion disk. Concurrently, visible changes occur in the size of the inner shadows and the light spots. It is noteworthy that under the condition of $\omega = 85^{\circ}$, the drift of the light spot is not significant with varying azimuth but slightly depends on the disk tilt, as clearly illustrated in the third row of figure 13. Additionally, it is found that the brightness of the image is correlated with observation azimuth, as $\varphi_{\textrm{obs}}$ determines the relative positions of the observer and the tilted accretion disk, subsequently affecting Doppler boosting. By comparing the images of $\varphi_{\textrm{obs}} = 180^{\circ}$ and $\varphi_{\textrm{obs}} = 0^{\circ}$, we observe that the two images exhibit morphological mirroring about the $x$-axis, albeit with slight differences in luminosity. This characteristic is evident when comparing figure 8 (j) and figure 13 (v). Furthermore, we confirm that the size of the critical curve remains independent of the observation azimuth. Considering that changing the azimuth of the observer is equivalent to altering the position of the accretion disk, this suggests that the movement of the light spot in the BH image may be attributed to the precession of the tilted accretion disk. This finding offers potential guidance for deducing the accretion disk's state through analyzing the shift of the light spot in time-averaged images obtained from sustained observations in the future.
\begin{figure*}
\center{
\includegraphics[width=3.7cm]{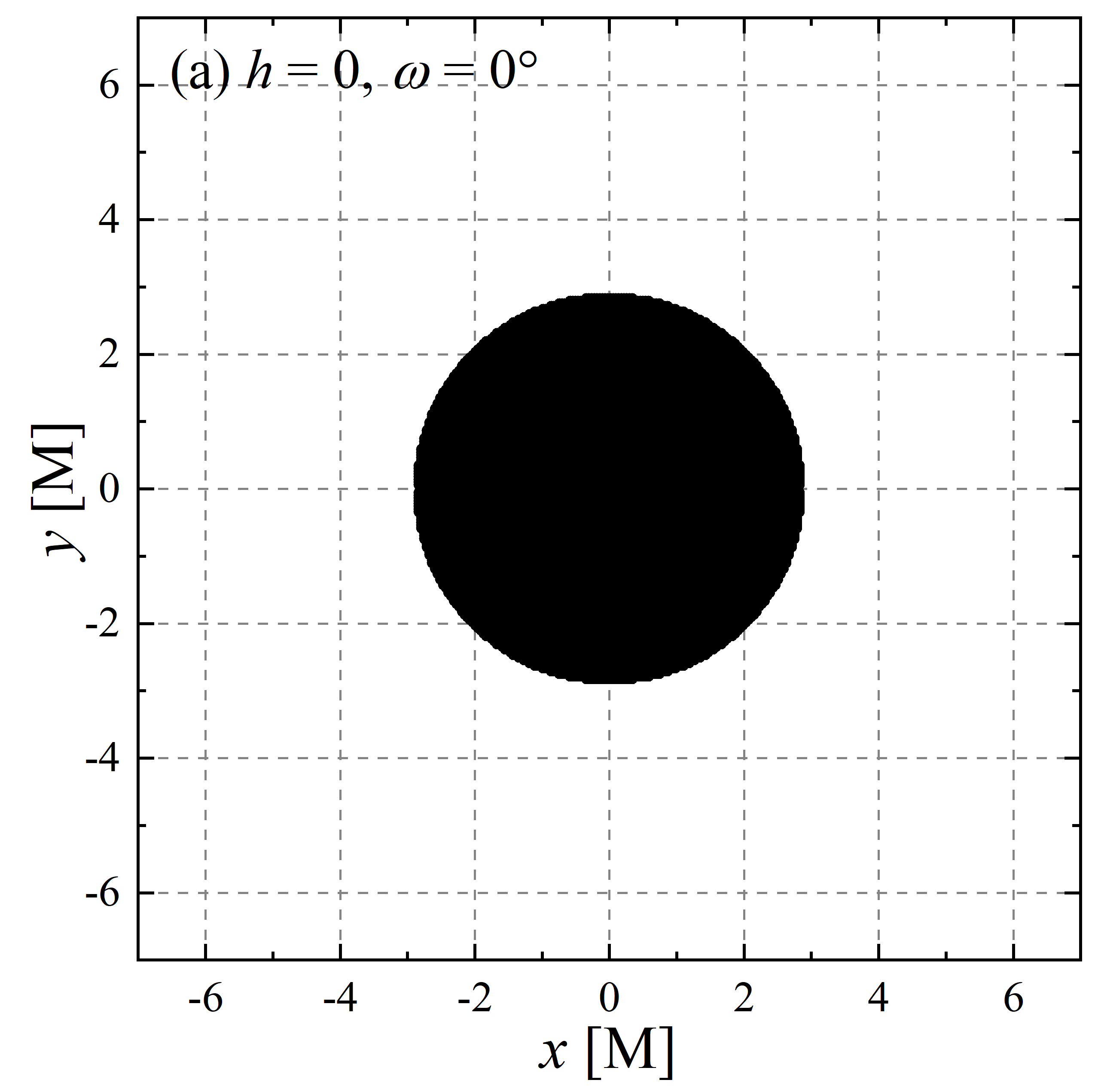}
\includegraphics[width=3.7cm]{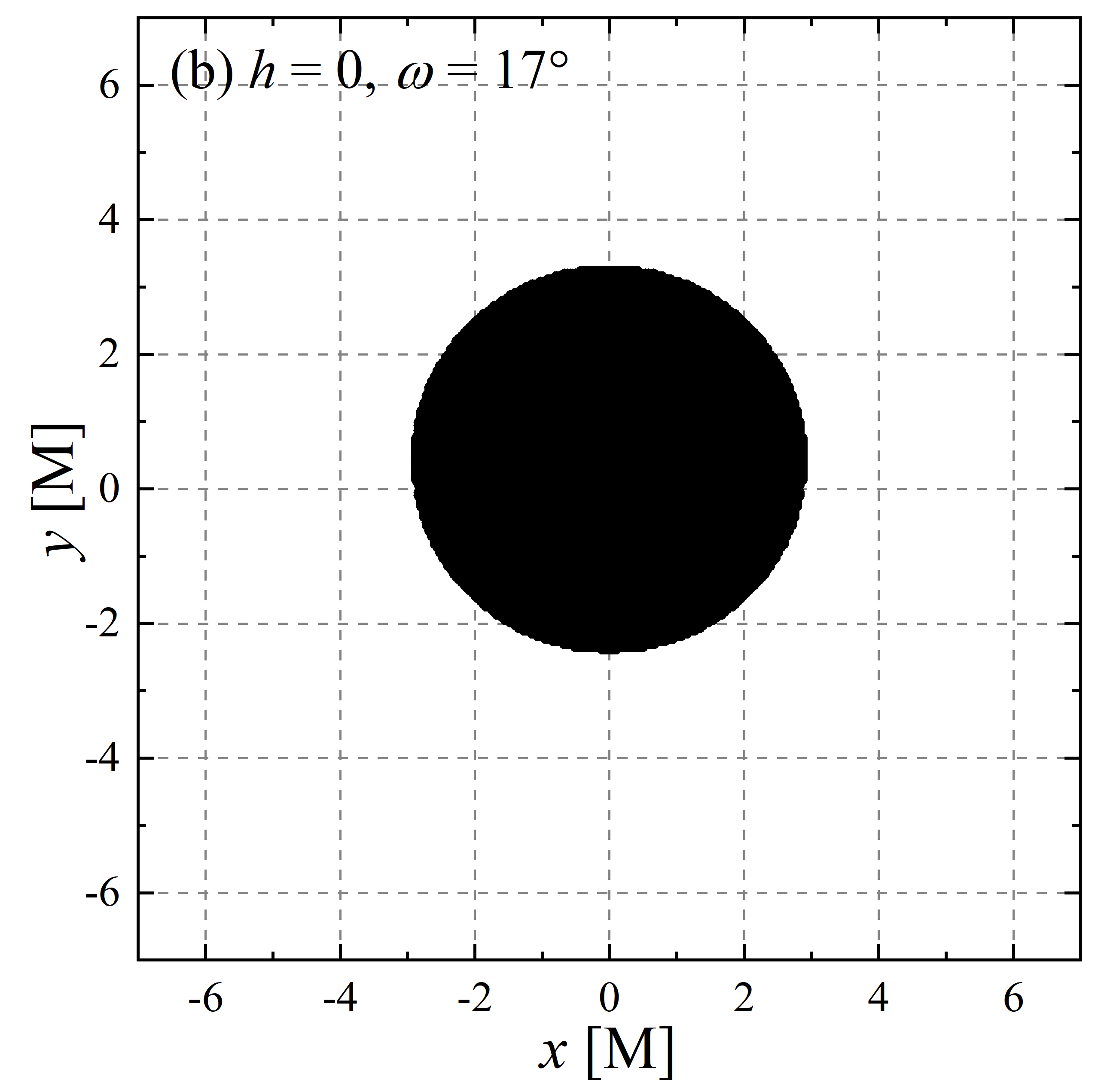}
\includegraphics[width=3.7cm]{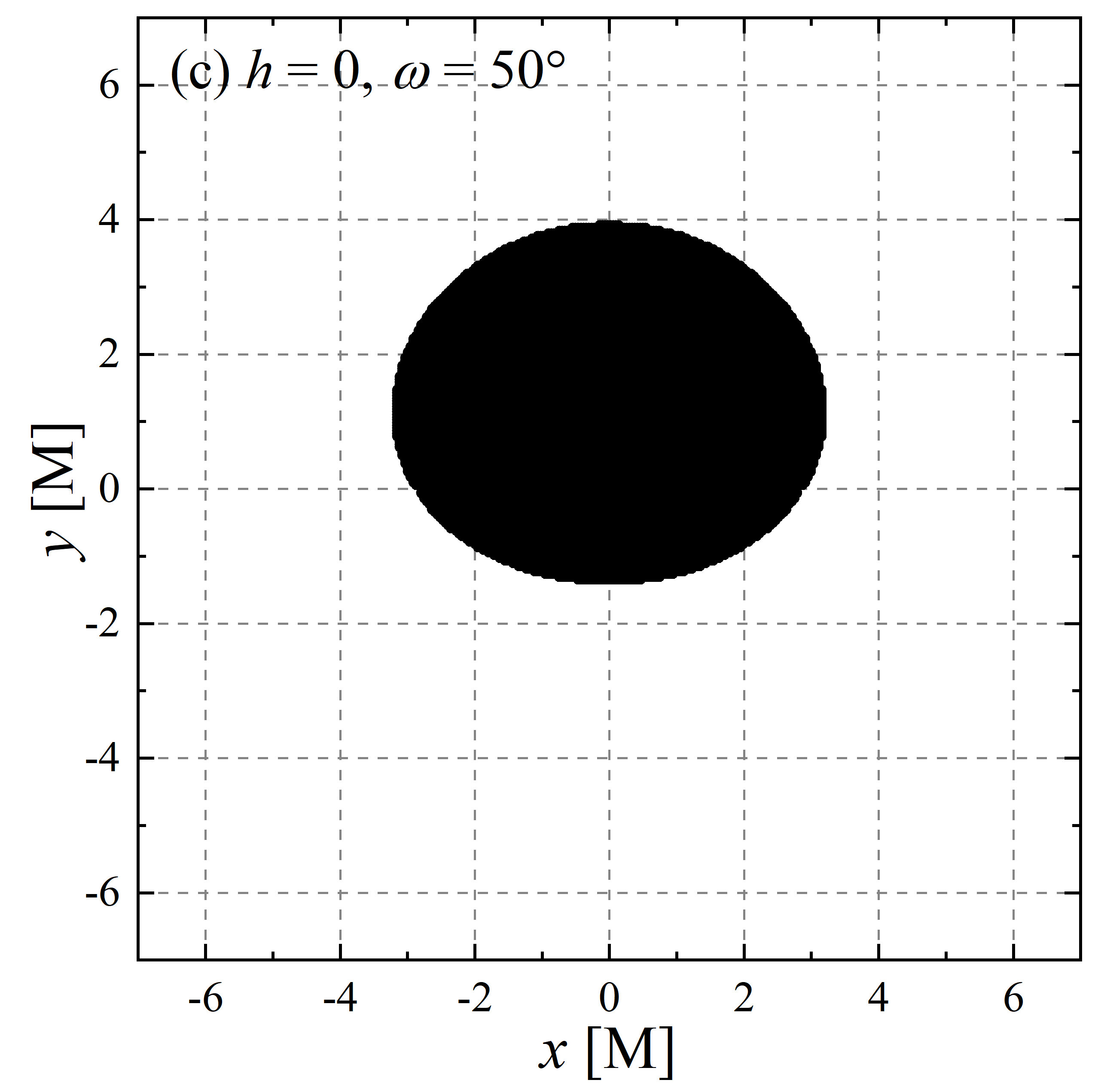}
\includegraphics[width=3.7cm]{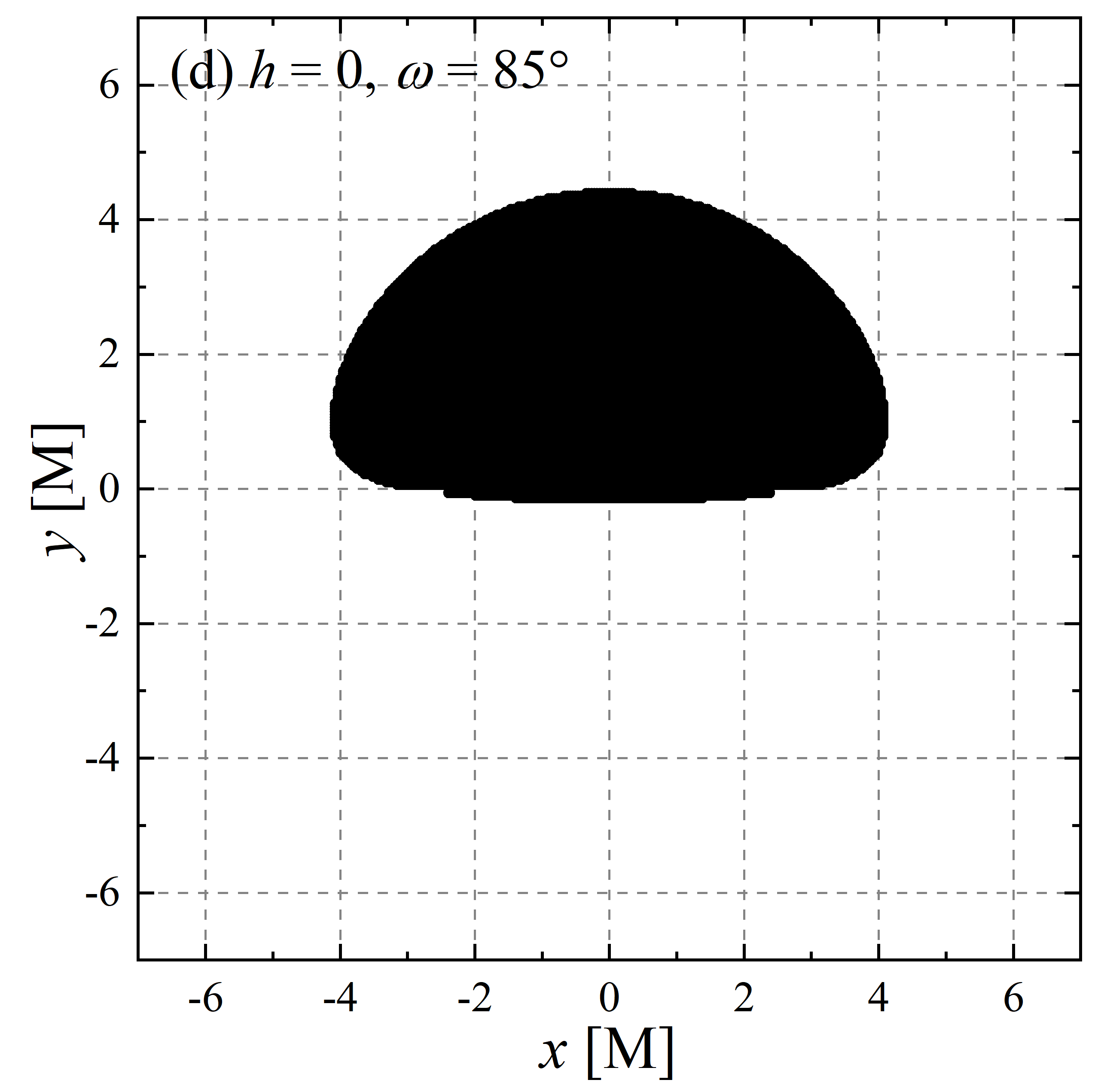}
\includegraphics[width=3.7cm]{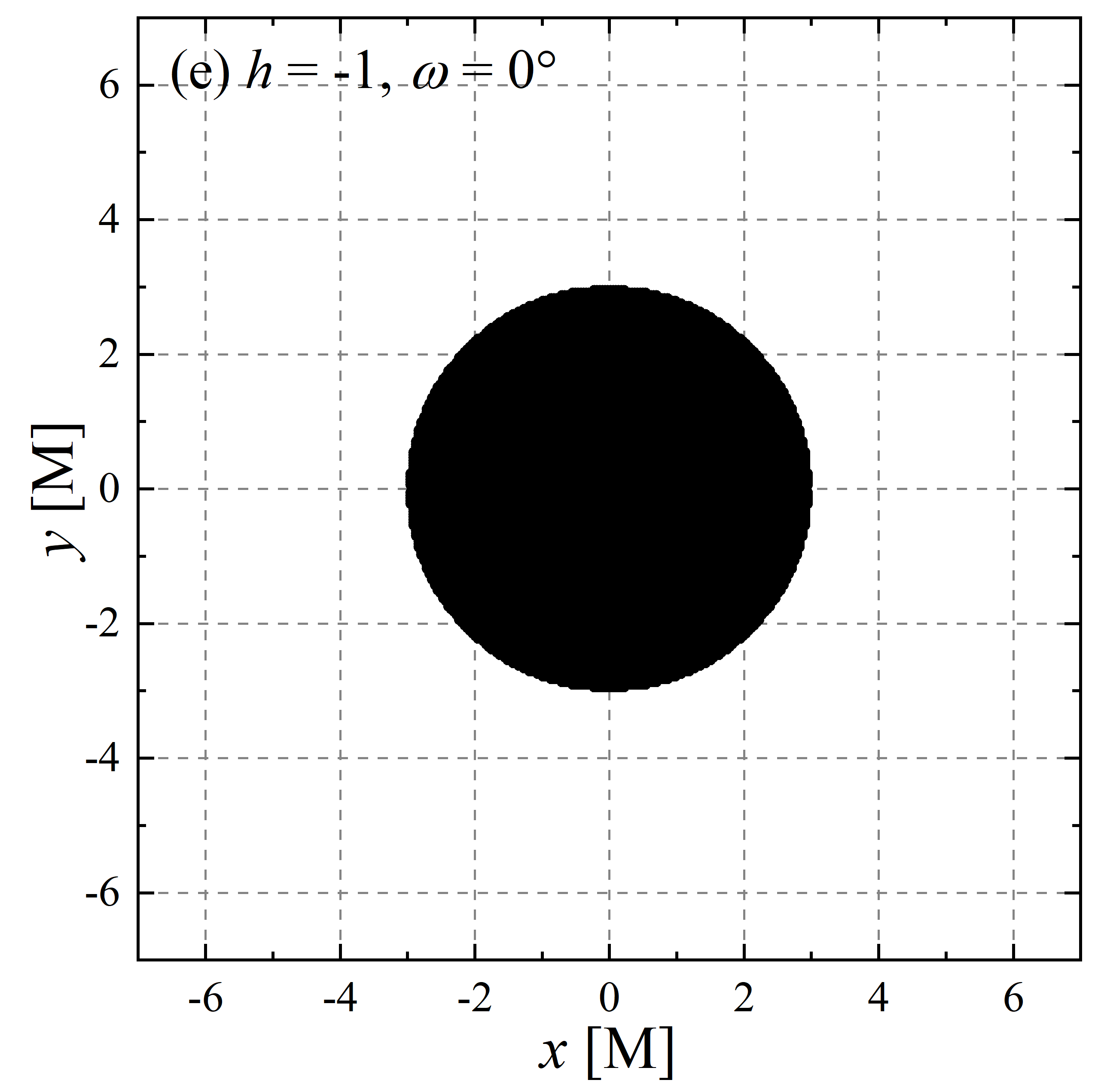}
\includegraphics[width=3.7cm]{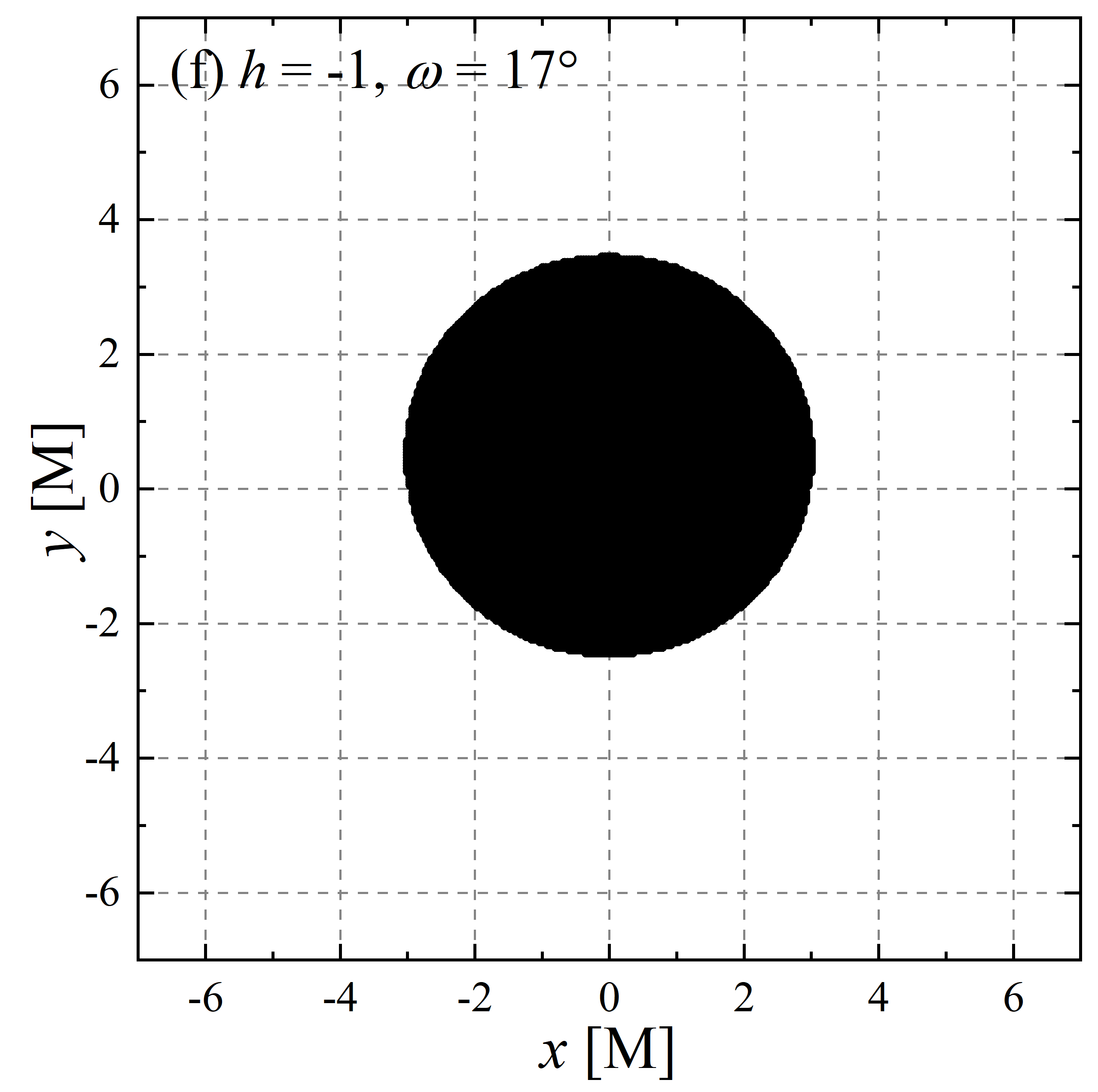}
\includegraphics[width=3.7cm]{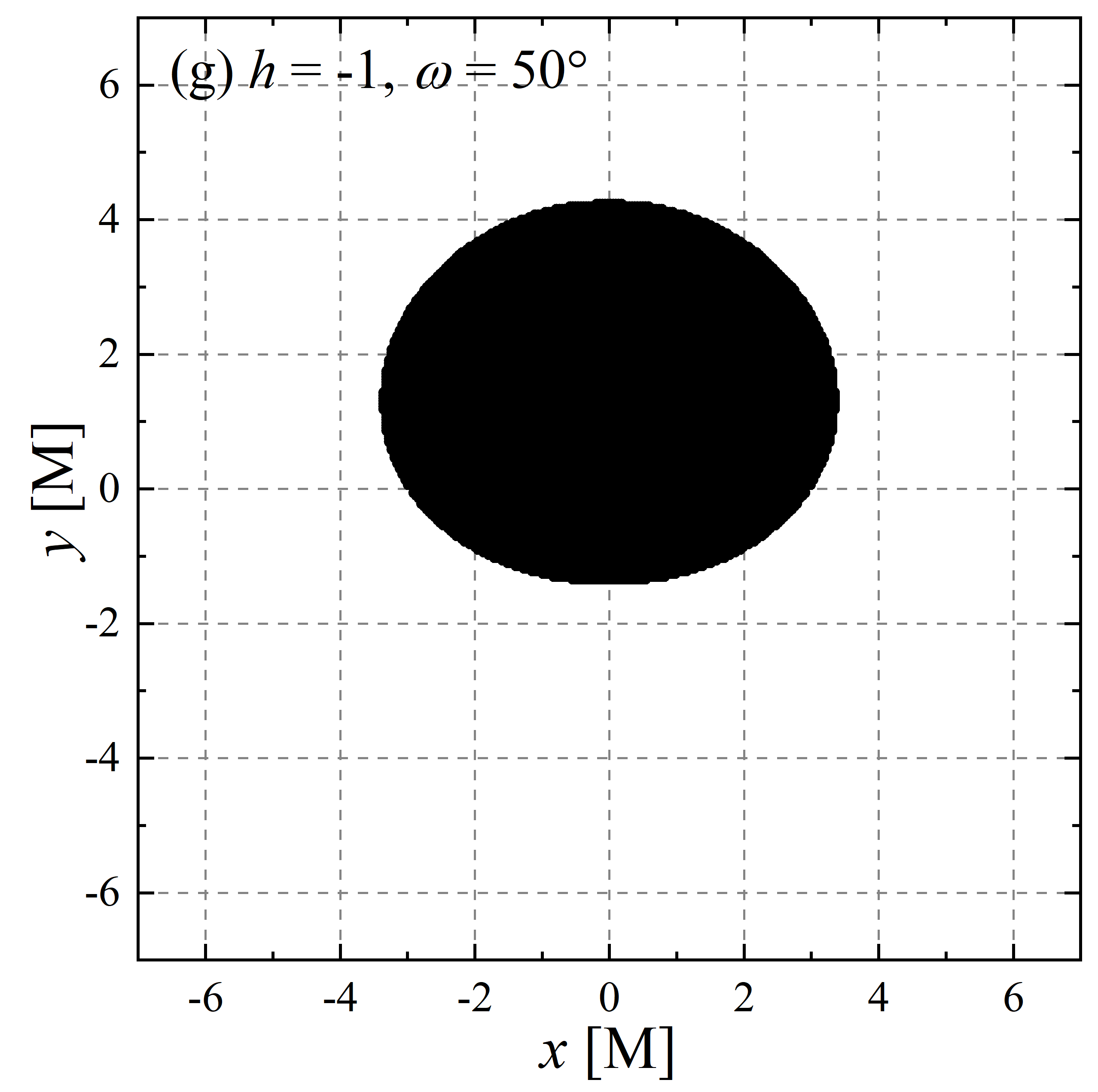}
\includegraphics[width=3.7cm]{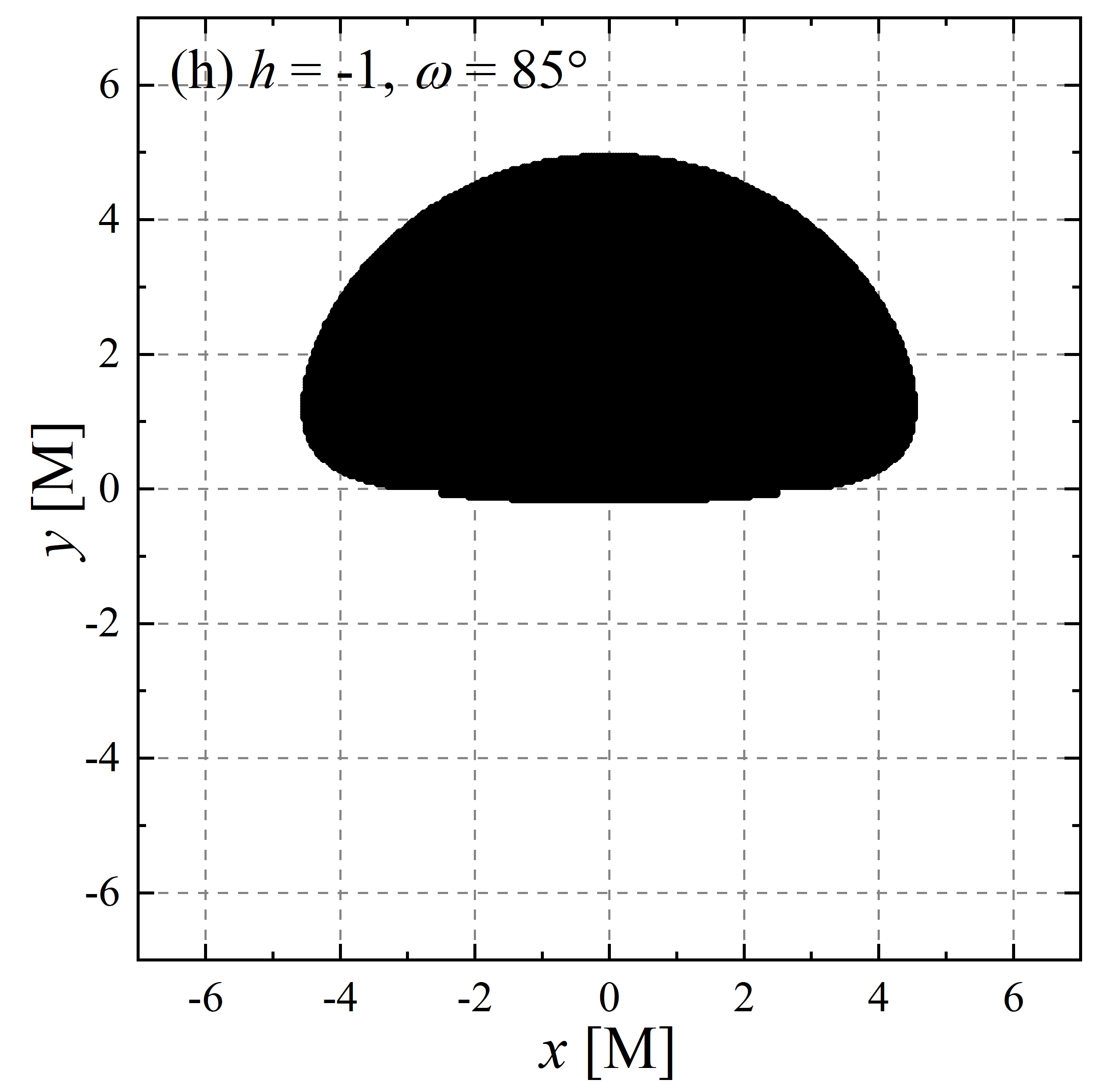}
\includegraphics[width=3.7cm]{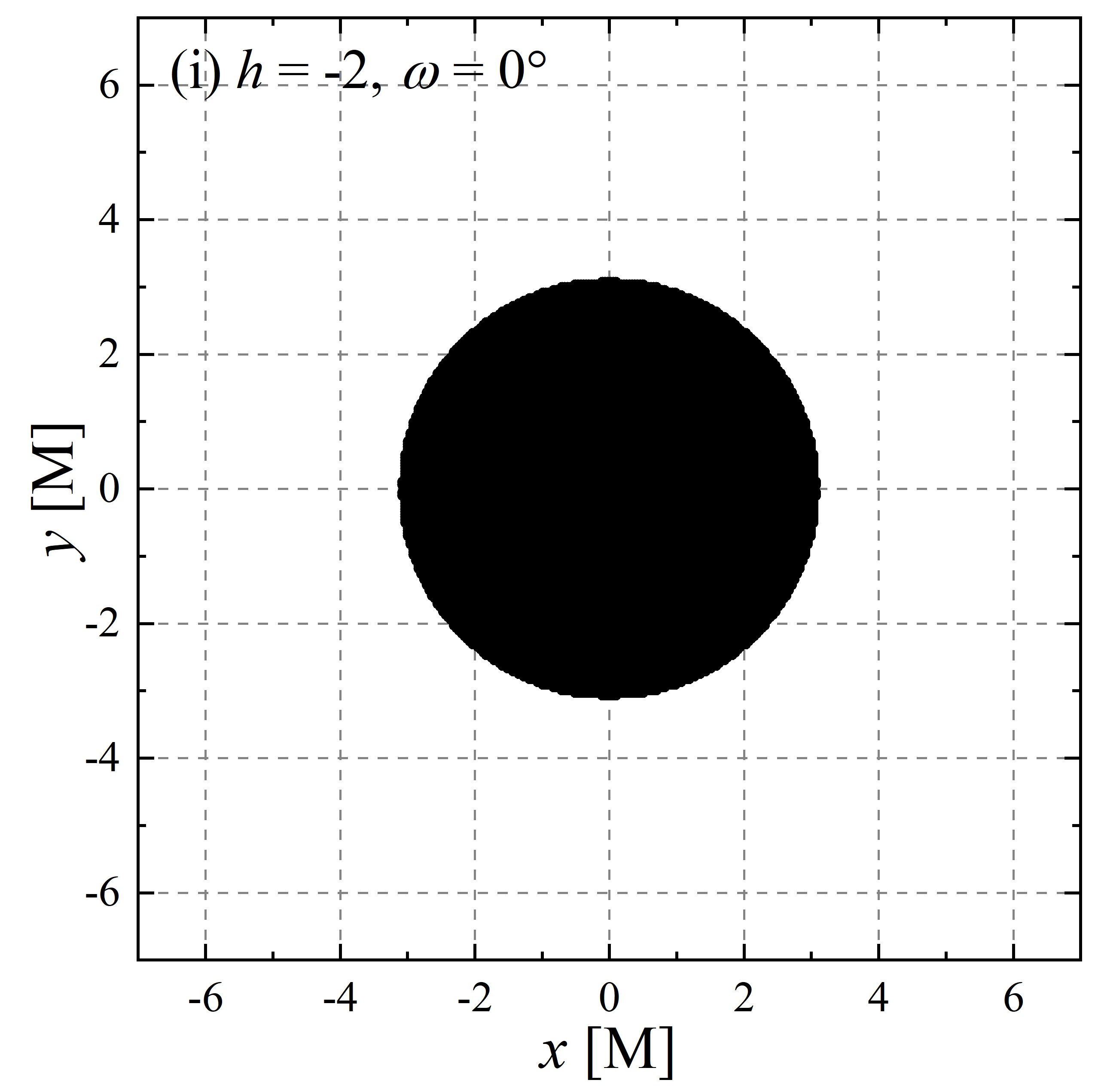}
\includegraphics[width=3.7cm]{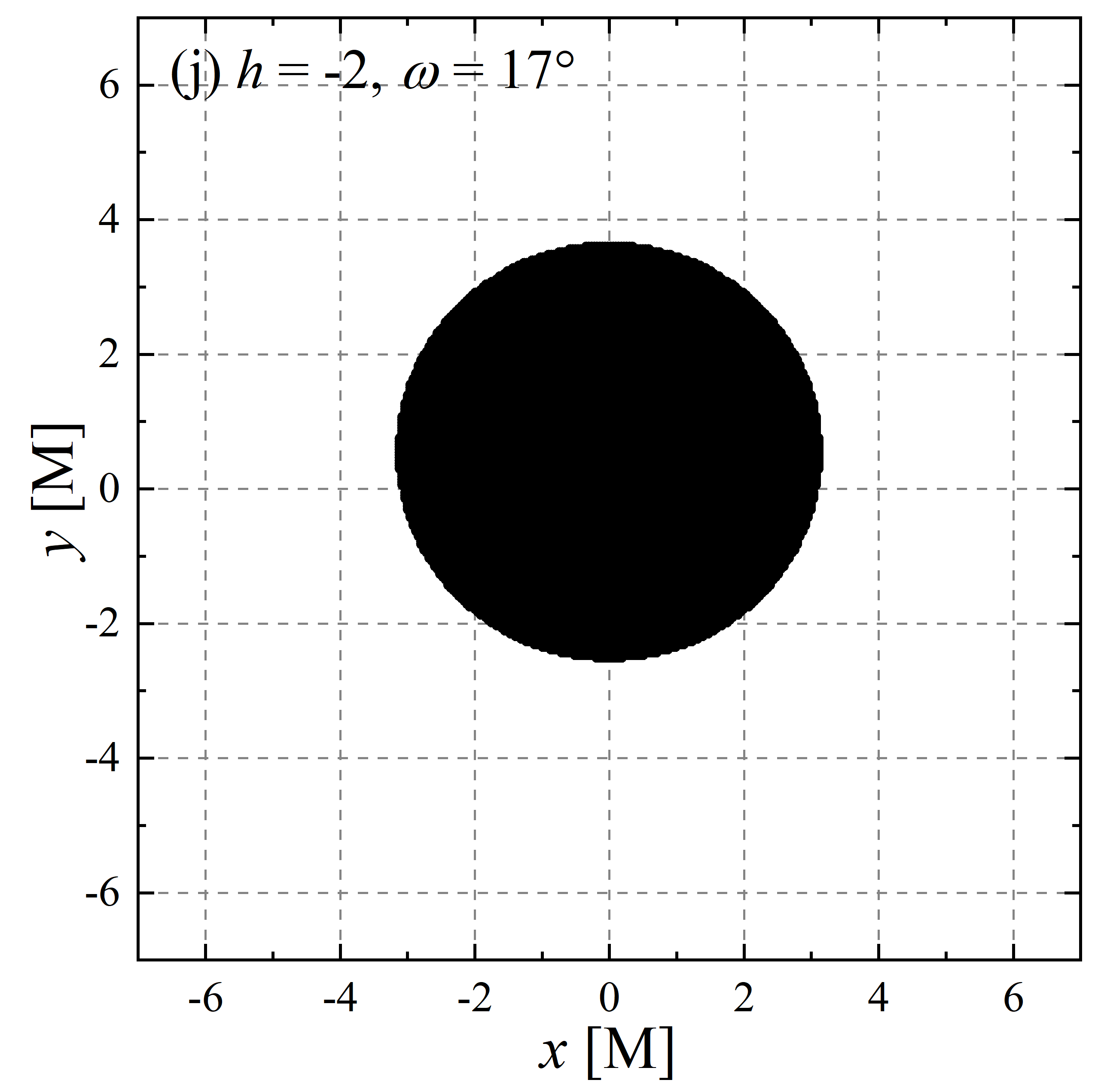}
\includegraphics[width=3.7cm]{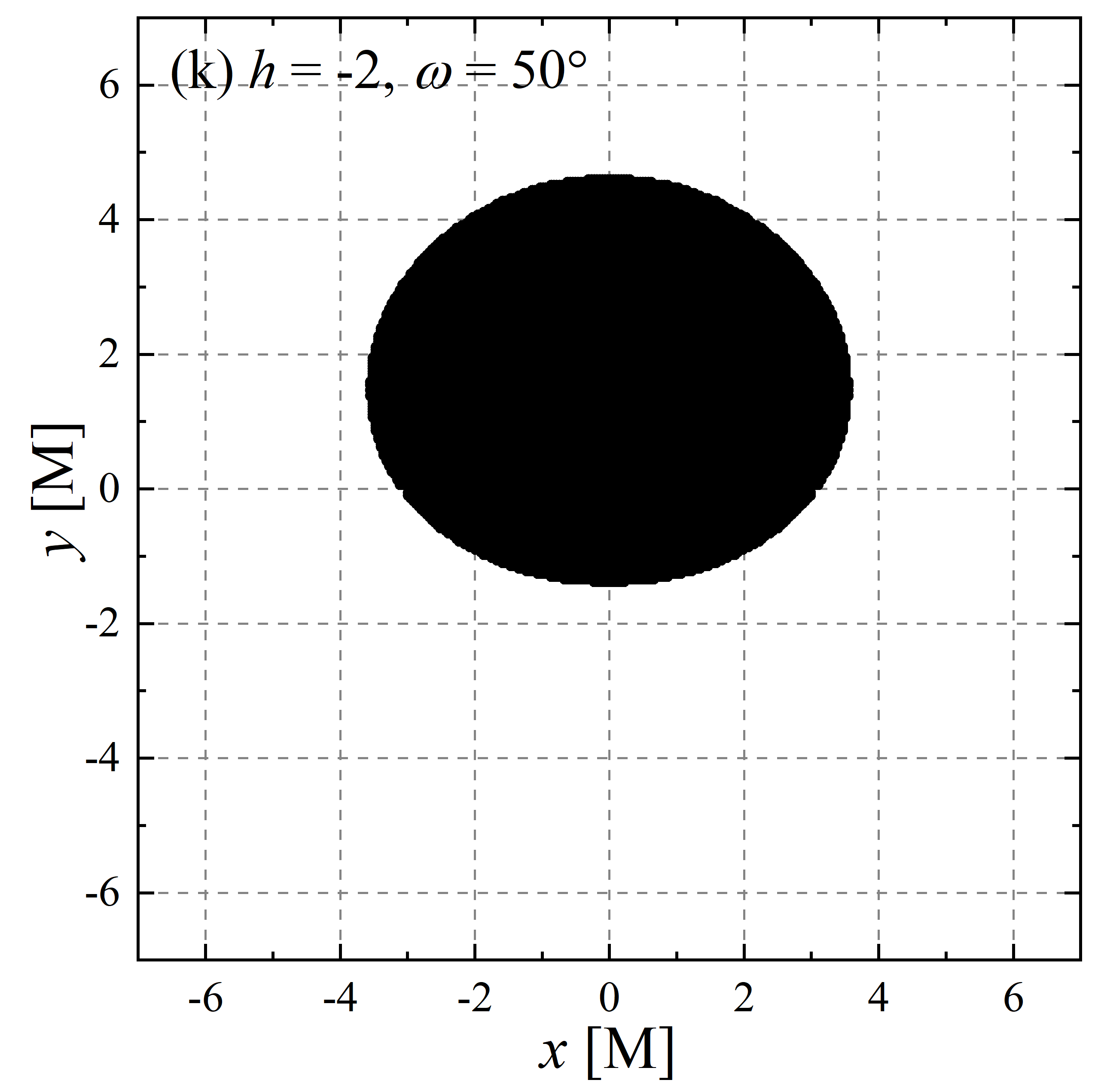}
\includegraphics[width=3.7cm]{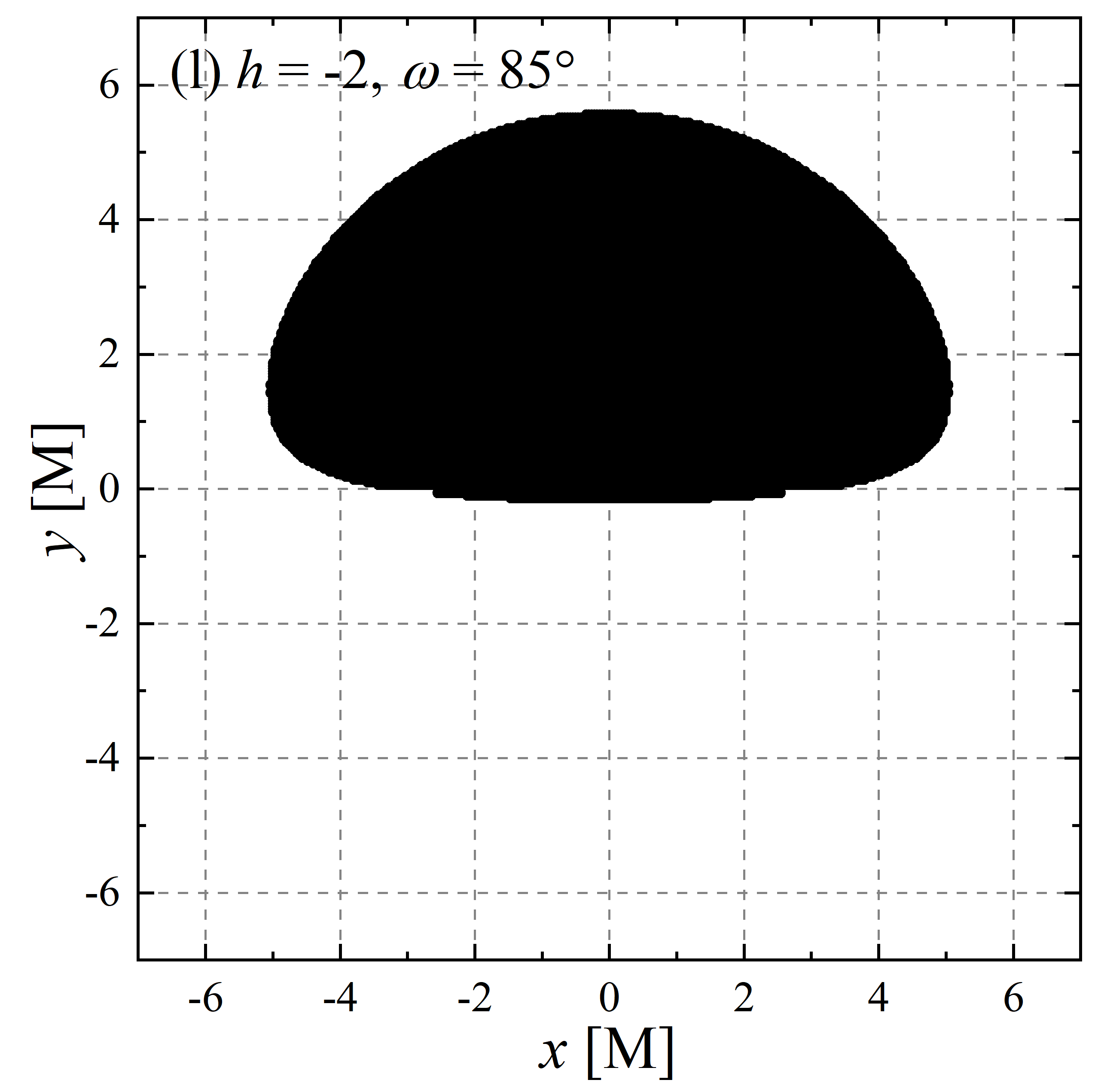}
\caption{Inner shadows of hairy BHs surrounded by an equatorial accretion disk. Evidently, the size of the inner shadow is affected by $h$, and its shape changes with $\omega$. The image resolution is $500$ $\times$ $500$ pixels.}}\label{fig11}
\end{figure*}

Next, to roughly simulate the resolution of the EHT array observing M$87^{*}$, we processed the images using Gaussian blurring with a $20$ $\upmu$as kernel to reveal the features of the $230$ GHz images of the hairy BH that the EHT array would actually capture. Figure 14 displays the blurred images of the Schwarzschild BH and the hairy BH illuminated by tilted accretion disks, as viewed from various inclinations. The image appears dimmer compared to the original, and the prominence of the light spot is also masked by the background. More importantly, the intricate details of the emission ring are washed out, and for small values of $\Theta$ the ring structure is even disrupted, leaving only the crescent visible, as depicted in the leftmost panel of the fourth row and the rightmost panel of the second row. Nevertheless, we can still recognize that the characteristic ring or crescent of the hairy BH is larger than that of the Schwarzschild BH. This suggests that the current EHT array may have the capability to capture the distinction between the two in certain parameter space. We also note that streaks contributed by inclined accretion disks survive the blurring operation, as seen in the second panel of the third column from the left. Although these features are faintly visible, they can still be used as a diagnostic to infer the inclination of the accretion disk when the viewing angle is known. Additionally, we explored whether the EHT array can detect the impact of the observer's azimuthal angle on the image by processing figures 12 and 13 with a Gaussian filter, as arranged in the left and right four columns of figure 15, respectively. The results show that when $\omega = 85^{\circ}$, the ring structure in the bulk parameter space is broken, leaving the picture with crescents and streaks. The location and morphology of the inner shadows are also challenging to identify. As a comparison, the ring structure remains intact after Gaussian blurring in the case of $\omega = 17^{\circ}$. Although the details of the image are blurred, it is still possible to track the drifting light spot caused by a change in observation azimuth. This is prominently demonstrated in the third and fourth columns from the left, strongly suggesting that information about the precession of the tilted accretion disk can be extracted through the analysis of the transition in the light spot.

However, it is necessary to clarify that our statements are based on a quasi-static accretion disk model, i.e., the evolution timescale of the accretion flow, which depends on the mass of the BH, is larger than a single observing night of the EHT campaign. For M87$^{*}$, the variability timescale at $230$ GHz ranges from $5$ days to $1$ month, depending on the value of spin, which affirms that the source structure remains unchanged during the course of an observing night \cite{Akiyama et al. (2019a),Akiyama et al. (2022a),Akiyama et al. (2019b),Akiyama et al. (2019c)}. Thus, the reconstructed images of M87$^{*}$ from $4$ days of observation are almost identical \cite{Akiyama et al. (2019a),Akiyama et al. (2019d)}, providing reliable observational features for deducing the physical parameters of M87$^{*}$ and its surroundings. In contrast, the range of variability timescale for Sagittarius A$^{*}$ is about $4-30$ minutes, leading to substantial variations in the accretion flow over an observing run. This violates the fundamental assumption for Earth-rotation VLBI aperture synthesis \cite{Akiyama et al. (2022a),Akiyama et al. (2022d)}. Therefore, the region of the second null in the observed data of Sagittarius A$^{*}$ is heavily affected by errors, resulting in highly uncertain reconstructed images that exhibit diverse azimuthal intensity distributions near the emission ring \cite{Akiyama et al. (2022a)}. Some features of the reconstructed image in this case, such as bright spots, are likely to be artefacts and are naturally unreliable for use in diagnosing accretion environments.
\section{Images of hairy black holes illuminated by multiple thin accretion disks}
In astrophysics, the accretion environment around a BH is highly intricate. As mentioned earlier, when the differential Lense-Thirring torques surpass the viscous torques, the tilted thin accretion disk can be torn into multiple sub-disks \cite{Liska et al. (2021),Liska et al. (2023b),Kaaz et al. (2023),Musoke et al. (2023)}. Among these, the angular momentum of the inner accretion disk can quickly align with the BH rotation due to the Bardeen-Petterson effect \cite{Liska et al. (2019),Liska et al. (2021),Bardeen $&$ Petterson (1975),Nealon et al. (2015)}, while the outer accretion disk can maintain its original inclination angle for a more extended period. Consequently, both equatorial and non-equatorial accretion flows coexist around the BH. It is imperative to elucidate the observational appearance of BHs in a multi-disk accretion scenario, as this can offer valuable insights into the structural evolution of the accretion flow through BH images. In this section, for simplicity, we adopt a toy model where the radiation regions of both accretion disks are extended to the BH event horizon to expose the observational signatures of the hairy BH.
\begin{figure*}
\center{
\includegraphics[width=3.5cm]{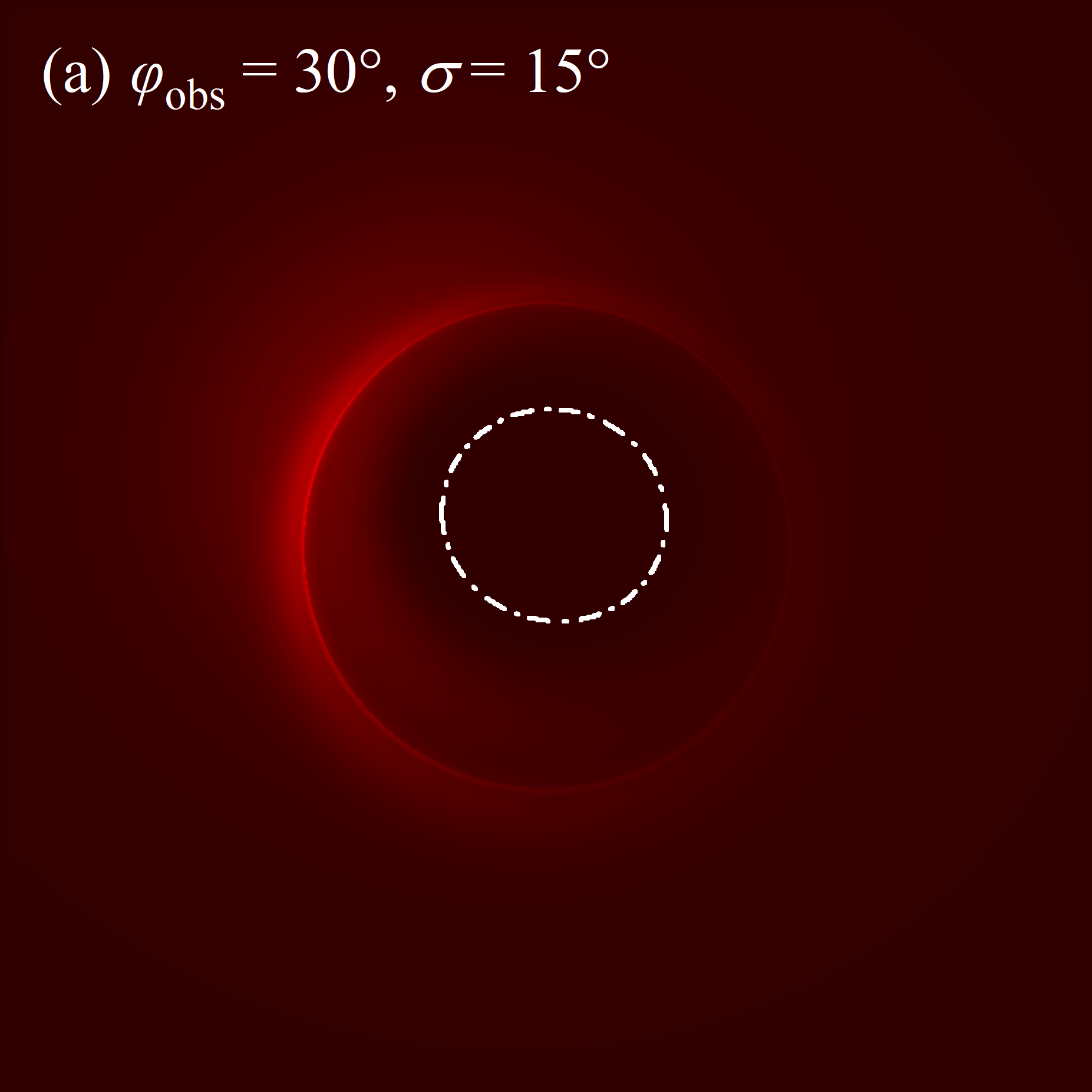}
\includegraphics[width=3.5cm]{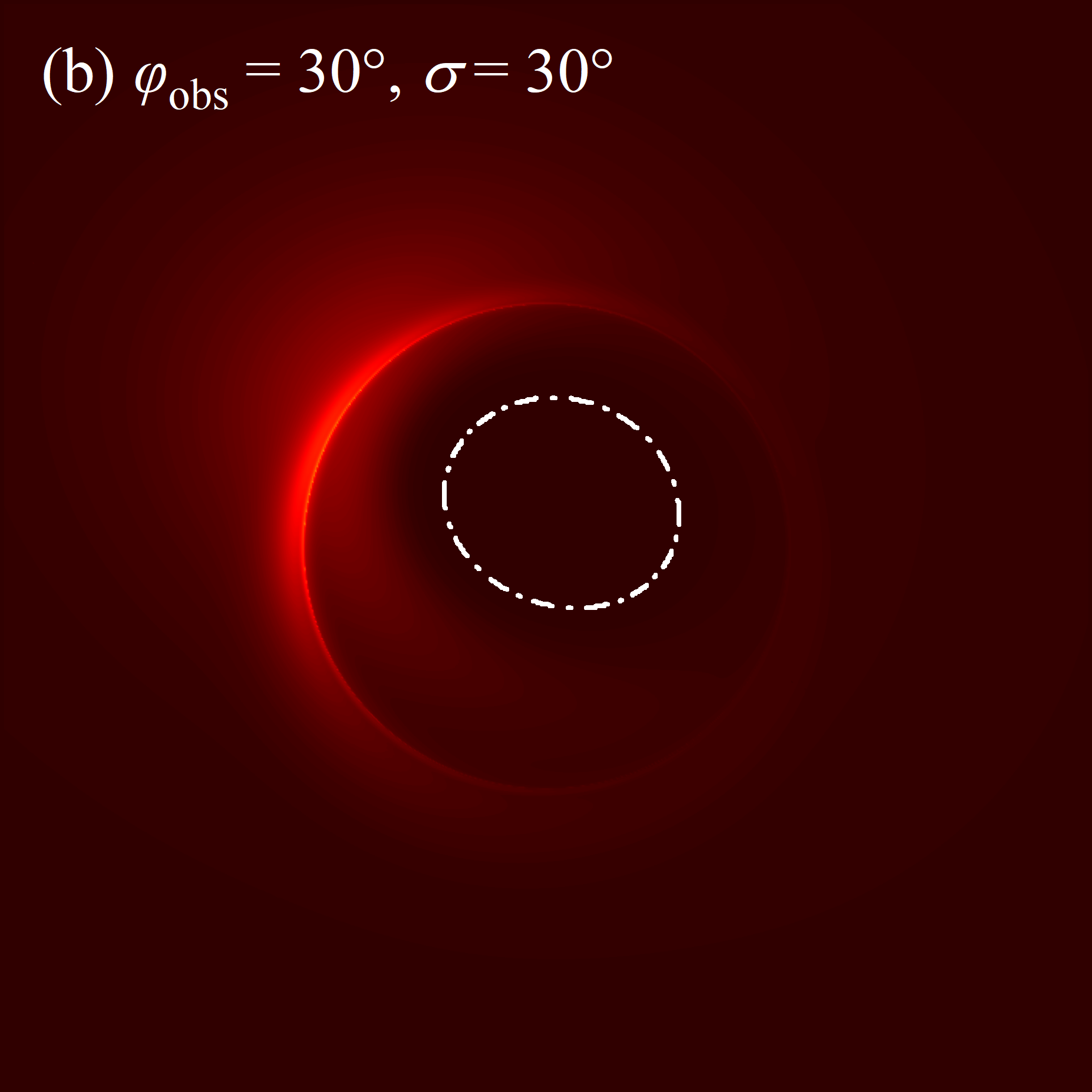}
\includegraphics[width=3.5cm]{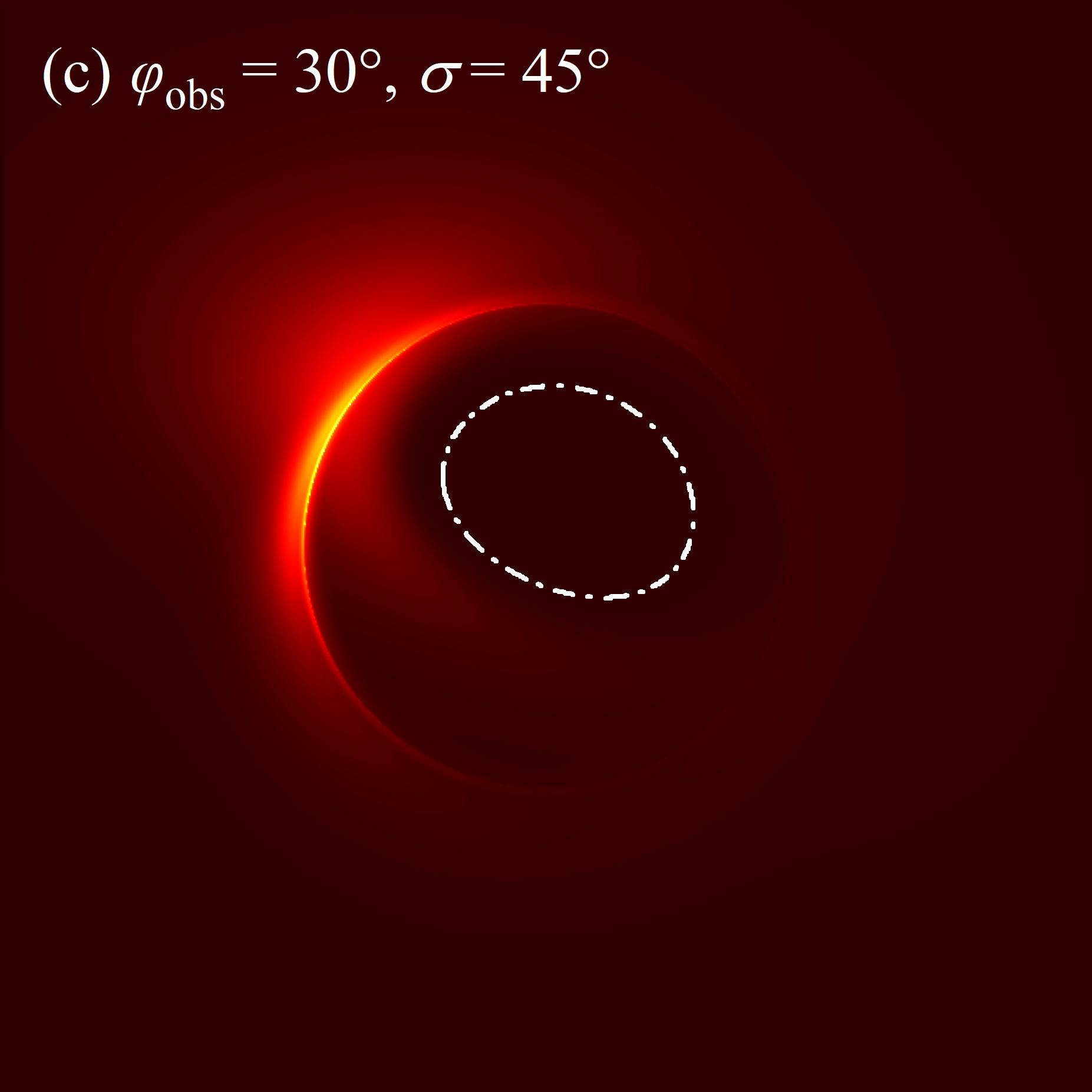}
\includegraphics[width=3.5cm]{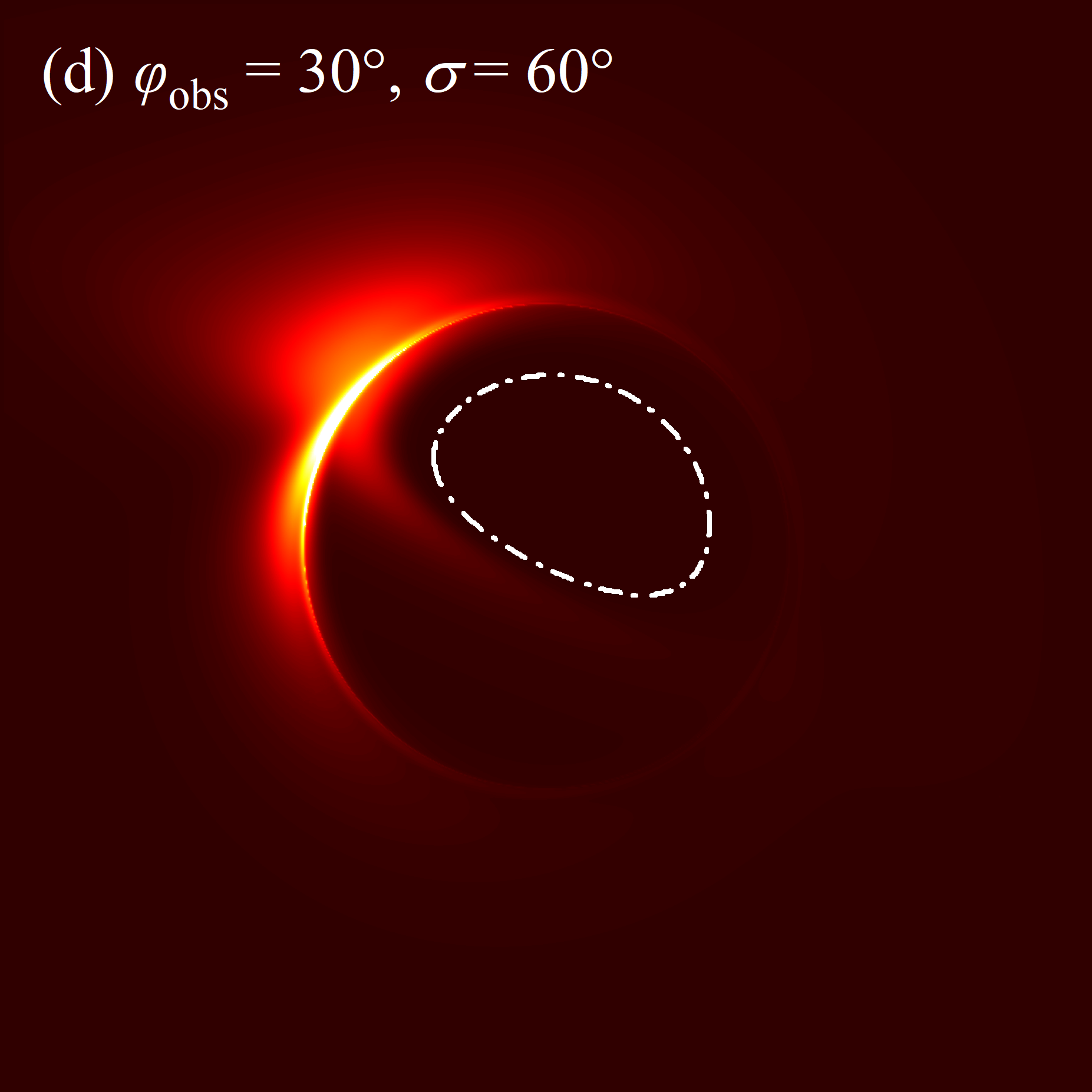}
\includegraphics[width=3.5cm]{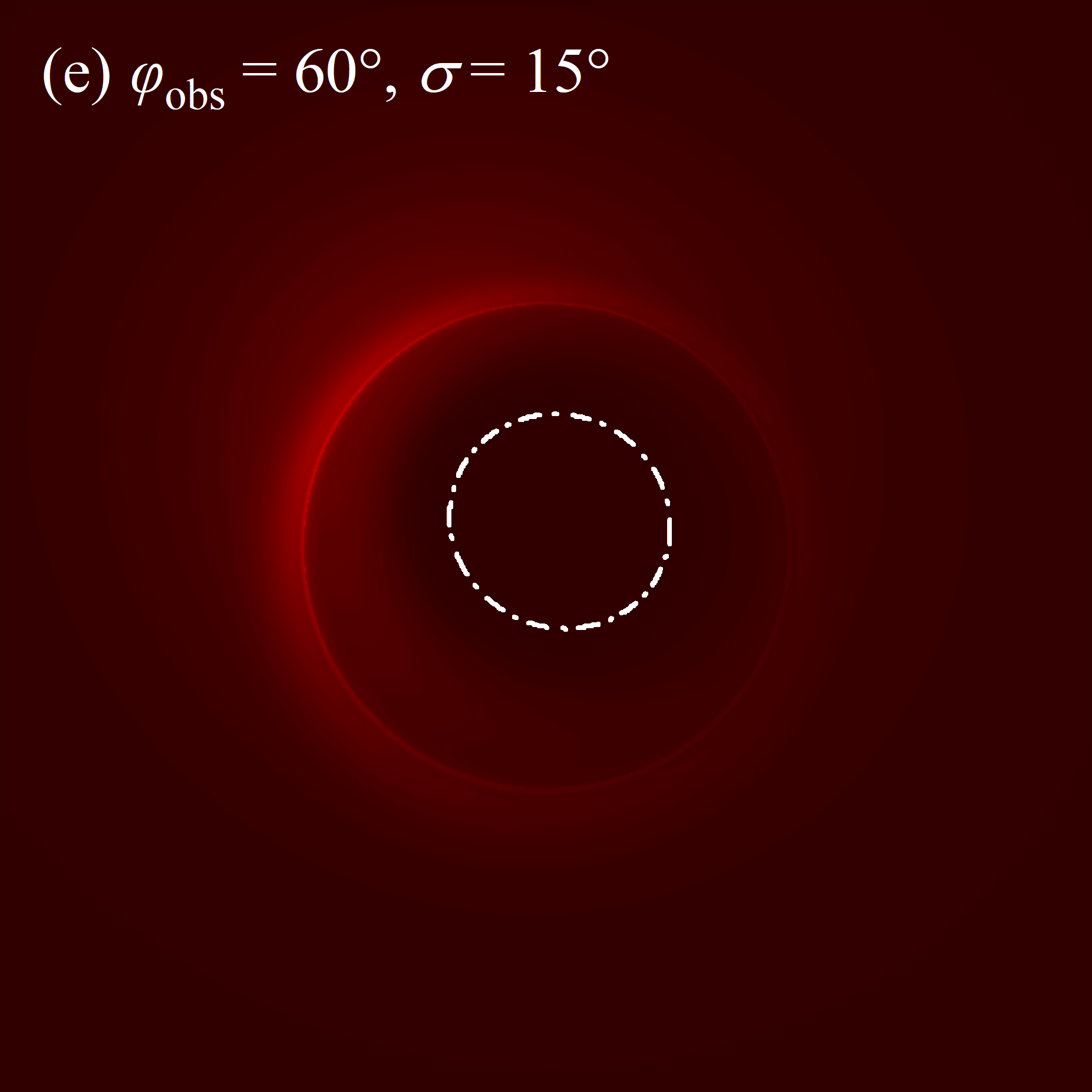}
\includegraphics[width=3.5cm]{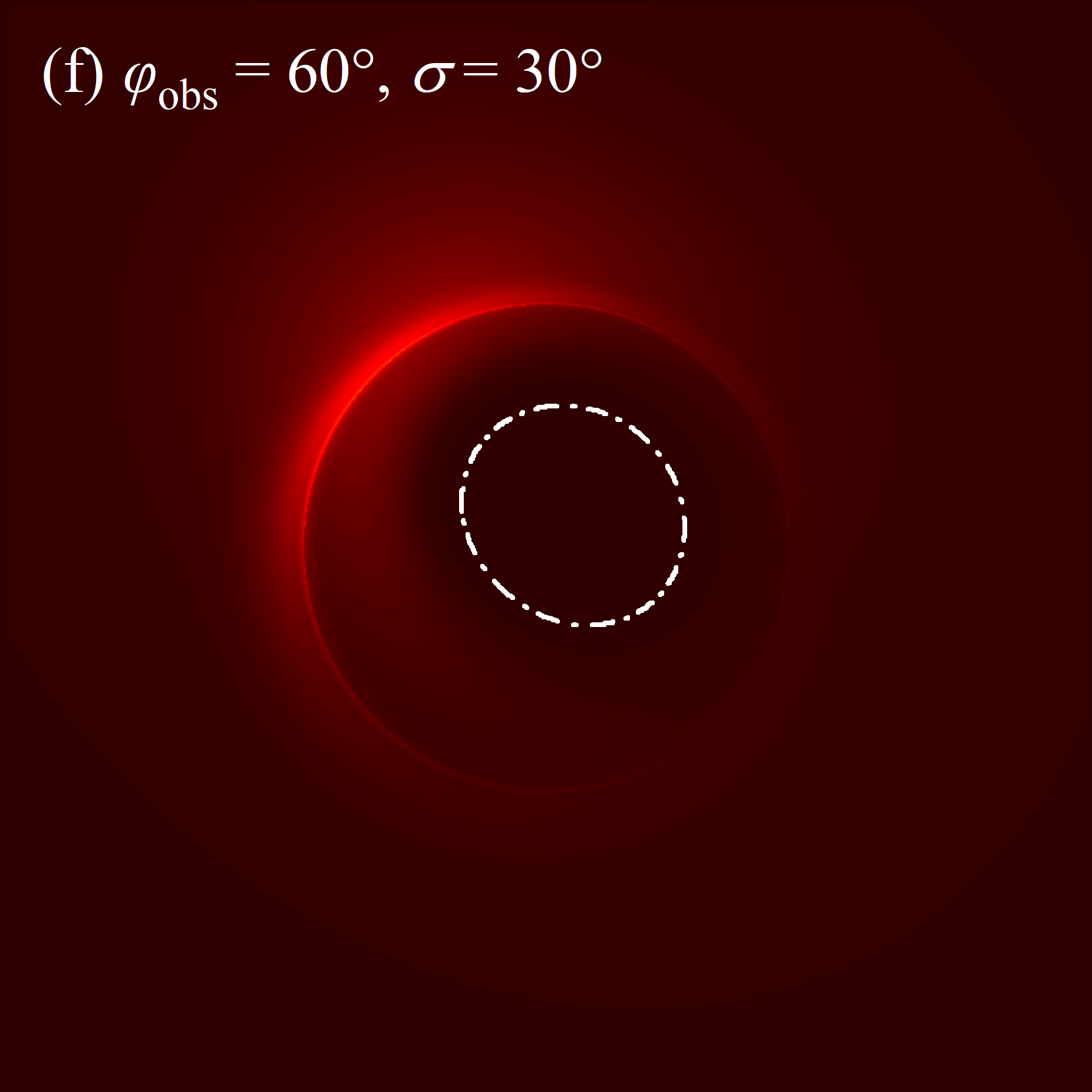}
\includegraphics[width=3.5cm]{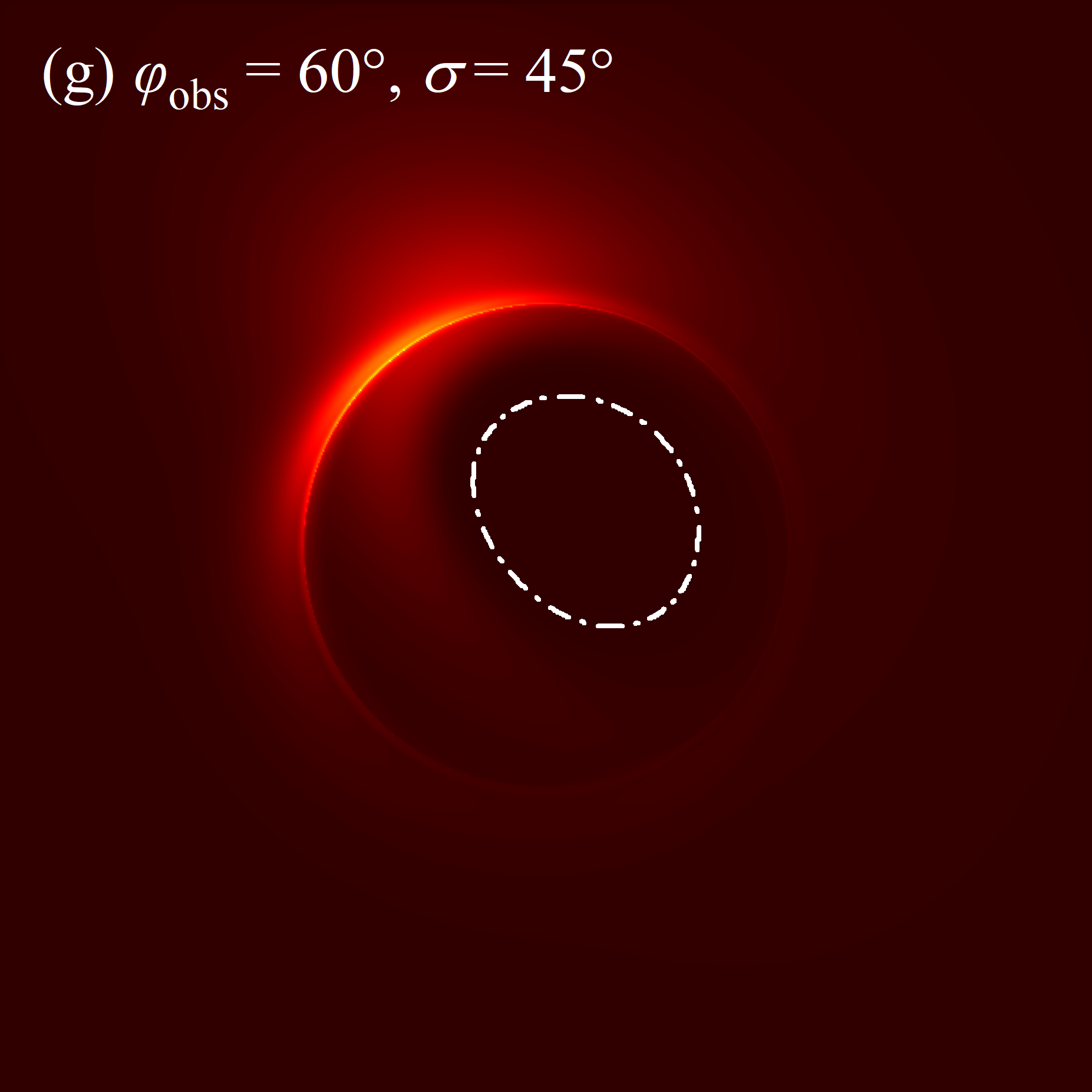}
\includegraphics[width=3.5cm]{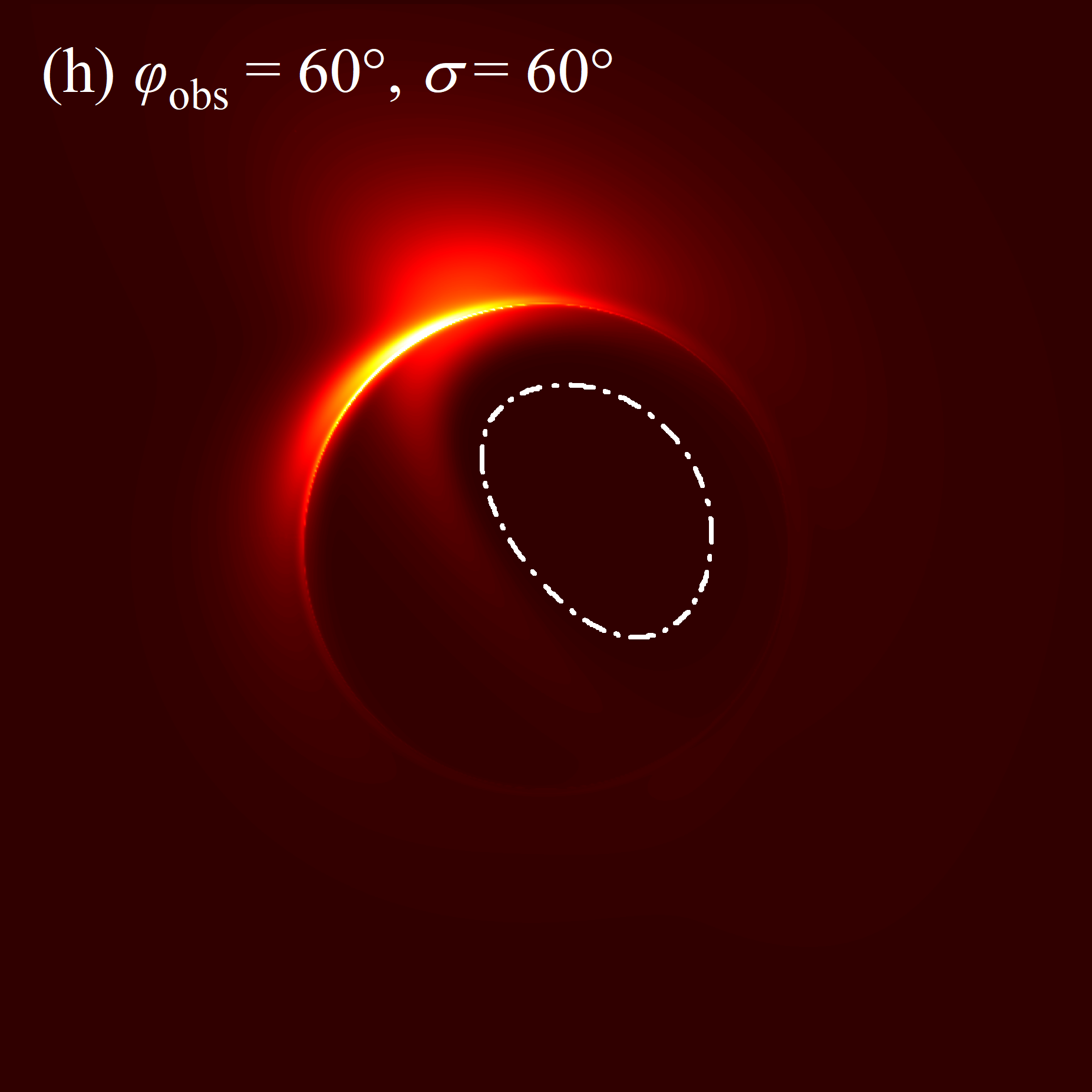}
\includegraphics[width=3.5cm]{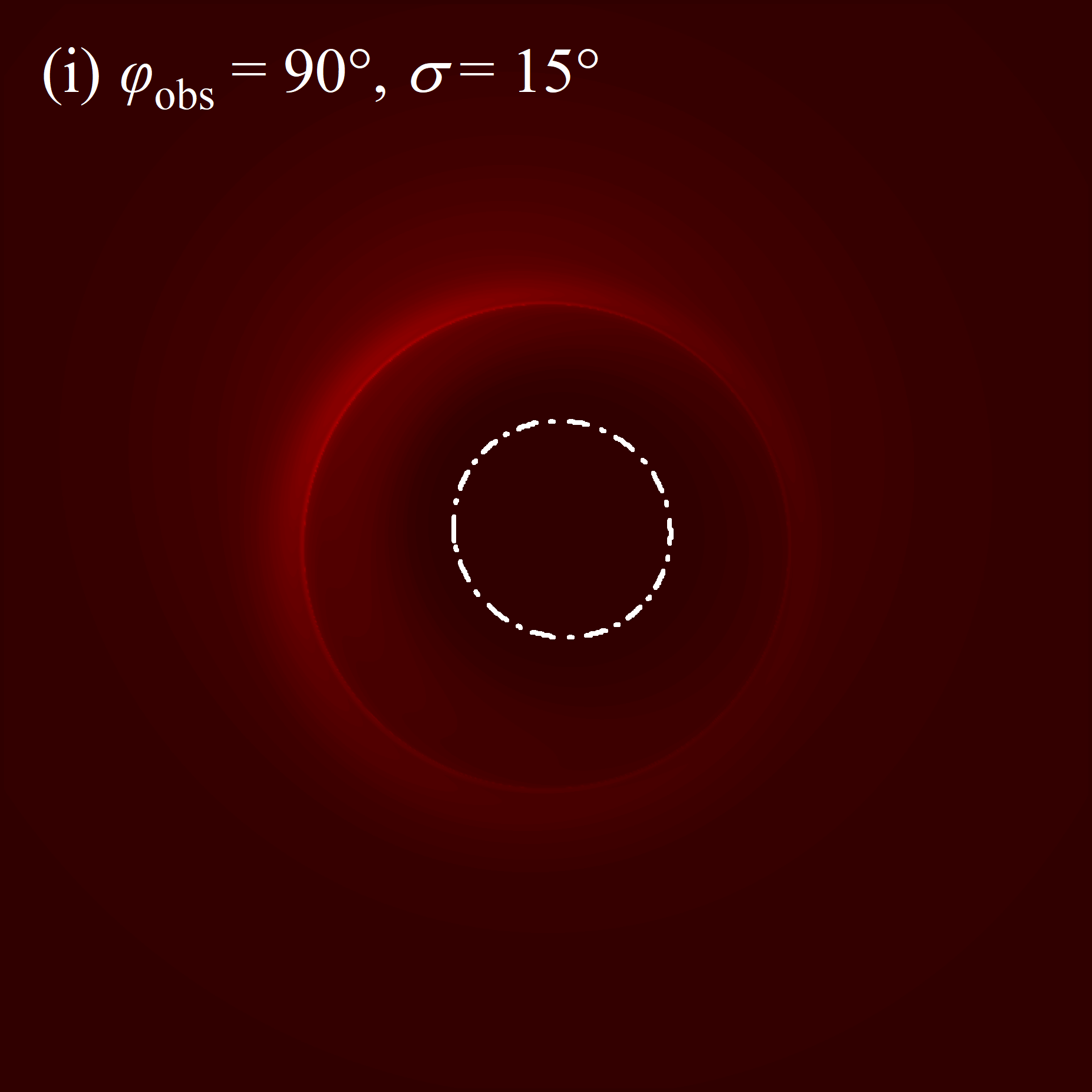}
\includegraphics[width=3.5cm]{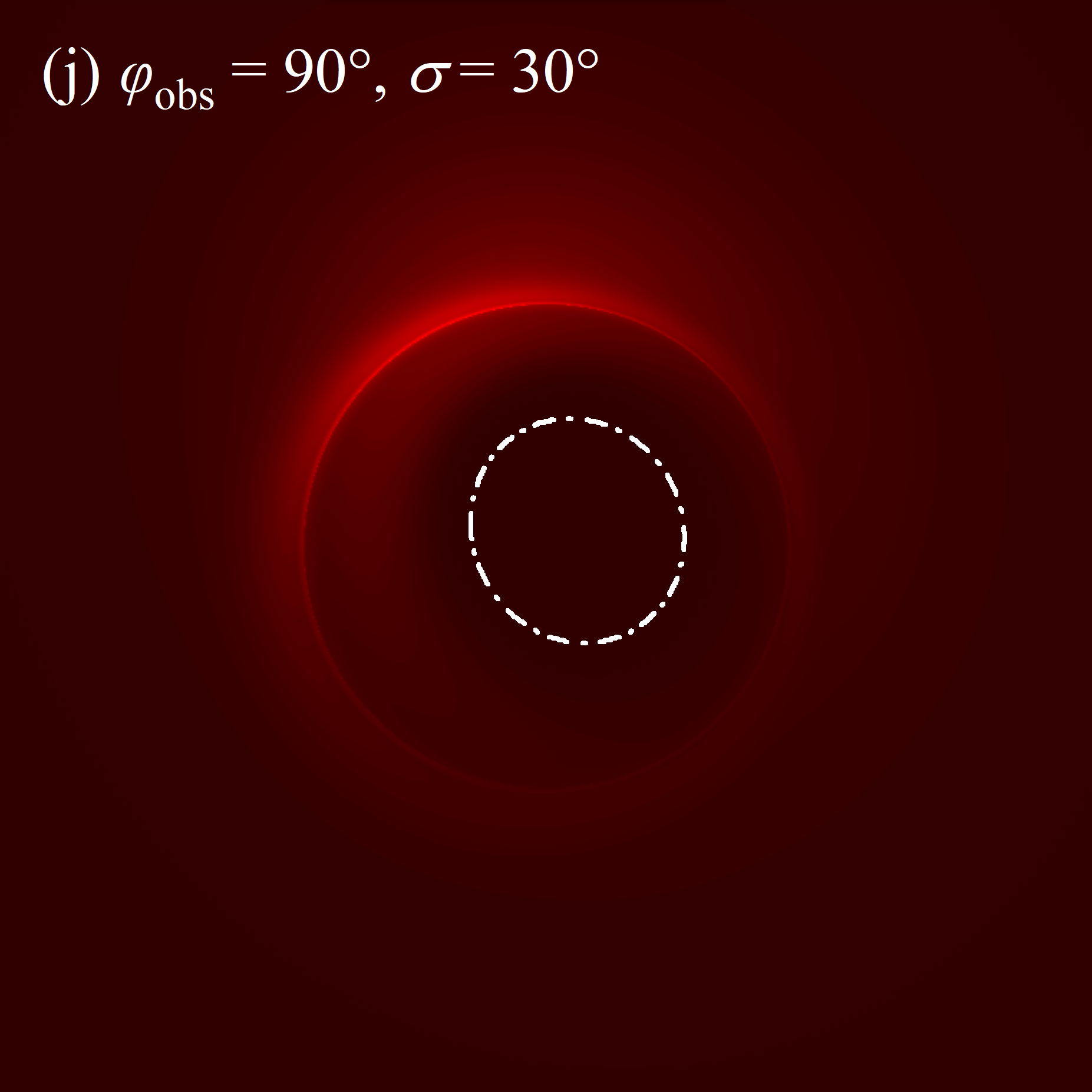}
\includegraphics[width=3.5cm]{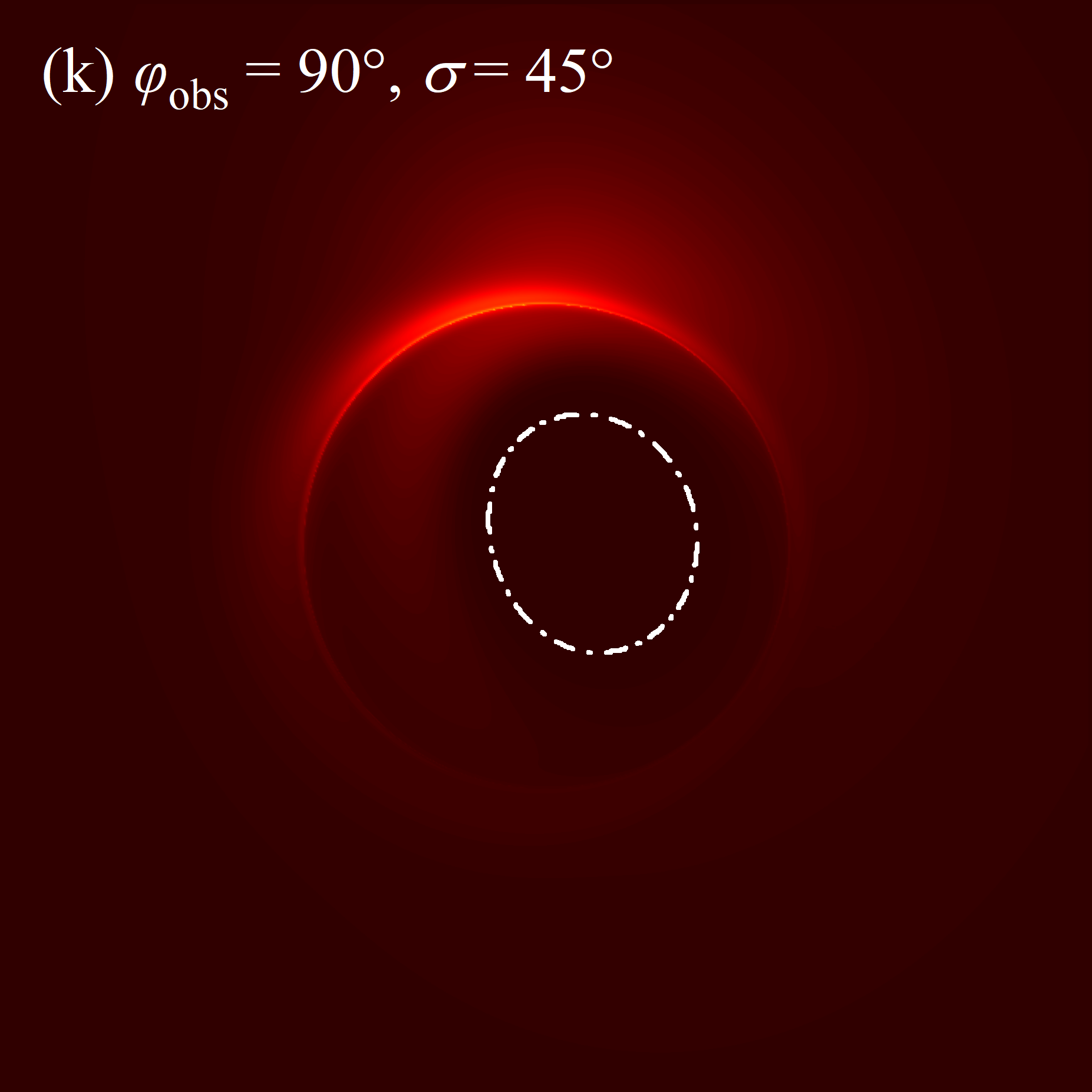}
\includegraphics[width=3.5cm]{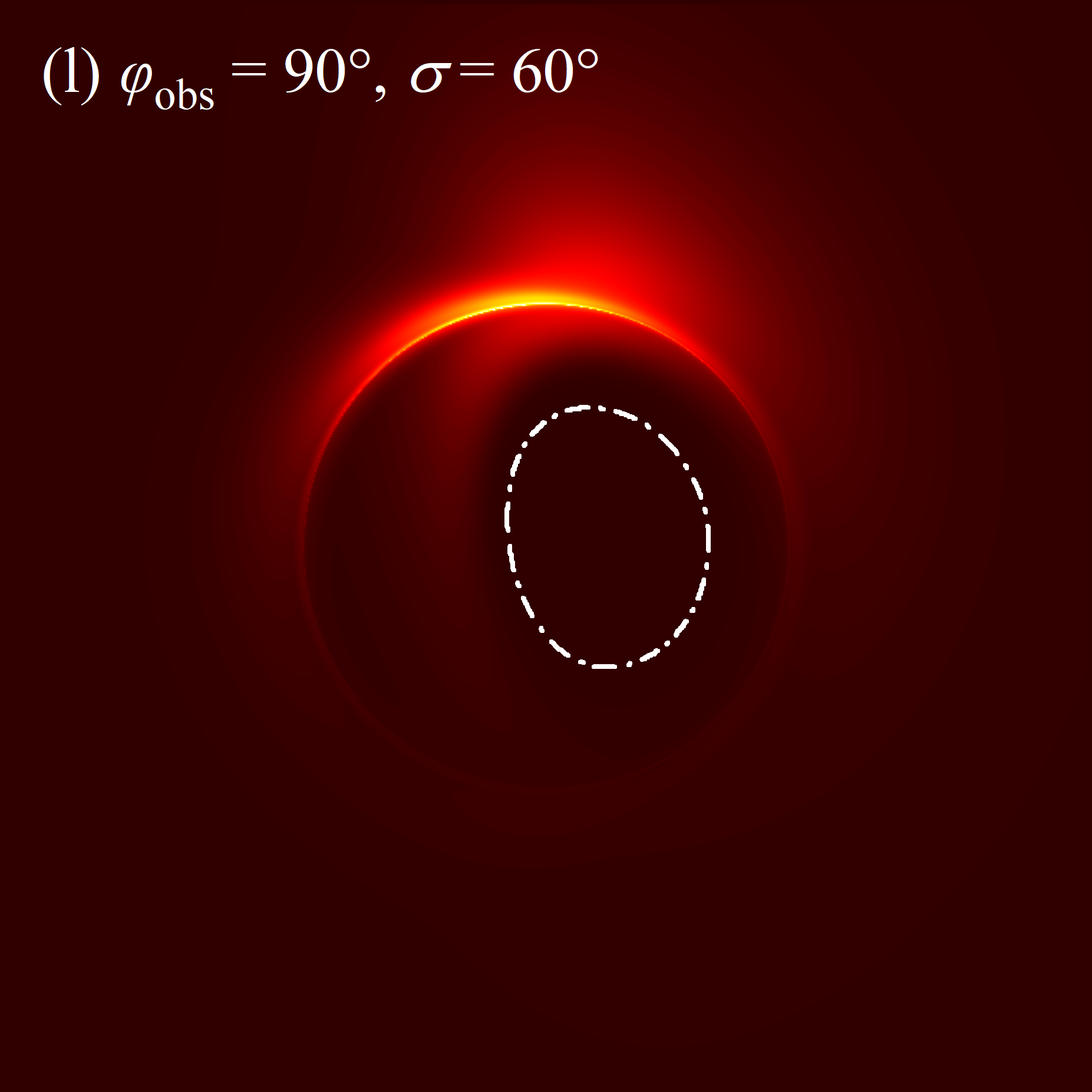}
\includegraphics[width=3.5cm]{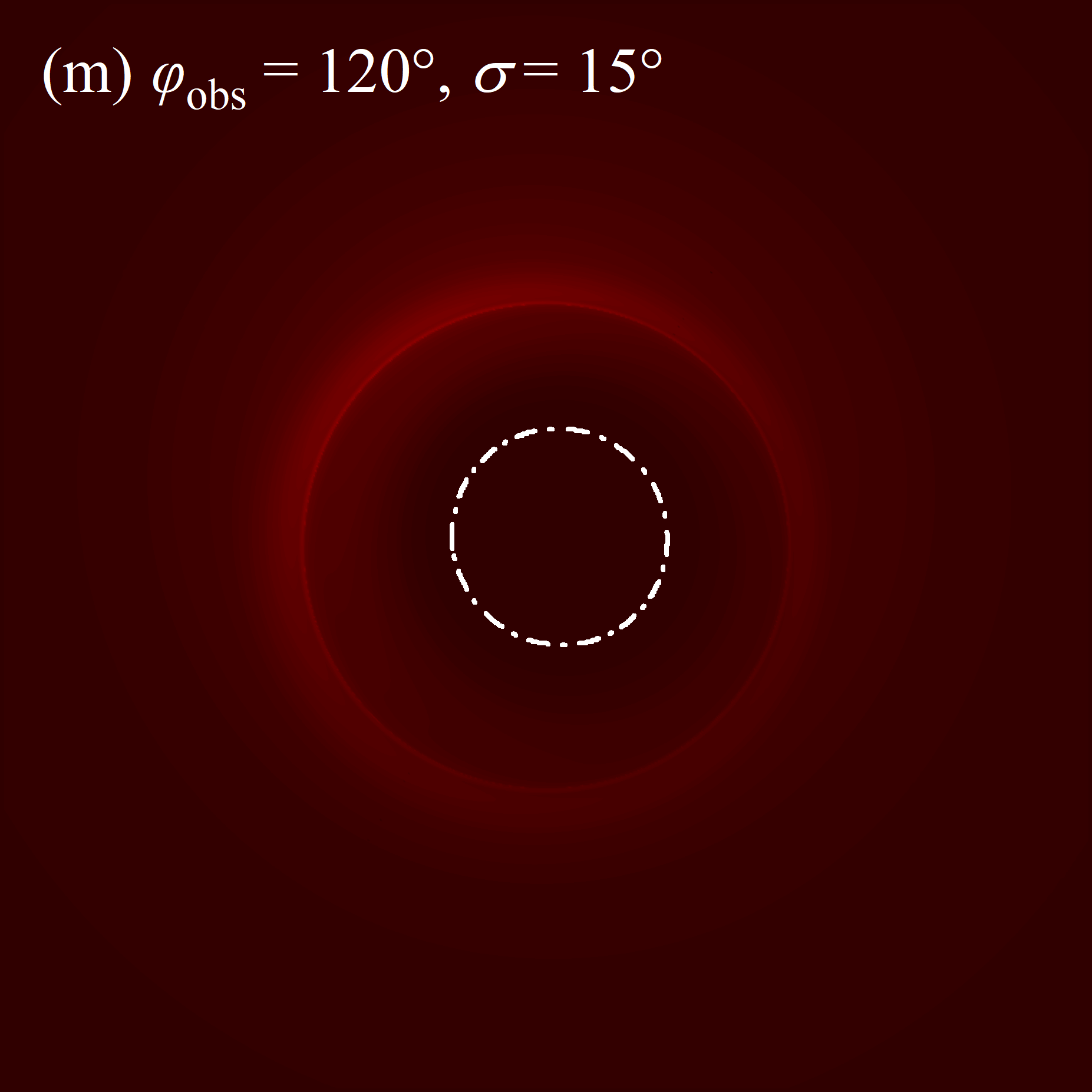}
\includegraphics[width=3.5cm]{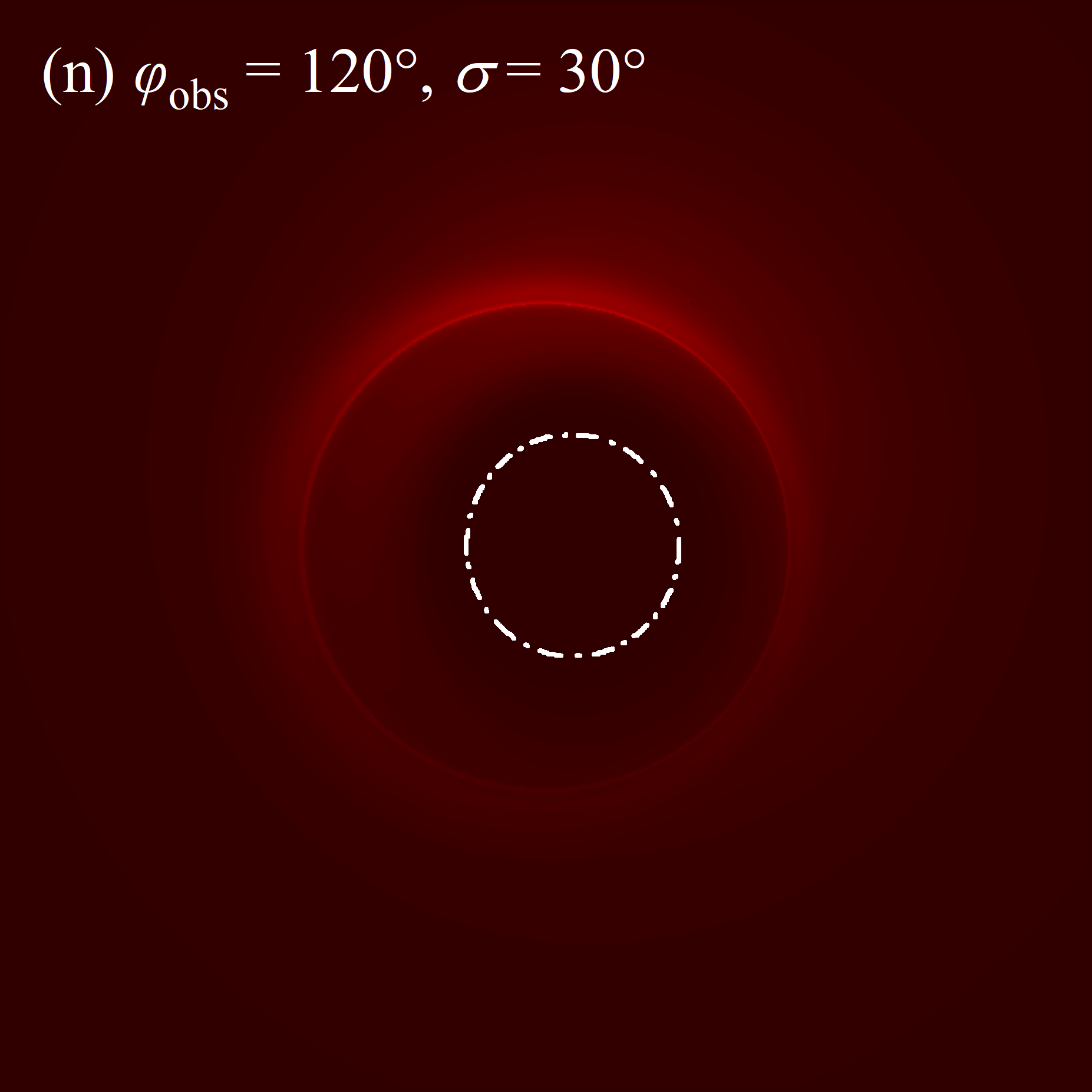}
\includegraphics[width=3.5cm]{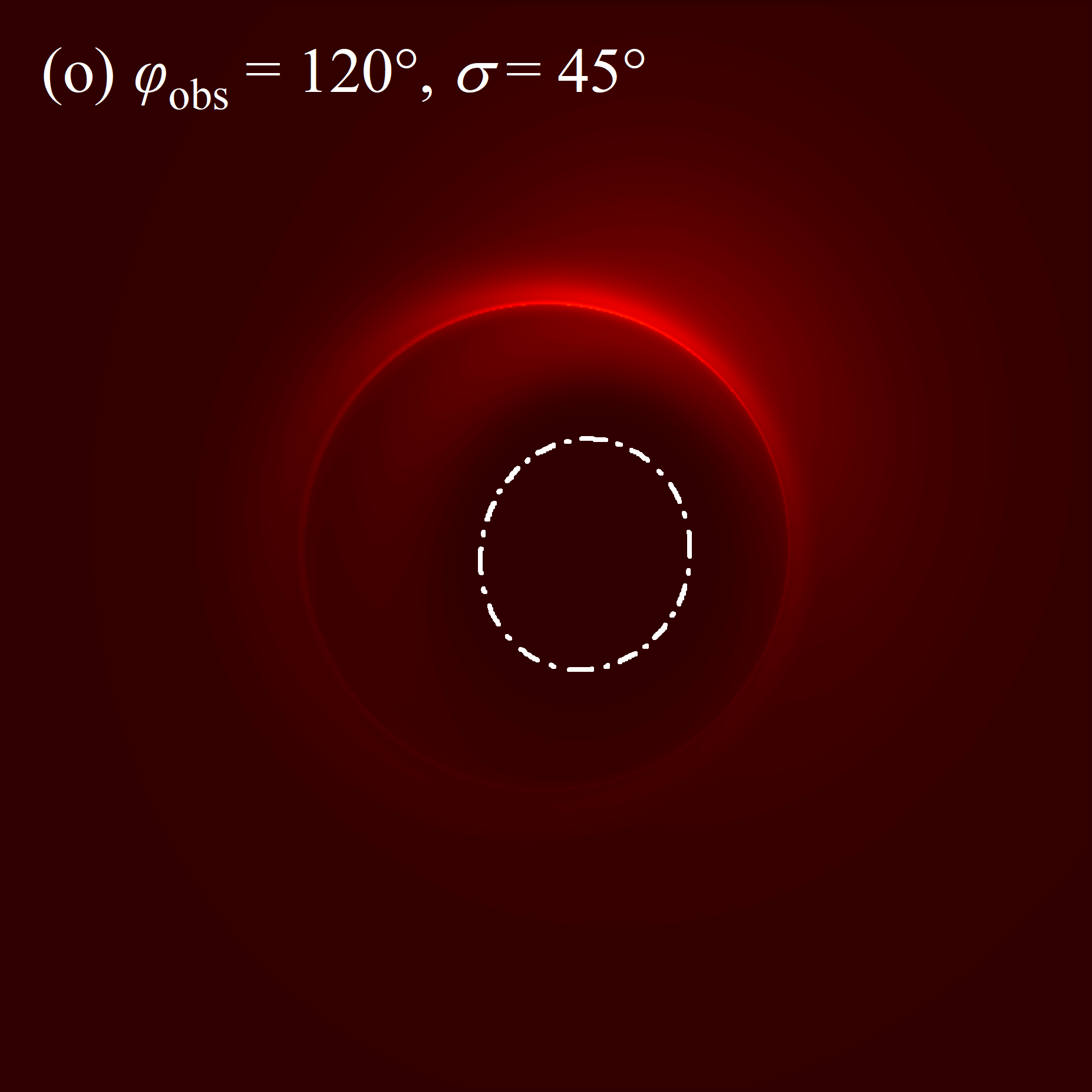}
\includegraphics[width=3.5cm]{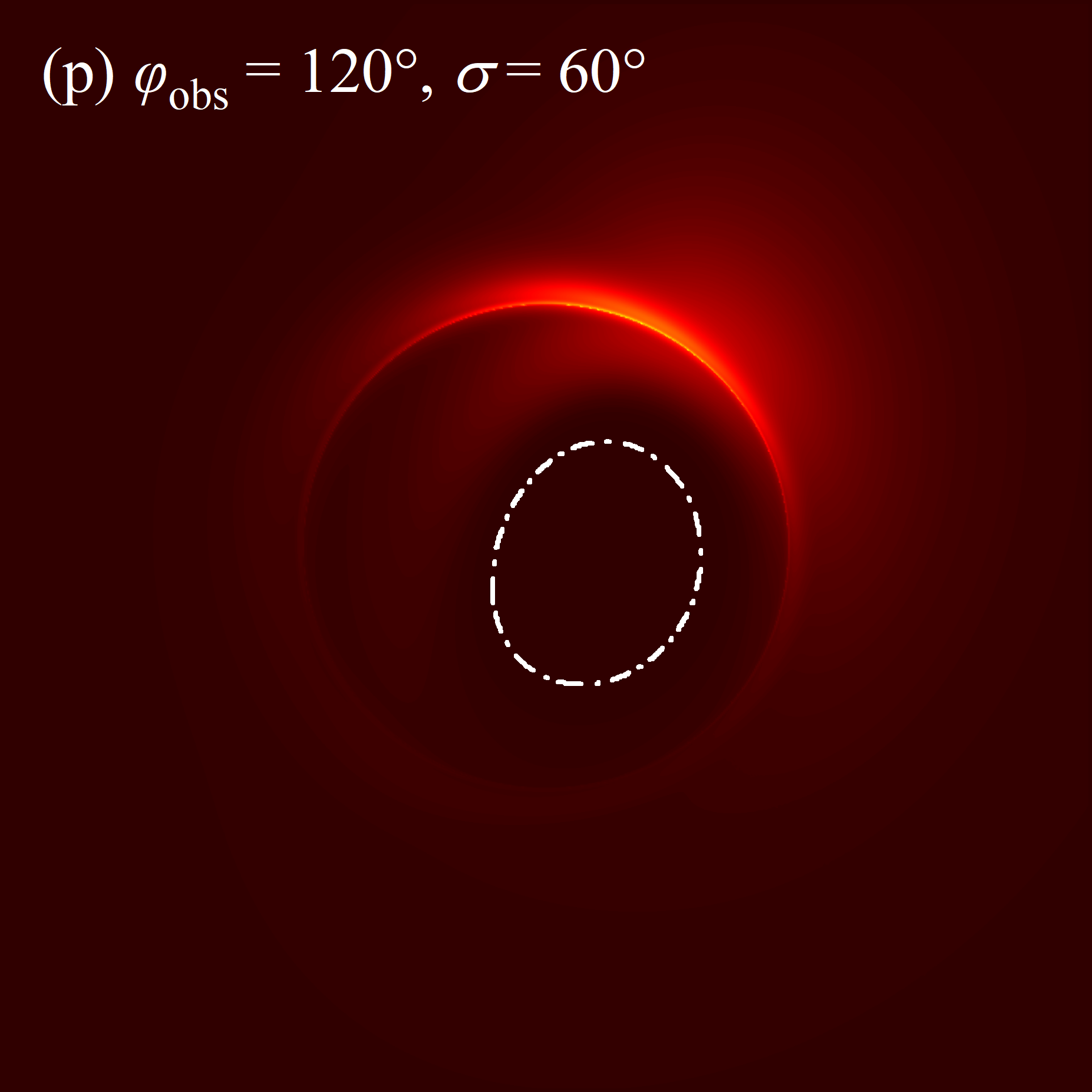}
\includegraphics[width=3.5cm]{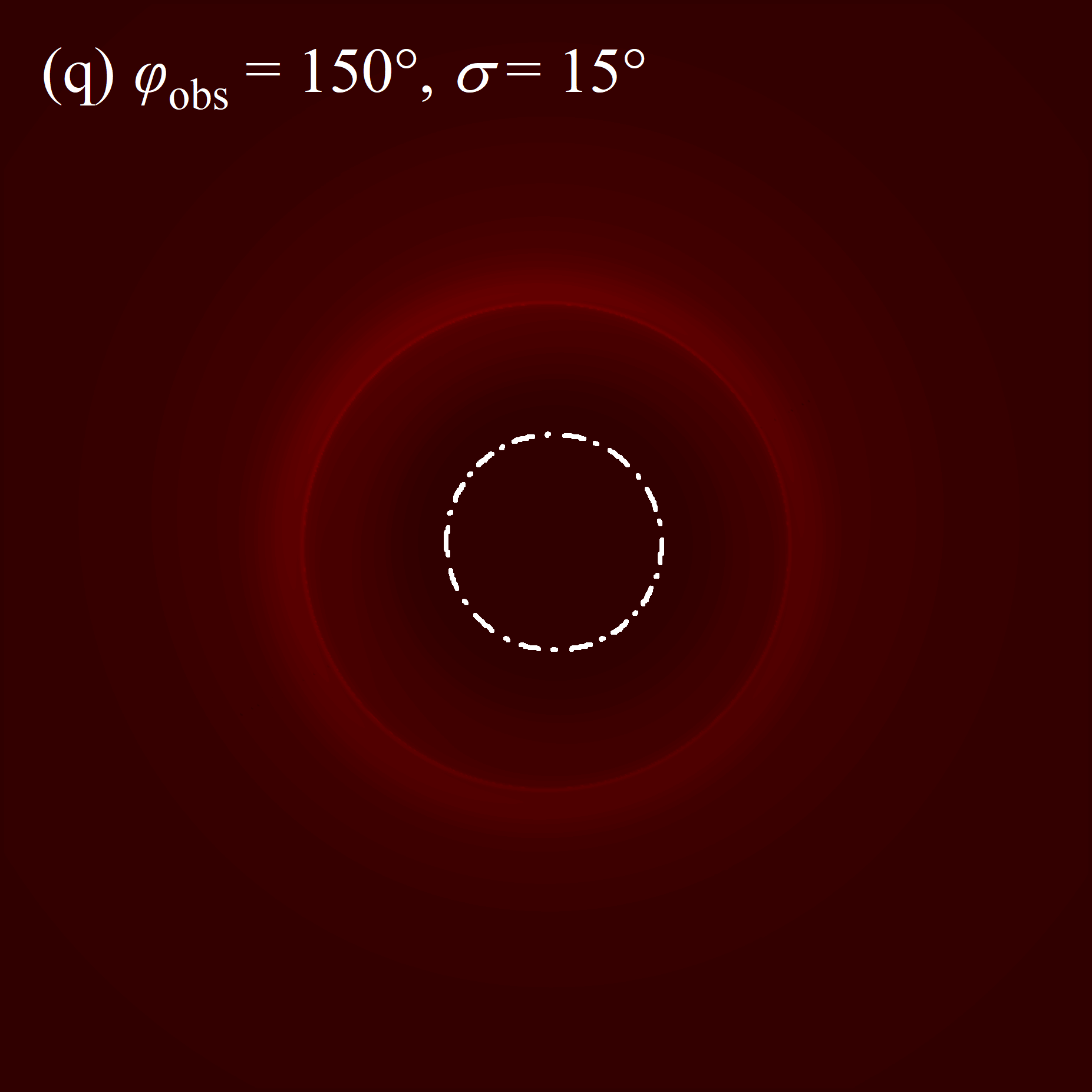}
\includegraphics[width=3.5cm]{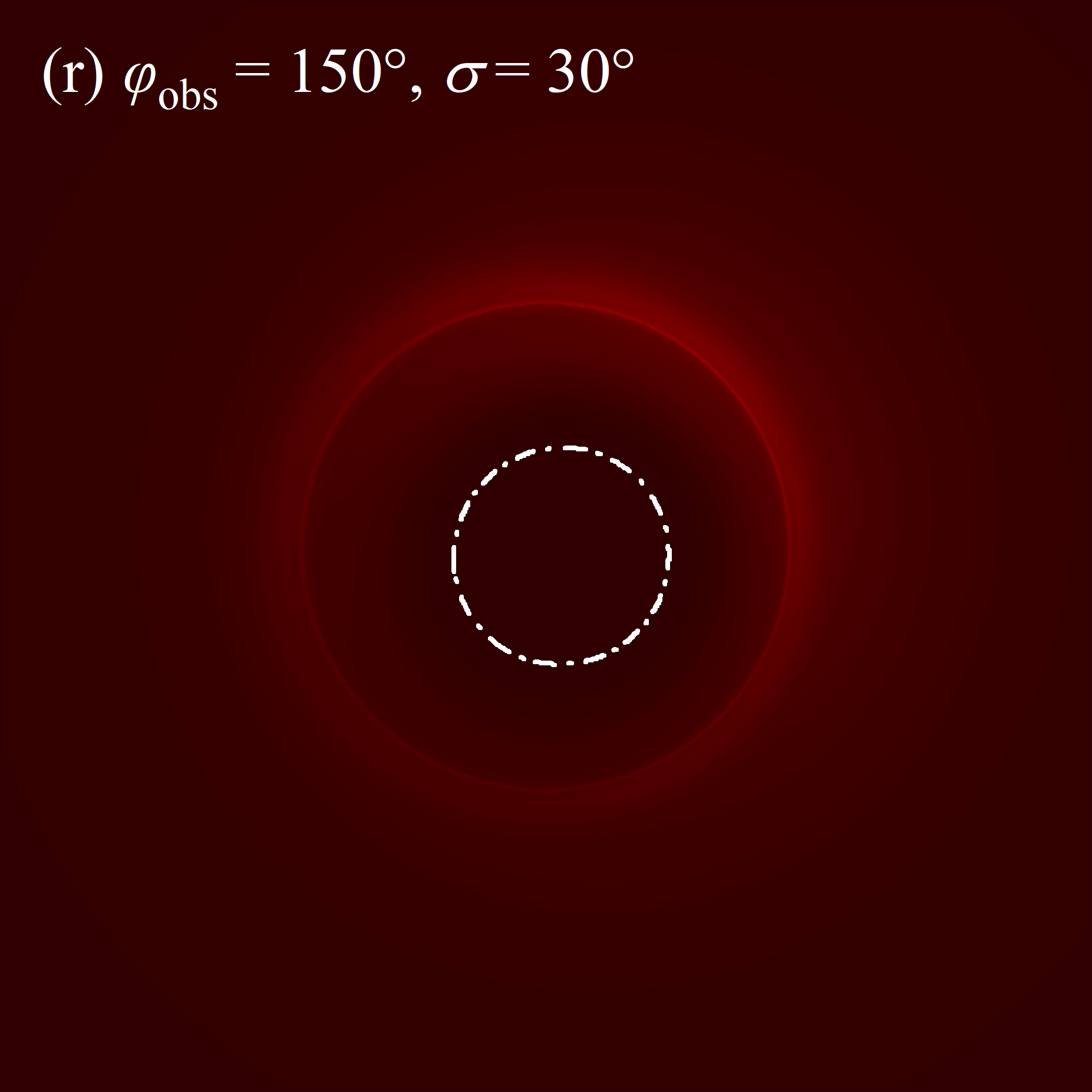}
\includegraphics[width=3.5cm]{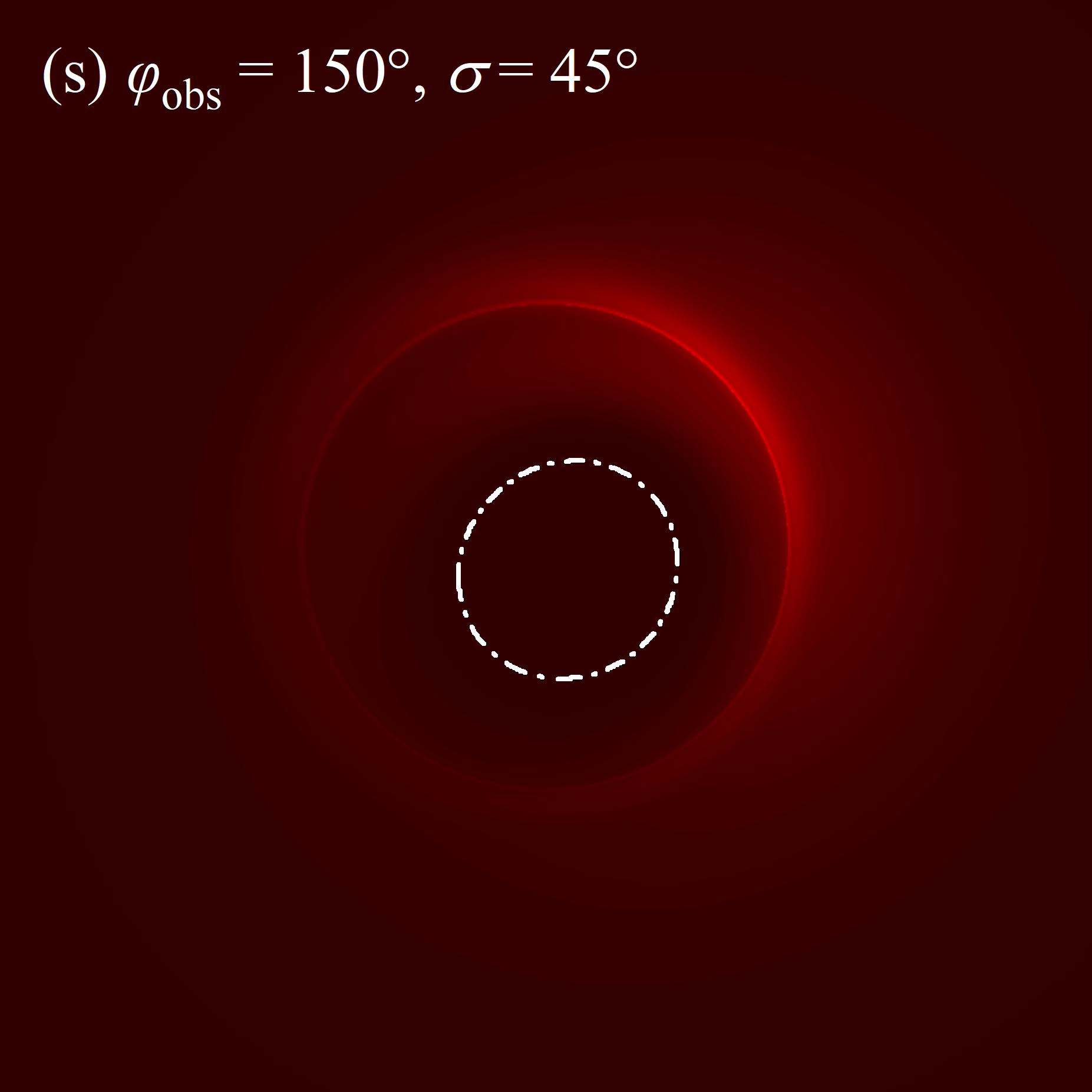}
\includegraphics[width=3.5cm]{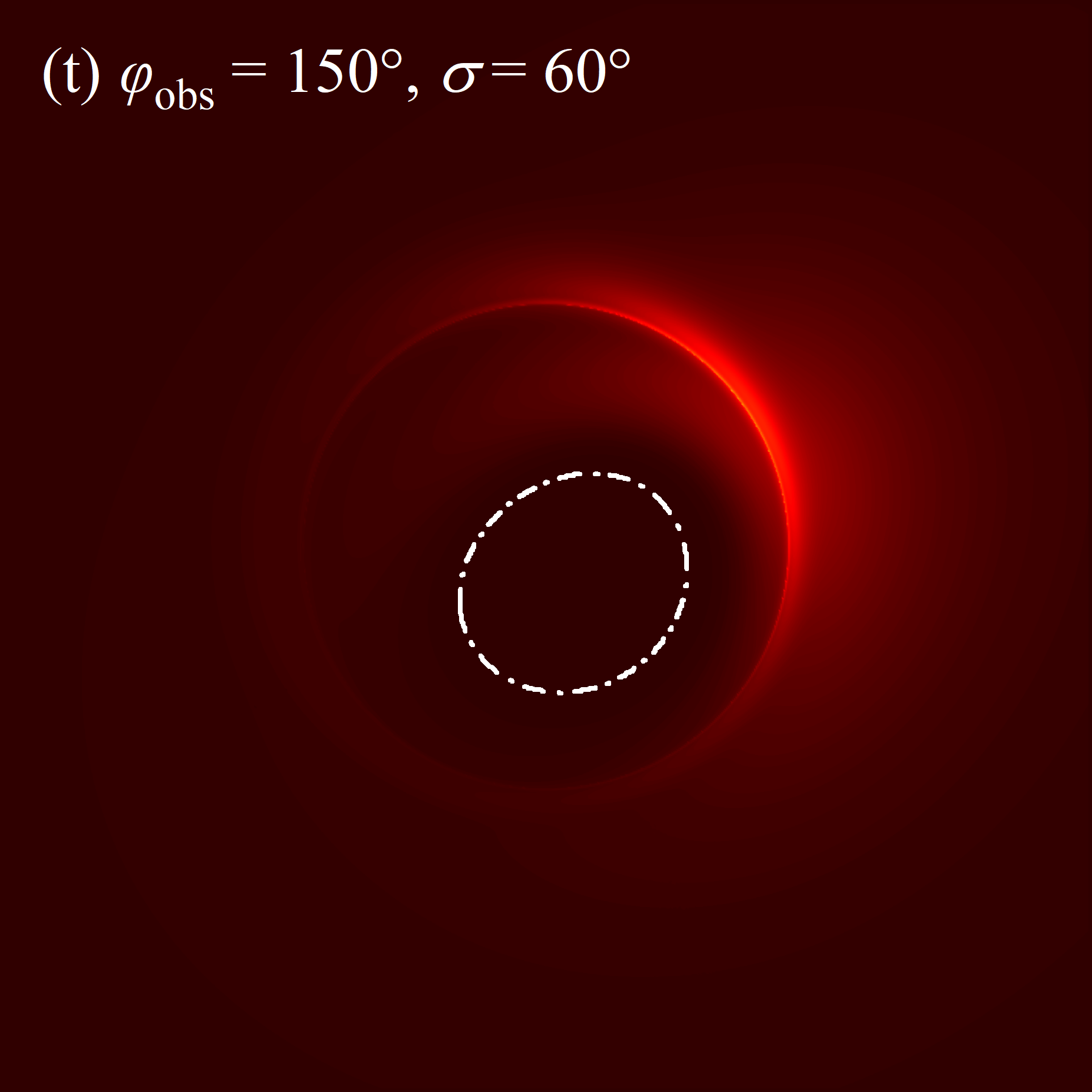}
\includegraphics[width=3.5cm]{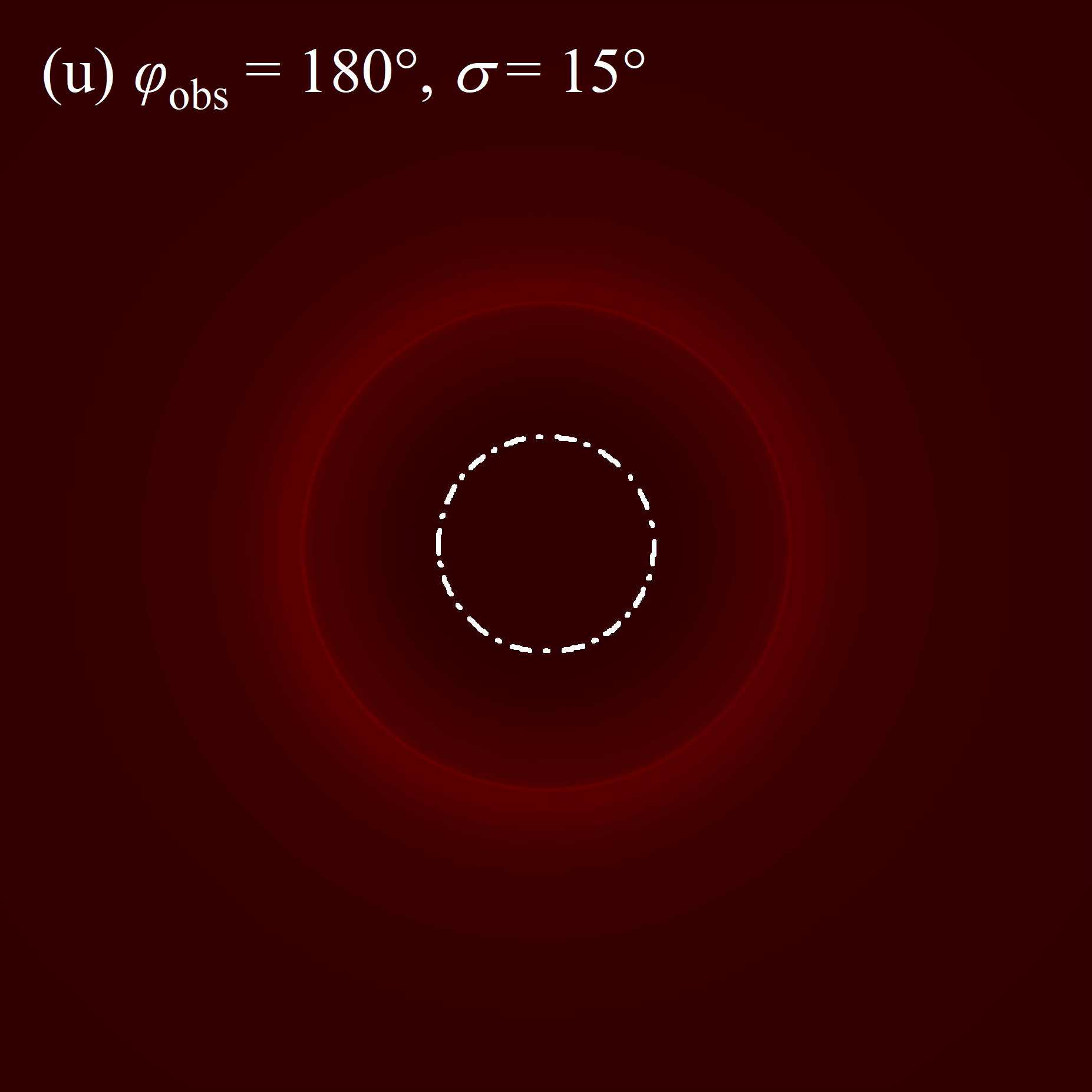}
\includegraphics[width=3.5cm]{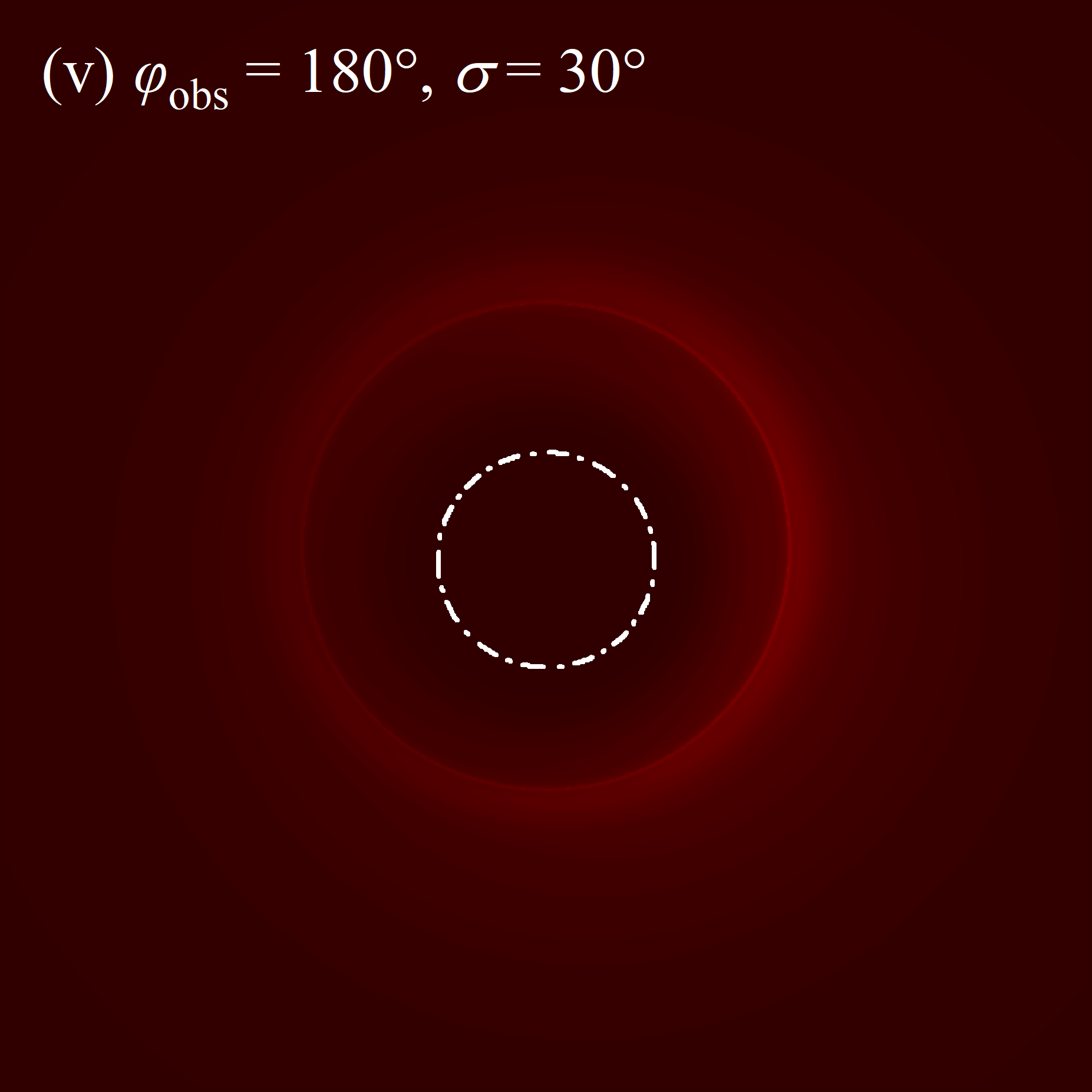}
\includegraphics[width=3.5cm]{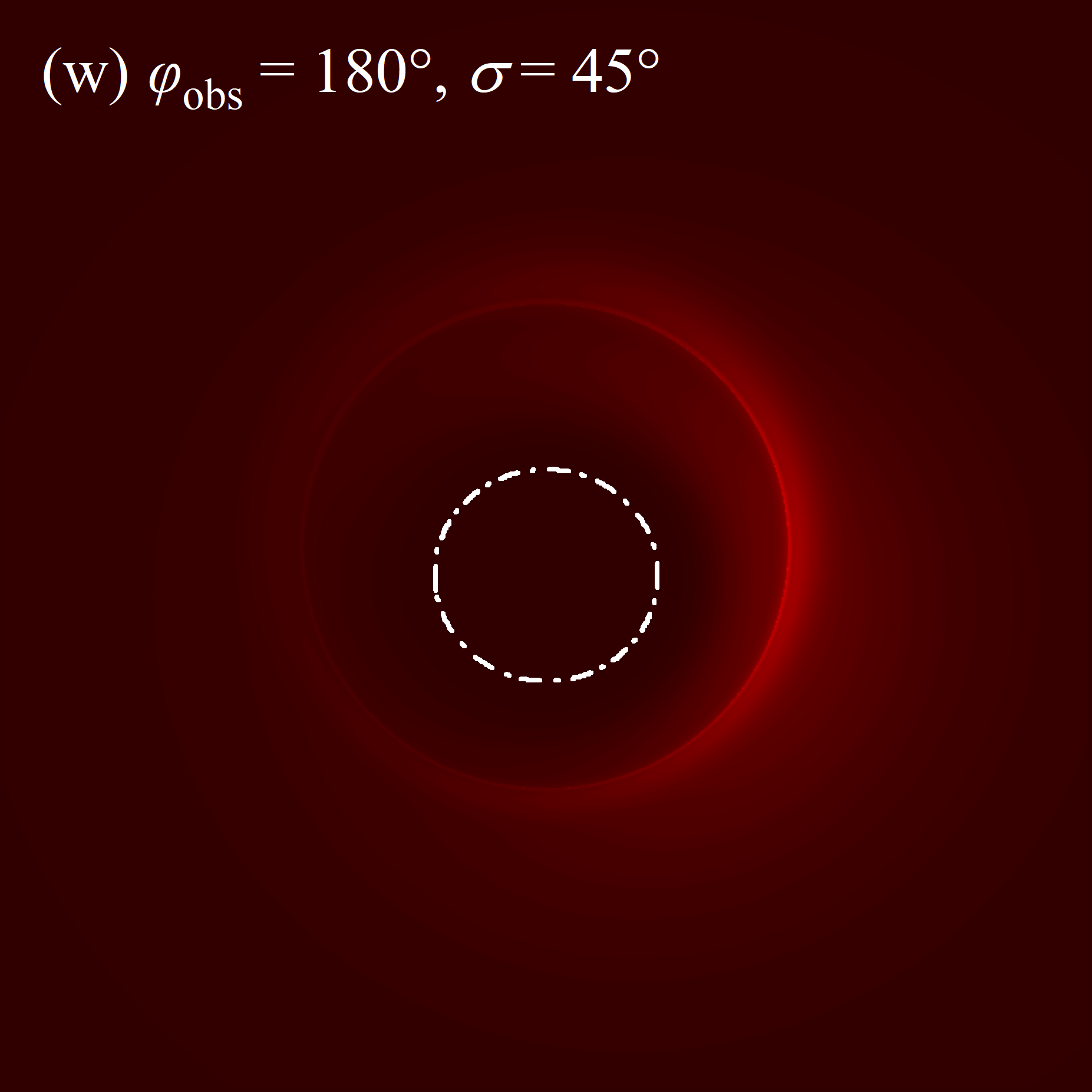}
\includegraphics[width=3.5cm]{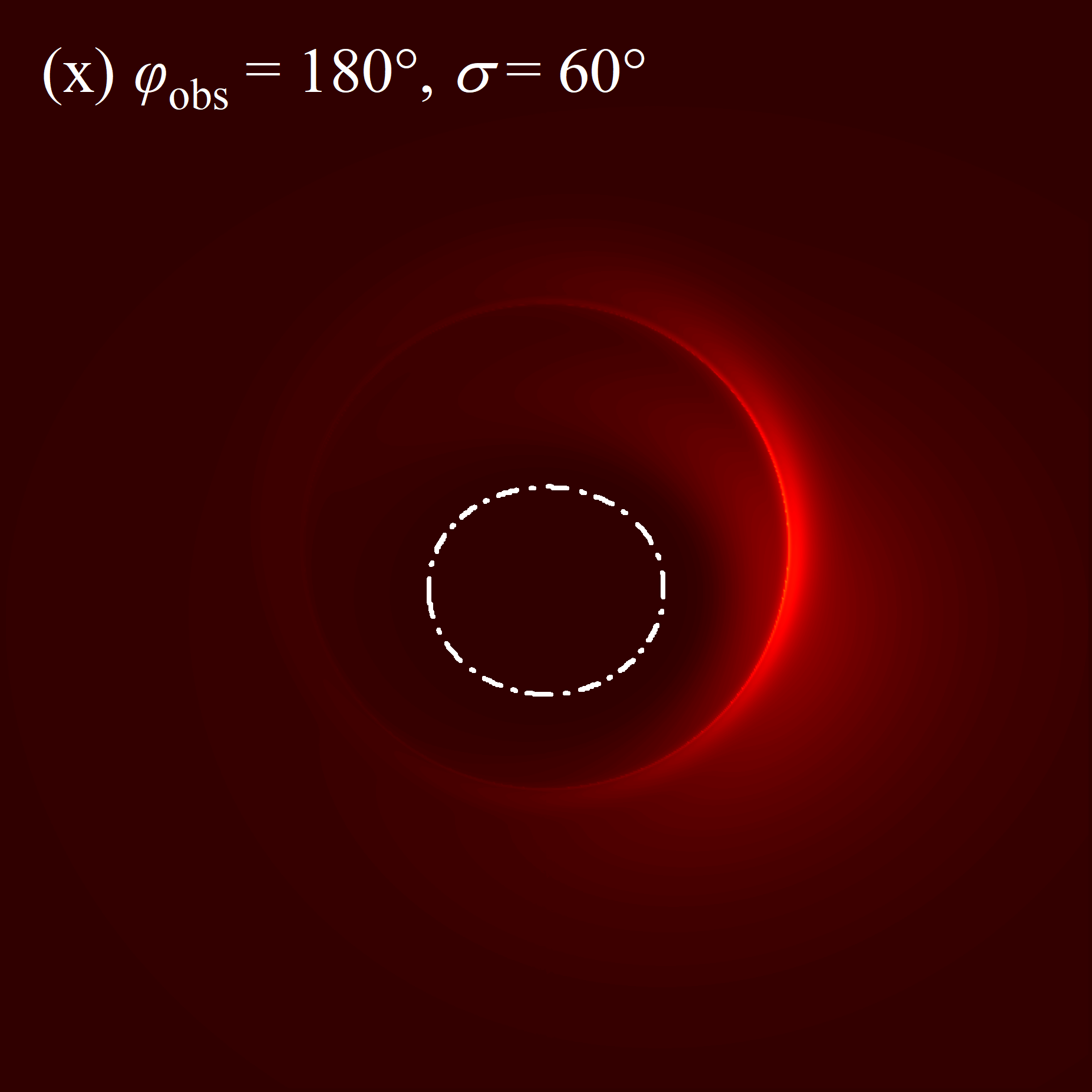}
\caption{Images of the hairy BH illuminated by different tilted thin accretion disks, as viewed from varying observation azimuths $\varphi_{\textrm{obs}}$. It is demonstrated that the positions of the inner shadow (white dashed line), as well as the flare, are changed due to the introduction of non-zero $\varphi_{\textrm{obs}}$. Here, we have the scalar hair parameter of $h = -1$ and the viewing angle of $\omega = 17^{\circ}$.}}\label{fig12}
\end{figure*}

\begin{figure*}
\center{
\includegraphics[width=3.5cm]{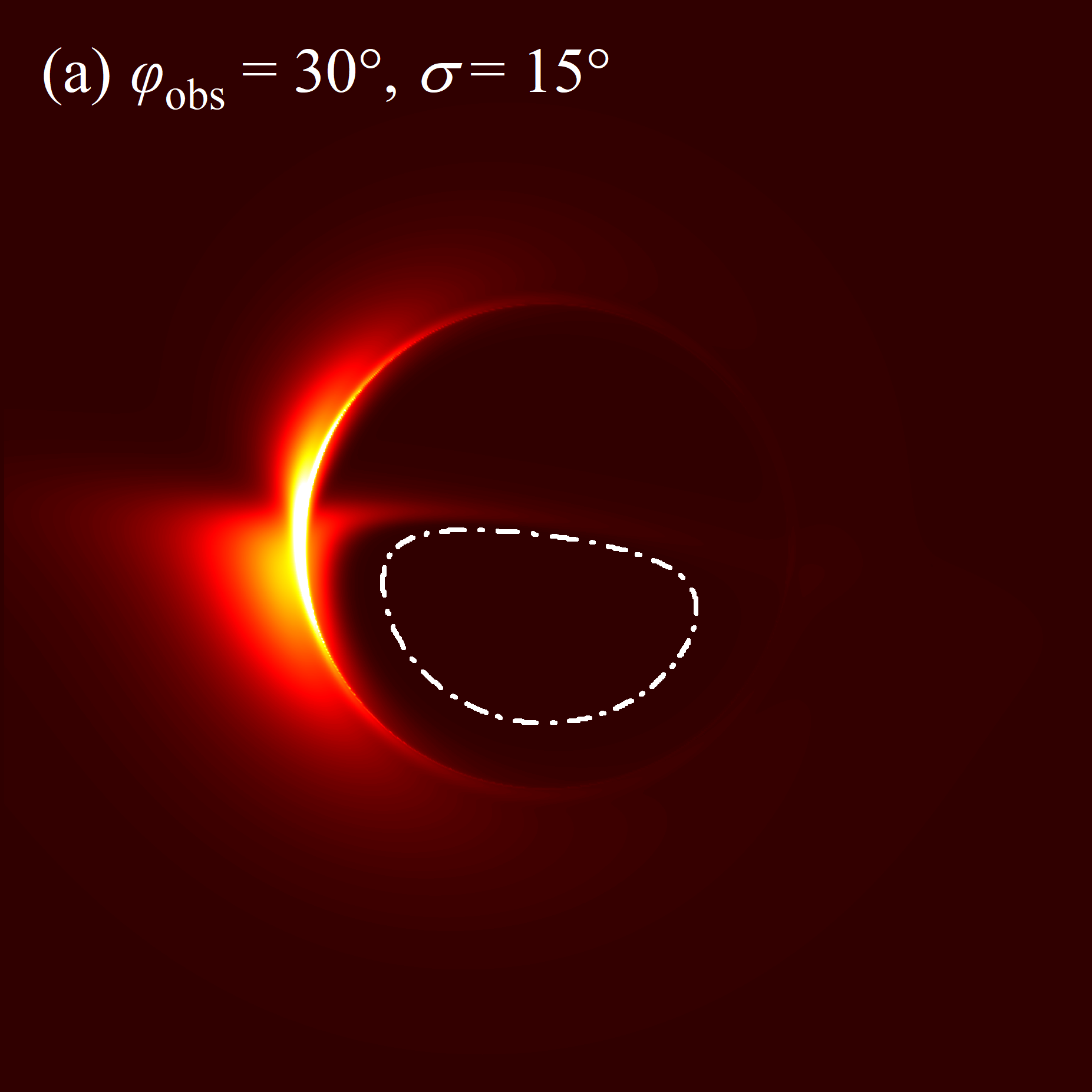}
\includegraphics[width=3.5cm]{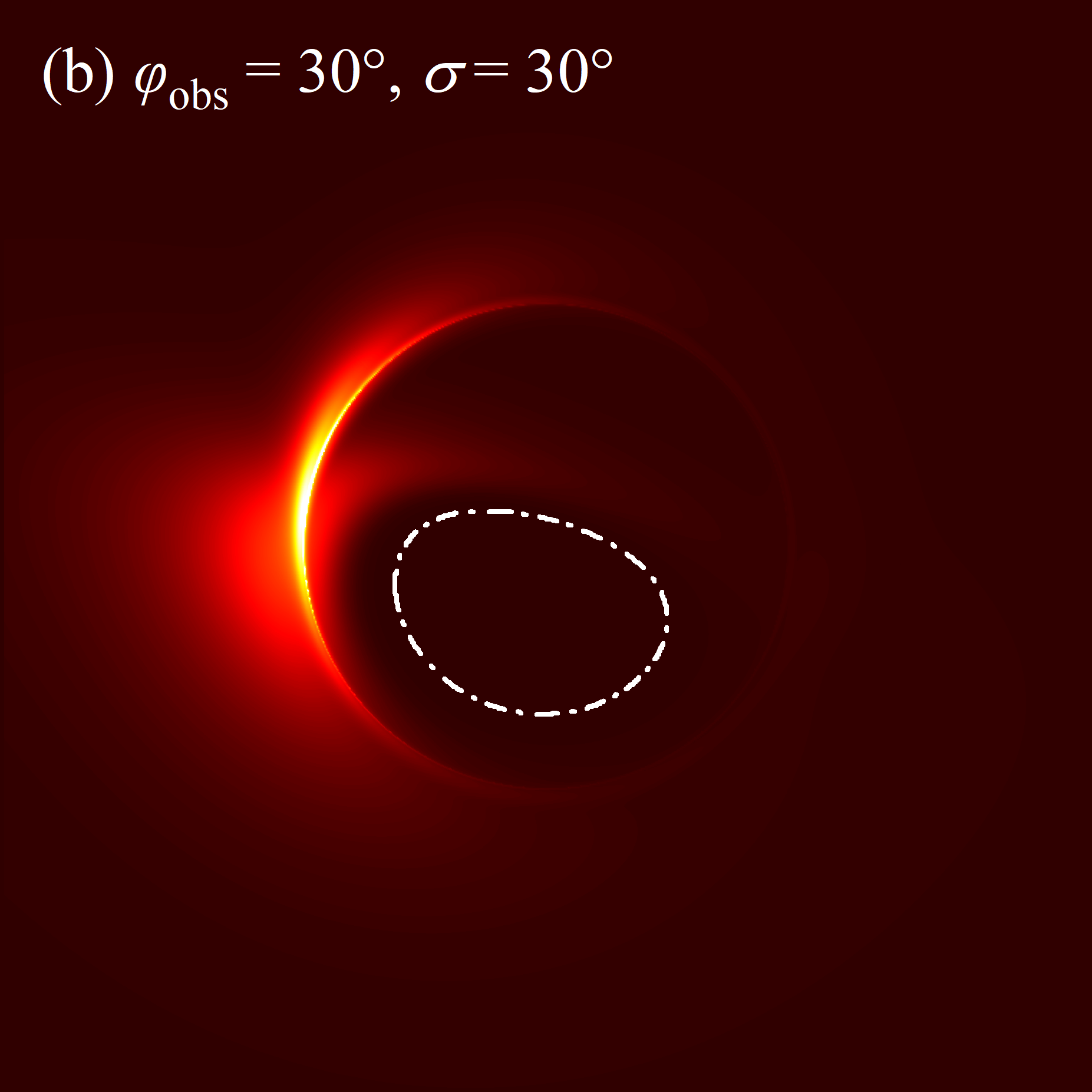}
\includegraphics[width=3.5cm]{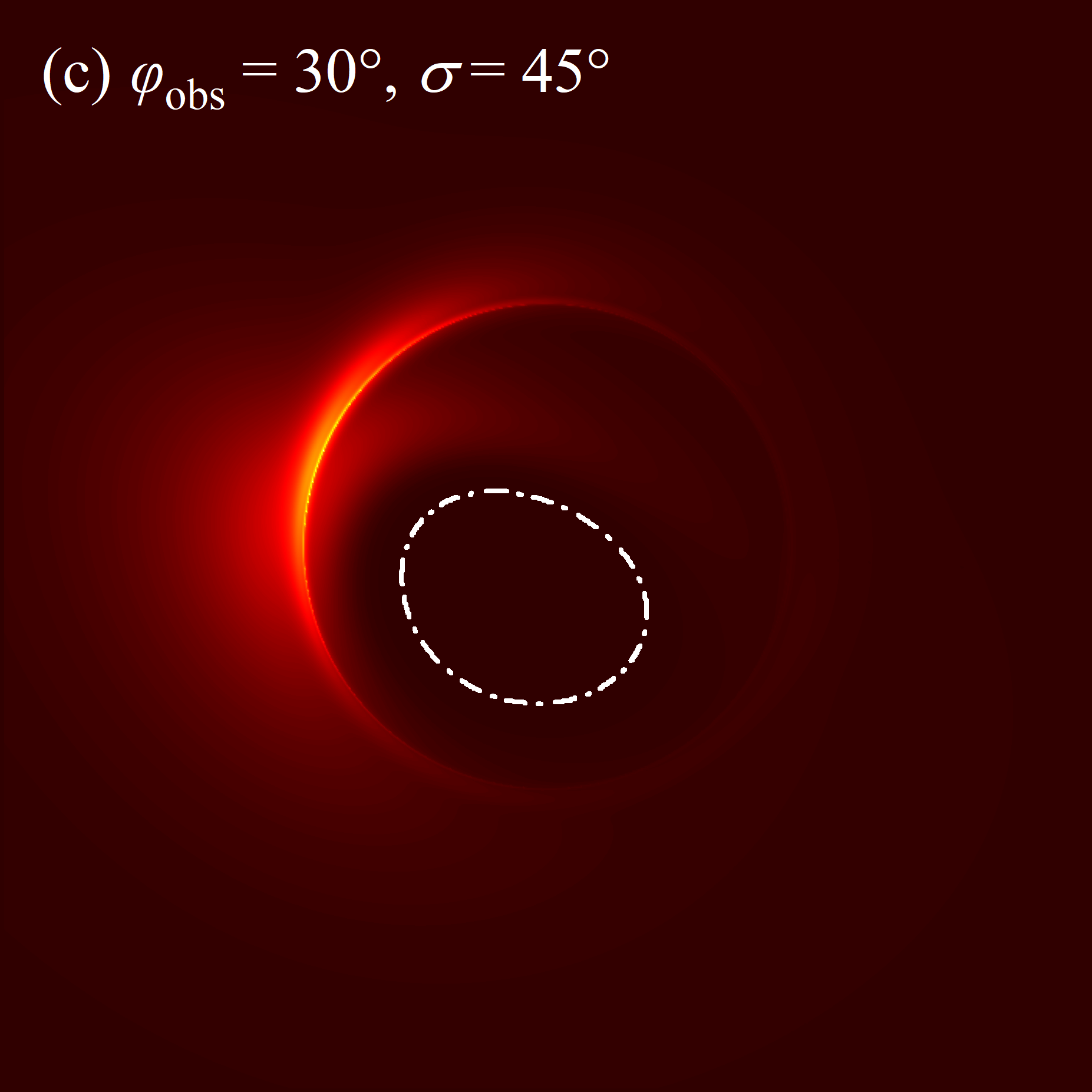}
\includegraphics[width=3.5cm]{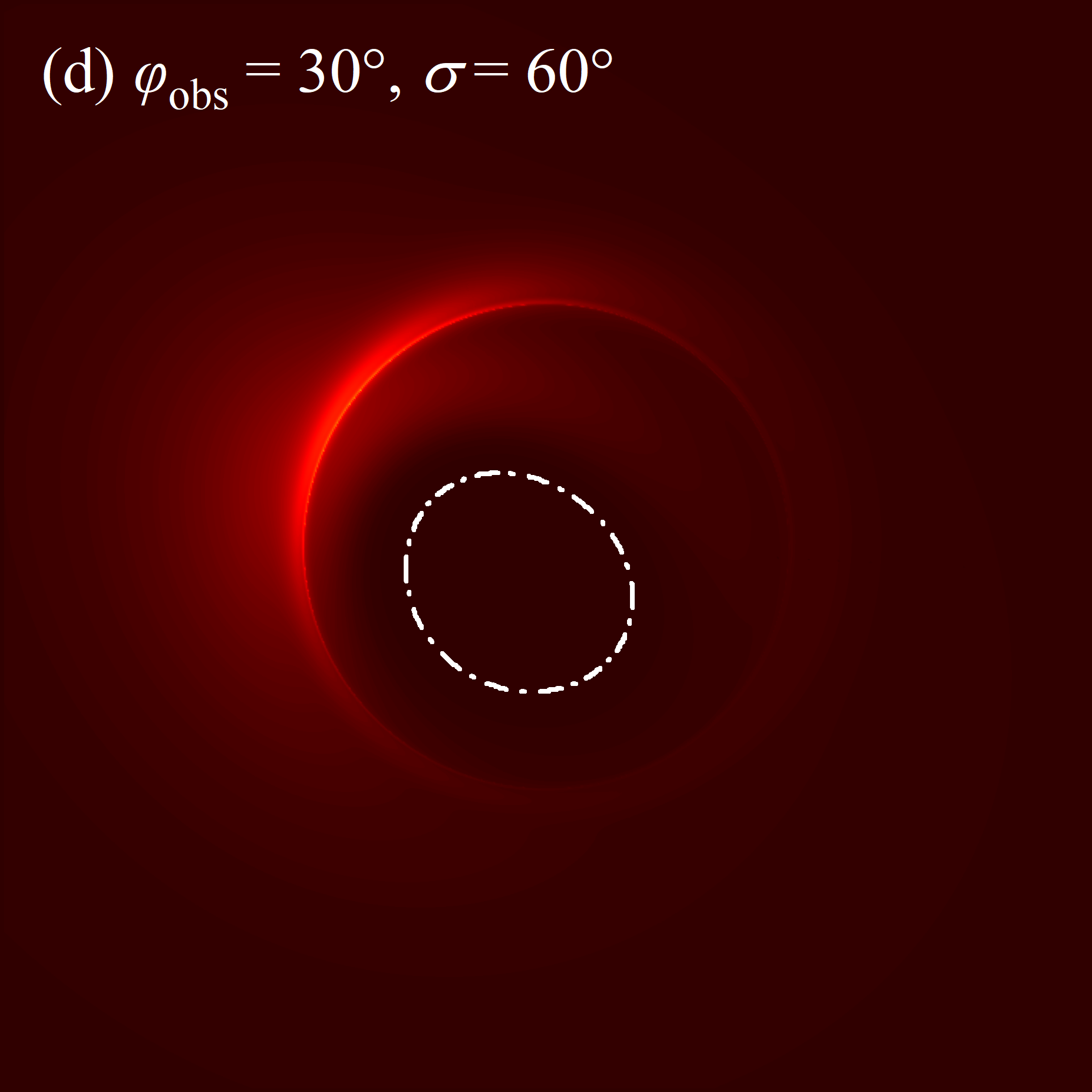}
\includegraphics[width=3.5cm]{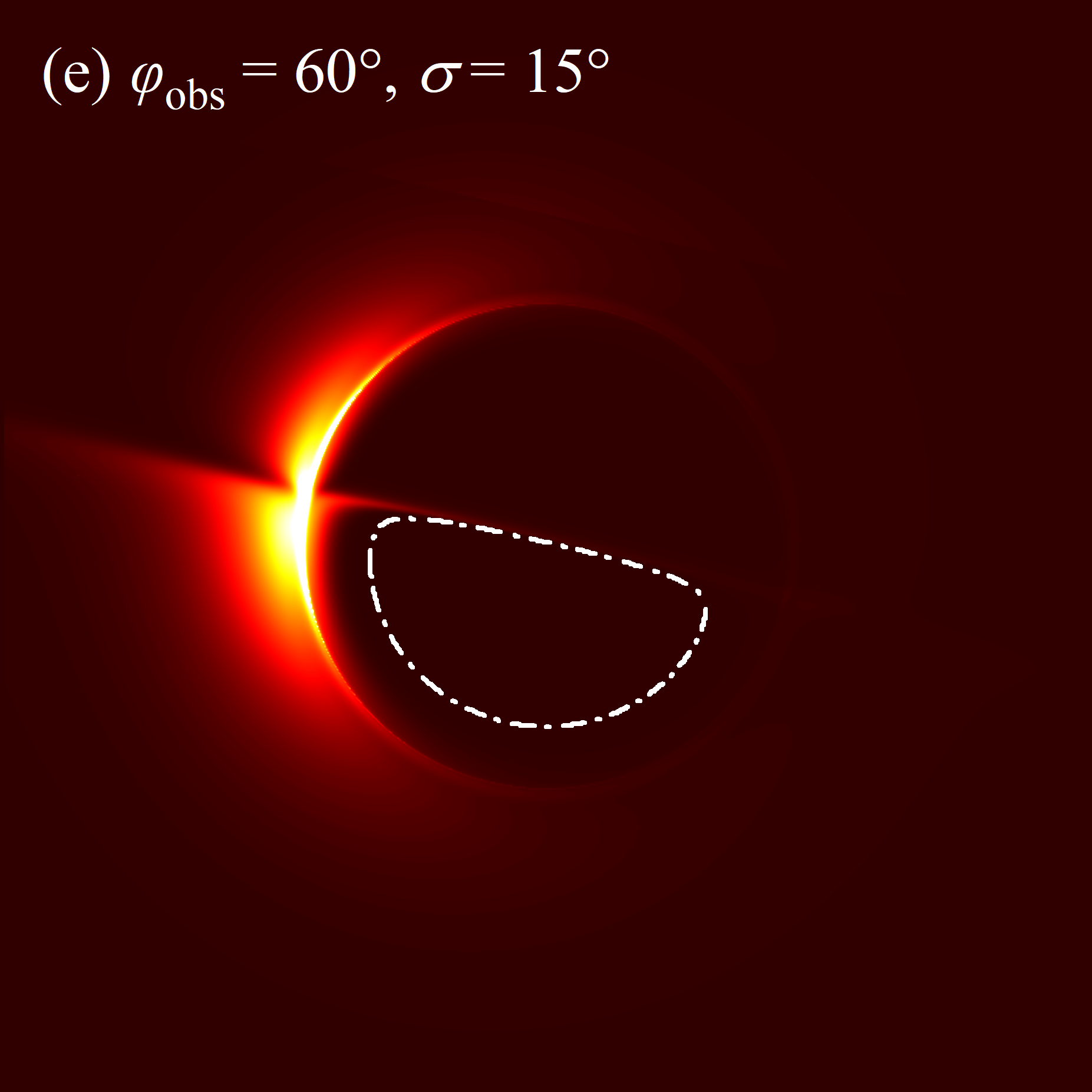}
\includegraphics[width=3.5cm]{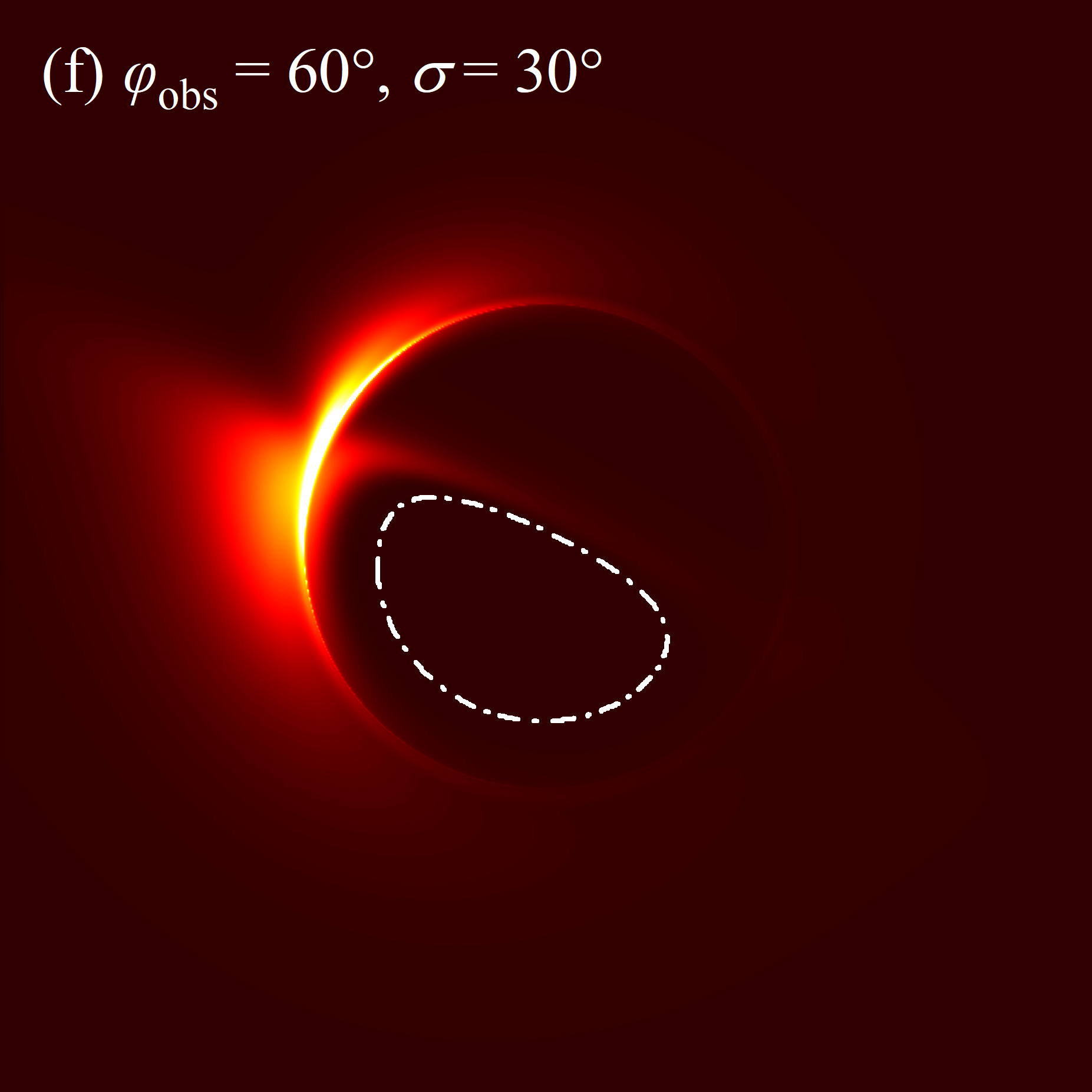}
\includegraphics[width=3.5cm]{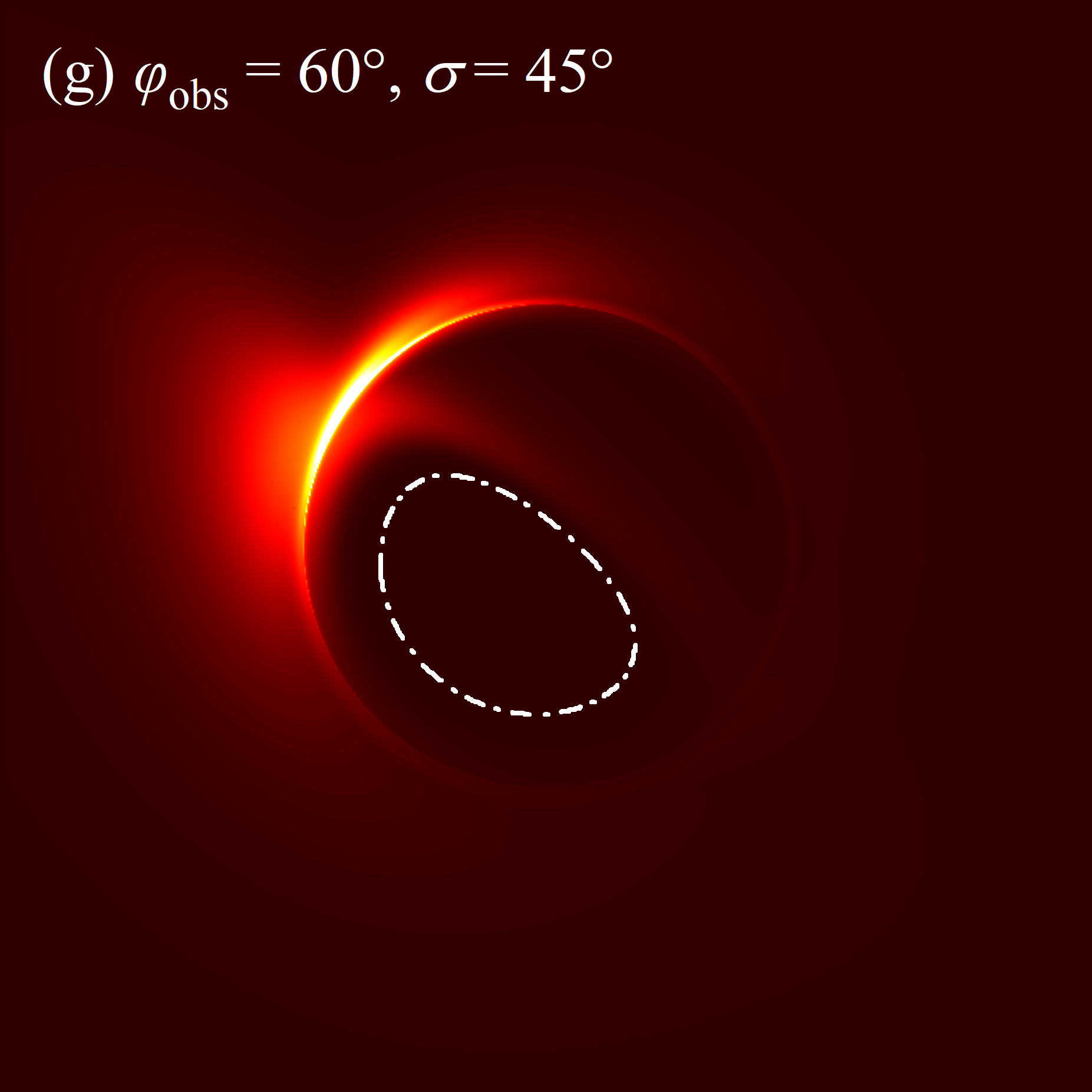}
\includegraphics[width=3.5cm]{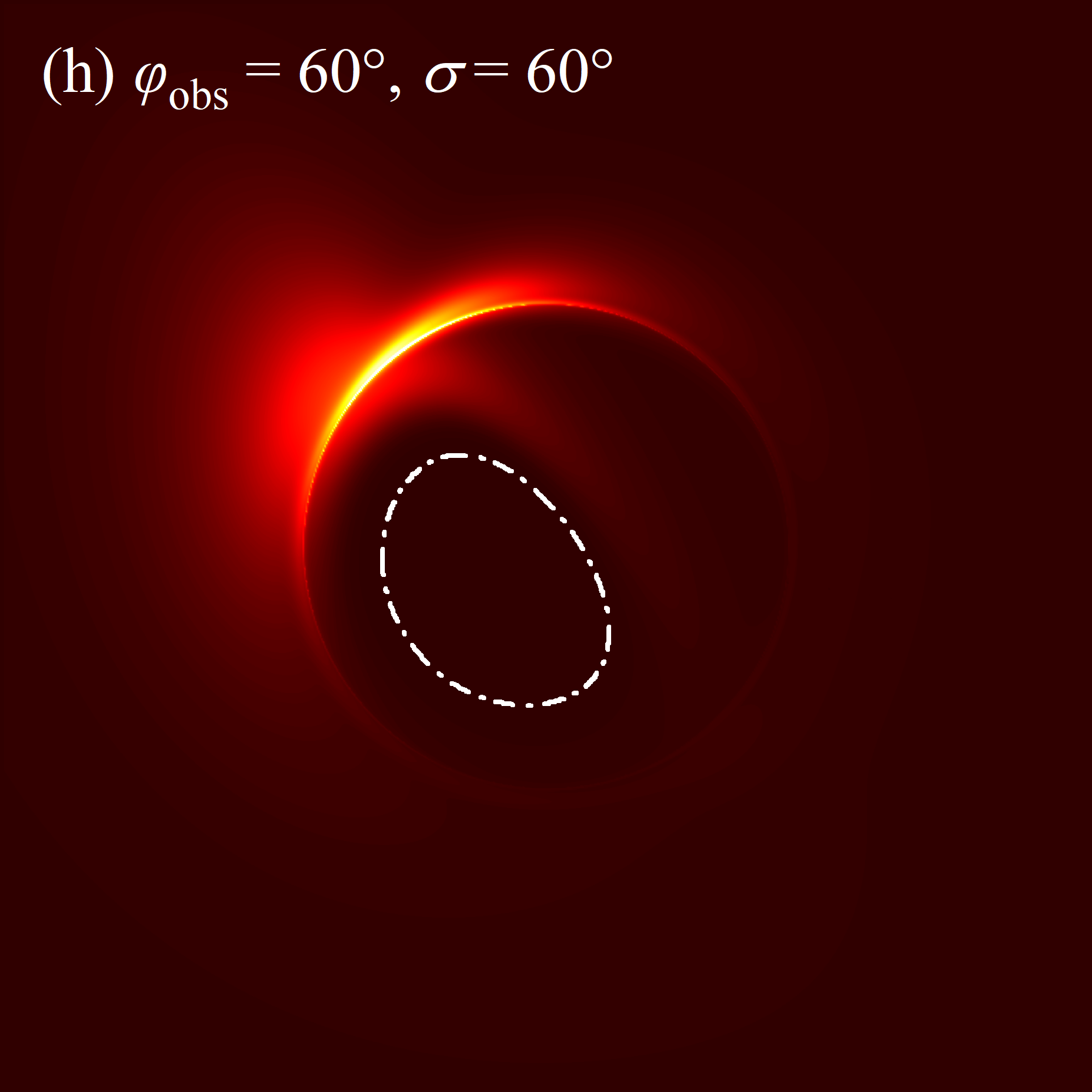}
\includegraphics[width=3.5cm]{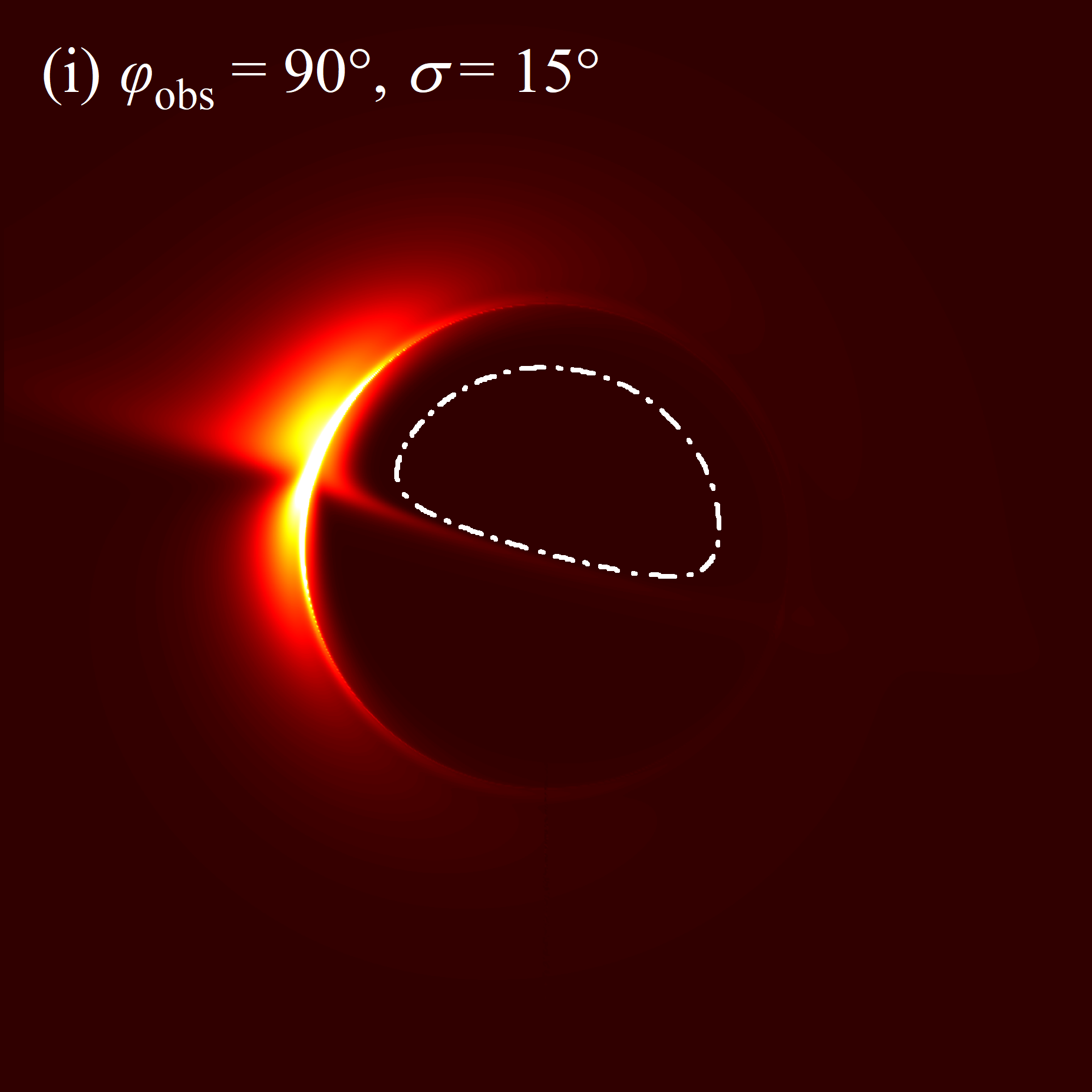}
\includegraphics[width=3.5cm]{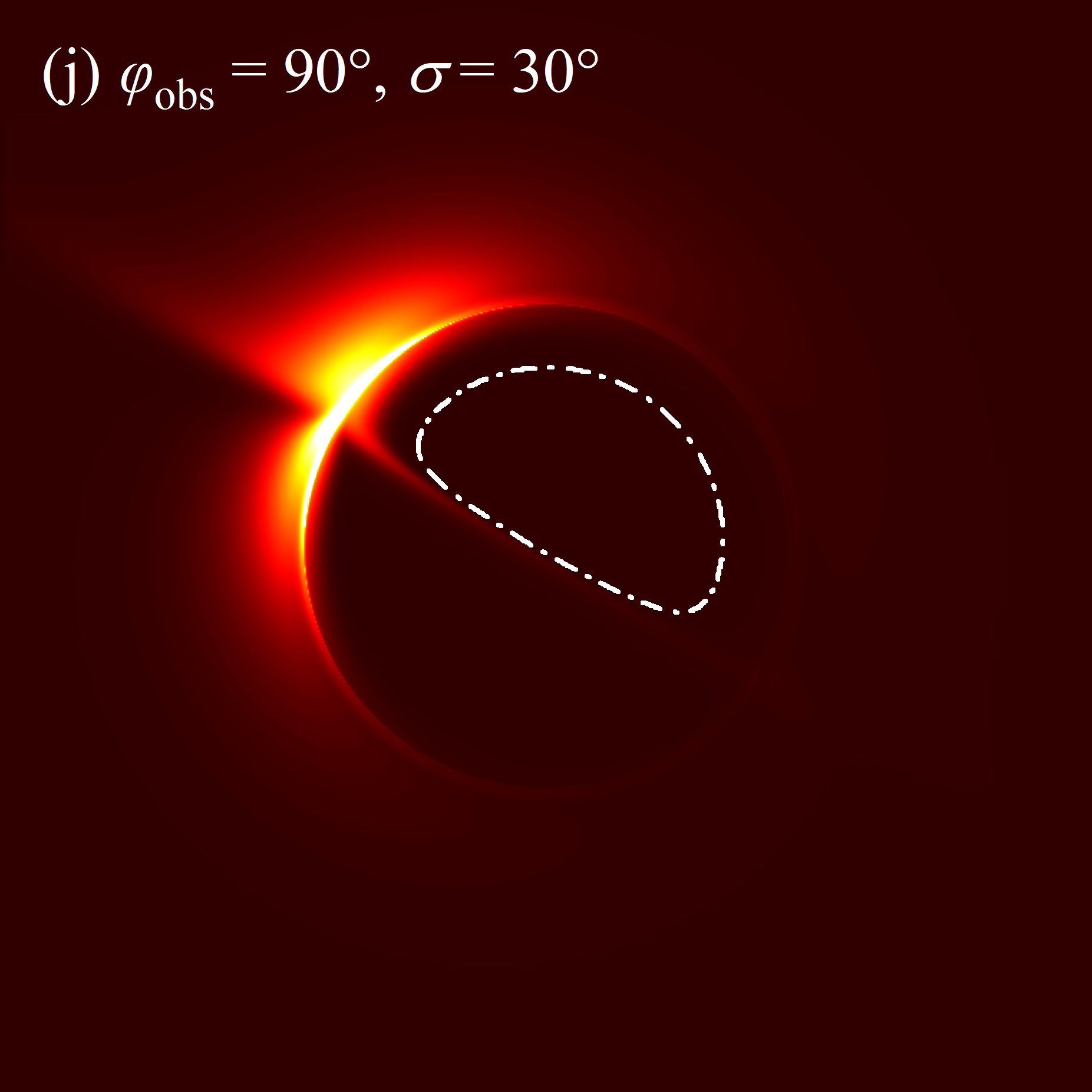}
\includegraphics[width=3.5cm]{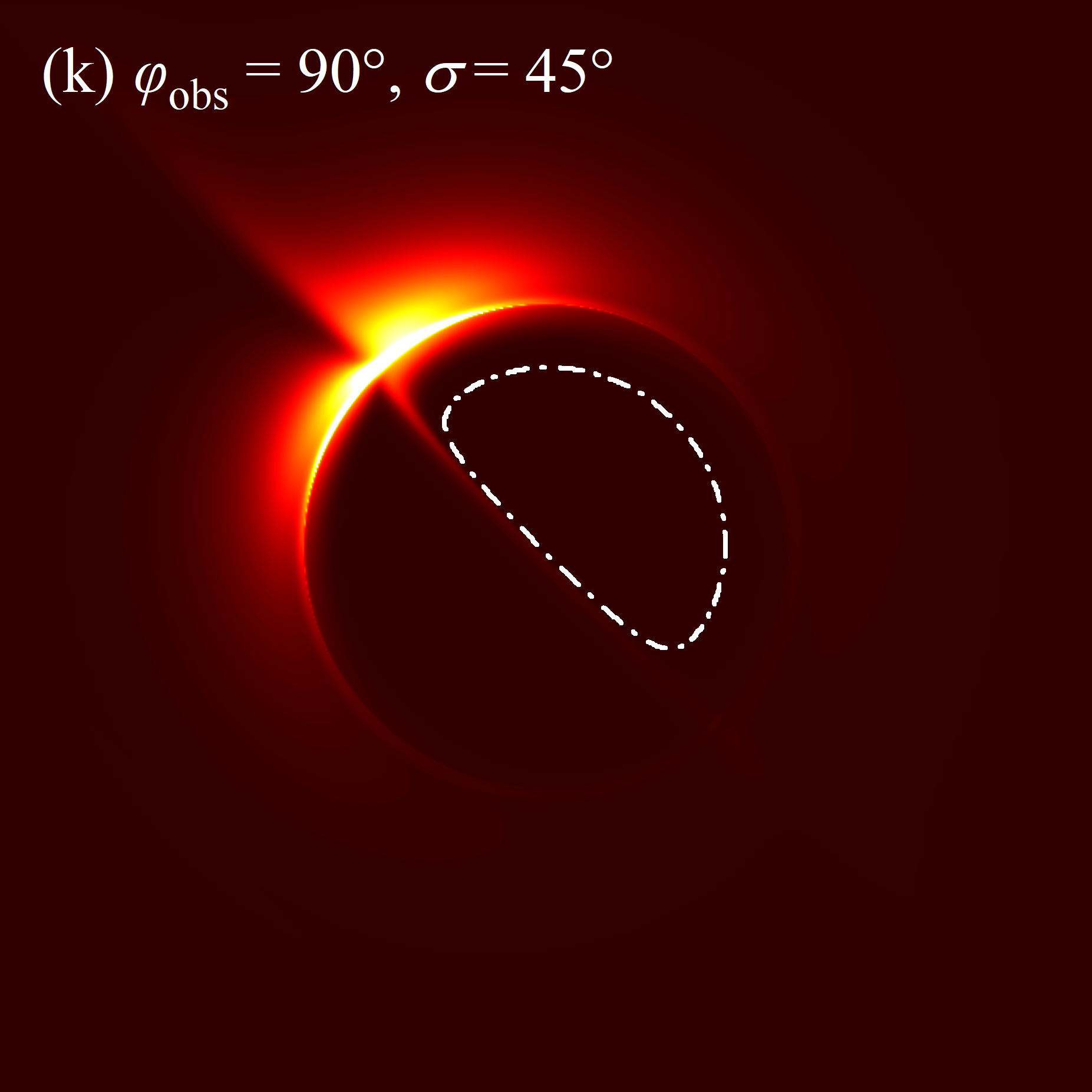}
\includegraphics[width=3.5cm]{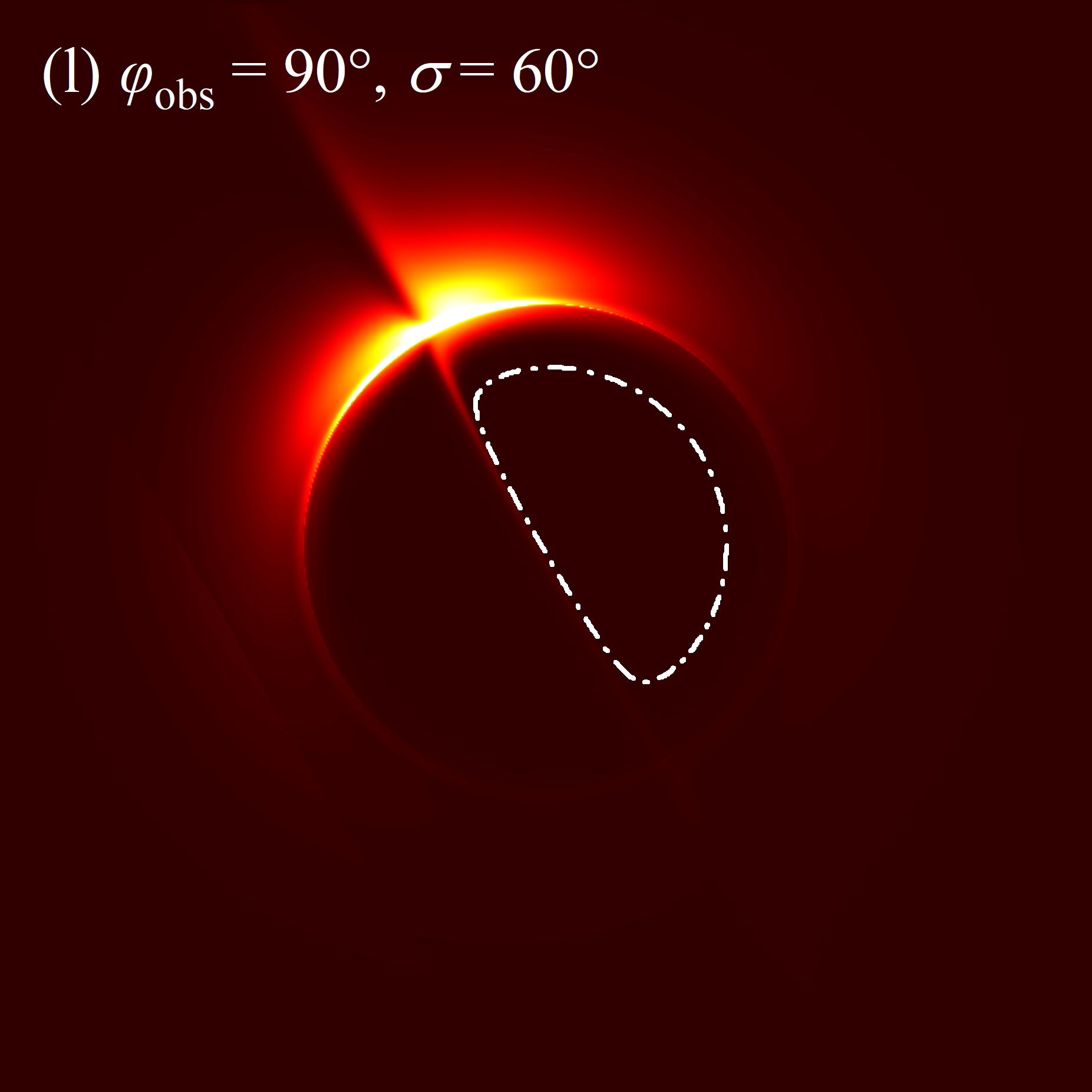}
\includegraphics[width=3.5cm]{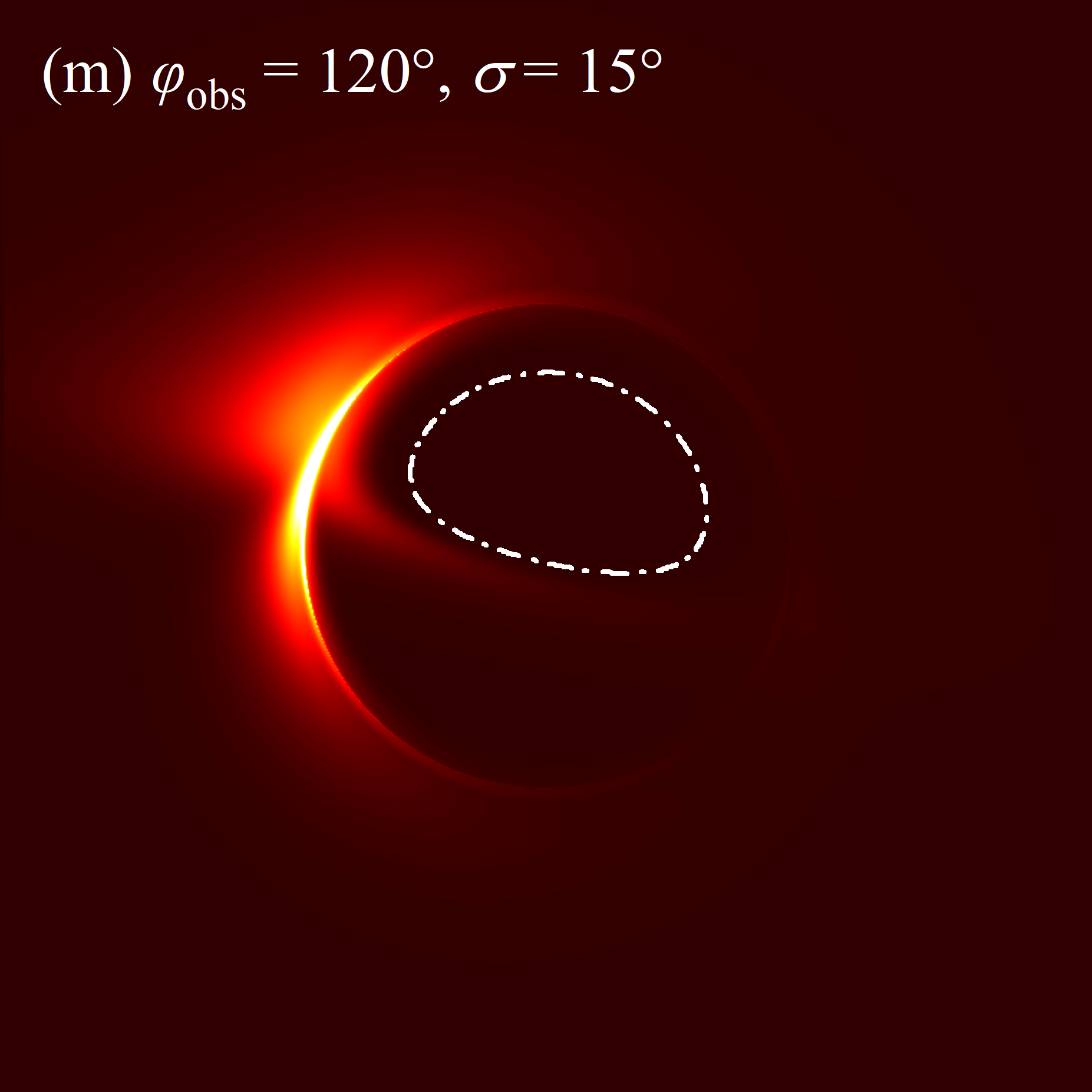}
\includegraphics[width=3.5cm]{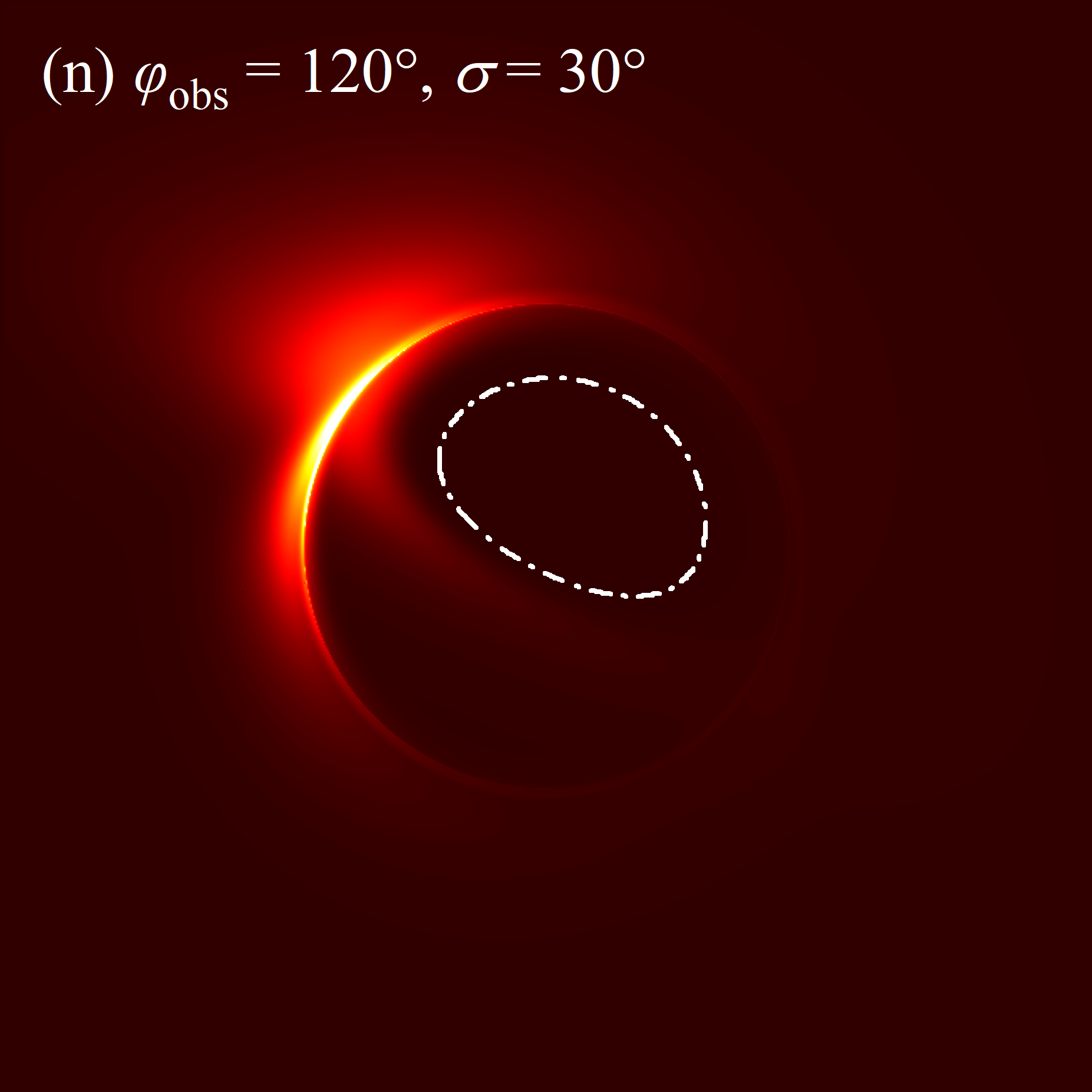}
\includegraphics[width=3.5cm]{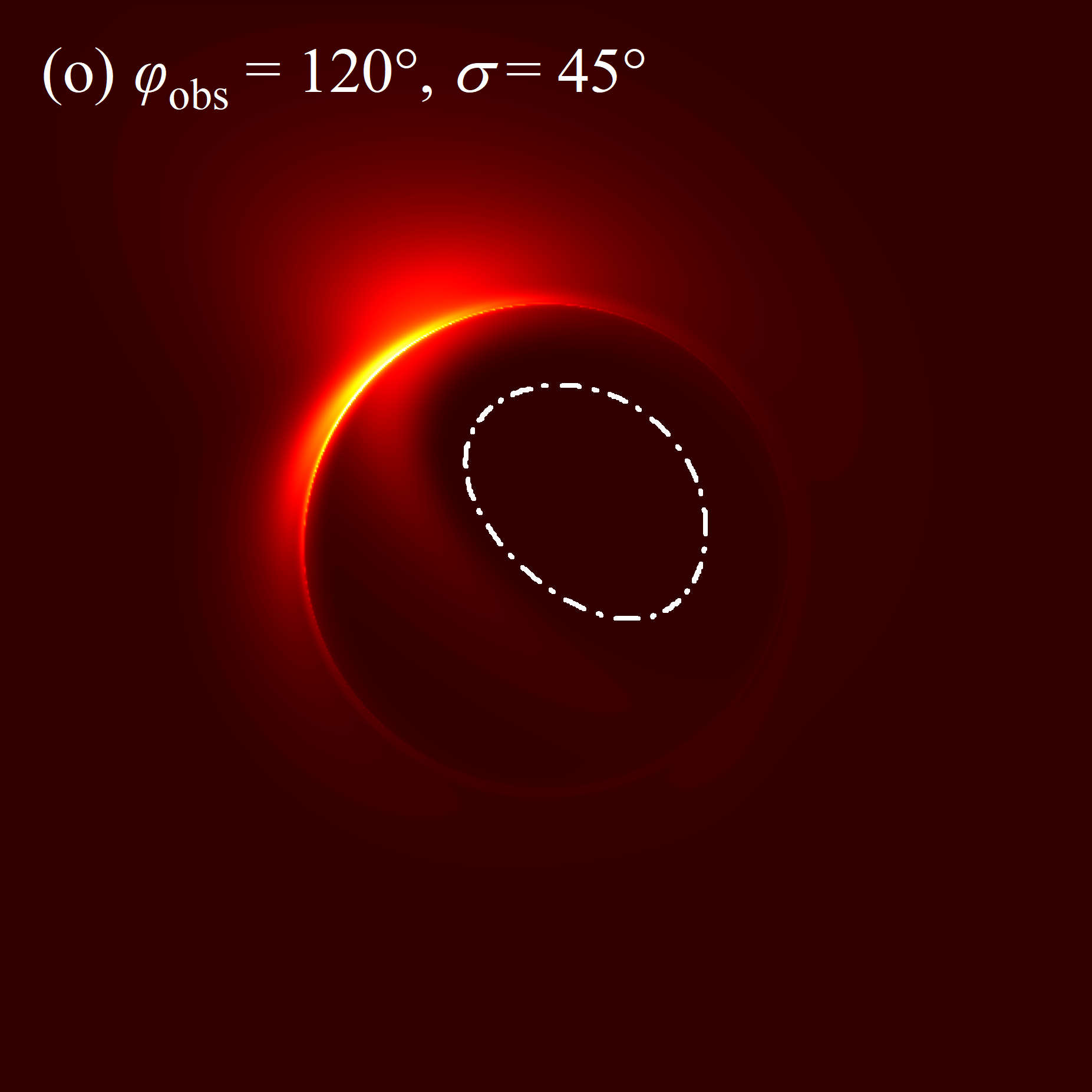}
\includegraphics[width=3.5cm]{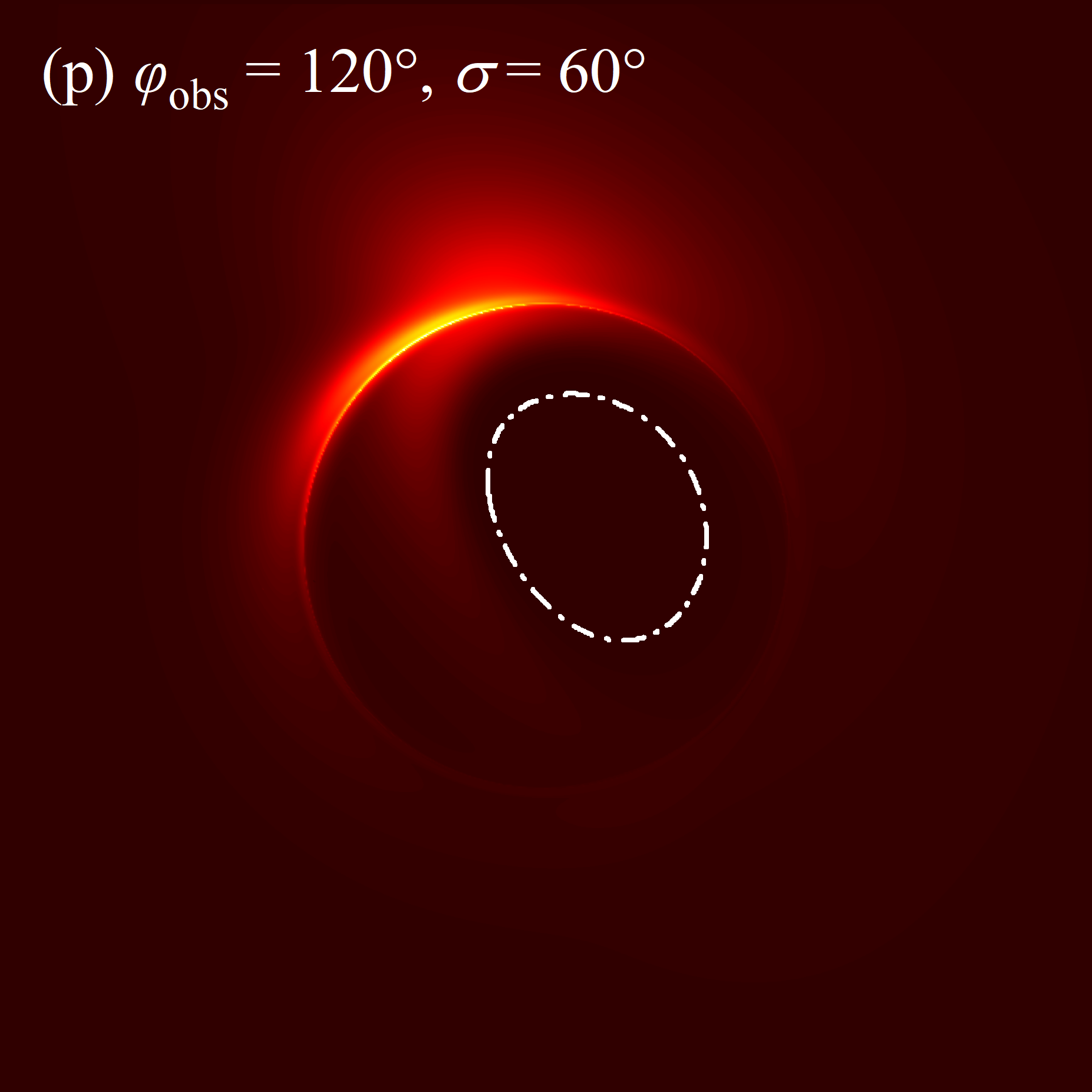}
\includegraphics[width=3.5cm]{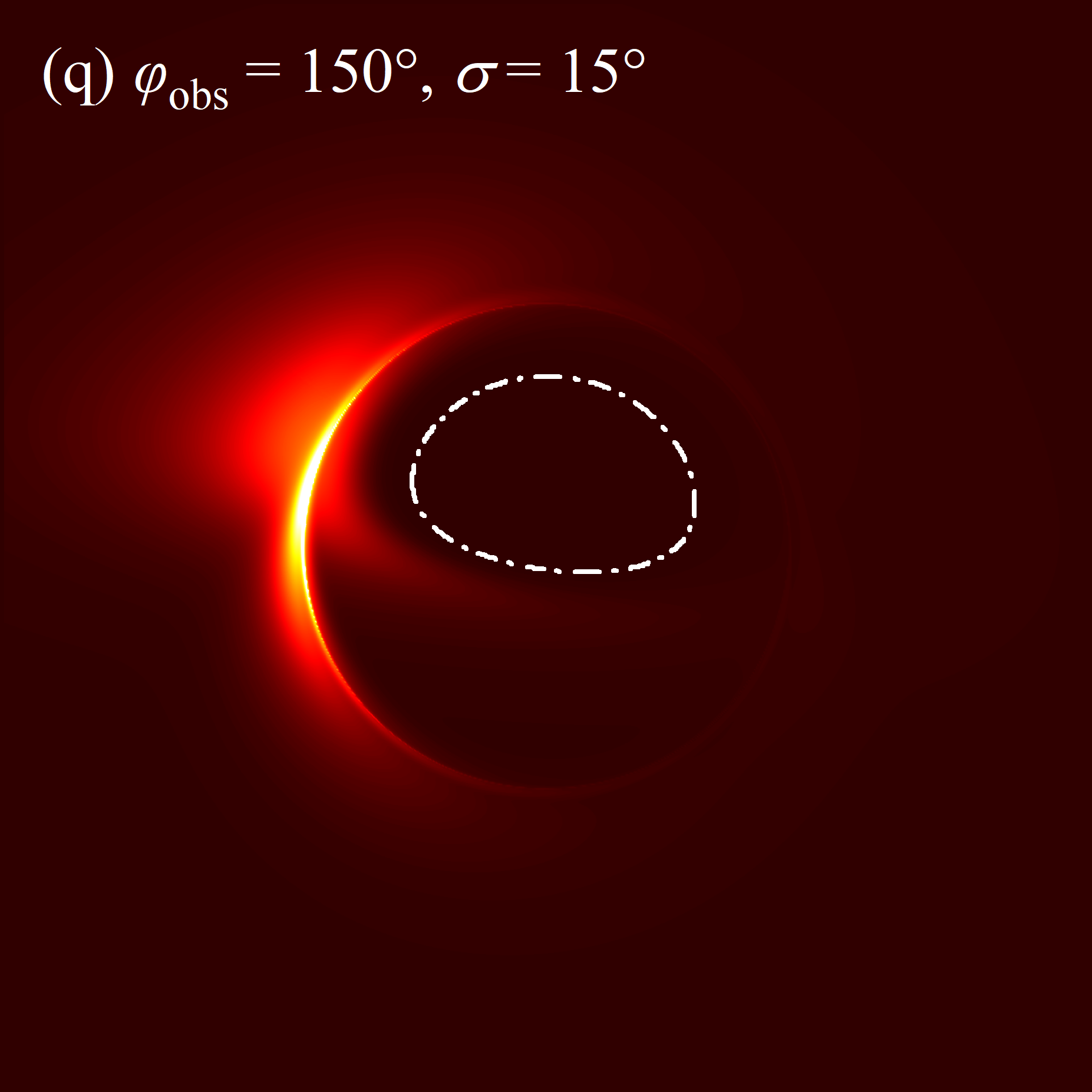}
\includegraphics[width=3.5cm]{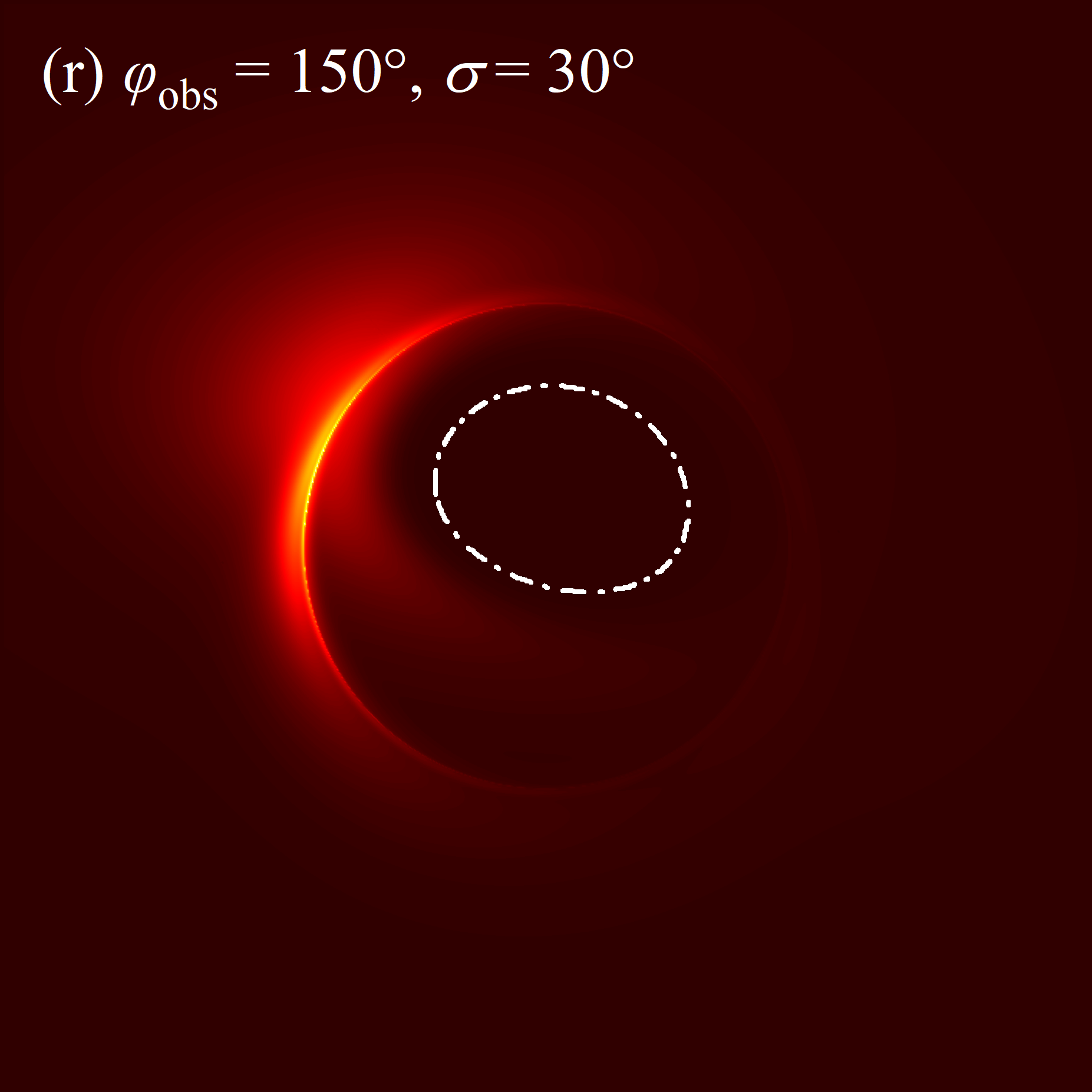}
\includegraphics[width=3.5cm]{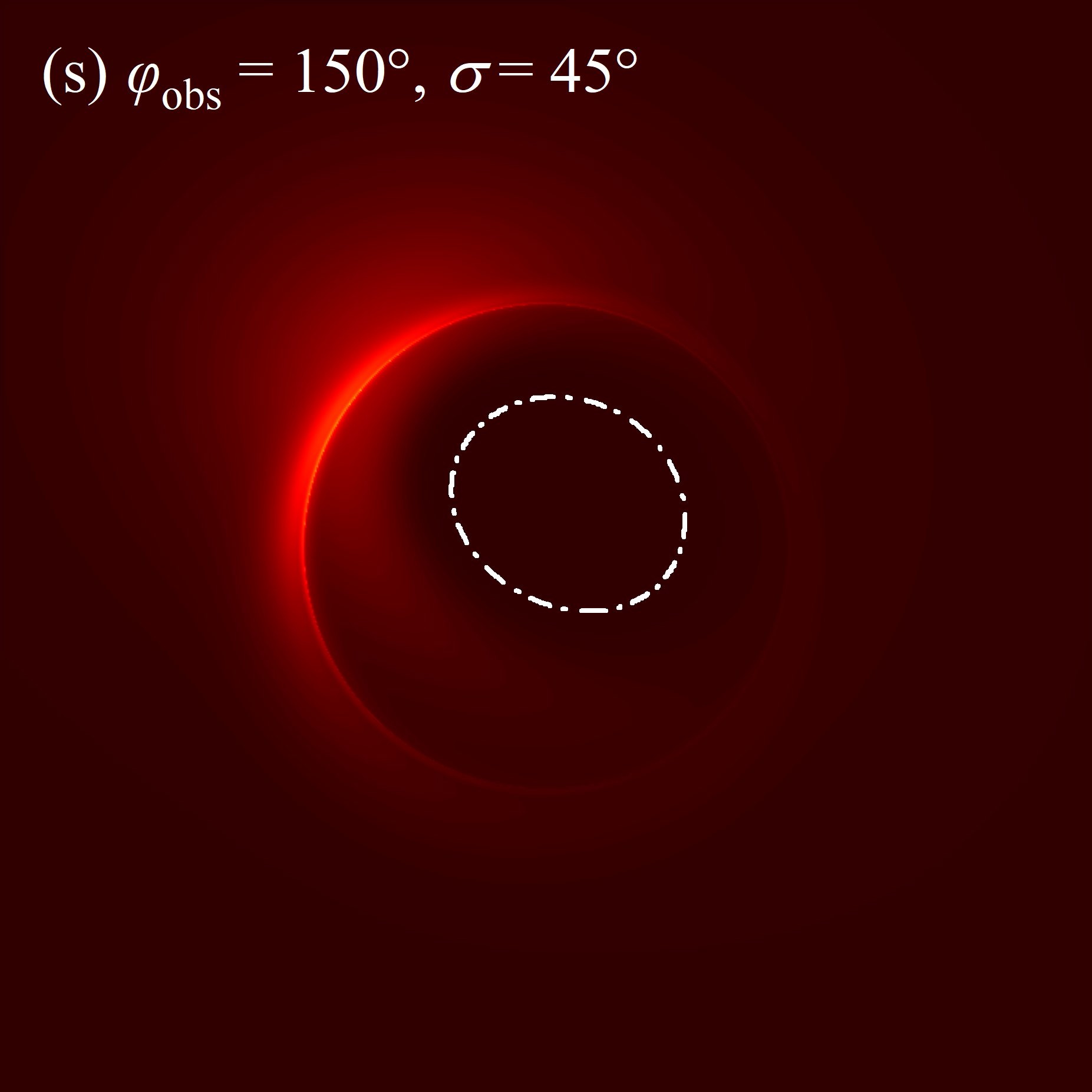}
\includegraphics[width=3.5cm]{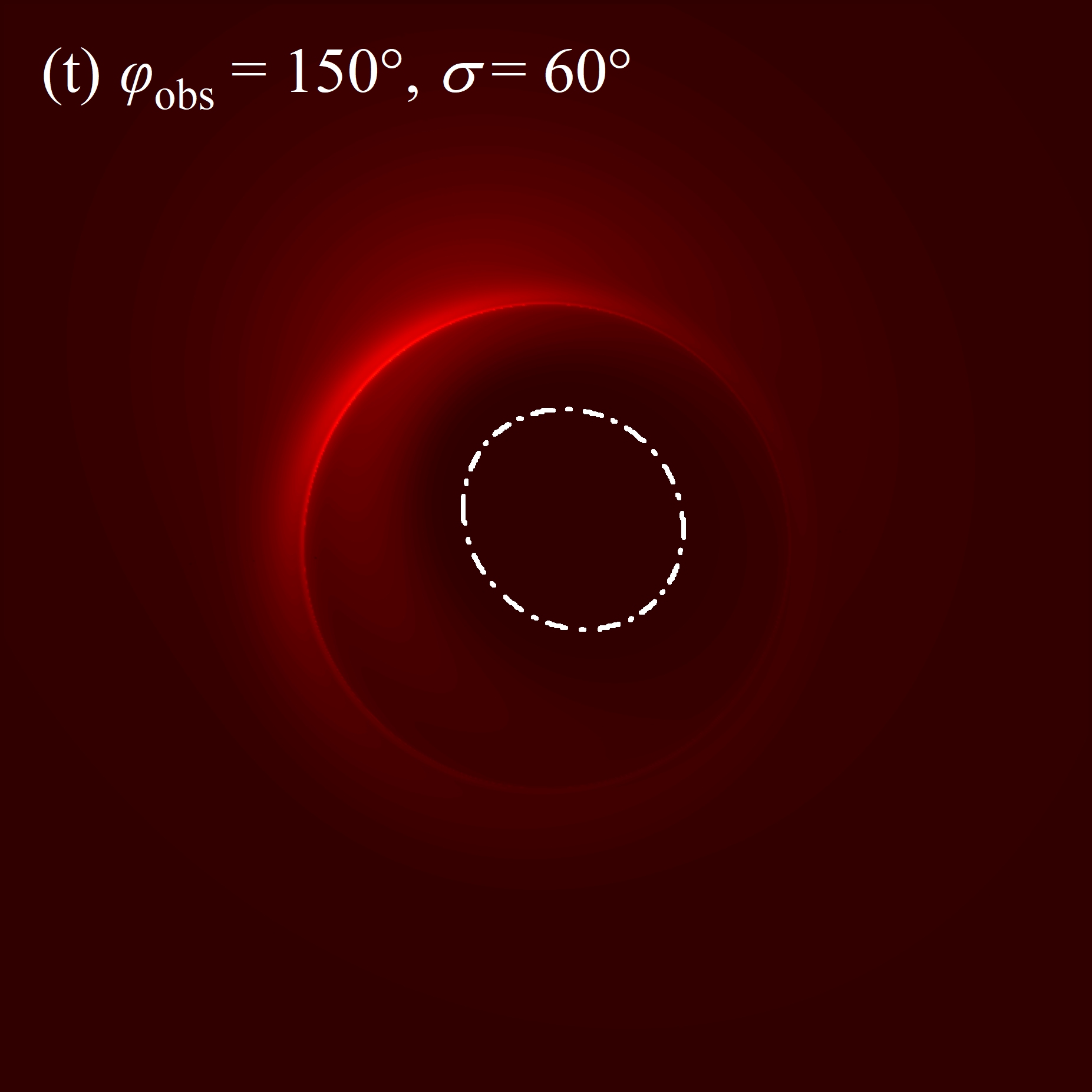}
\includegraphics[width=3.5cm]{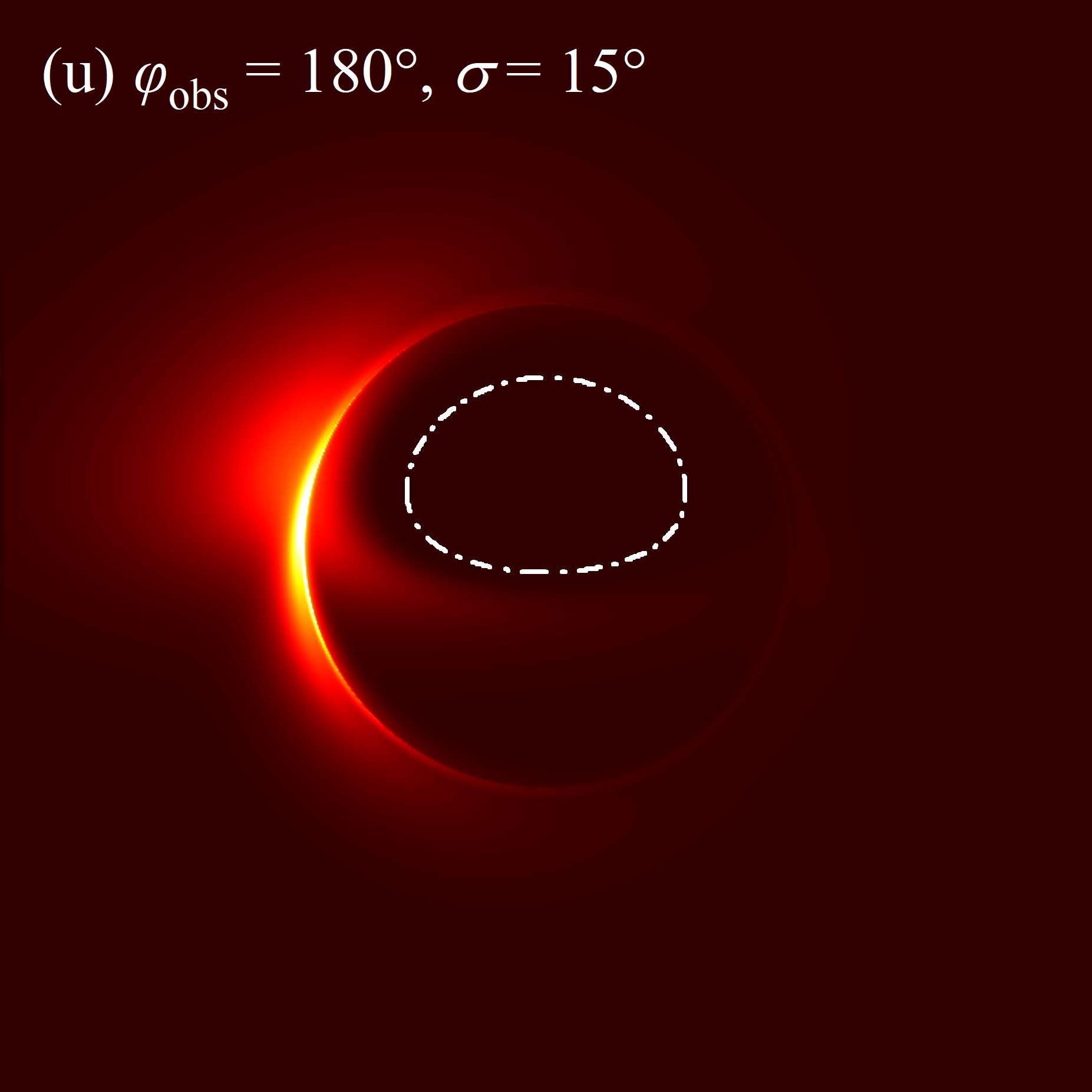}
\includegraphics[width=3.5cm]{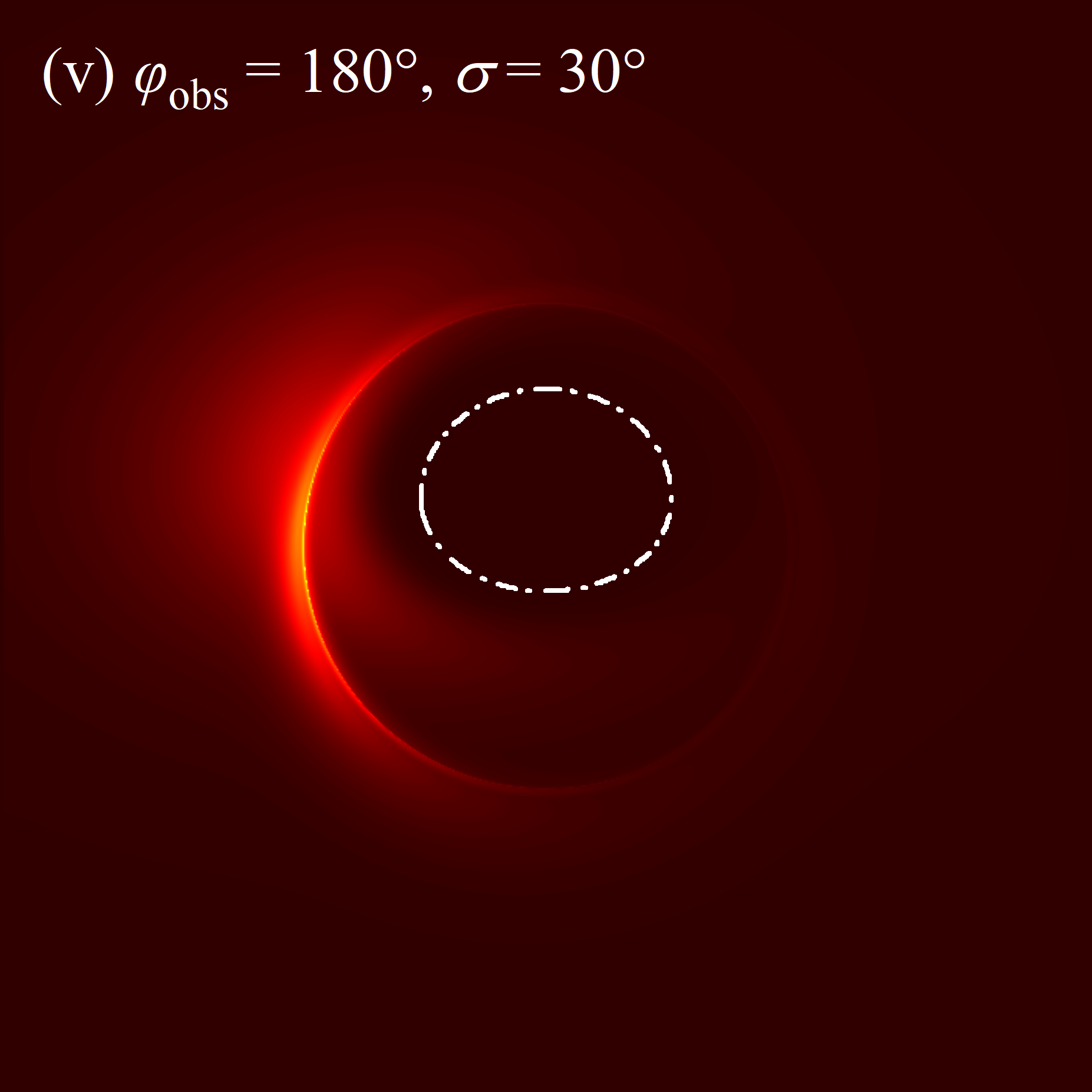}
\includegraphics[width=3.5cm]{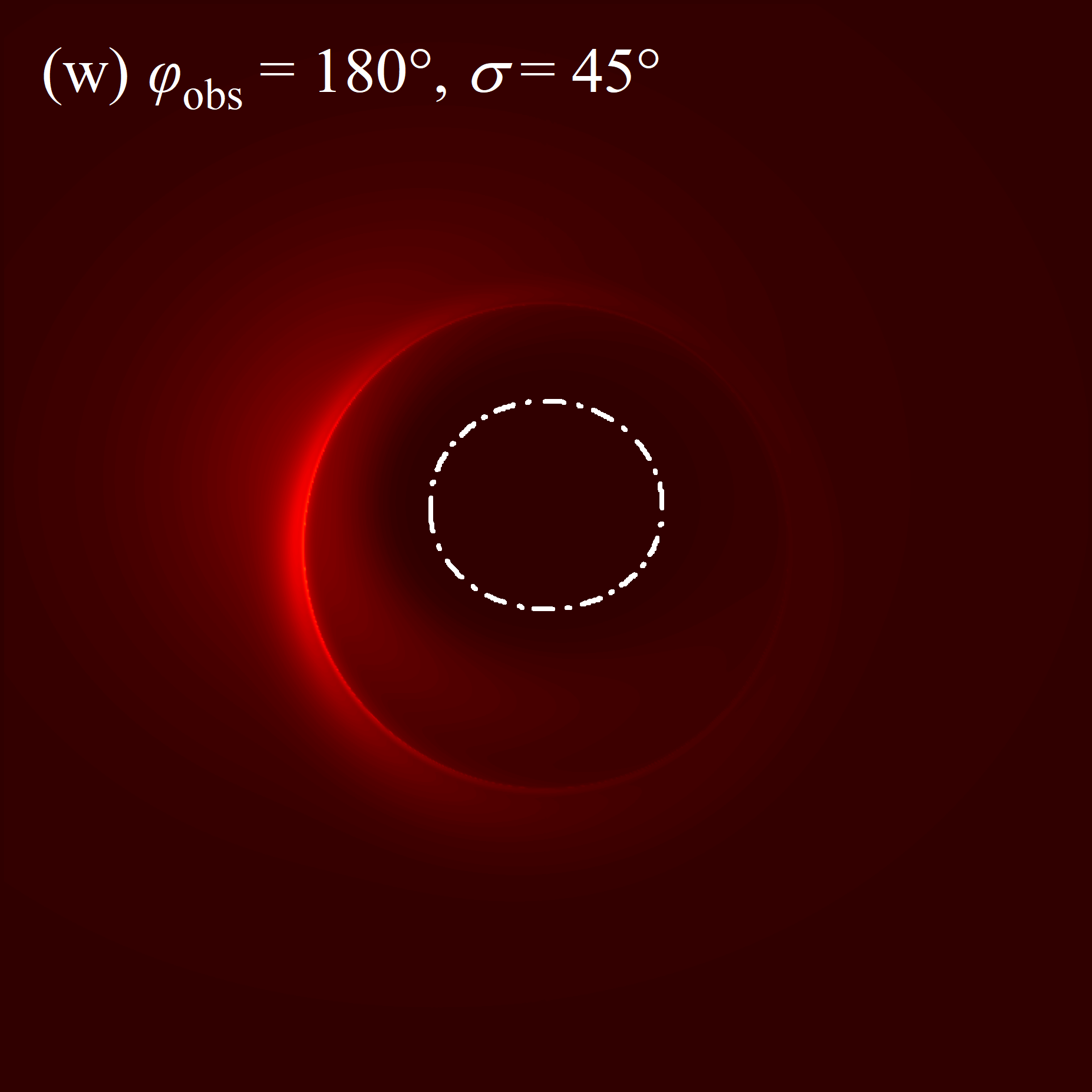}
\includegraphics[width=3.5cm]{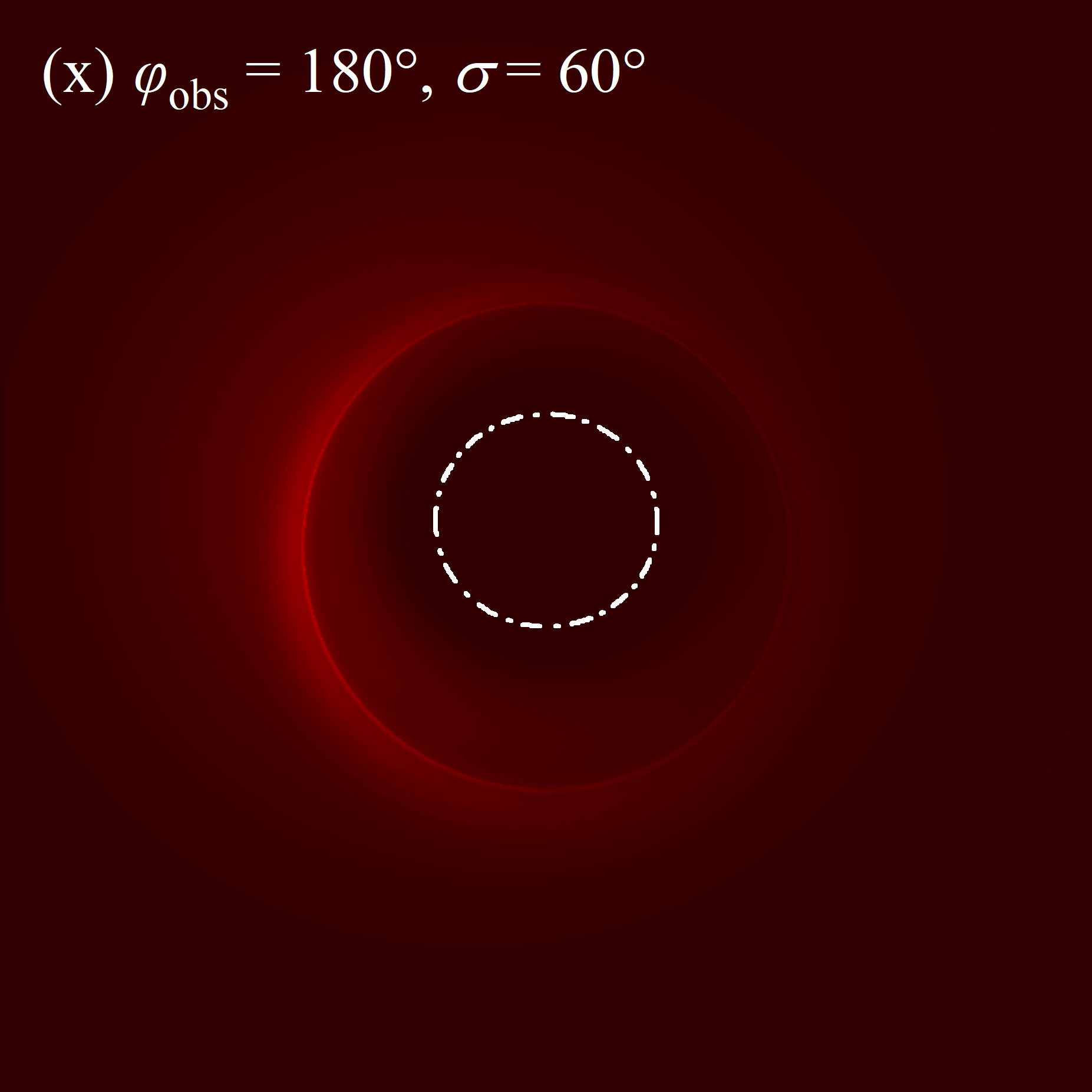}
\caption{Similar to figure 12, but for the viewing angle of $\omega = 85^{\circ}$.}}\label{fig13}
\end{figure*}
\begin{figure*}
\center{
\includegraphics[width=3.5cm]{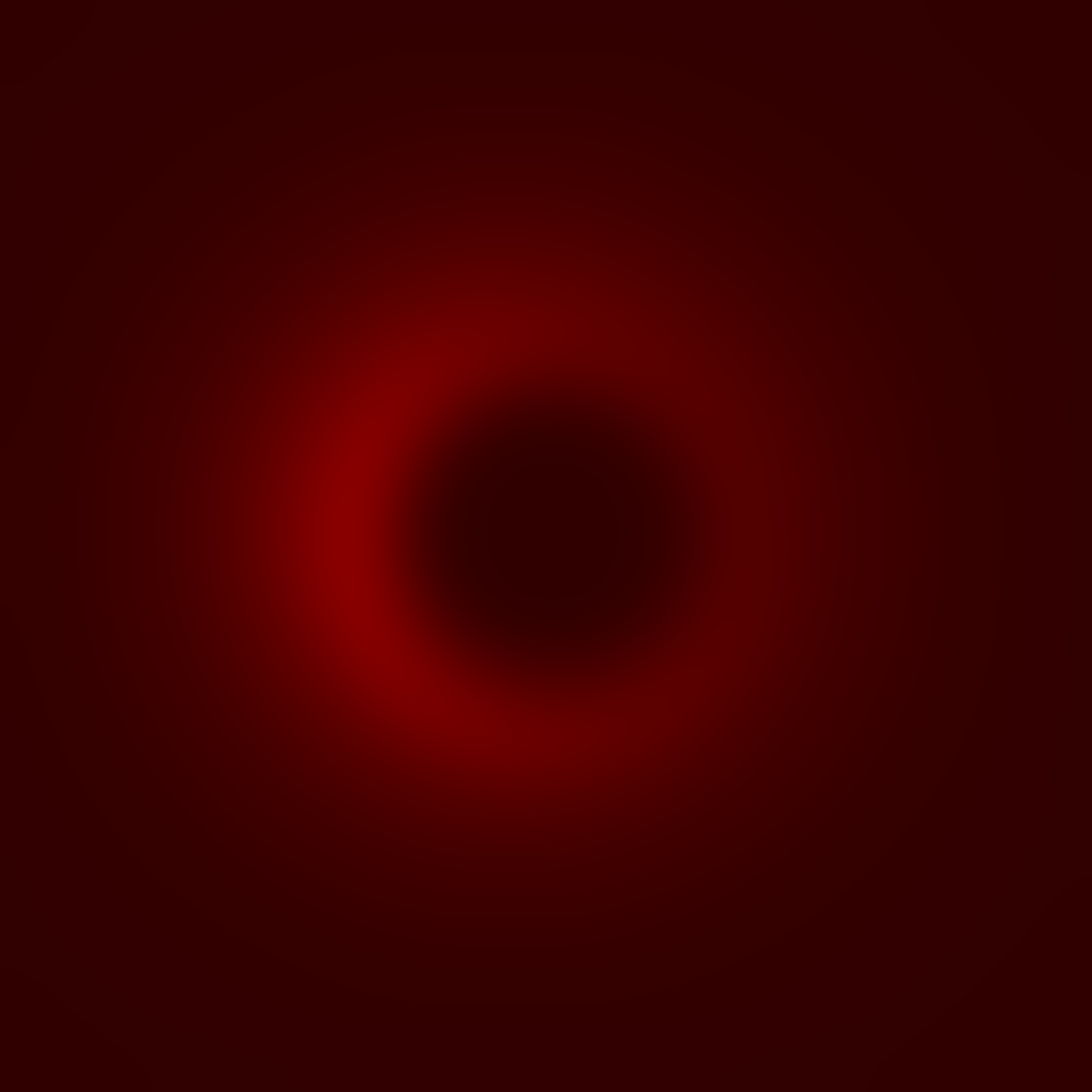}
\includegraphics[width=3.5cm]{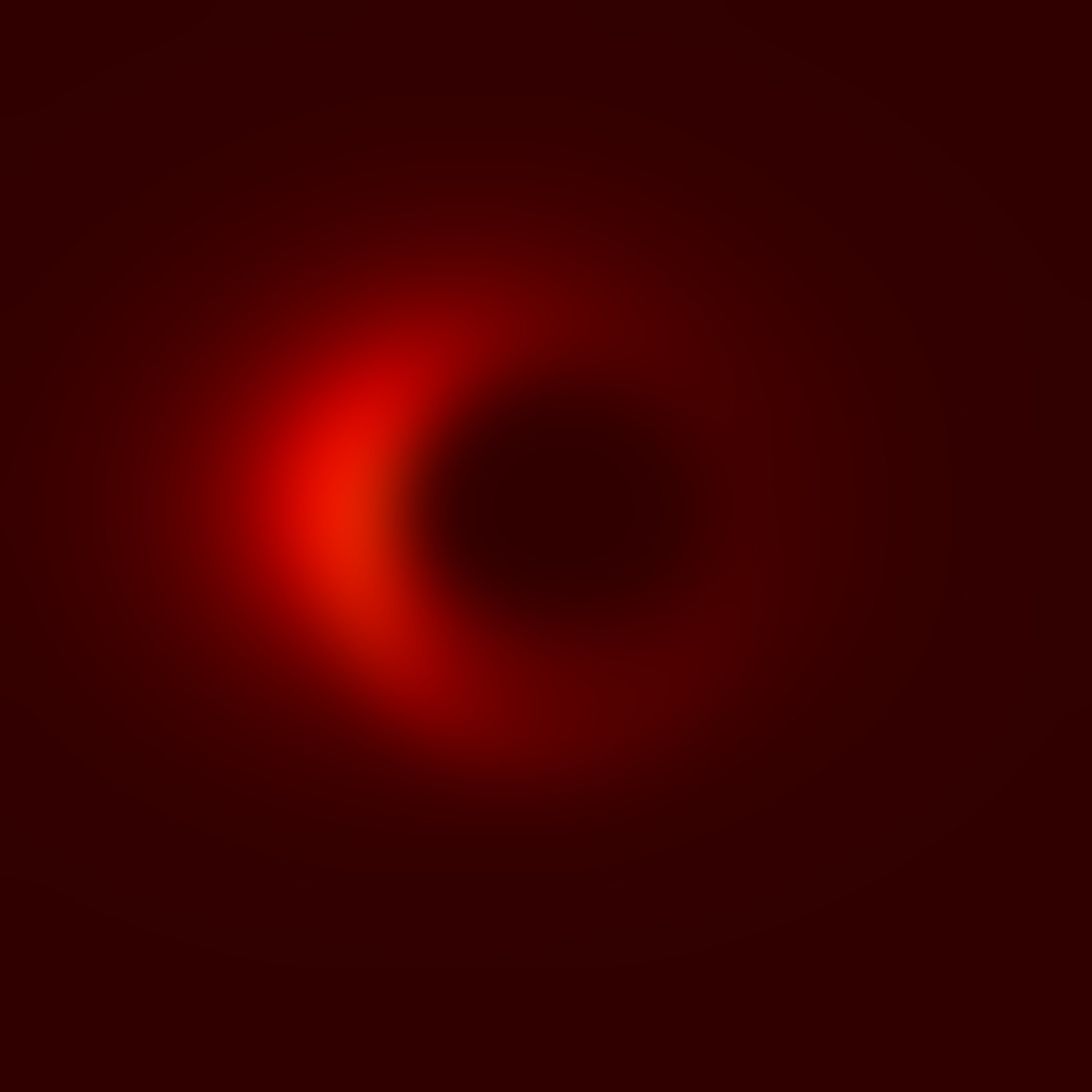}
\includegraphics[width=3.5cm]{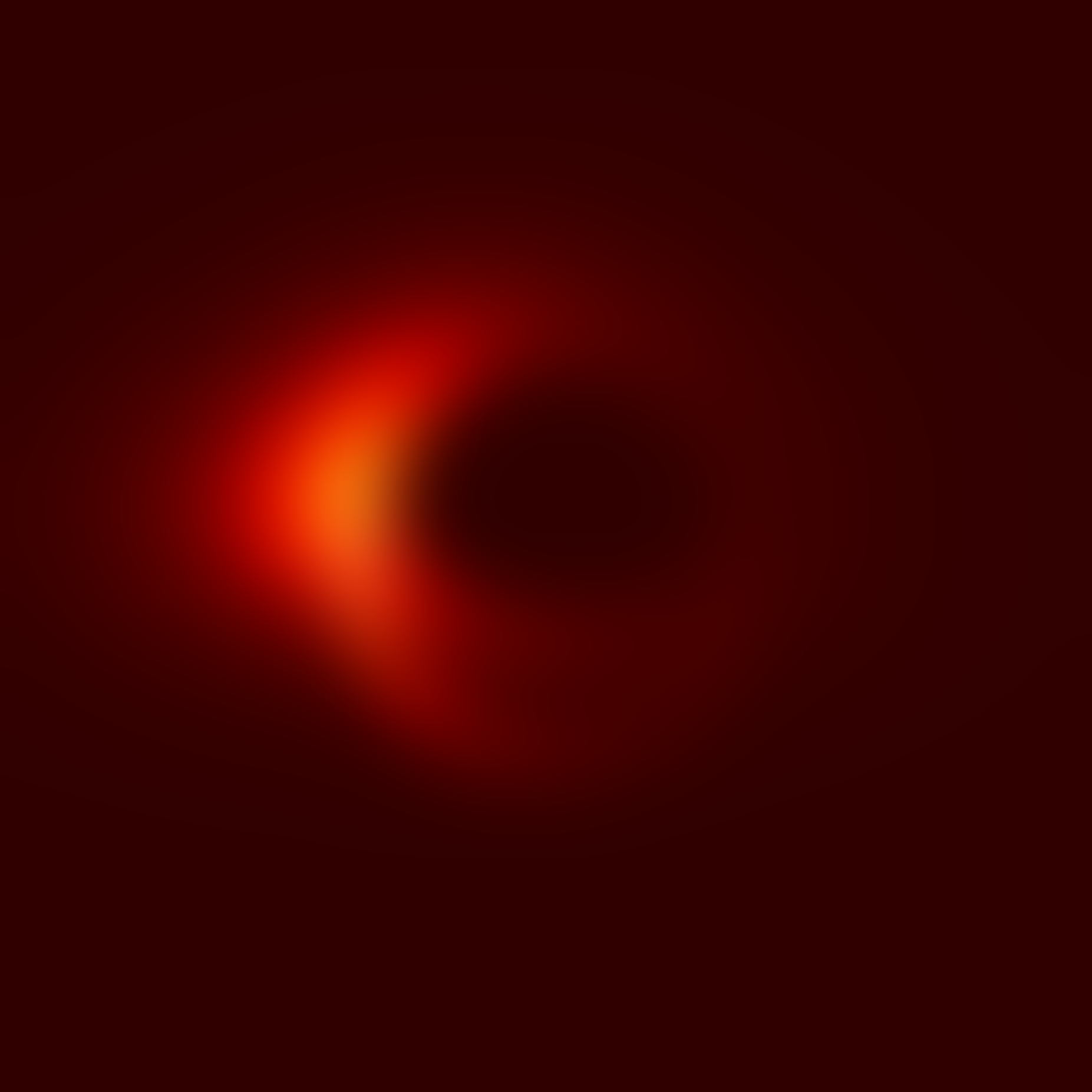}
\includegraphics[width=3.5cm]{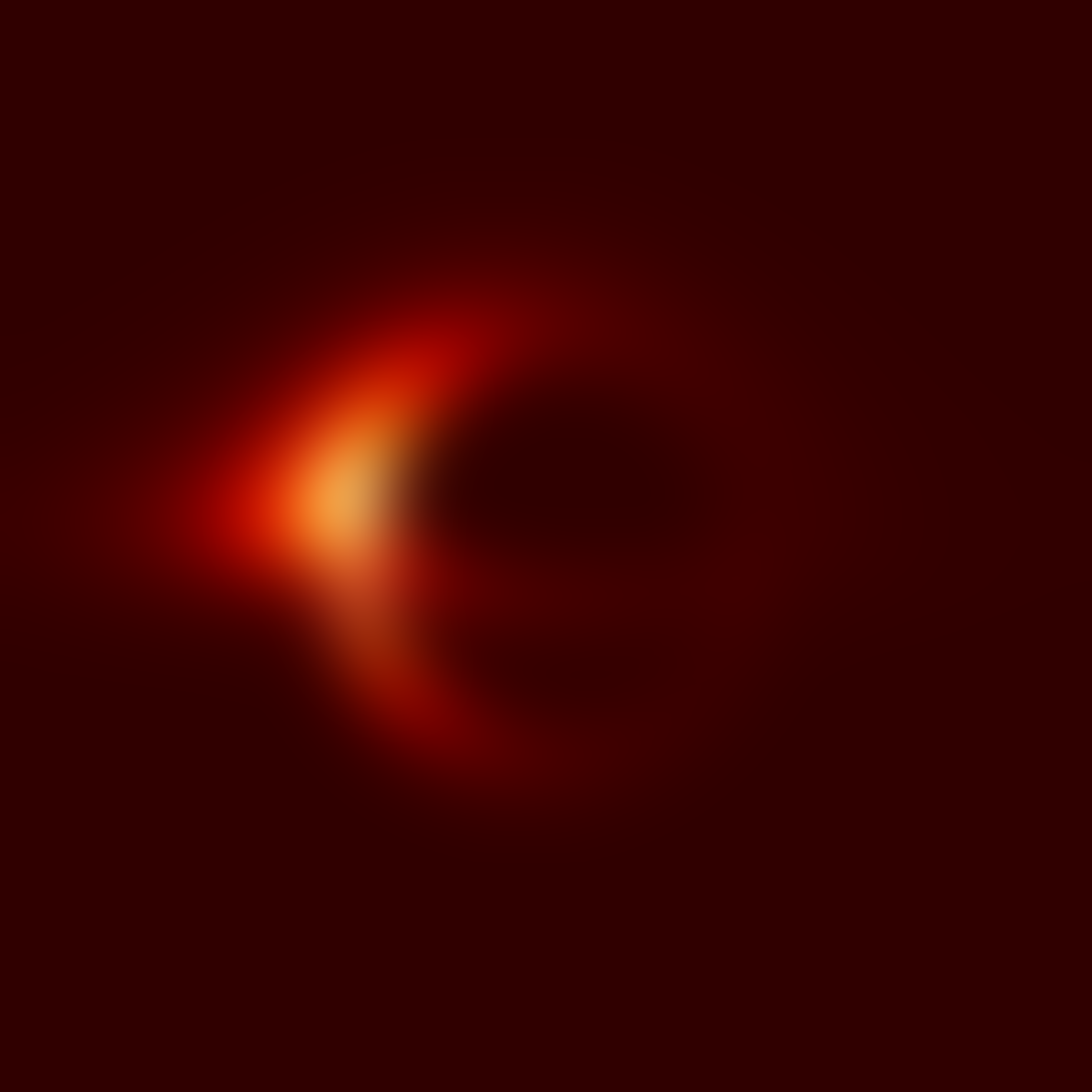}
\includegraphics[width=3.5cm]{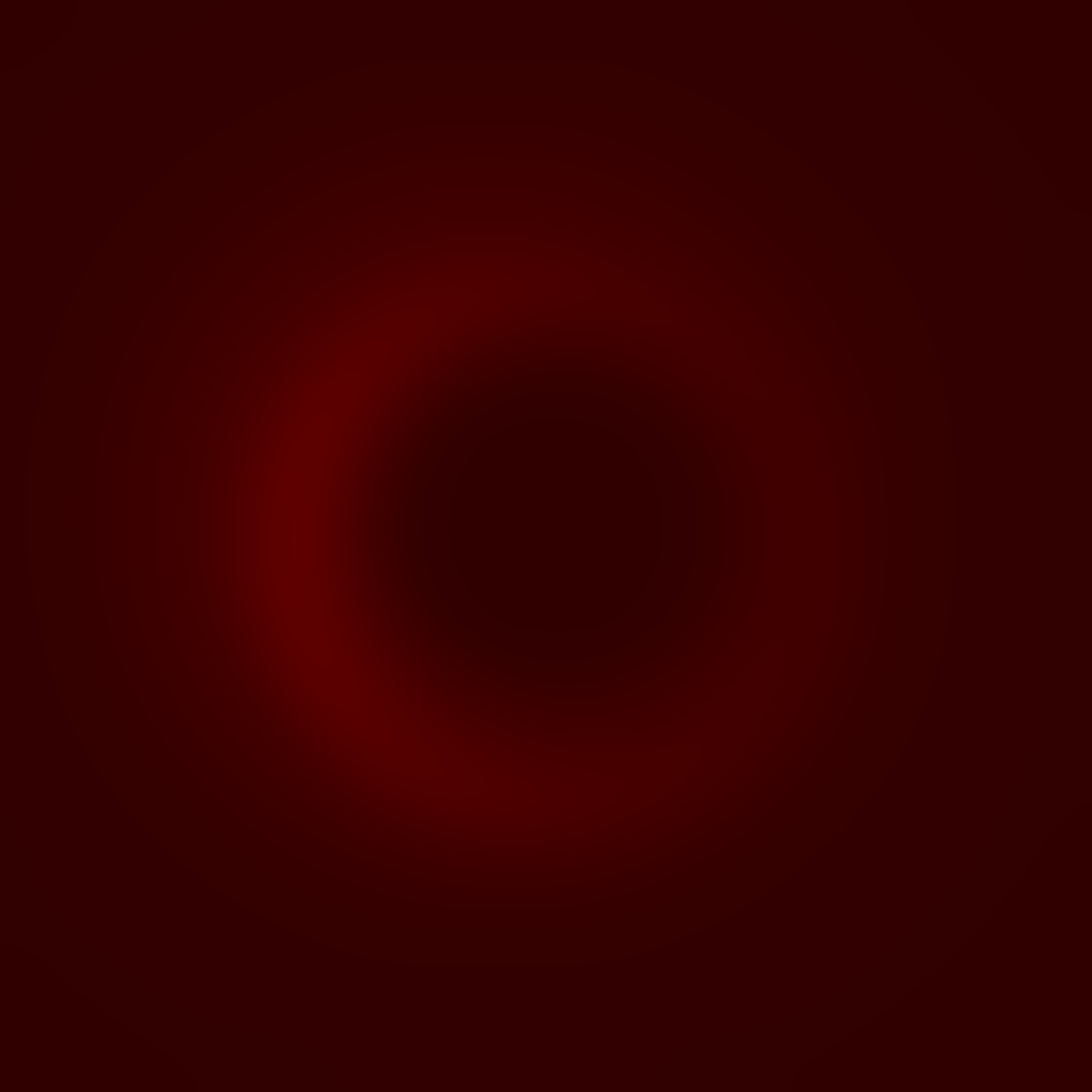}
\includegraphics[width=3.5cm]{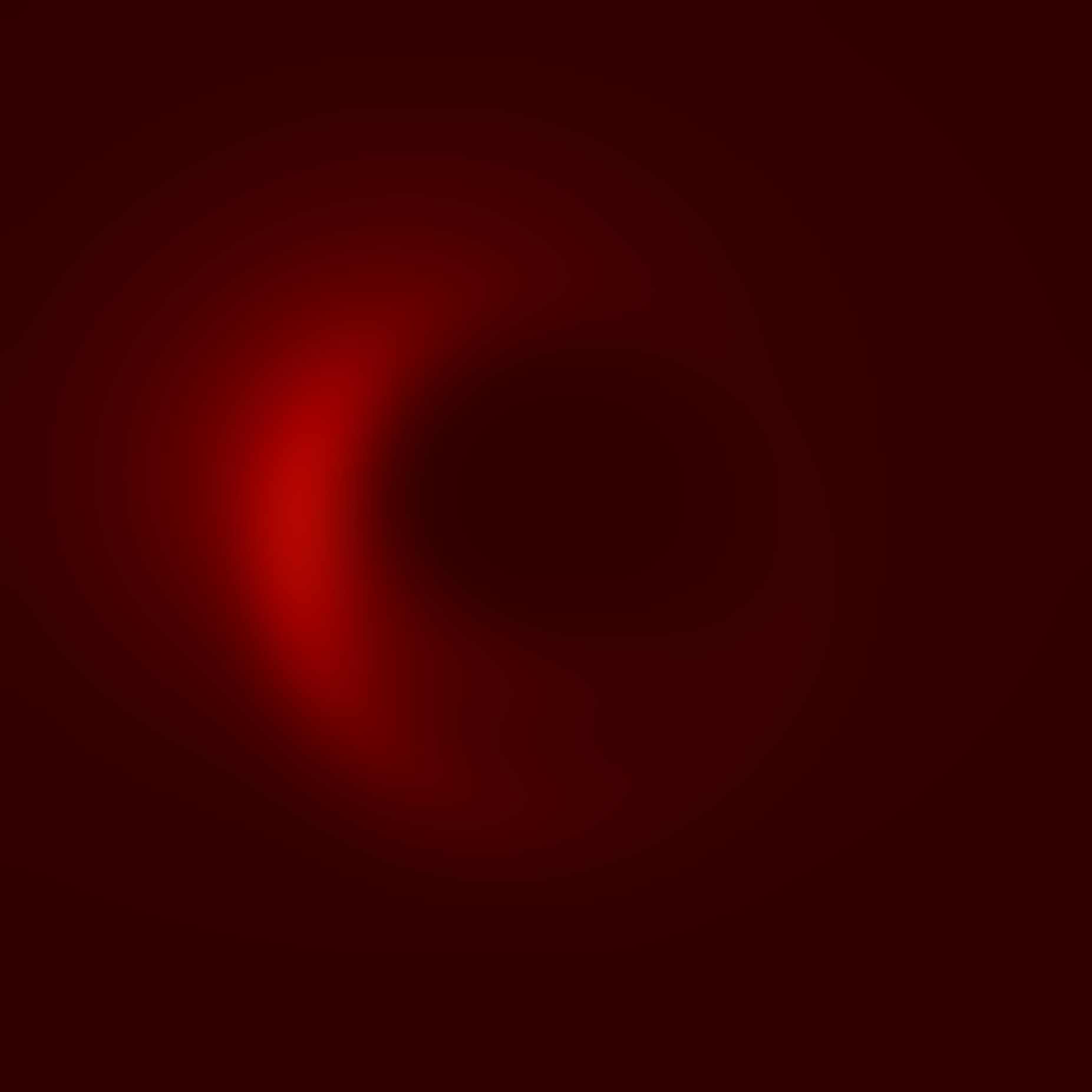}
\includegraphics[width=3.5cm]{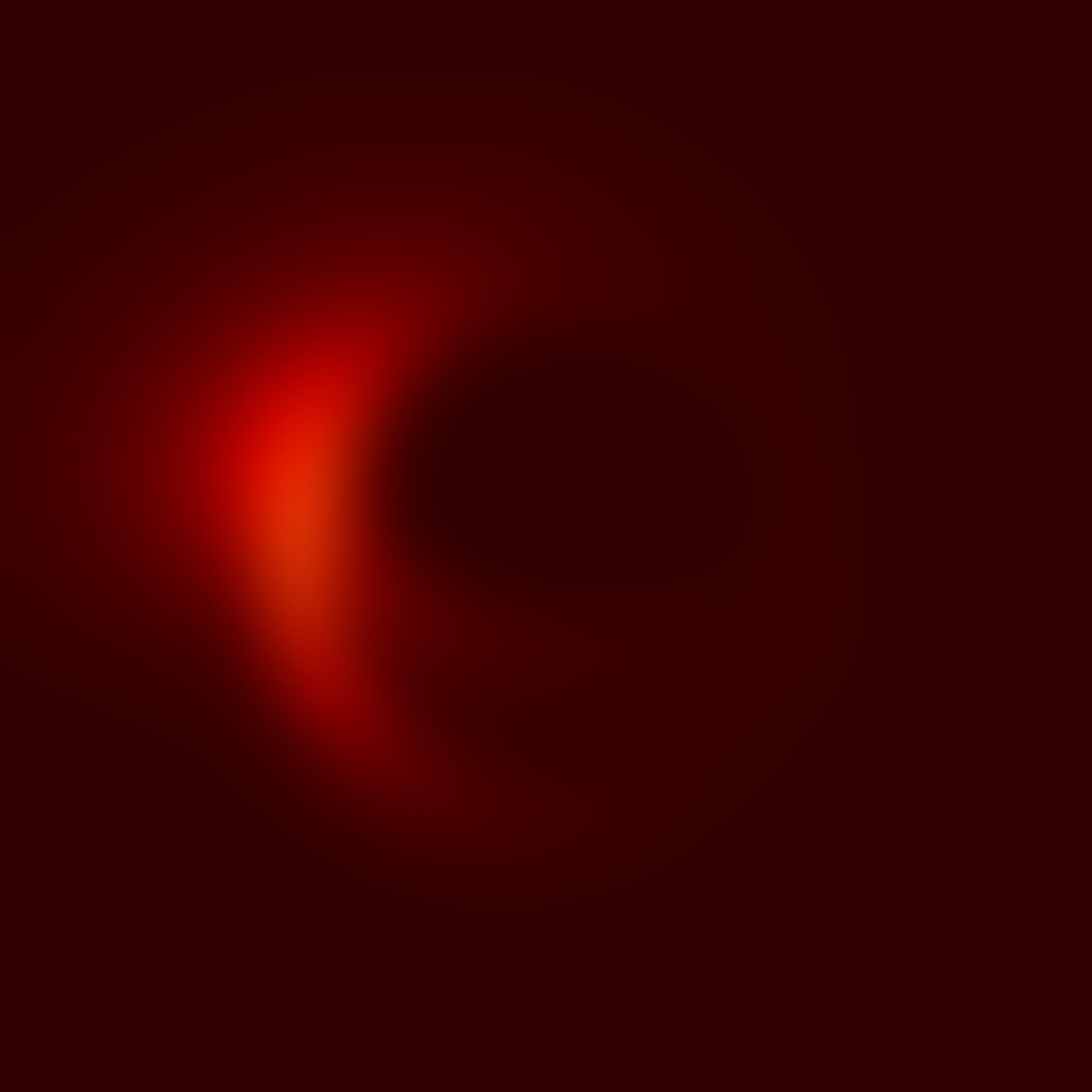}
\includegraphics[width=3.5cm]{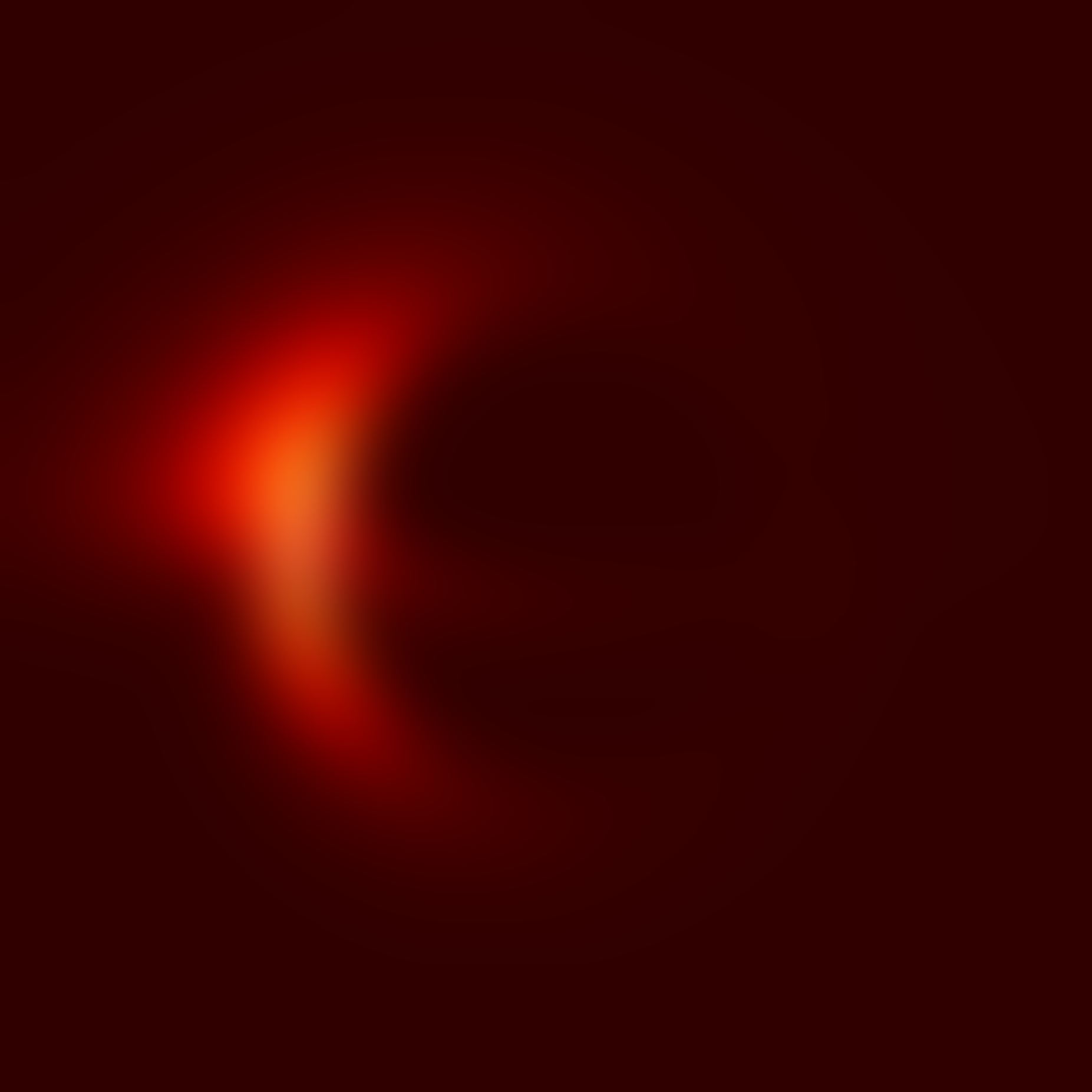}
\includegraphics[width=3.5cm]{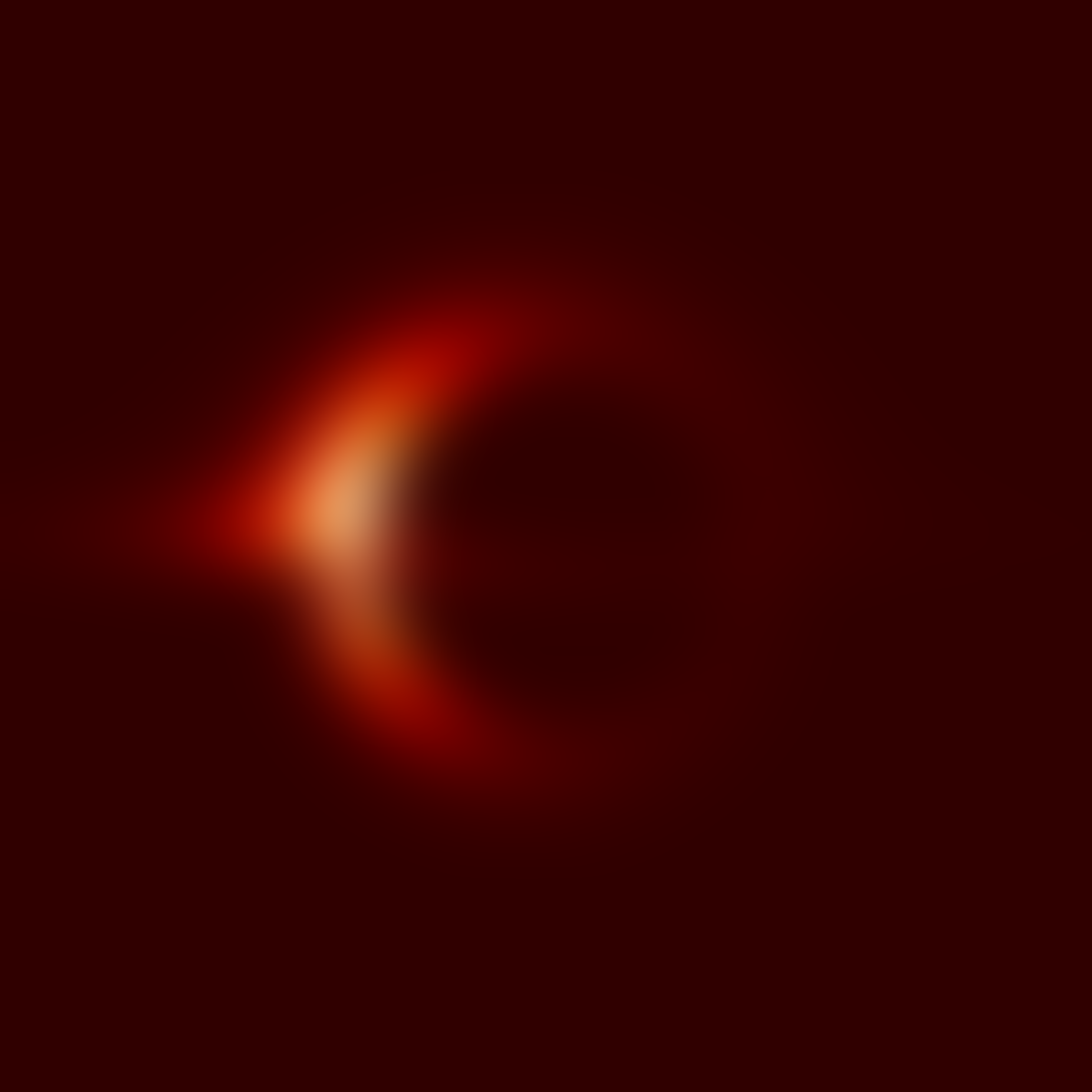}
\includegraphics[width=3.5cm]{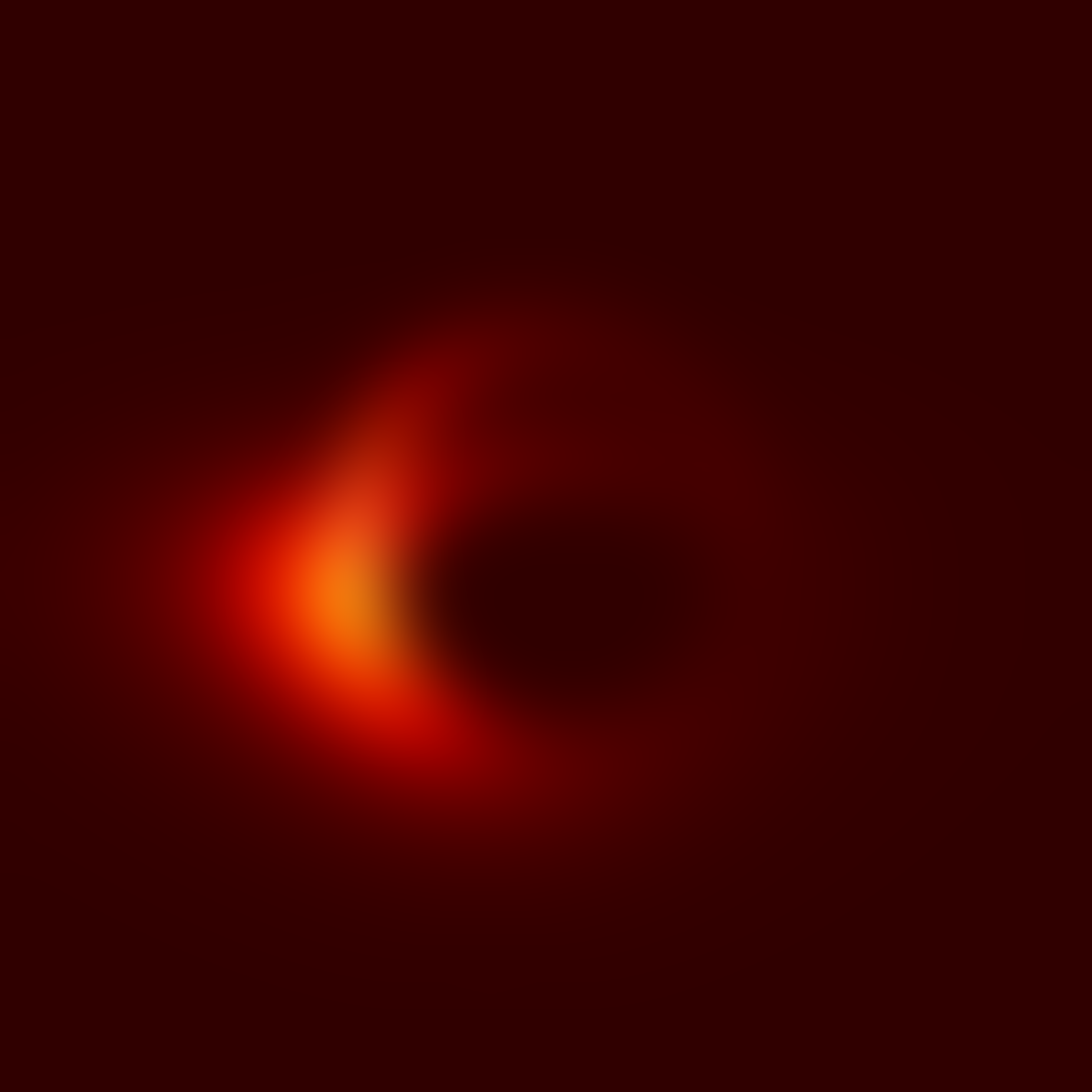}
\includegraphics[width=3.5cm]{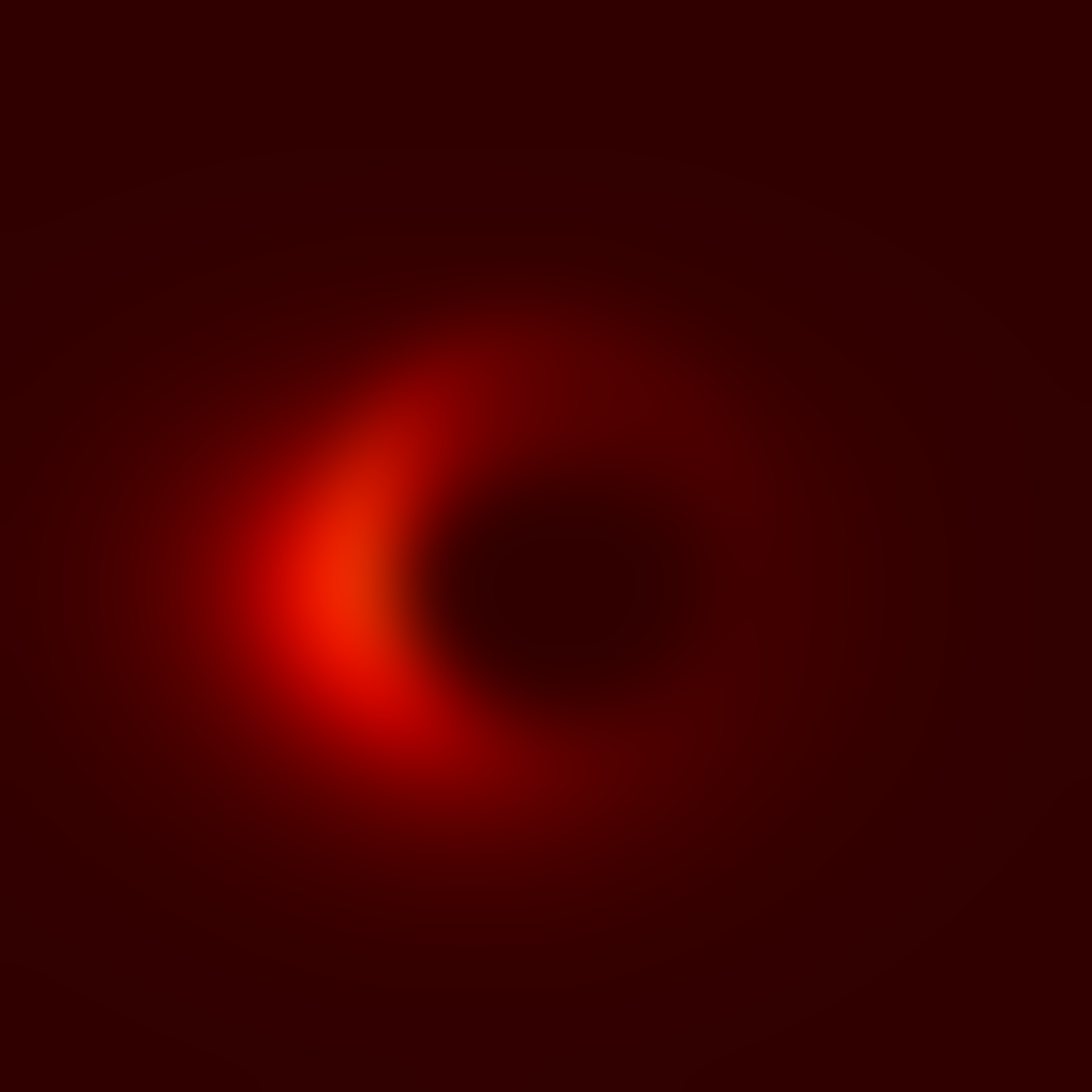}
\includegraphics[width=3.5cm]{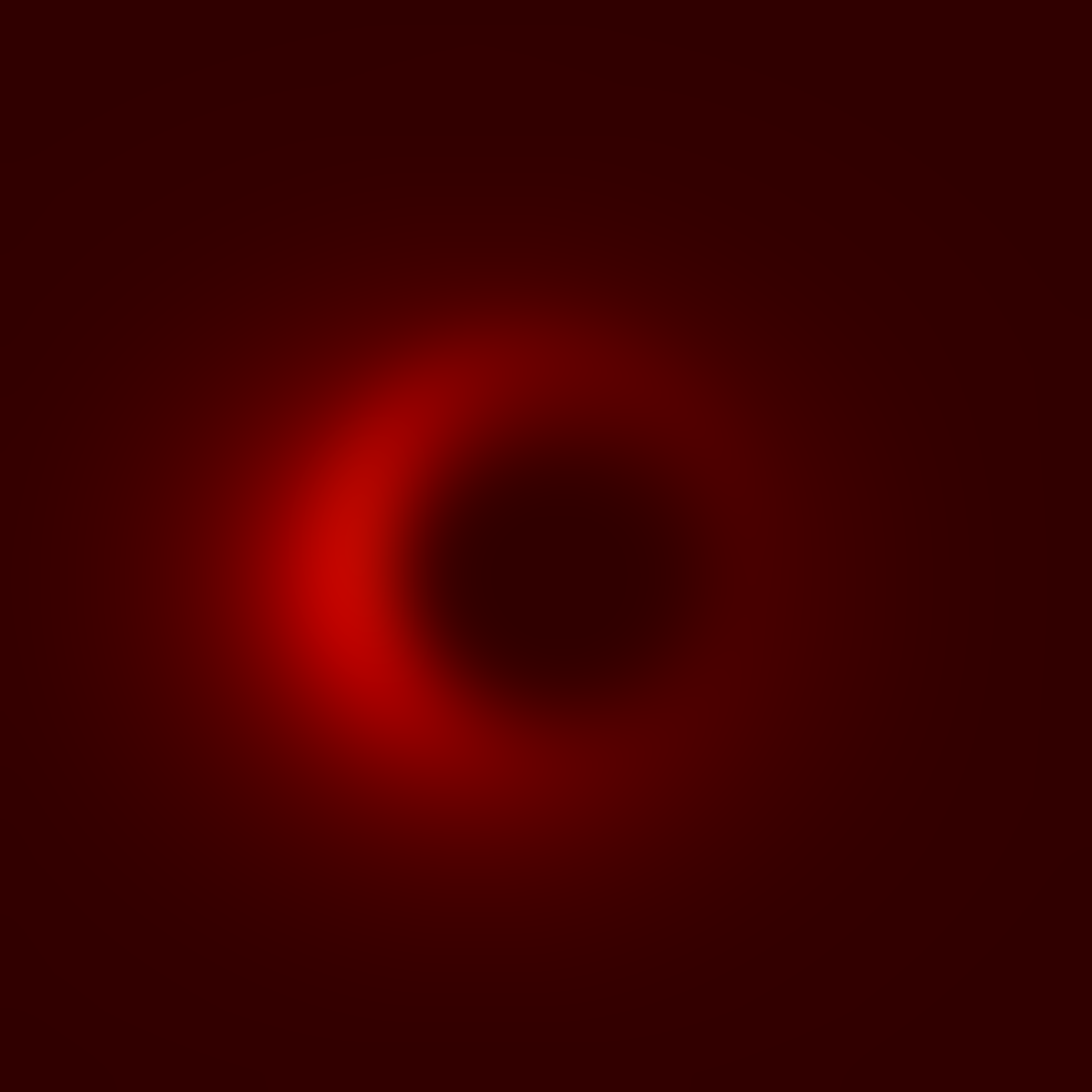}
\includegraphics[width=3.5cm]{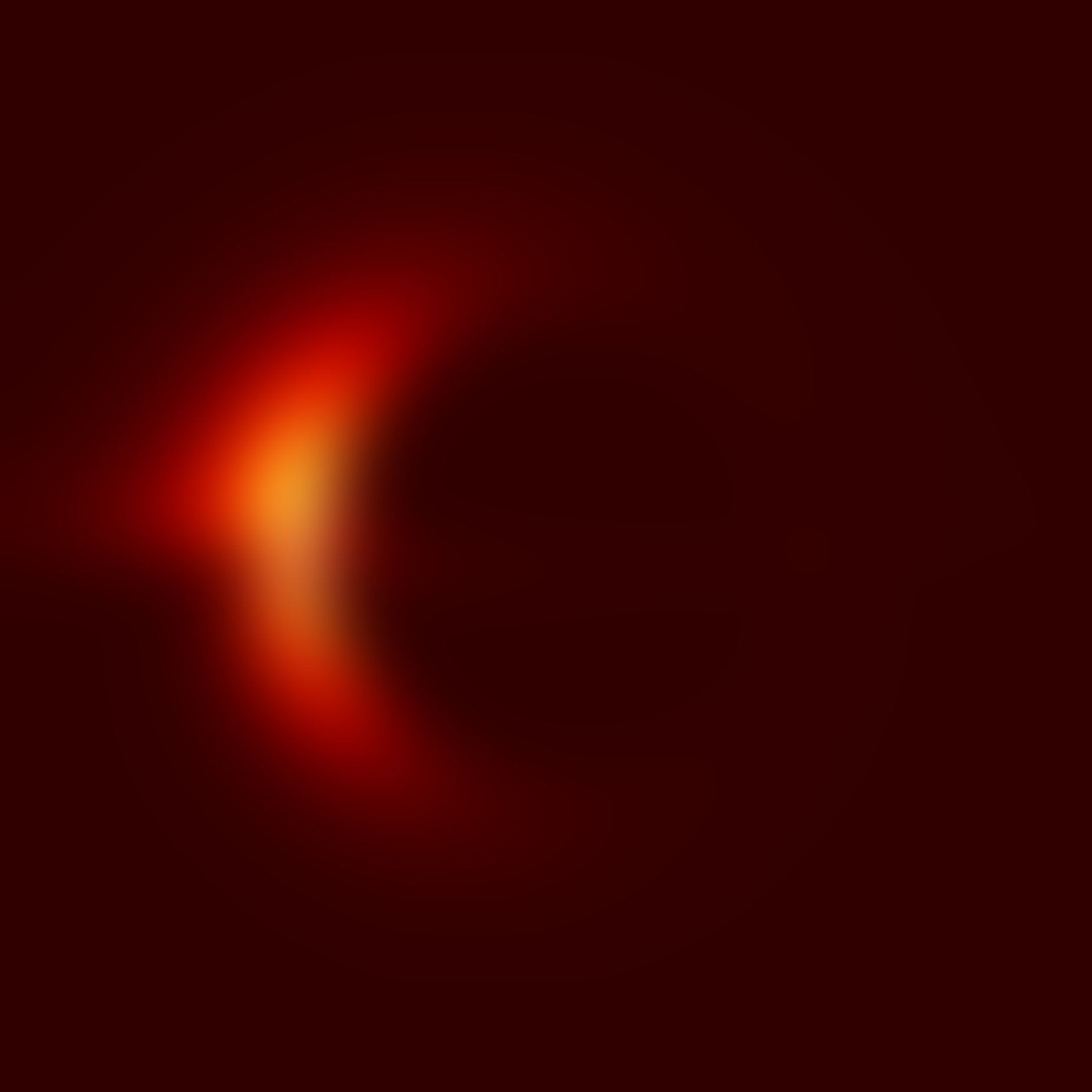}
\includegraphics[width=3.5cm]{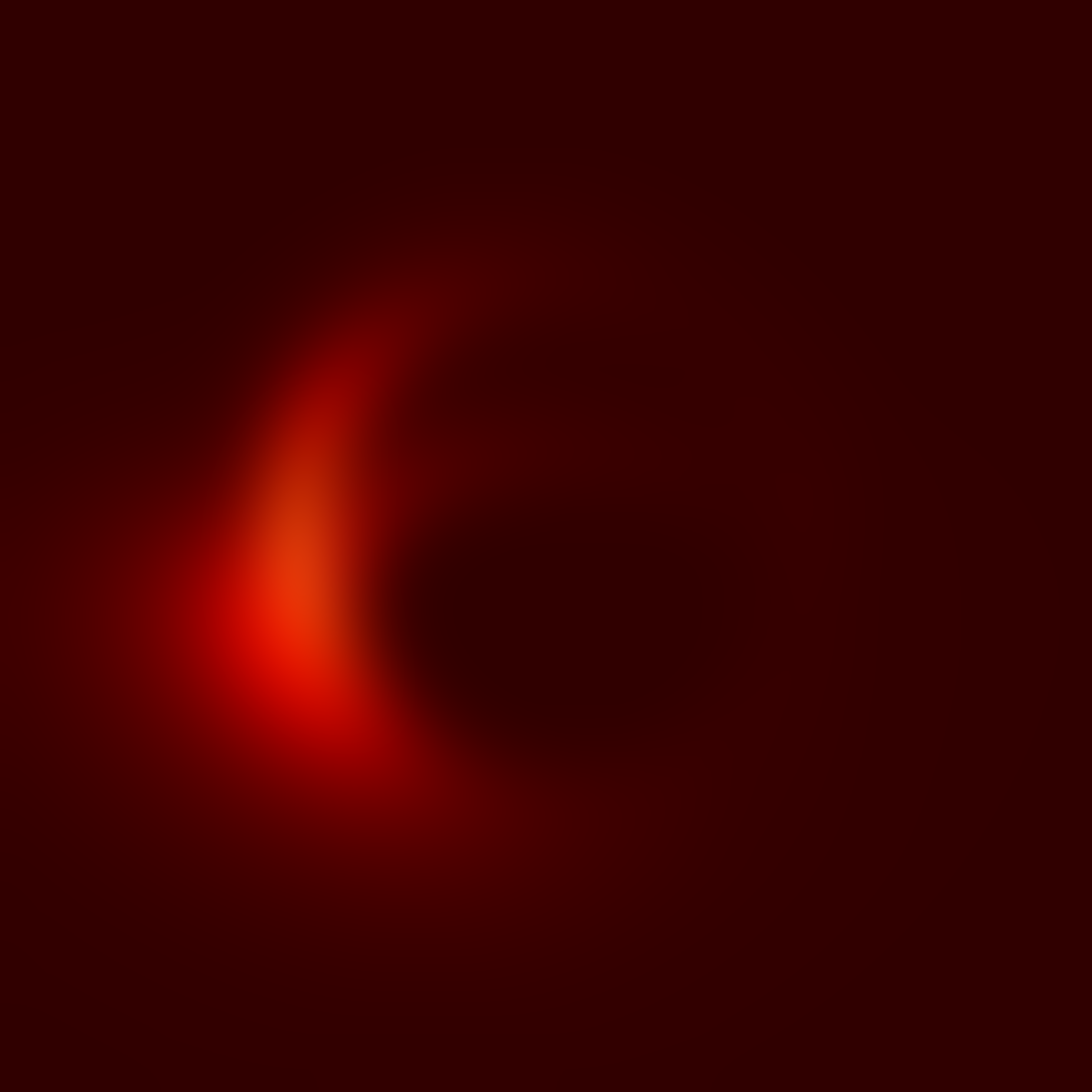}
\includegraphics[width=3.5cm]{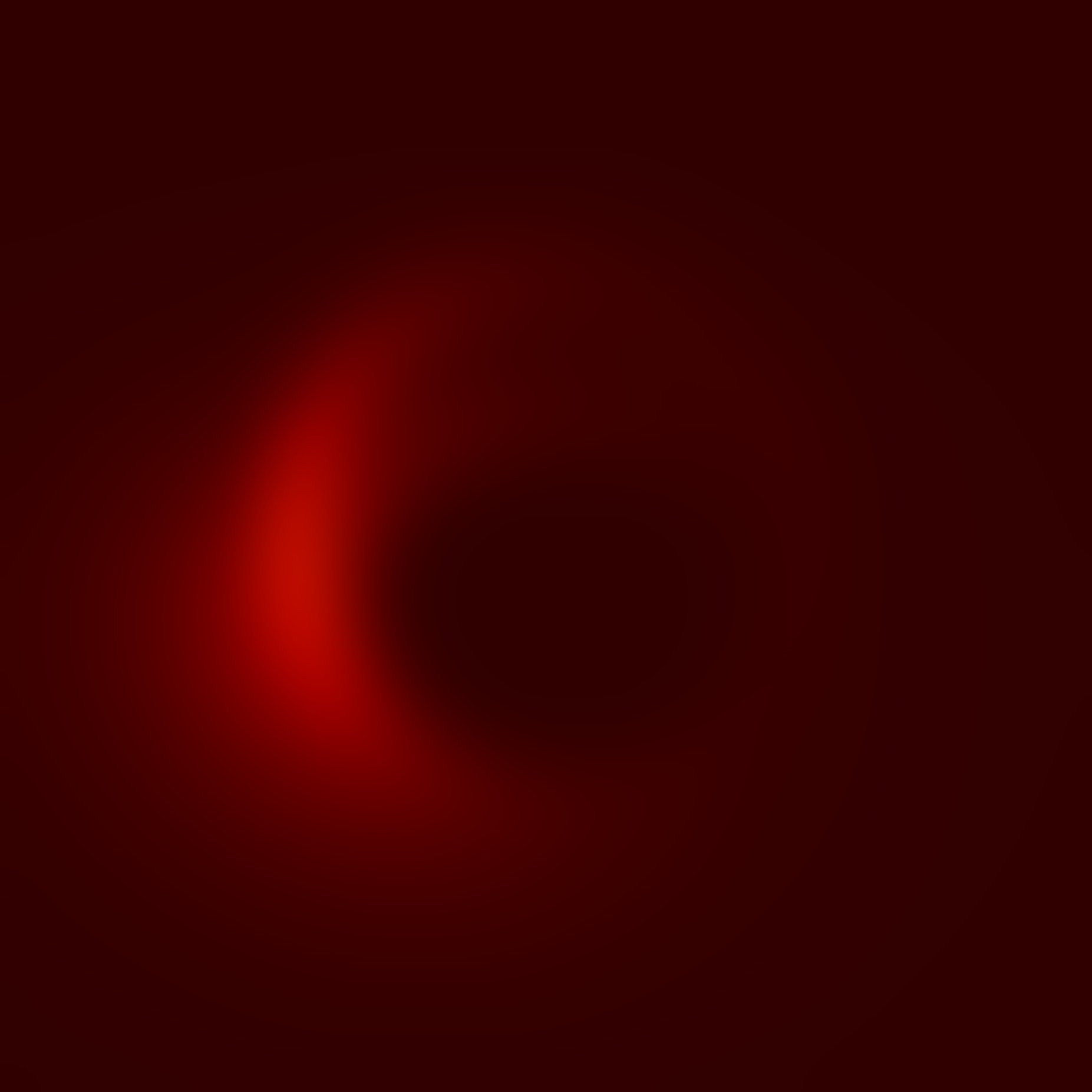}
\includegraphics[width=3.5cm]{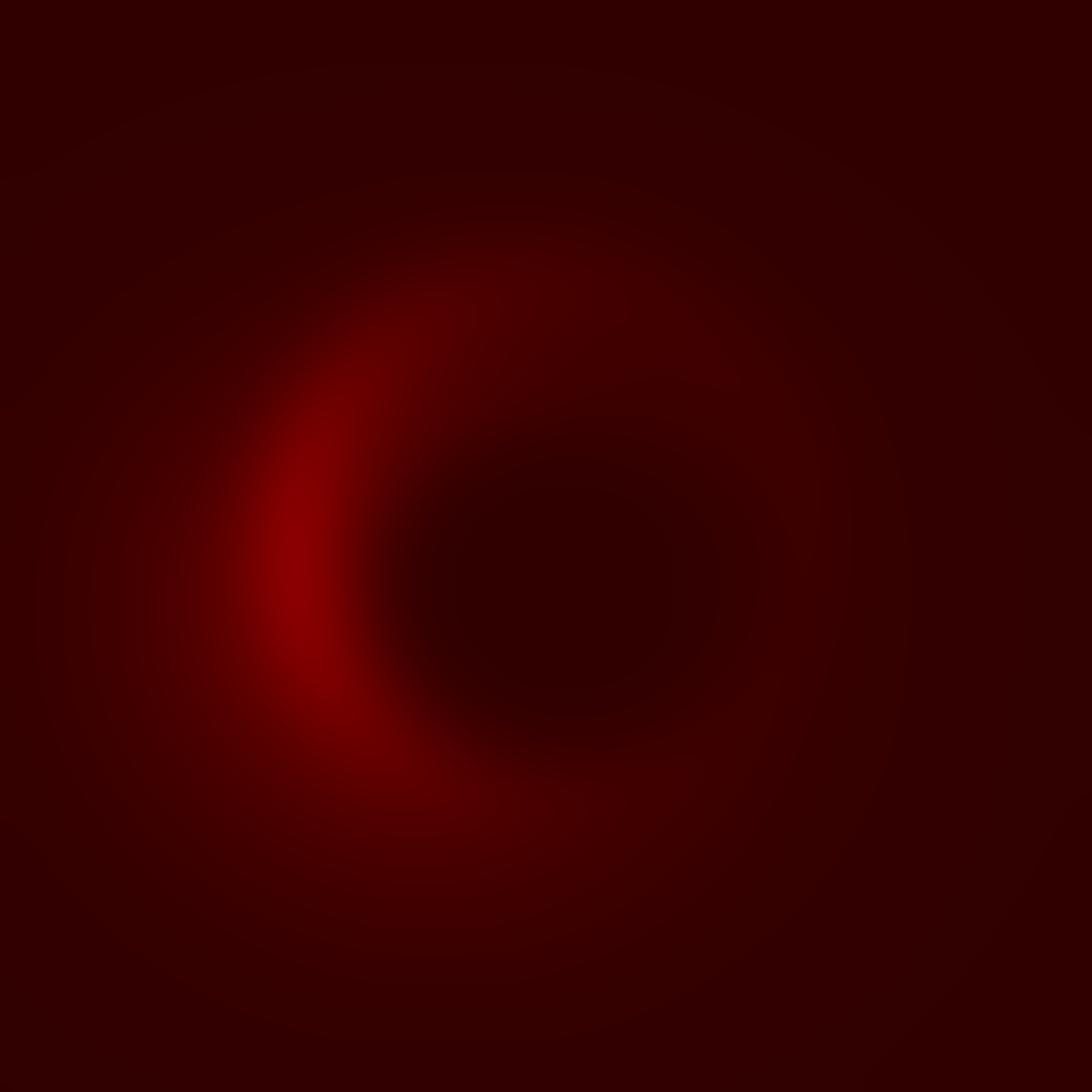}
\caption{Images of the Schwarzschild BH (first and third row) and the hairy BH of scalar hair parameter of $h = -1$ (second and fourth row) after being blurred with a Gaussian filter using a $20$ $\upmu$as kernel. The disk inclinations from left to right are $0^{\circ}$, $30^{\circ}$, $45^{\circ}$, and $60^{\circ}$, respectively, and are consistent in each column. The observation angle is $17^{\circ}$ for the first two rows and $85^{\circ}$ for the last two rows. It is observed that the precise position of the photon ring can no longer be accurately determined from the blurred image, even in certain parameter spaces where the emission ring structure is disrupted and only the crescent shape remains.}}\label{fig14}
\end{figure*}
\begin{figure*}
\center{
\includegraphics[width=1.8cm]{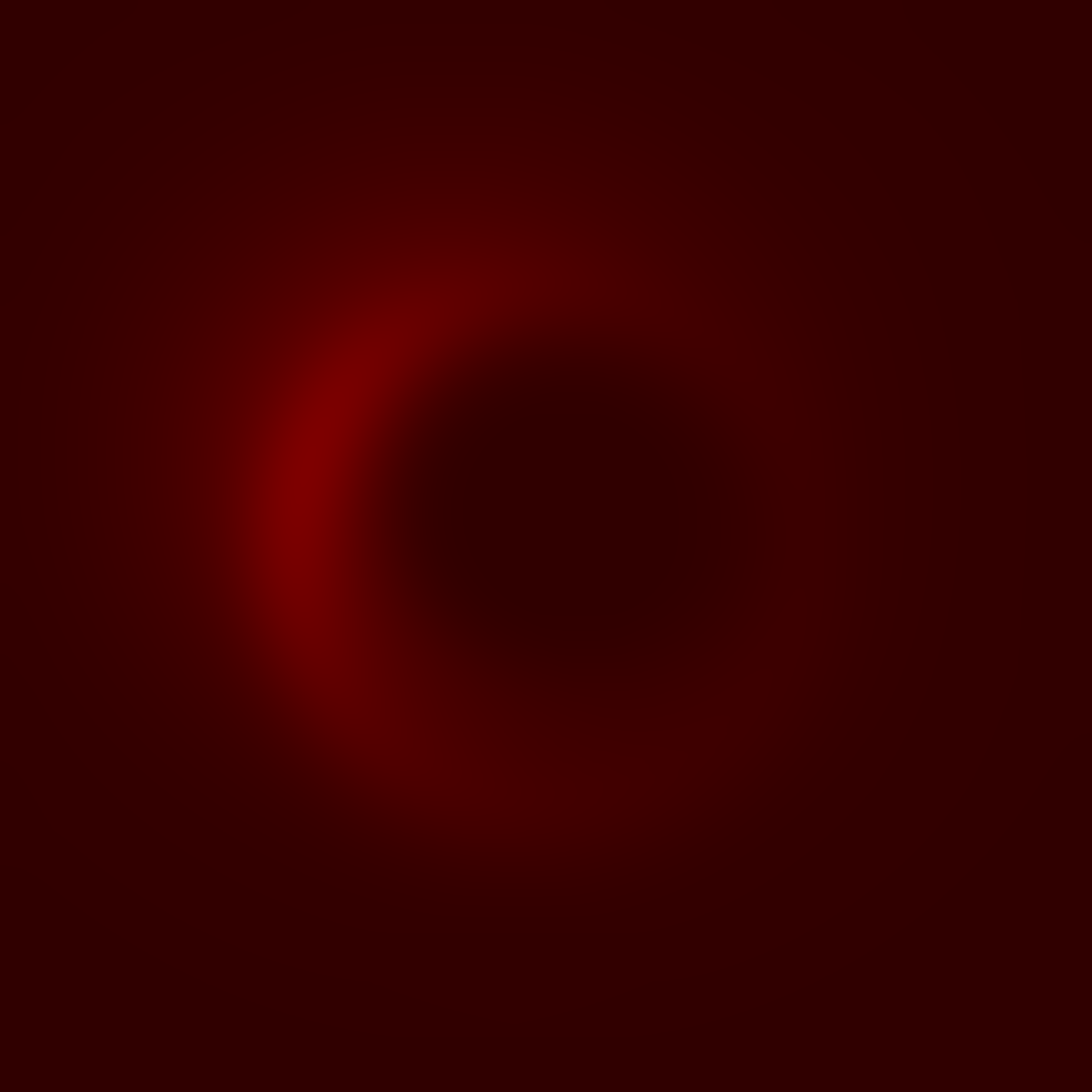}
\includegraphics[width=1.8cm]{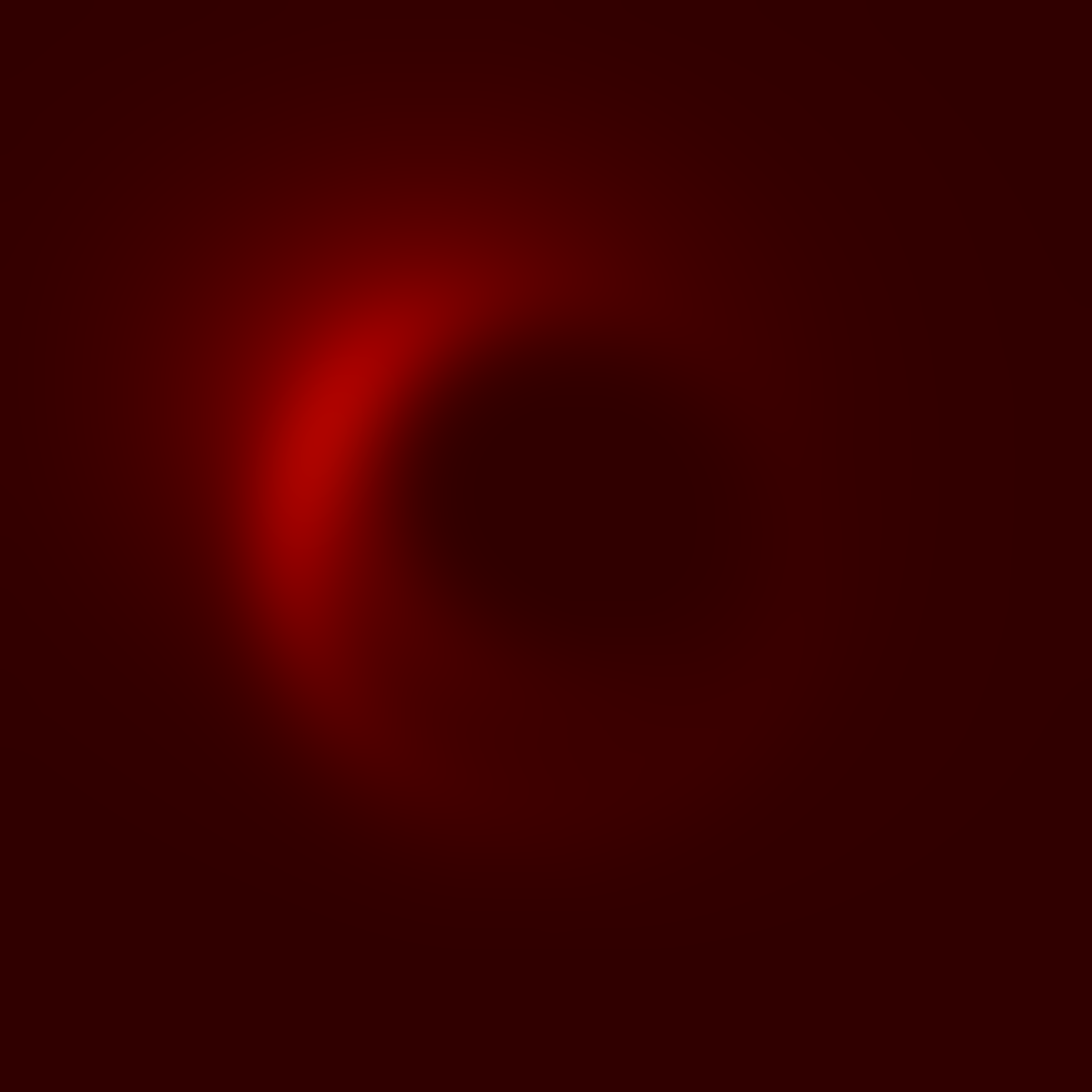}
\includegraphics[width=1.8cm]{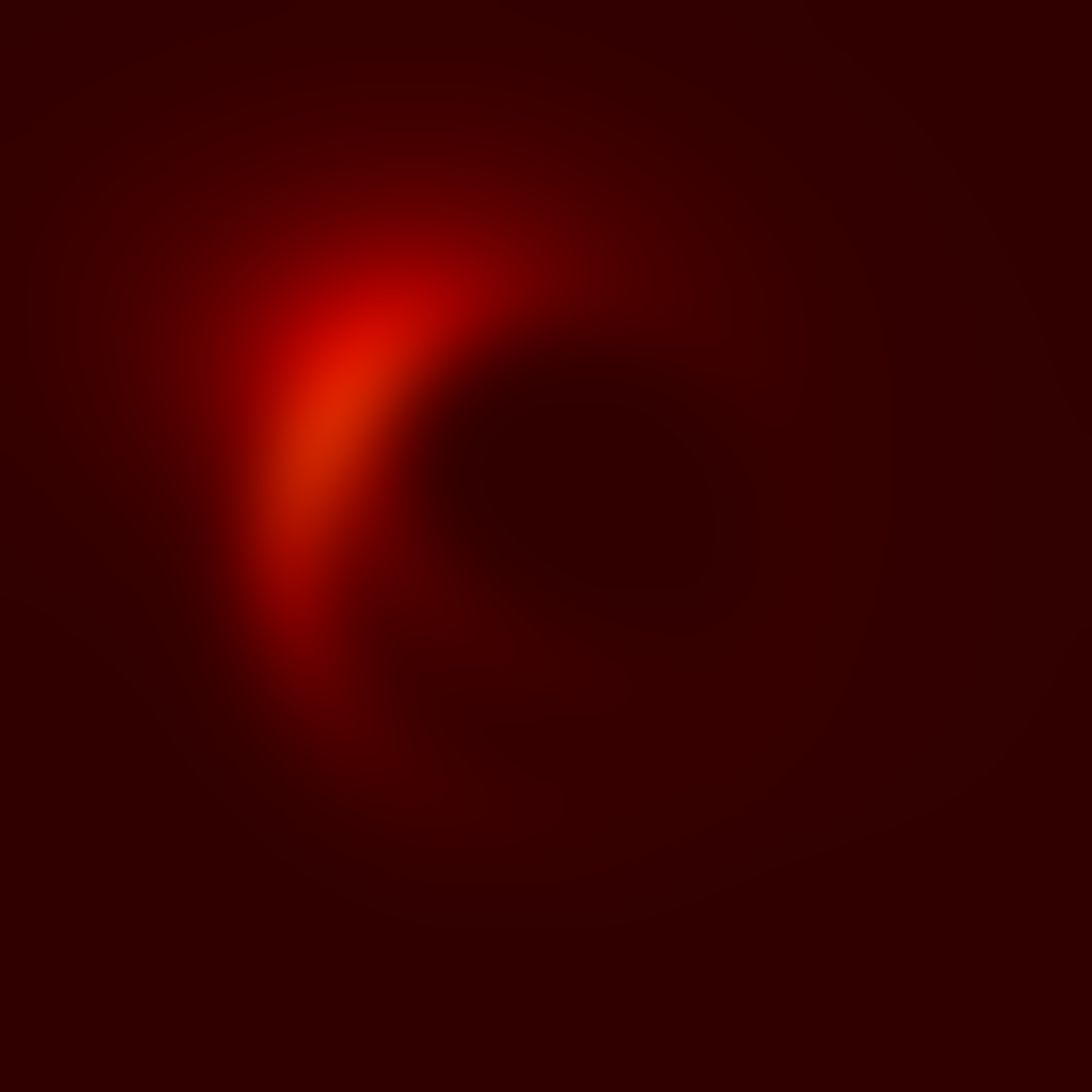}
\includegraphics[width=1.8cm]{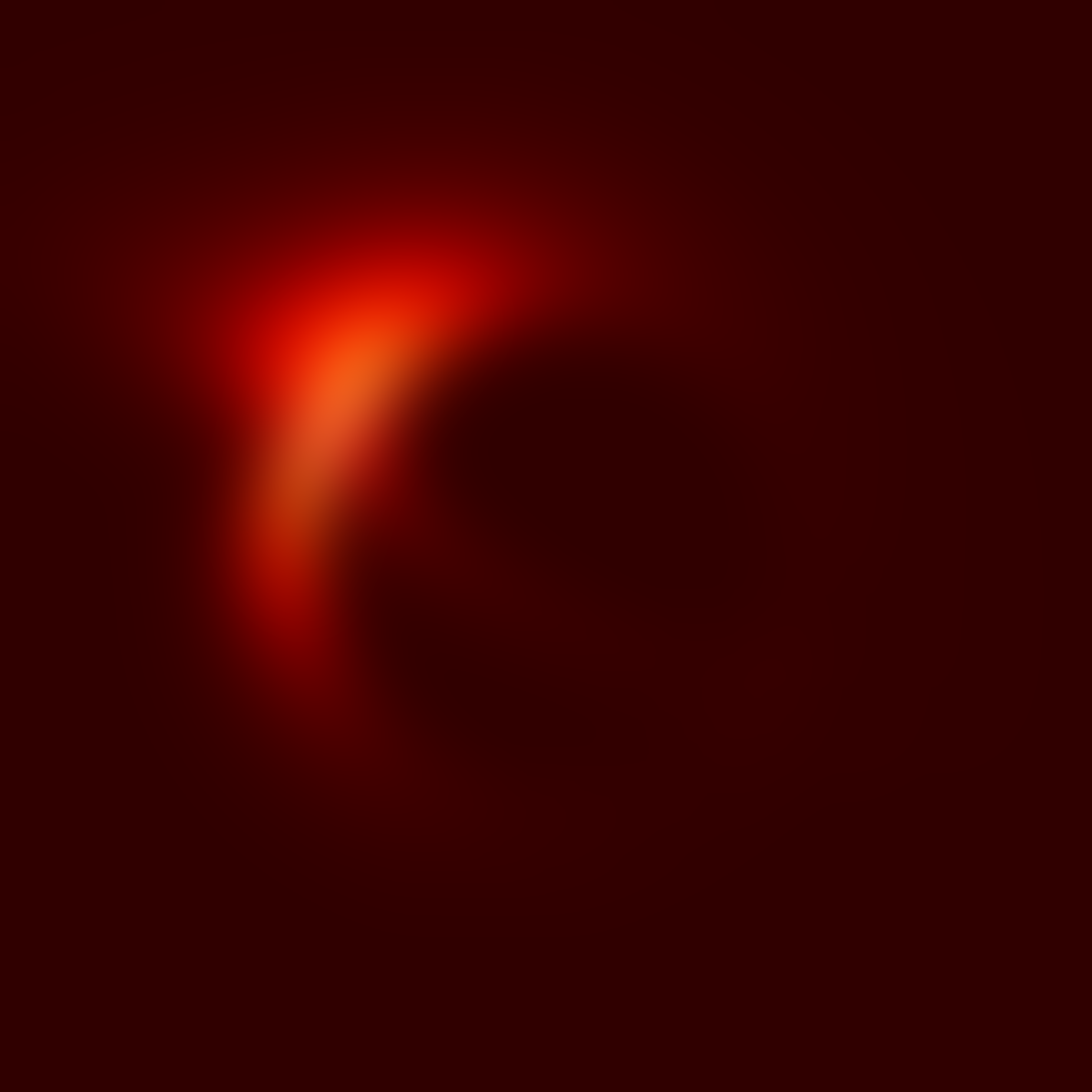}
\includegraphics[width=1.8cm]{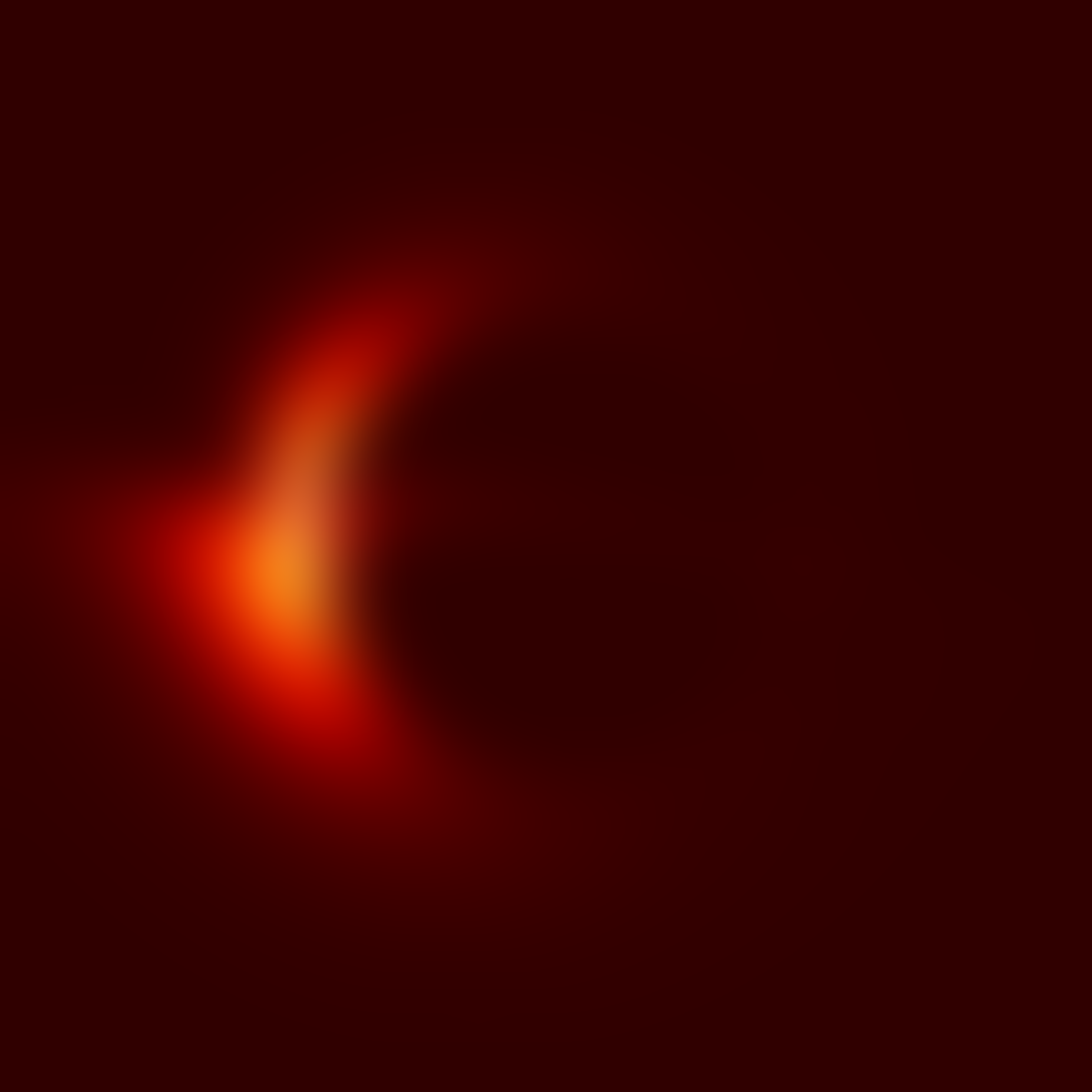}
\includegraphics[width=1.8cm]{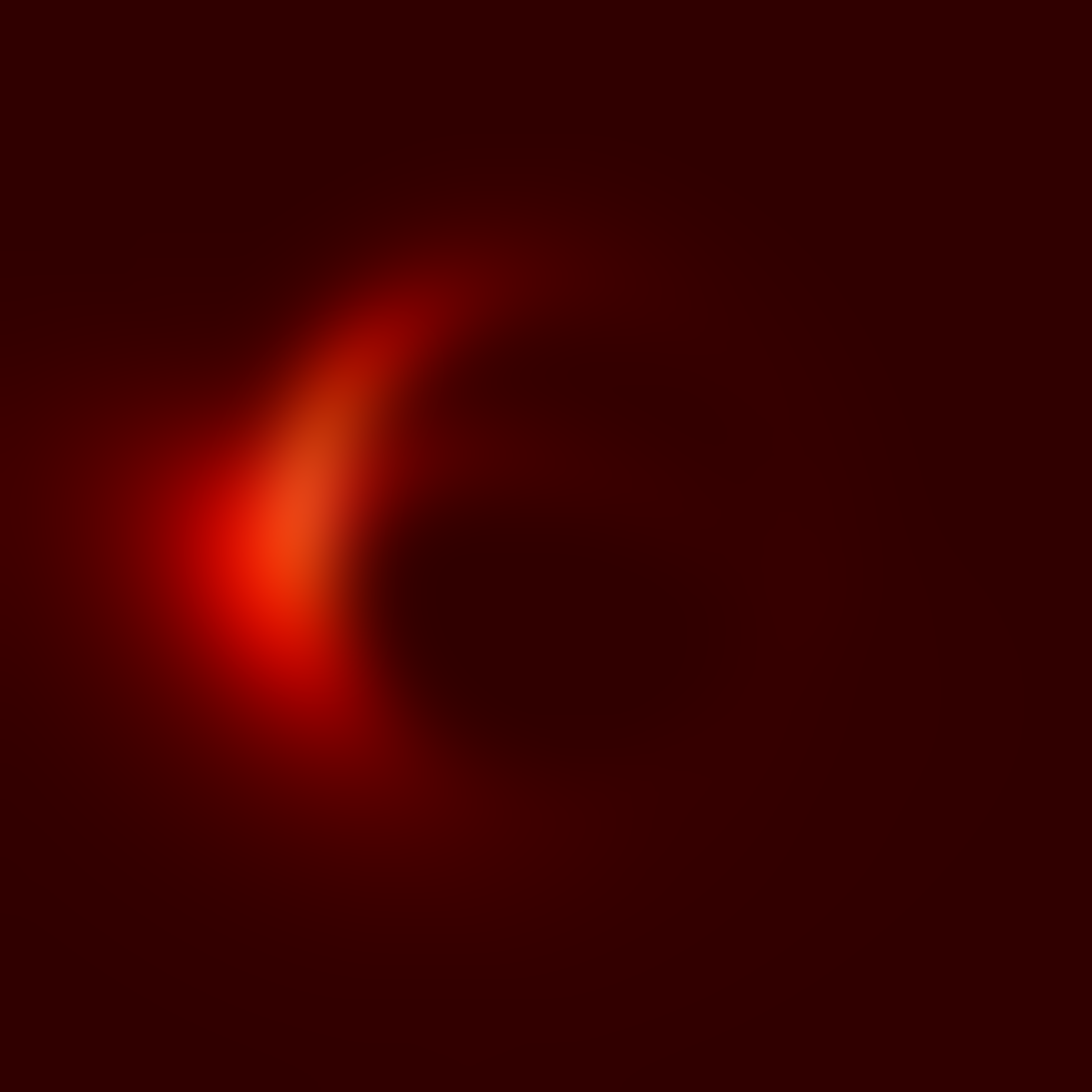}
\includegraphics[width=1.8cm]{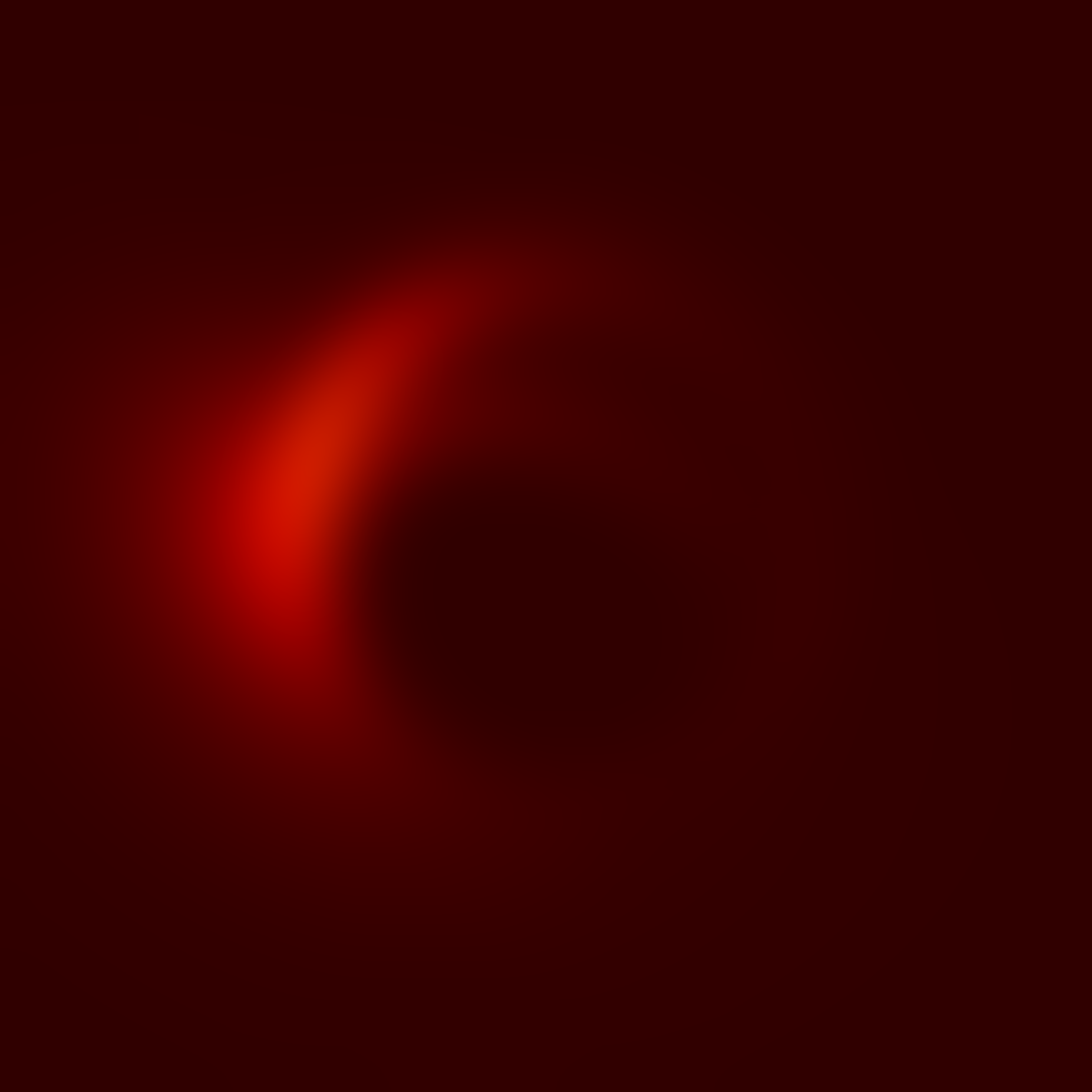}
\includegraphics[width=1.8cm]{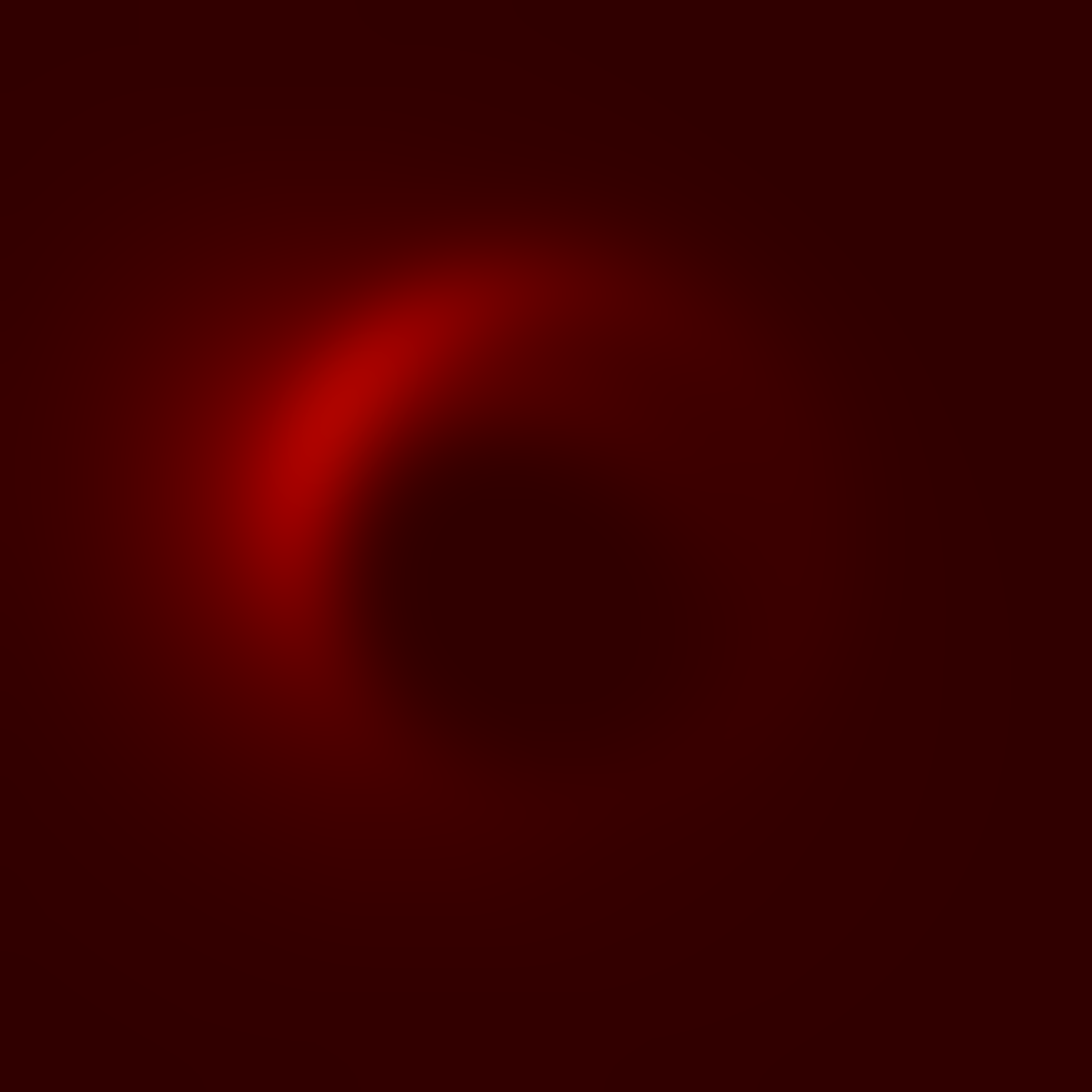}
\includegraphics[width=1.8cm]{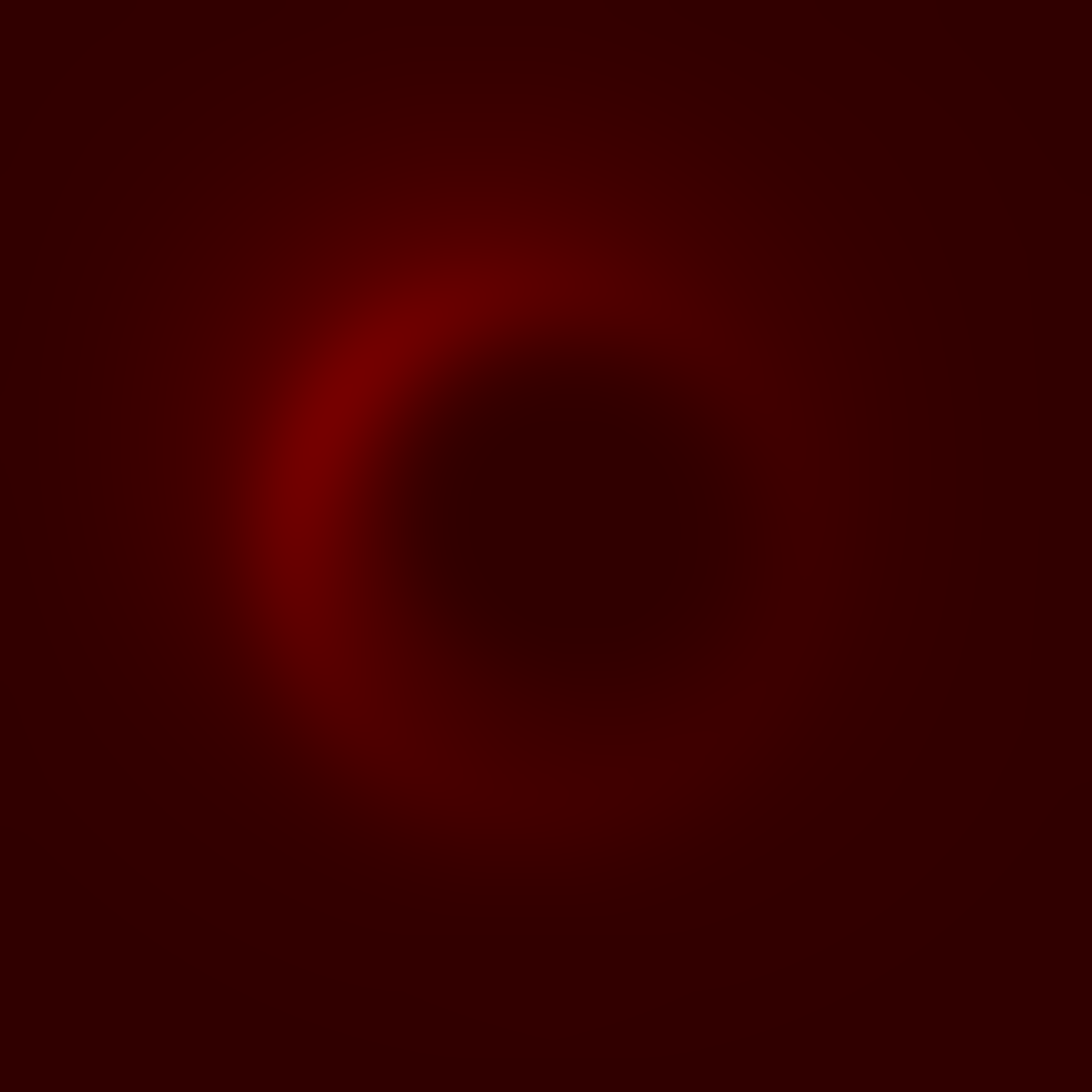}
\includegraphics[width=1.8cm]{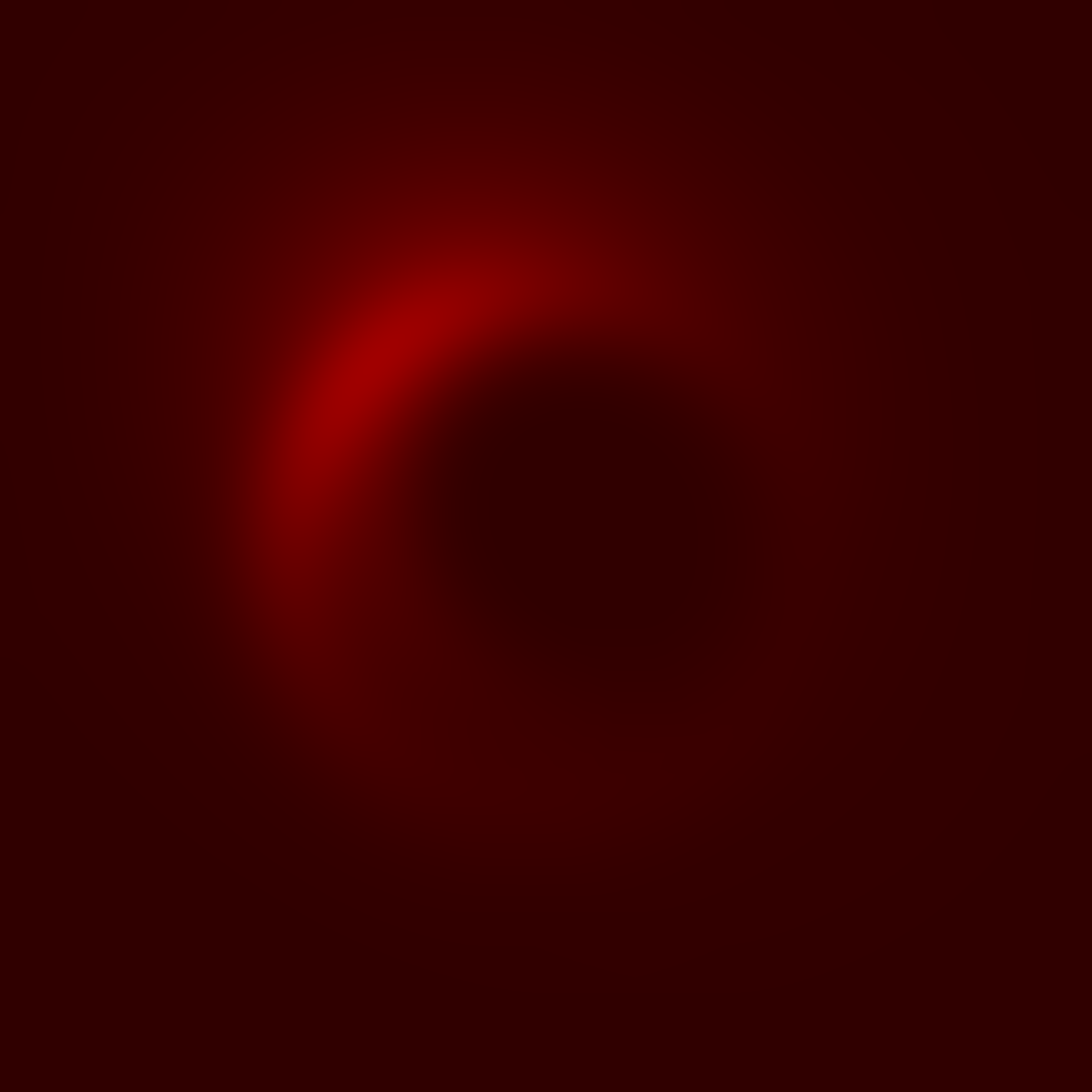}
\includegraphics[width=1.8cm]{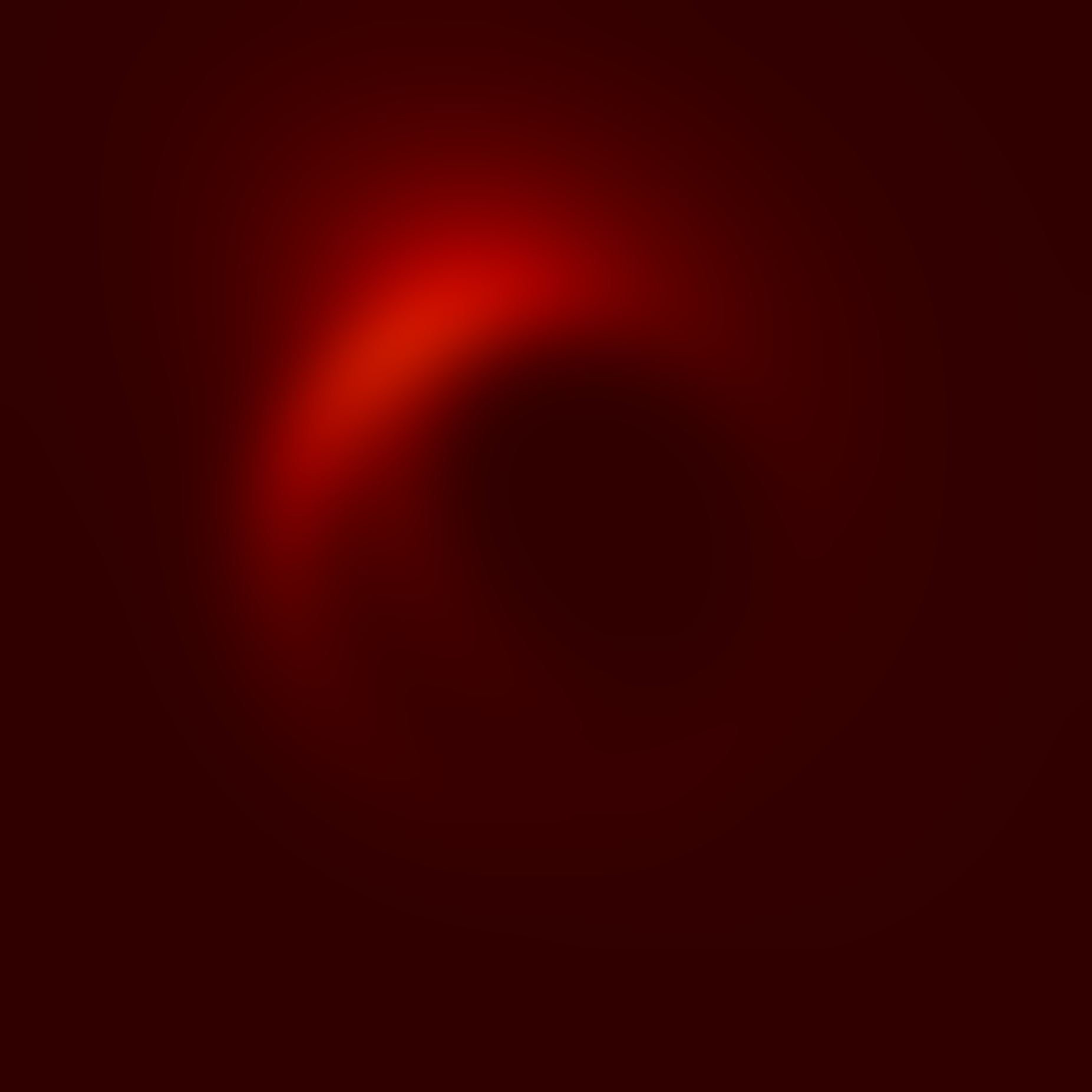}
\includegraphics[width=1.8cm]{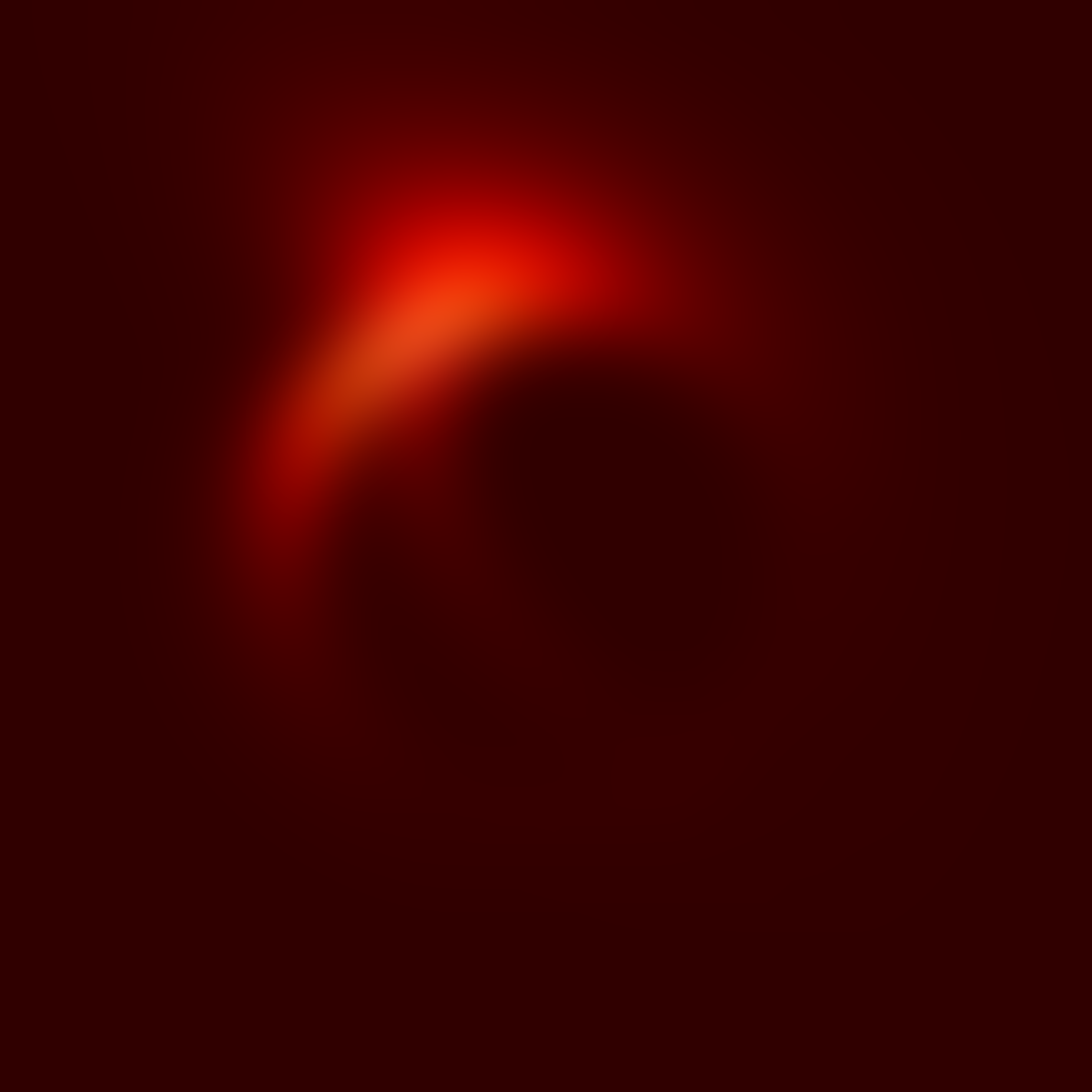}
\includegraphics[width=1.8cm]{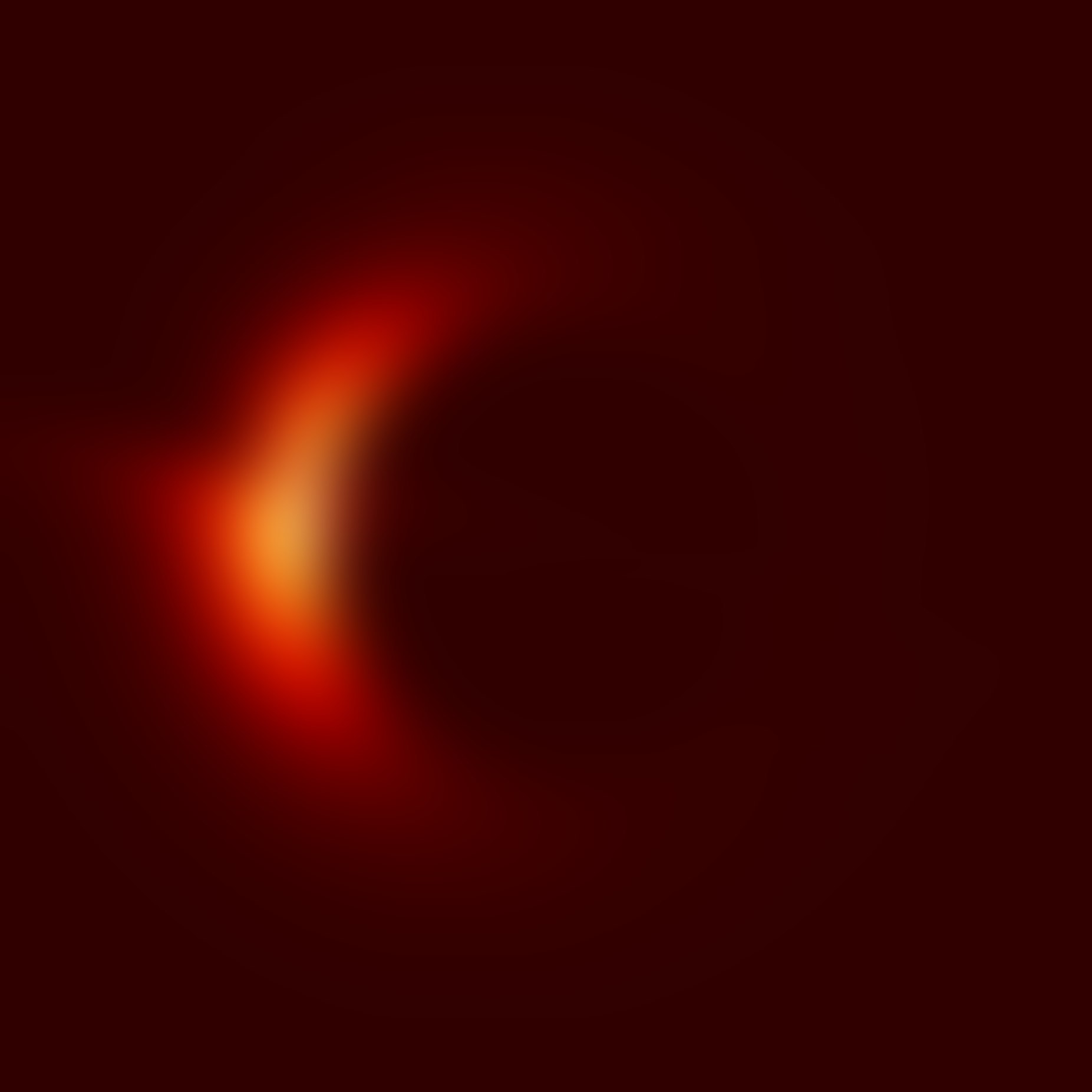}
\includegraphics[width=1.8cm]{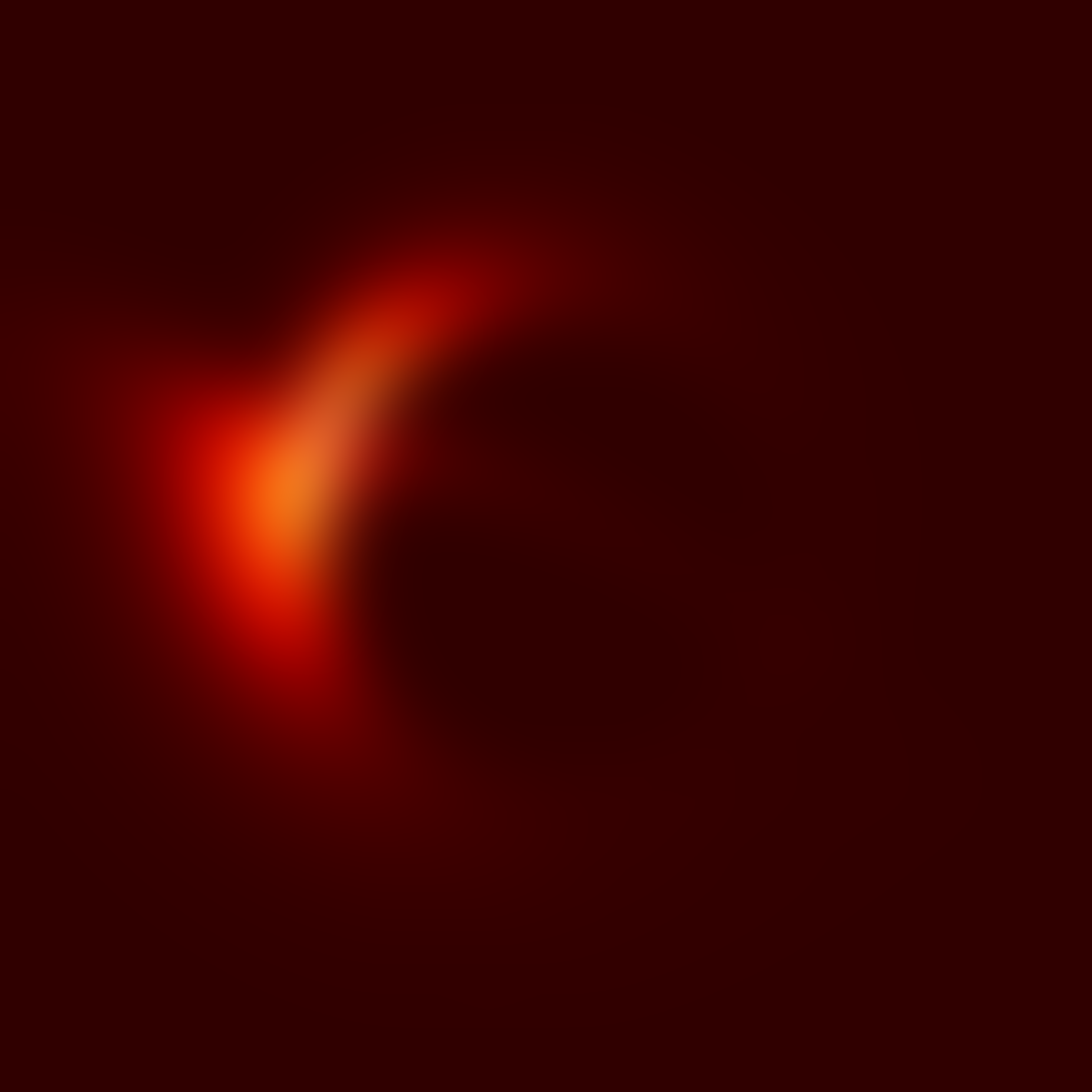}
\includegraphics[width=1.8cm]{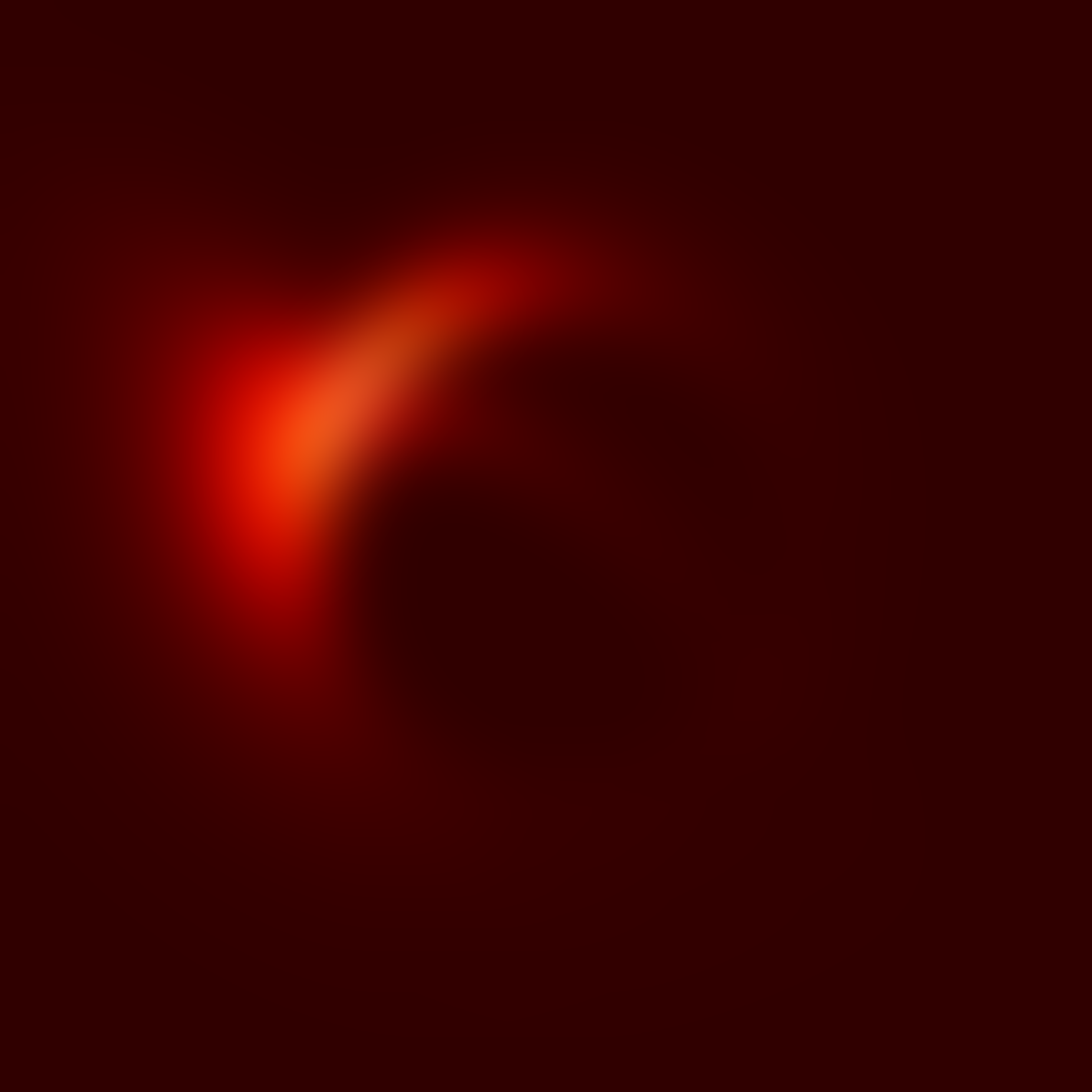}
\includegraphics[width=1.8cm]{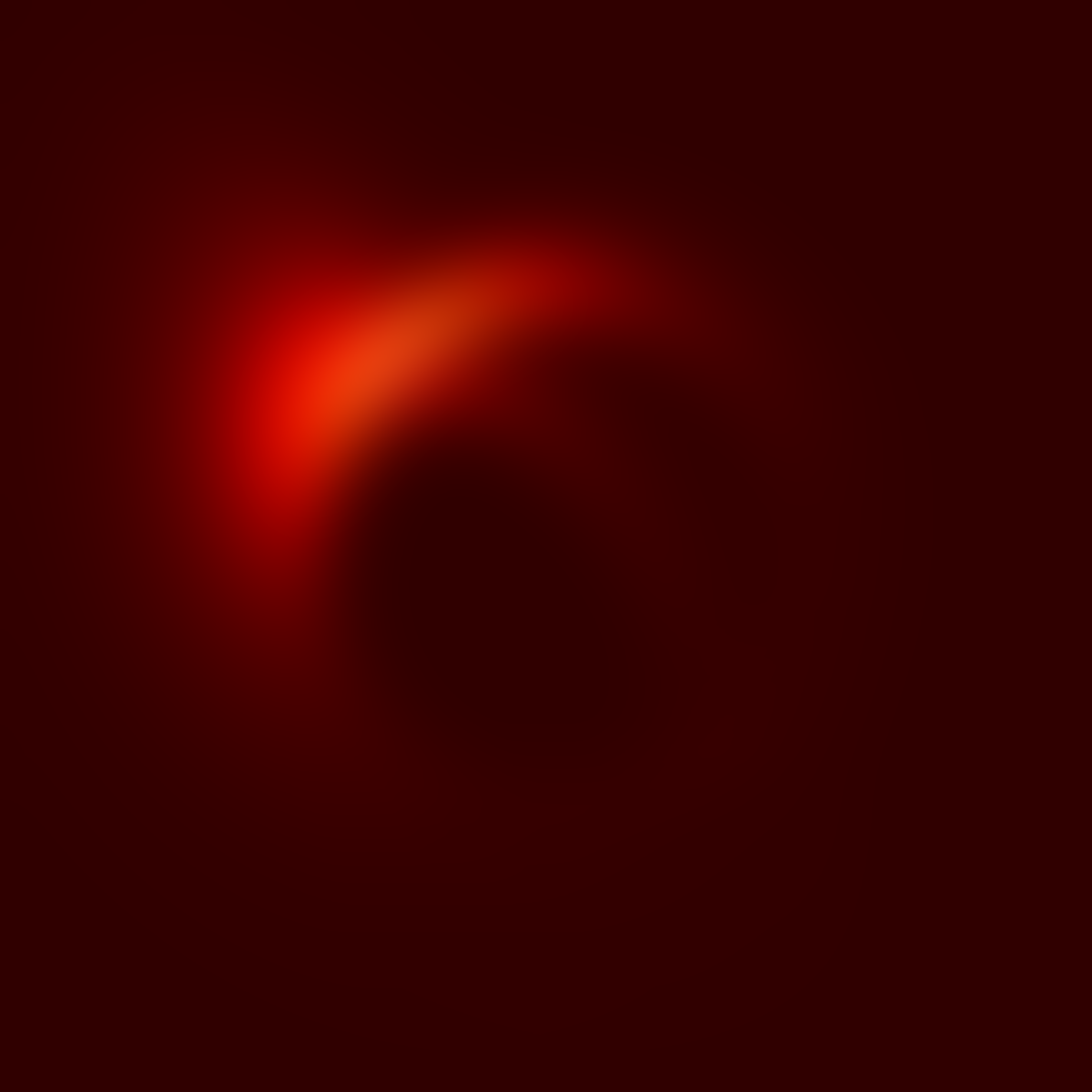}
\includegraphics[width=1.8cm]{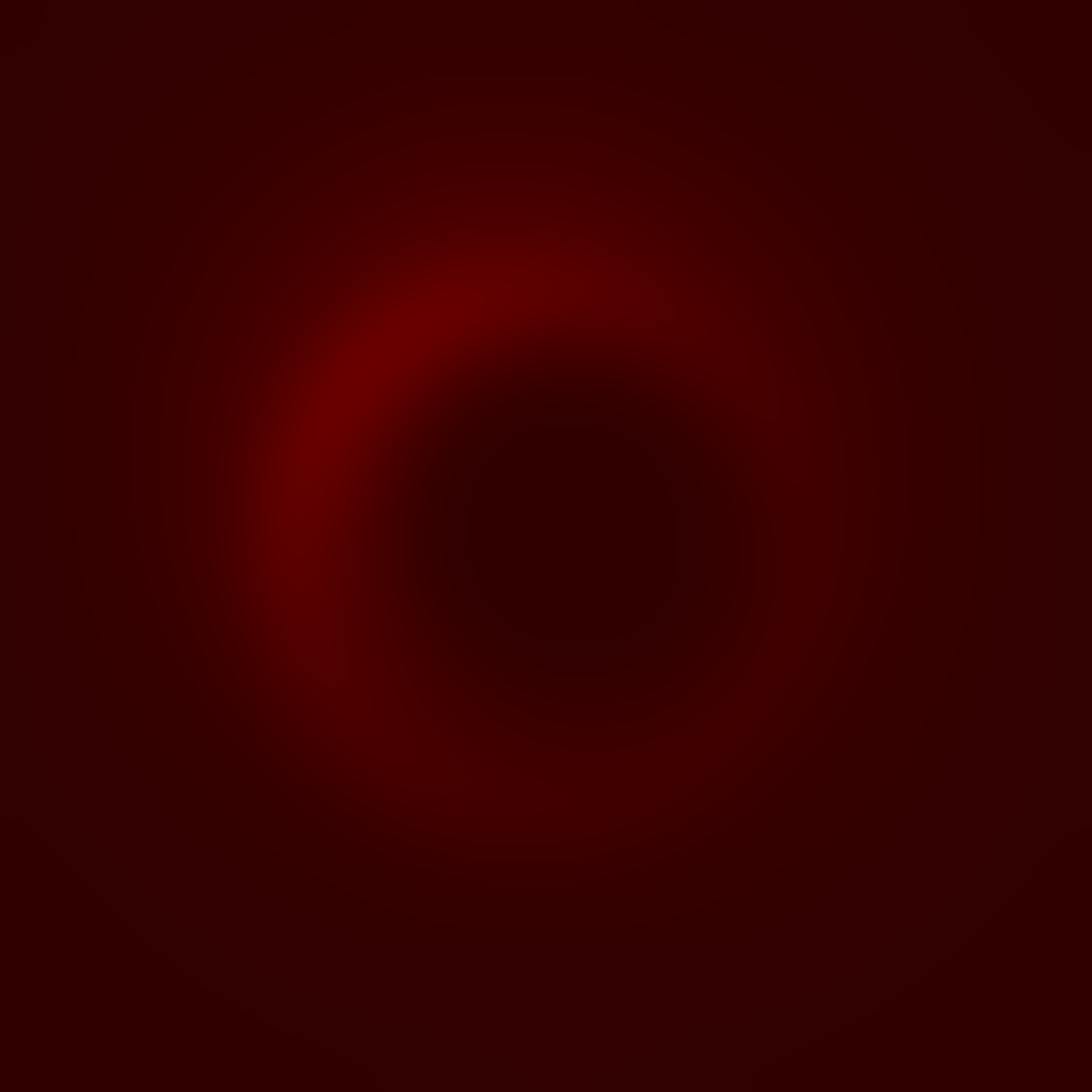}
\includegraphics[width=1.8cm]{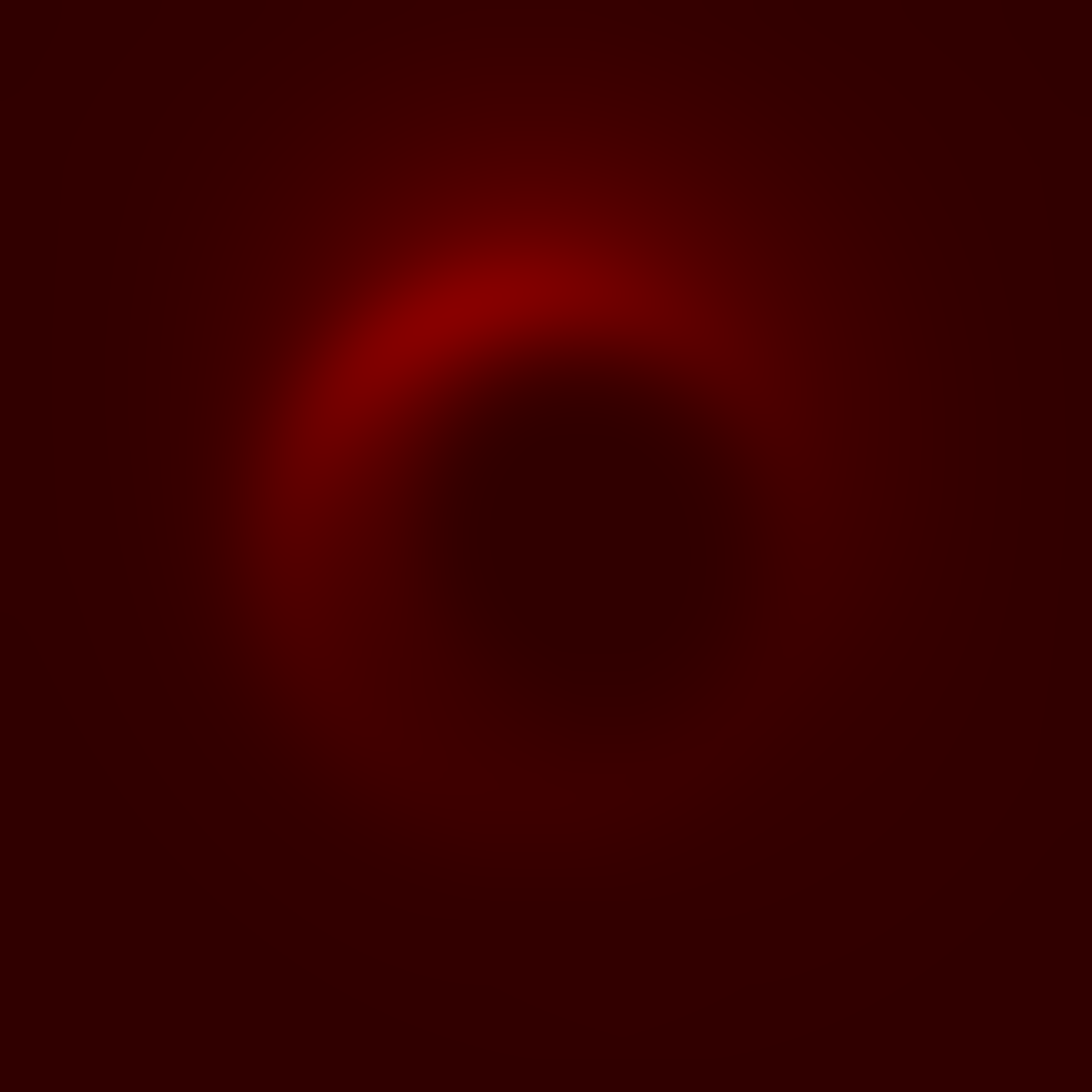}
\includegraphics[width=1.8cm]{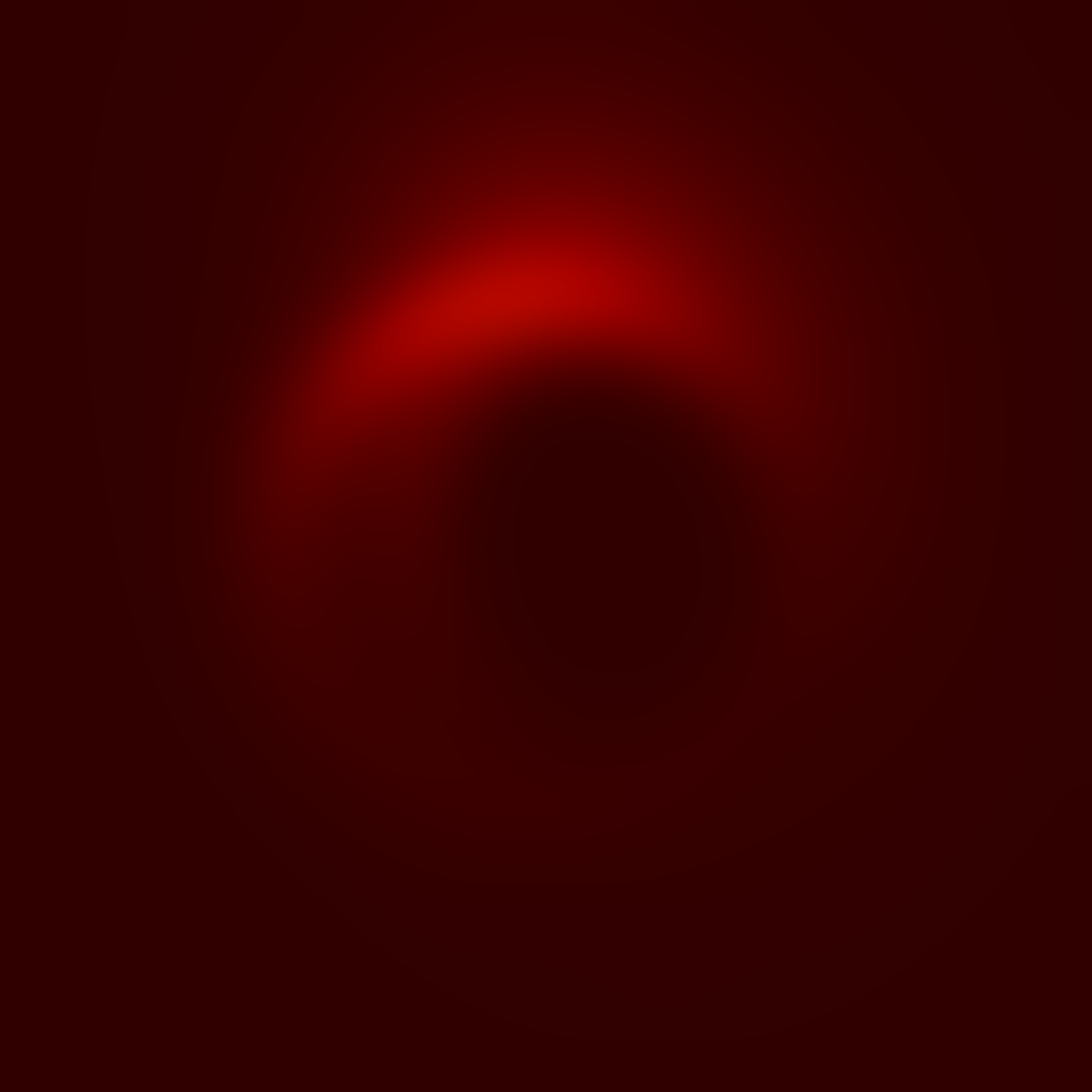}
\includegraphics[width=1.8cm]{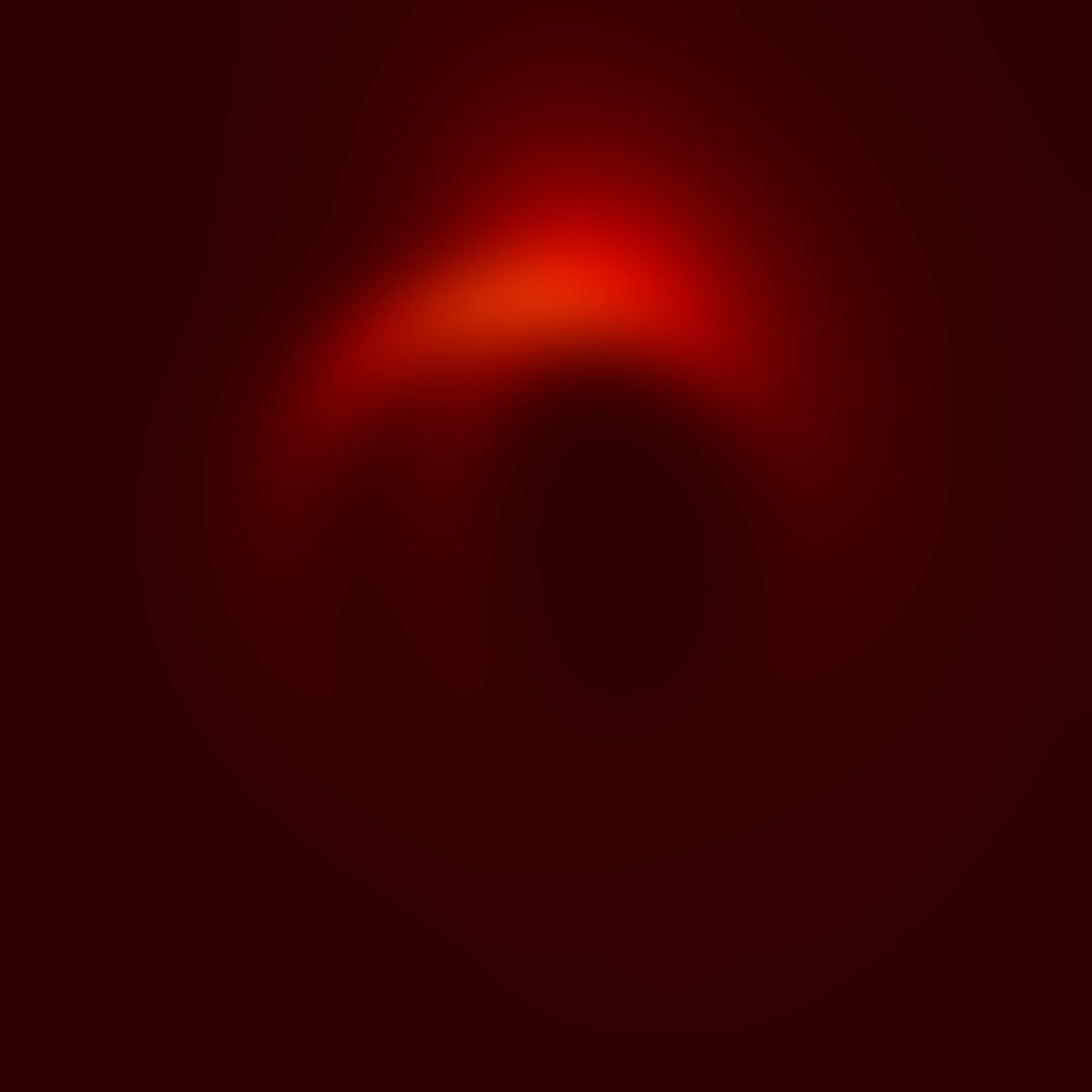}
\includegraphics[width=1.8cm]{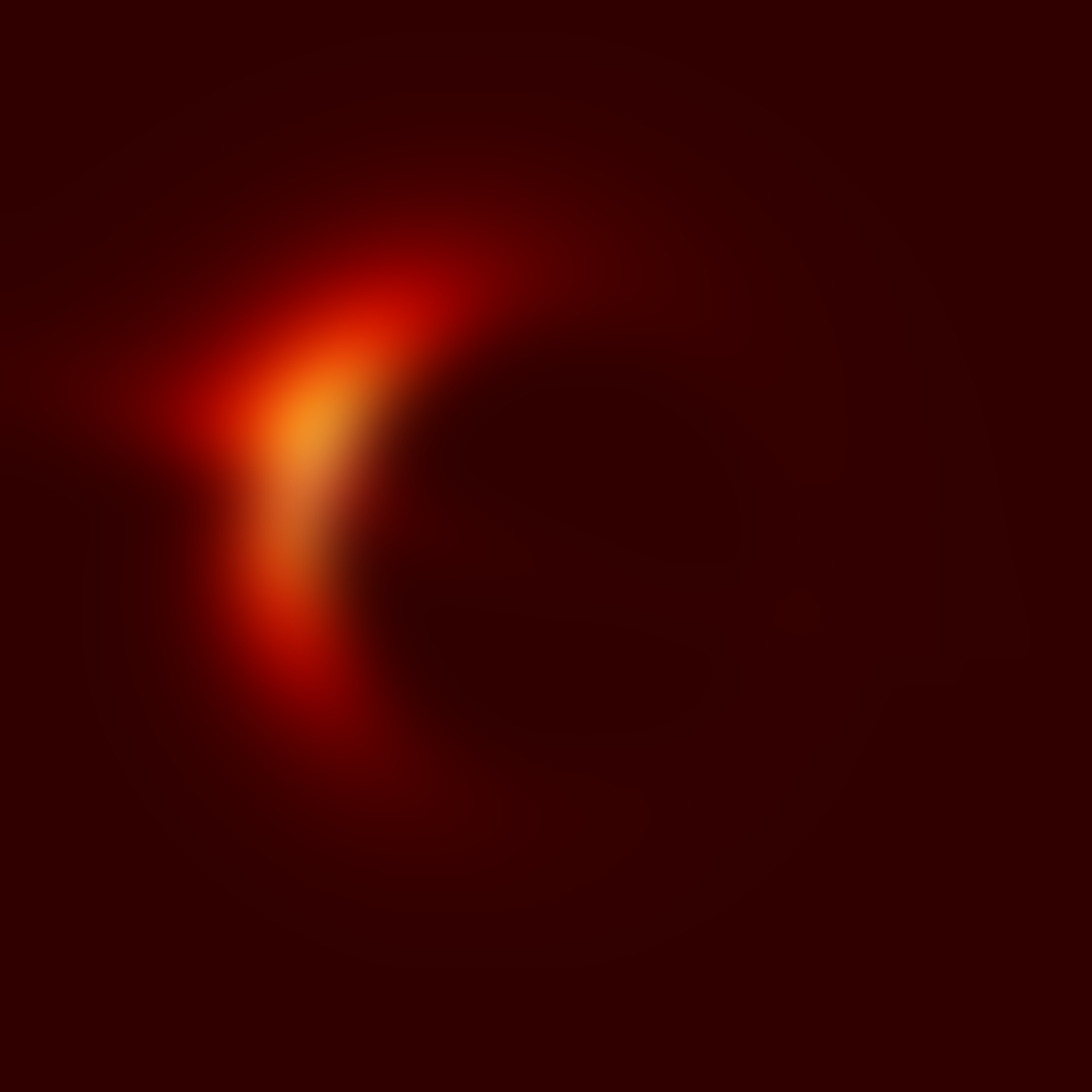}
\includegraphics[width=1.8cm]{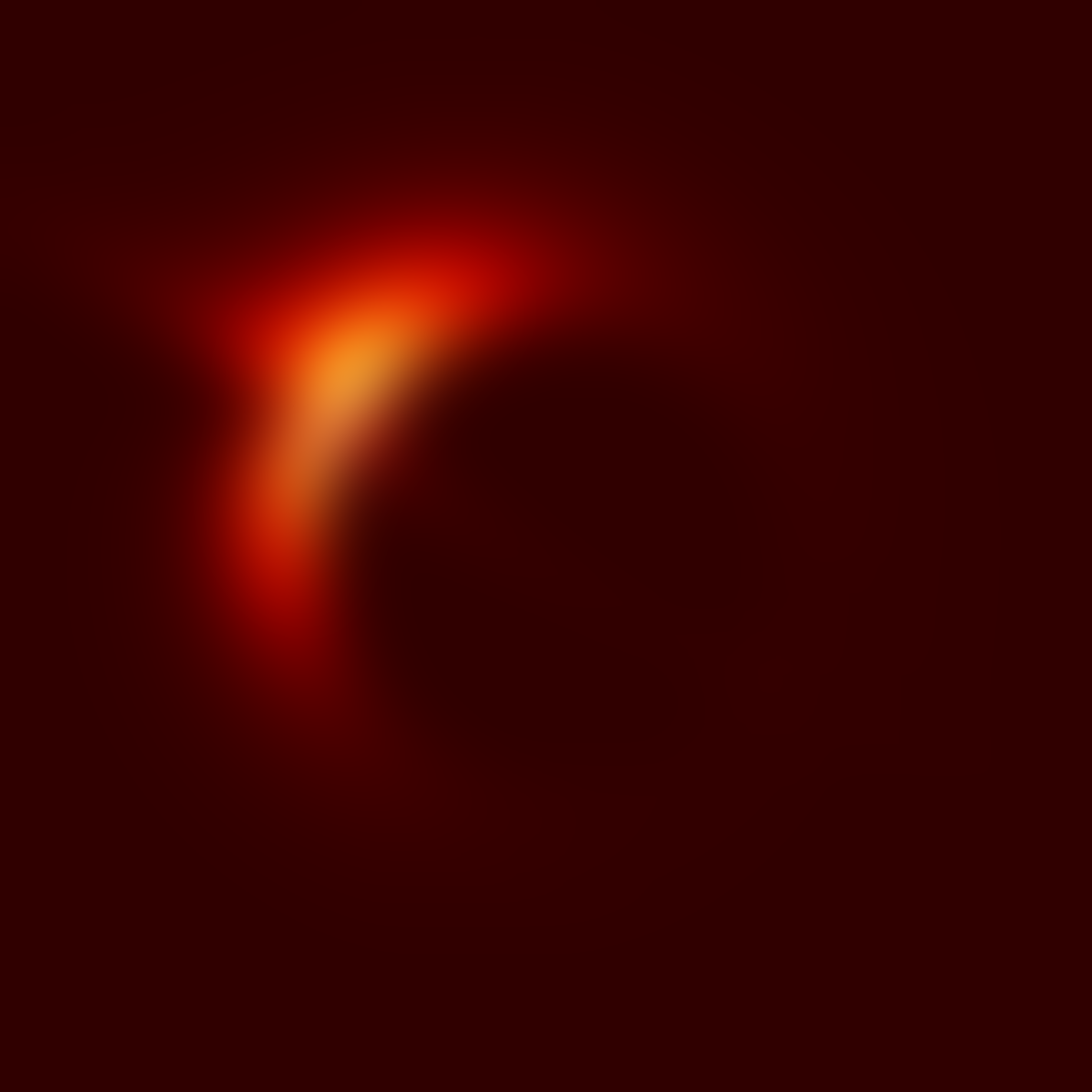}
\includegraphics[width=1.8cm]{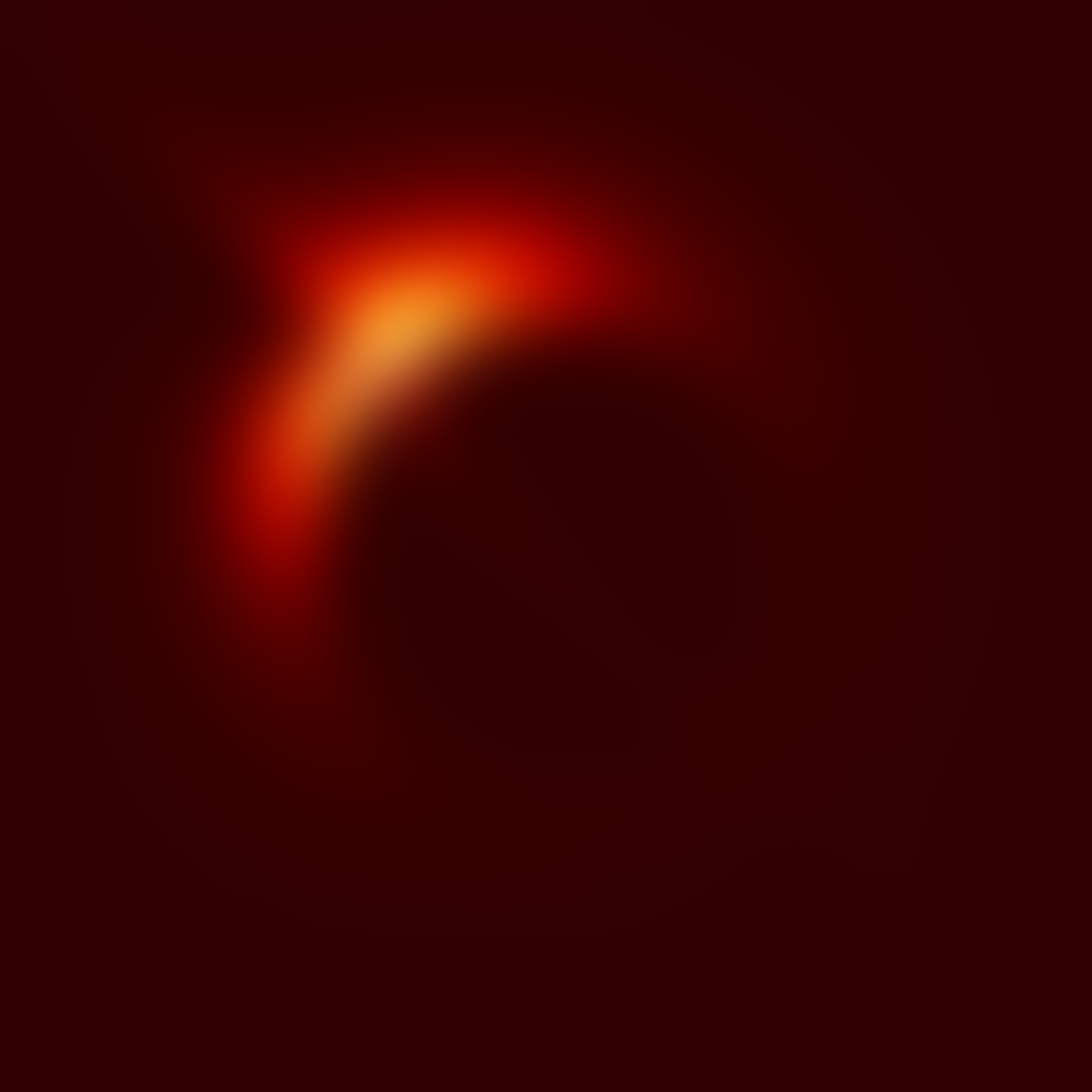}
\includegraphics[width=1.8cm]{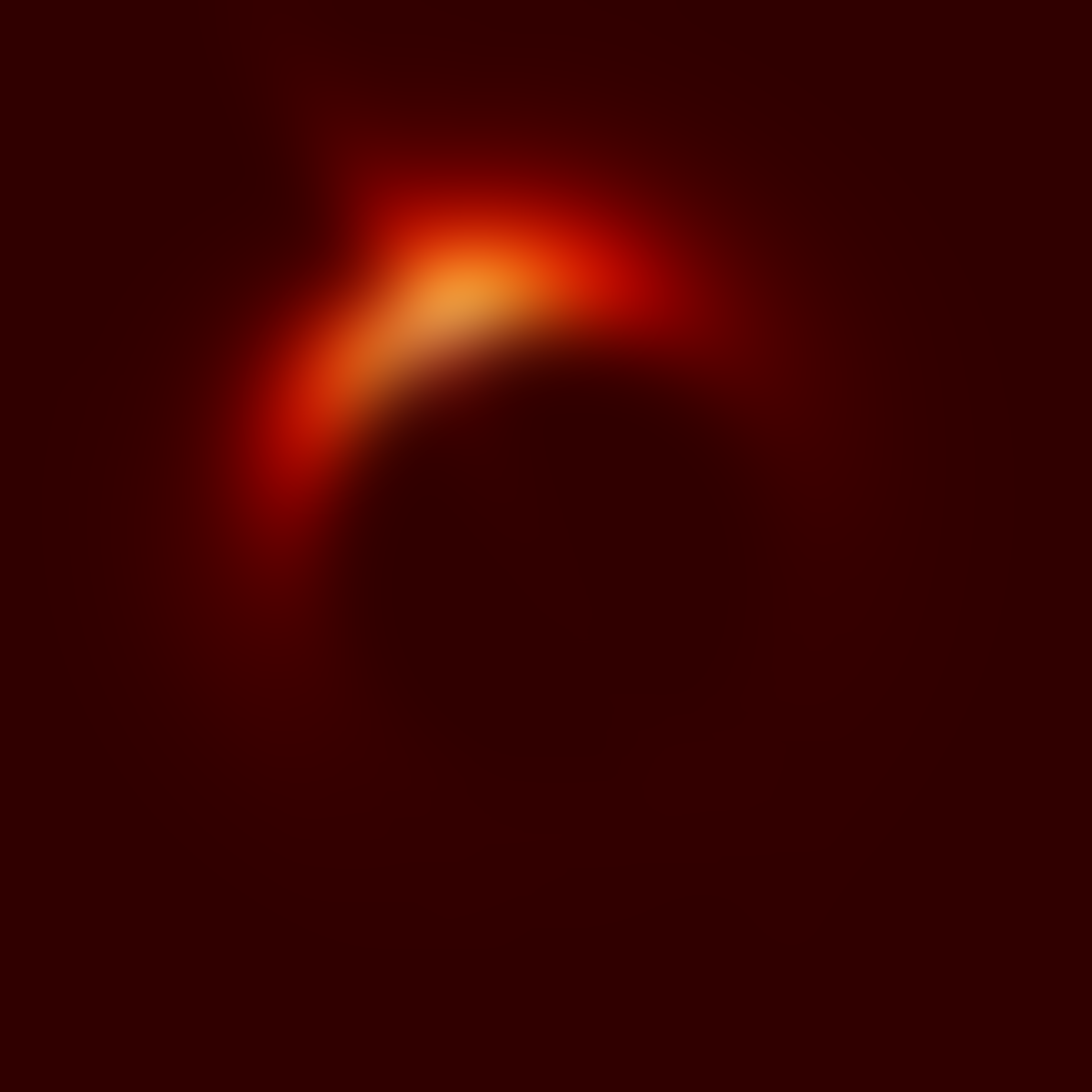}
\includegraphics[width=1.8cm]{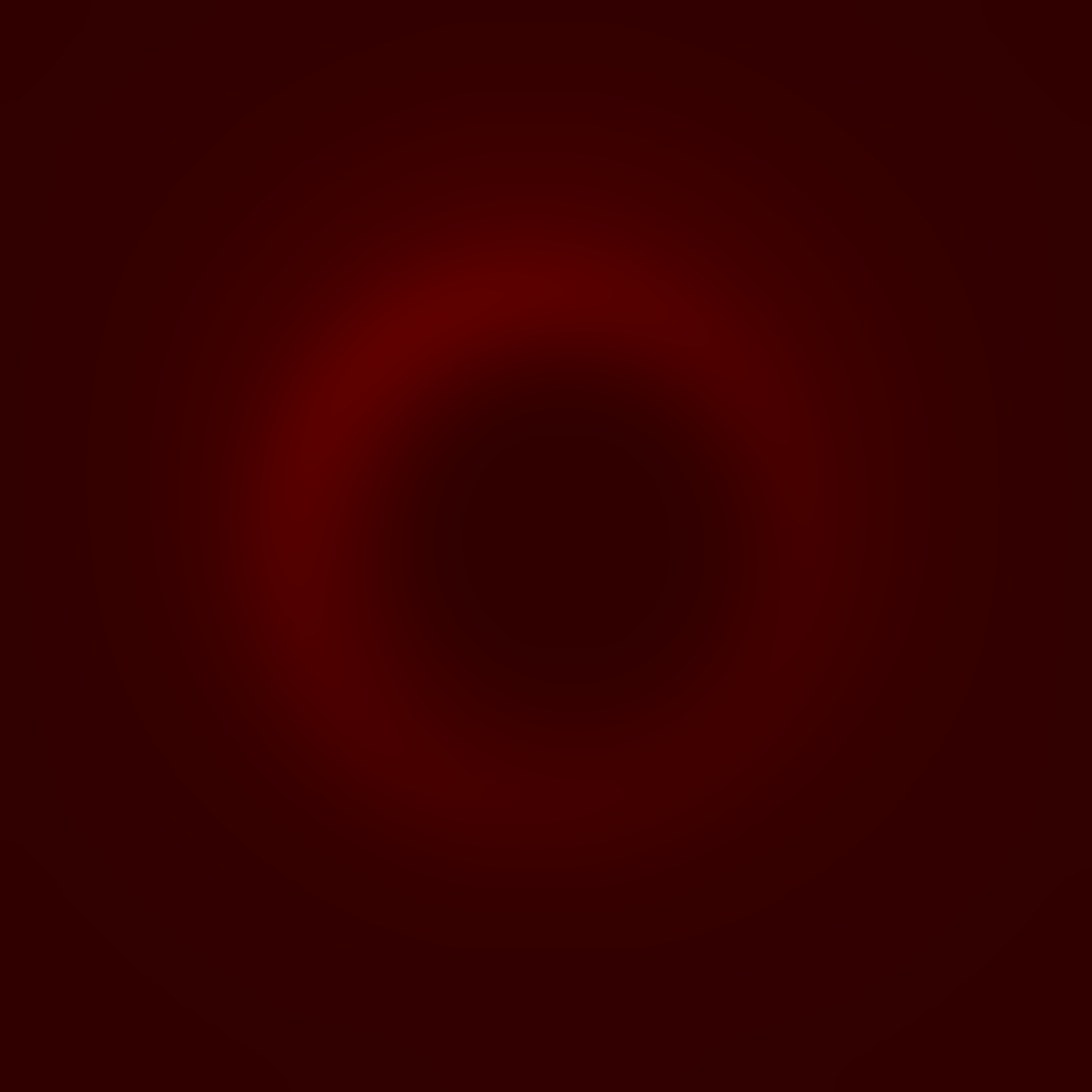}
\includegraphics[width=1.8cm]{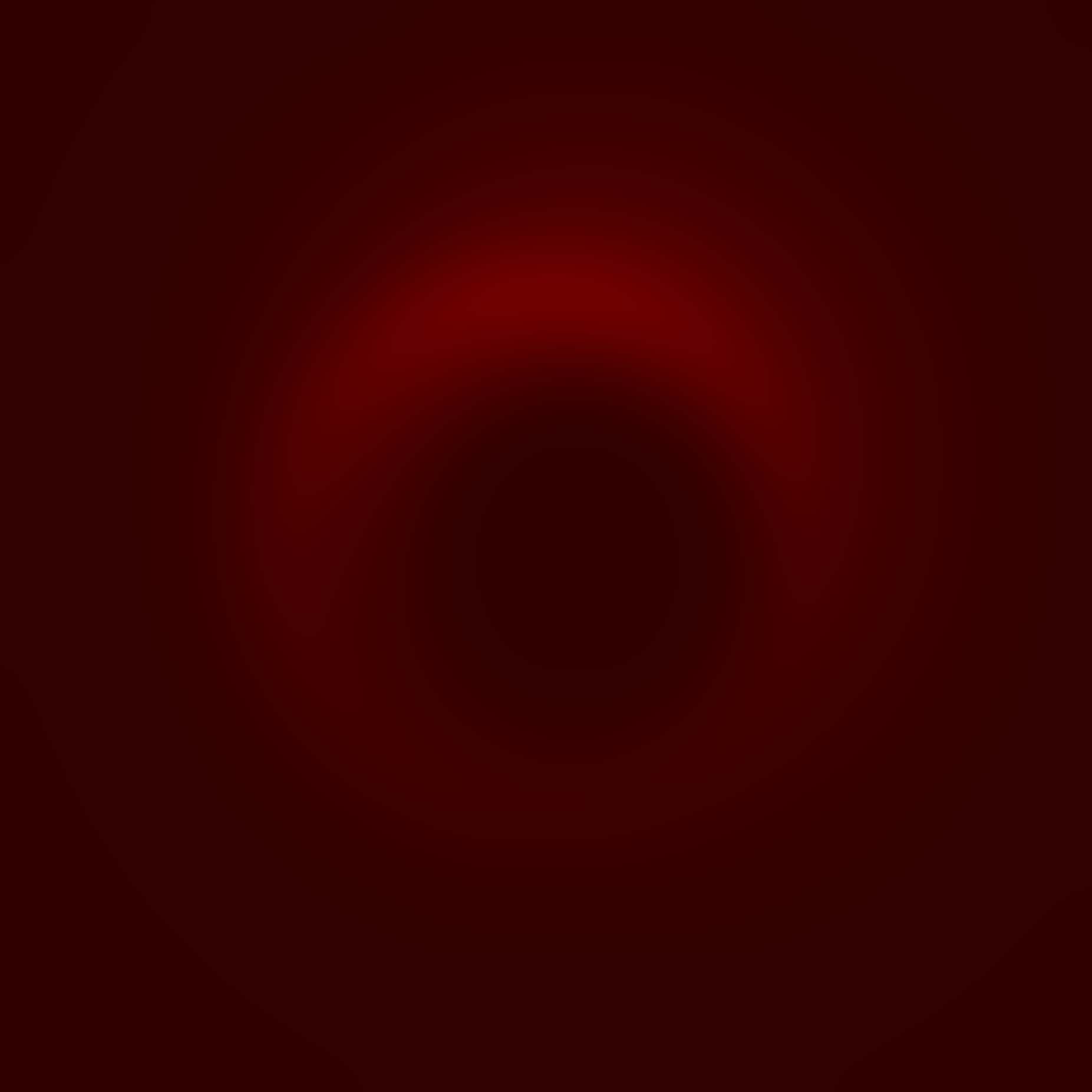}
\includegraphics[width=1.8cm]{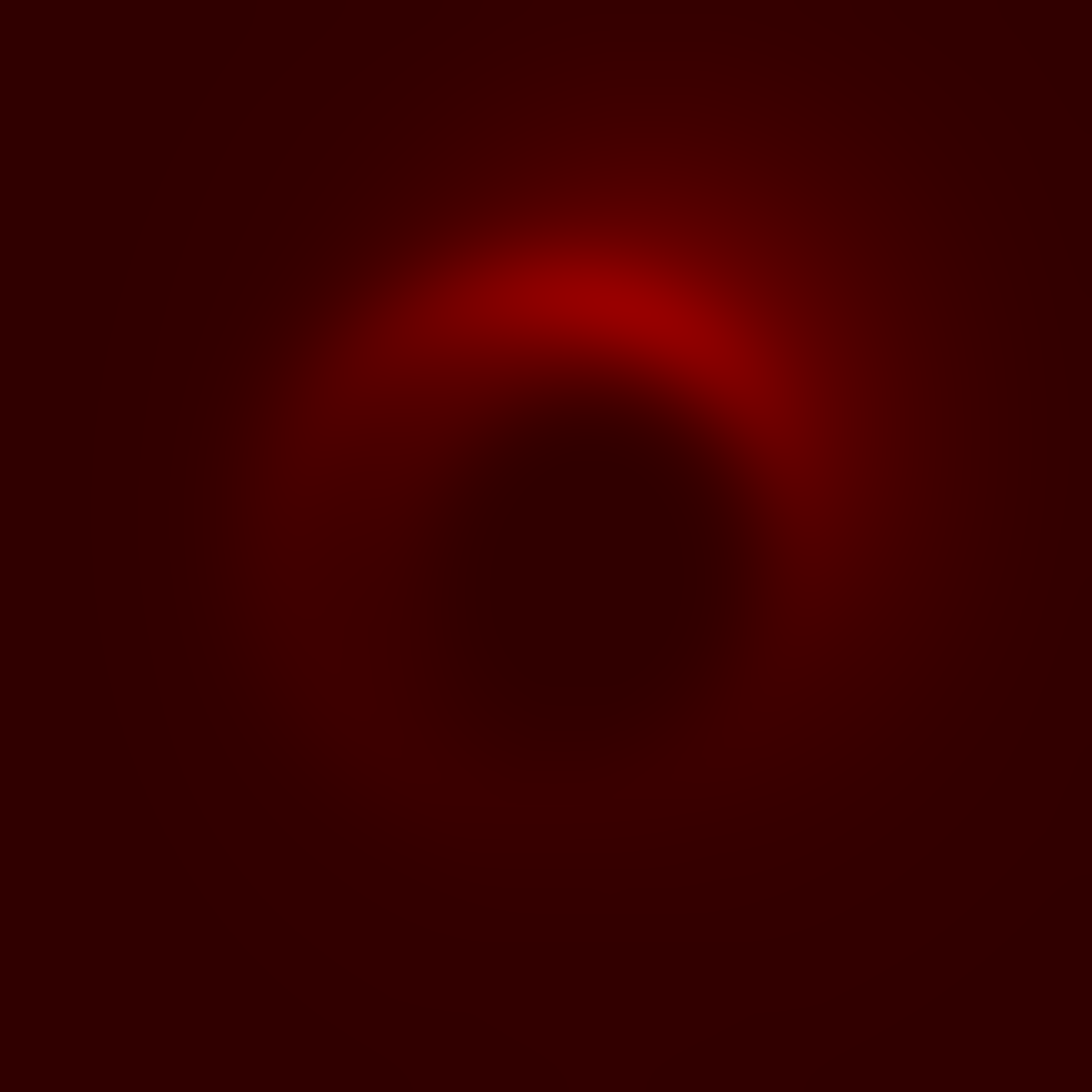}
\includegraphics[width=1.8cm]{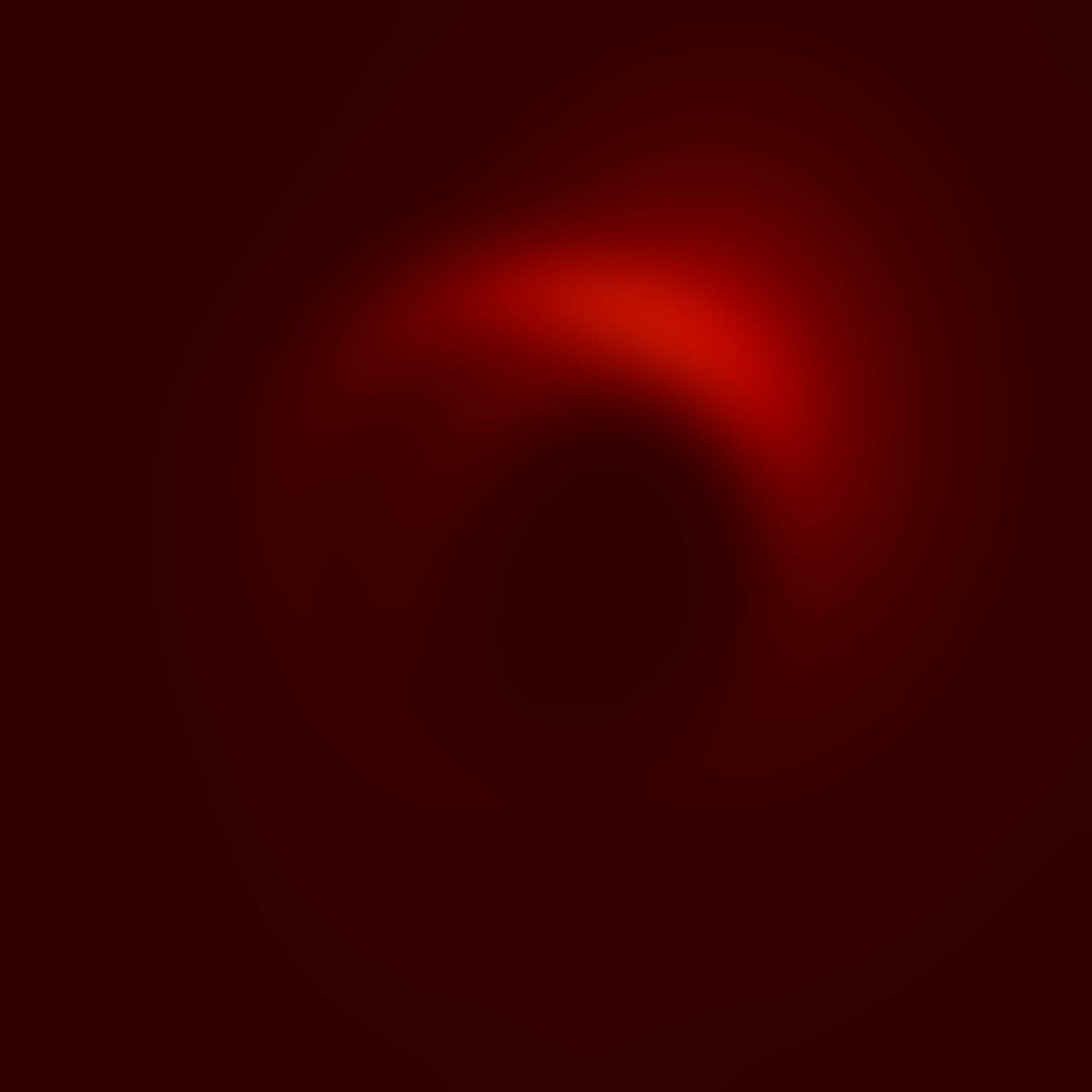}
\includegraphics[width=1.8cm]{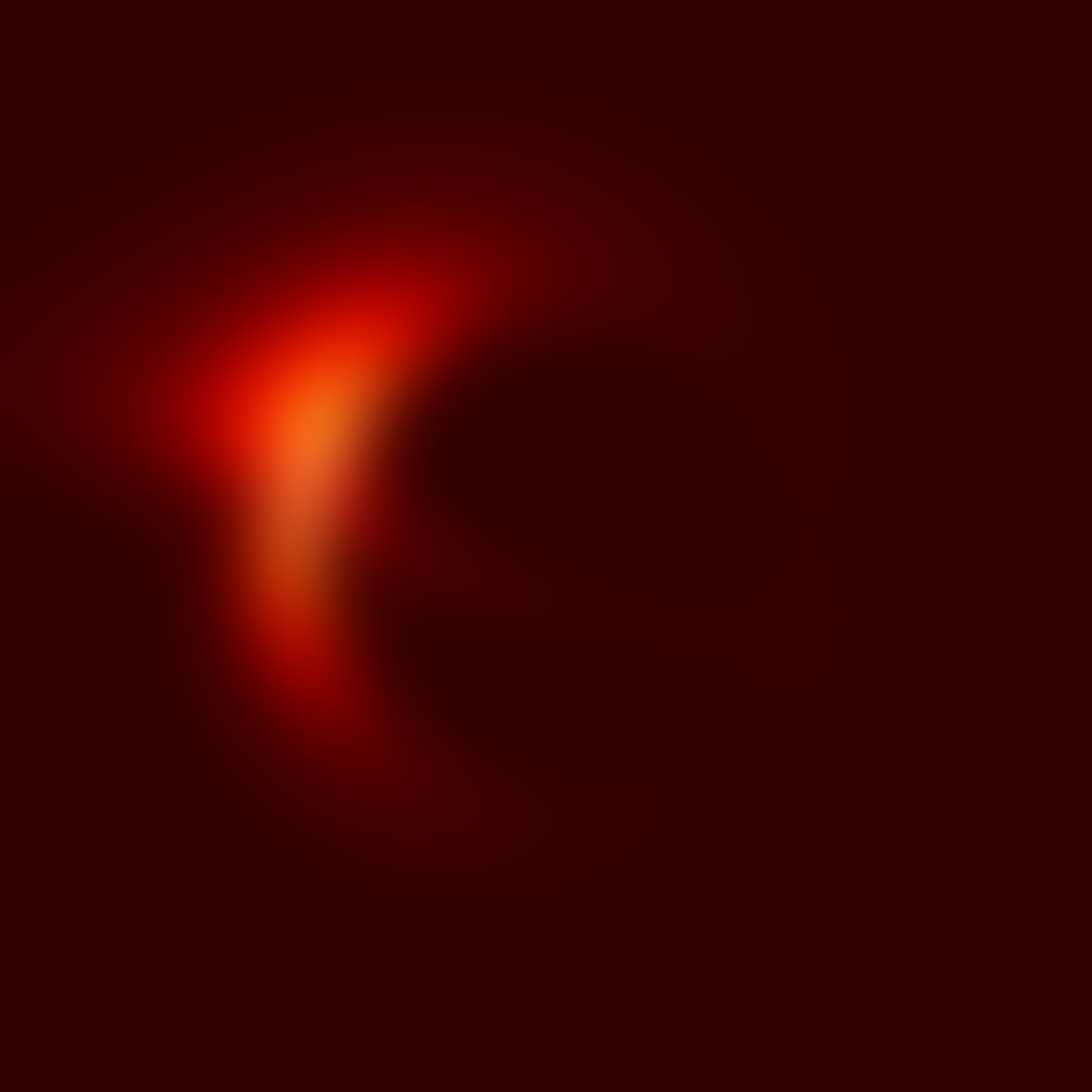}
\includegraphics[width=1.8cm]{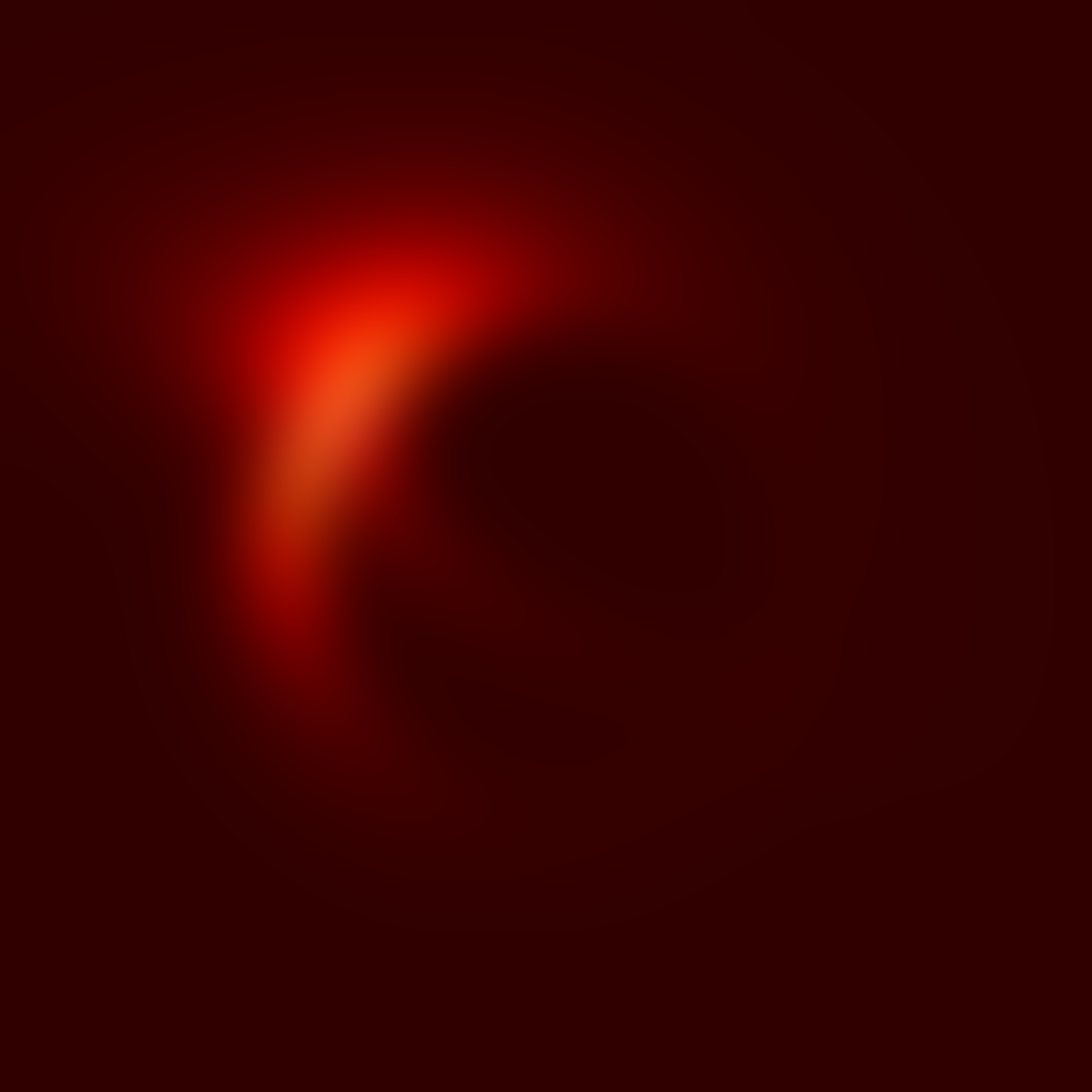}
\includegraphics[width=1.8cm]{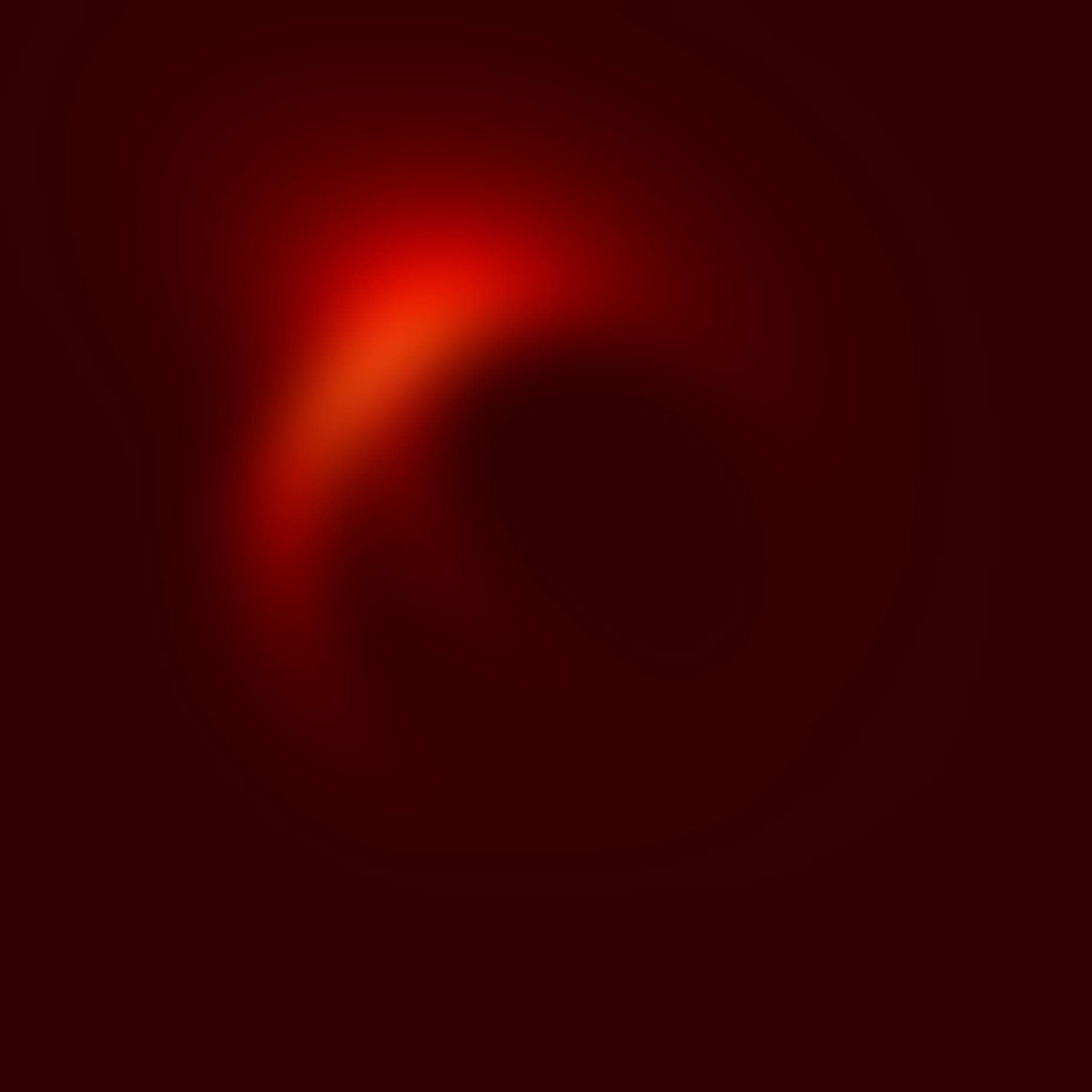}
\includegraphics[width=1.8cm]{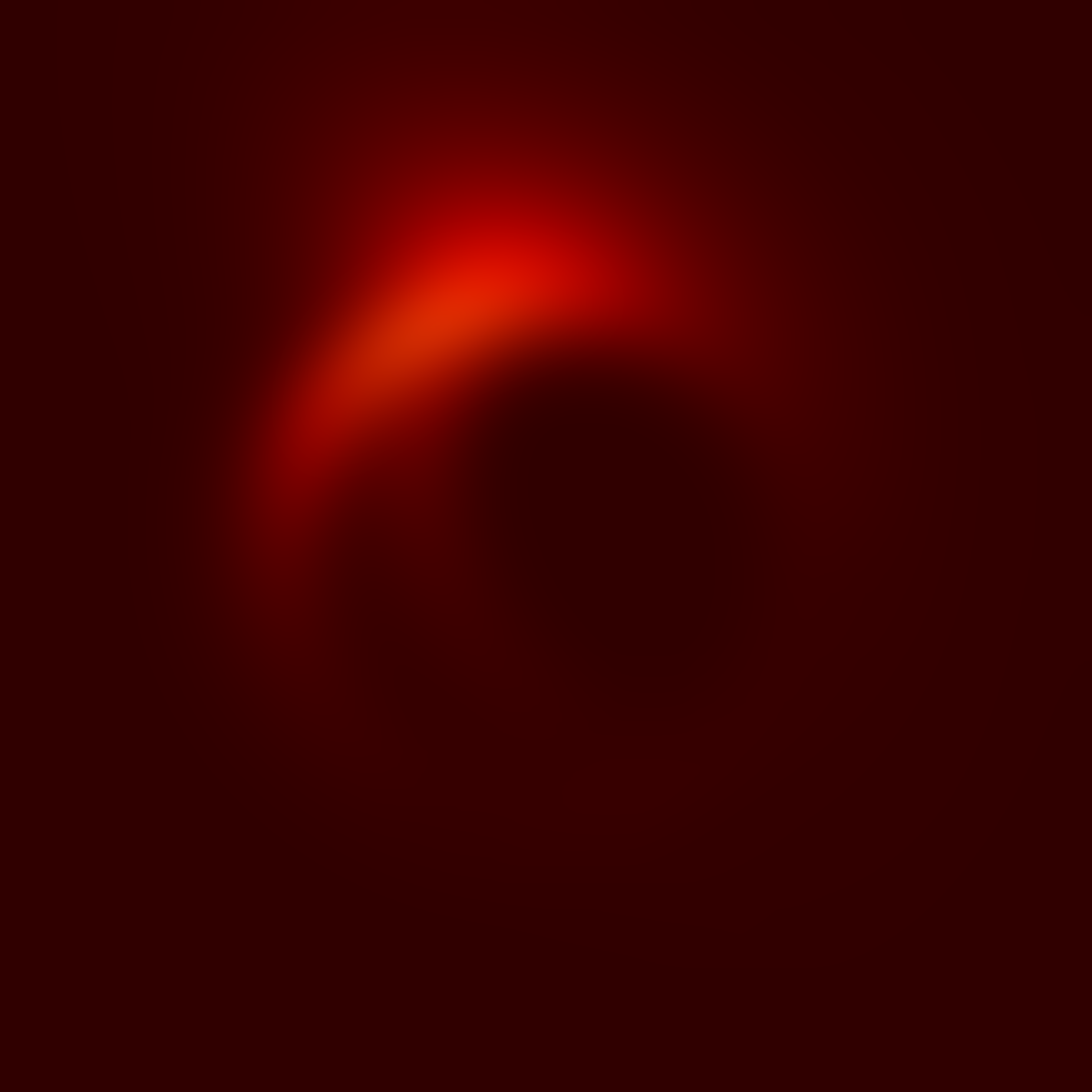}
\includegraphics[width=1.8cm]{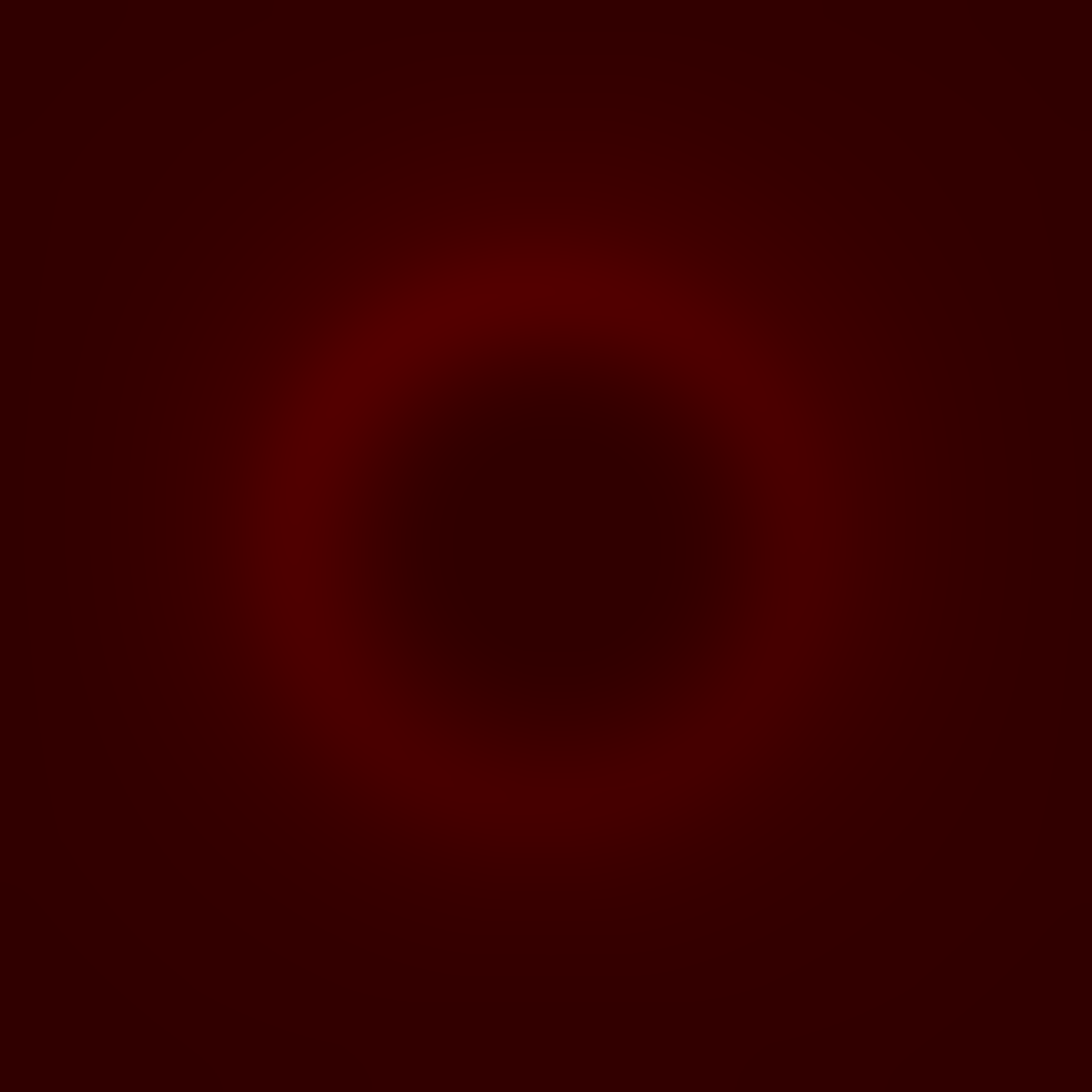}
\includegraphics[width=1.8cm]{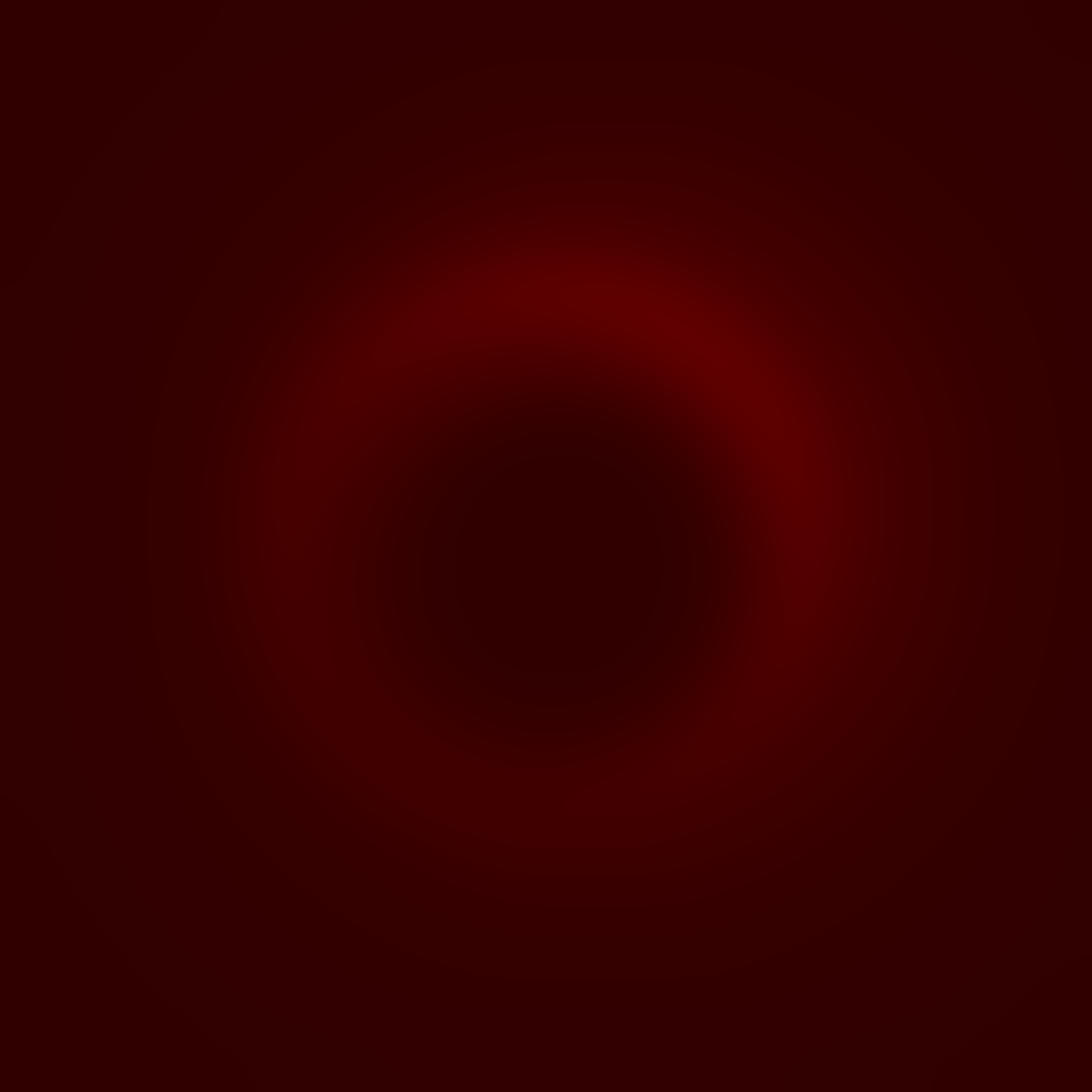}
\includegraphics[width=1.8cm]{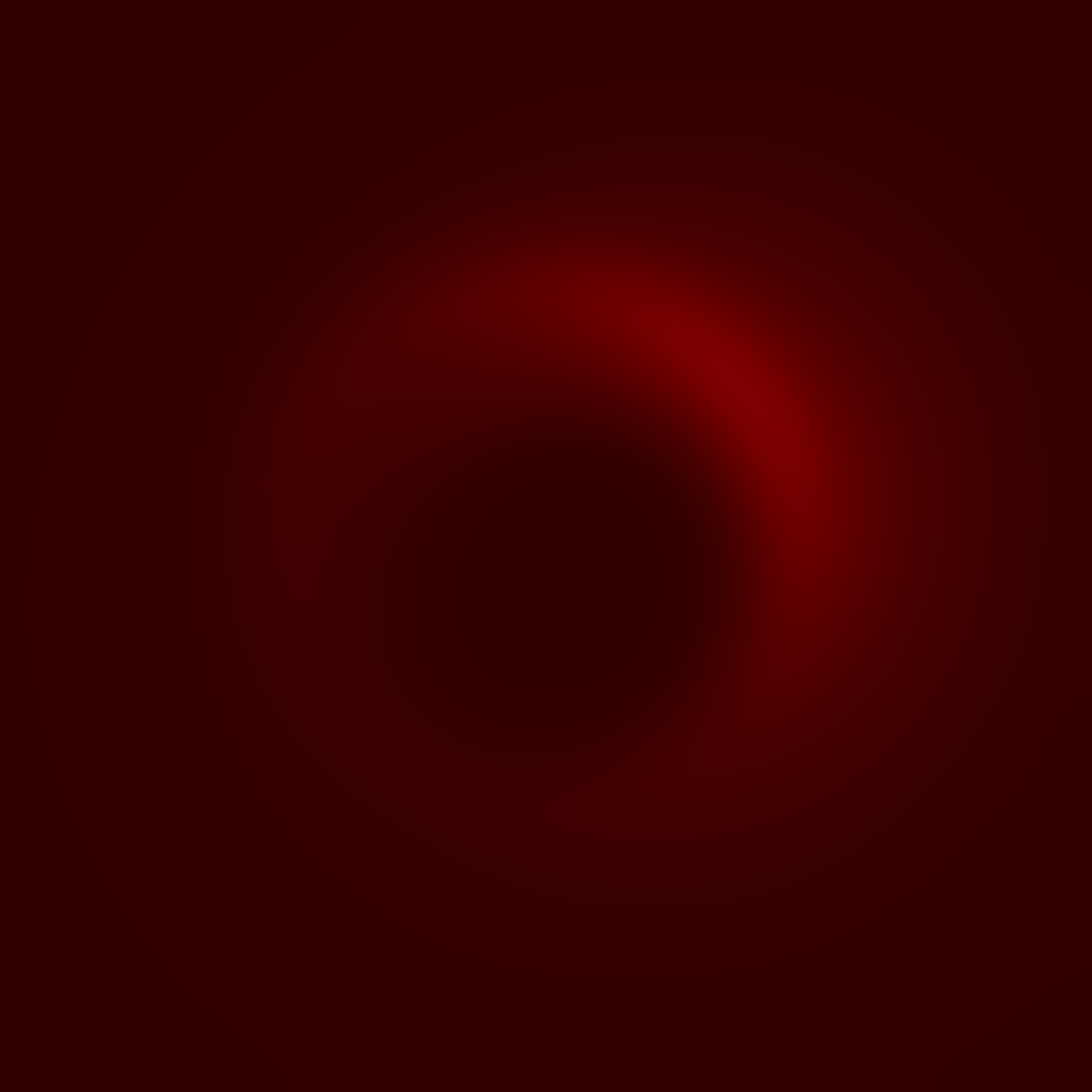}
\includegraphics[width=1.8cm]{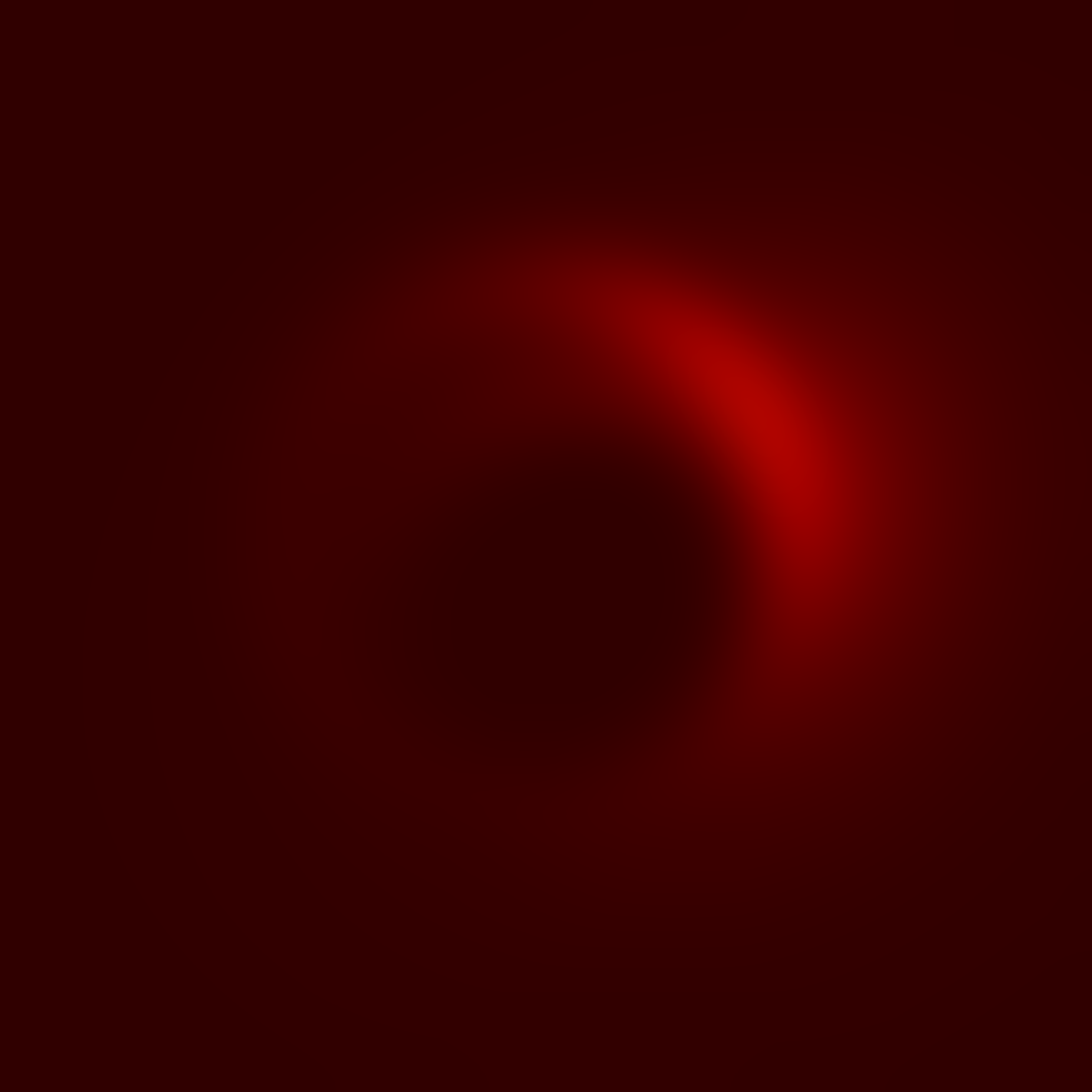}
\includegraphics[width=1.8cm]{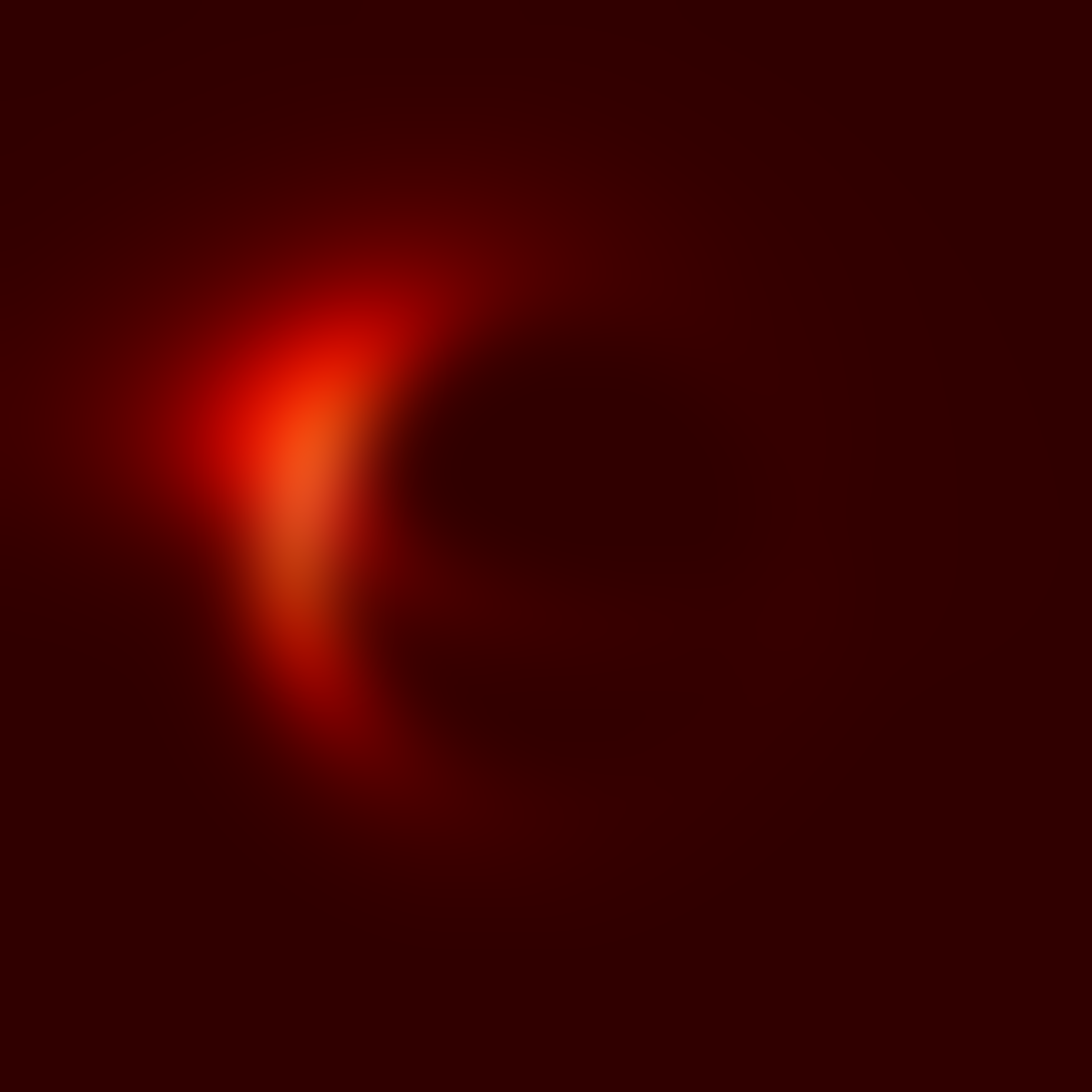}
\includegraphics[width=1.8cm]{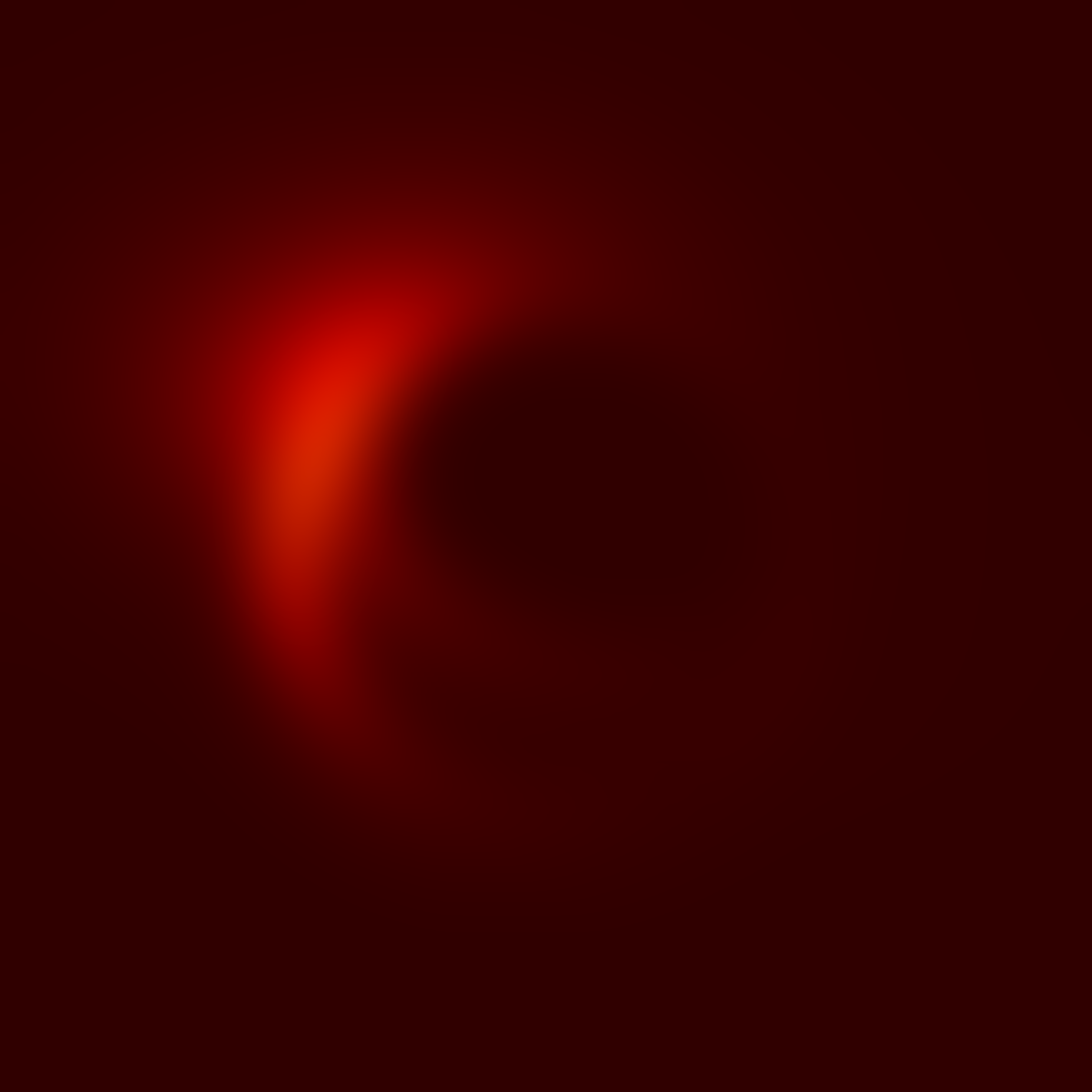}
\includegraphics[width=1.8cm]{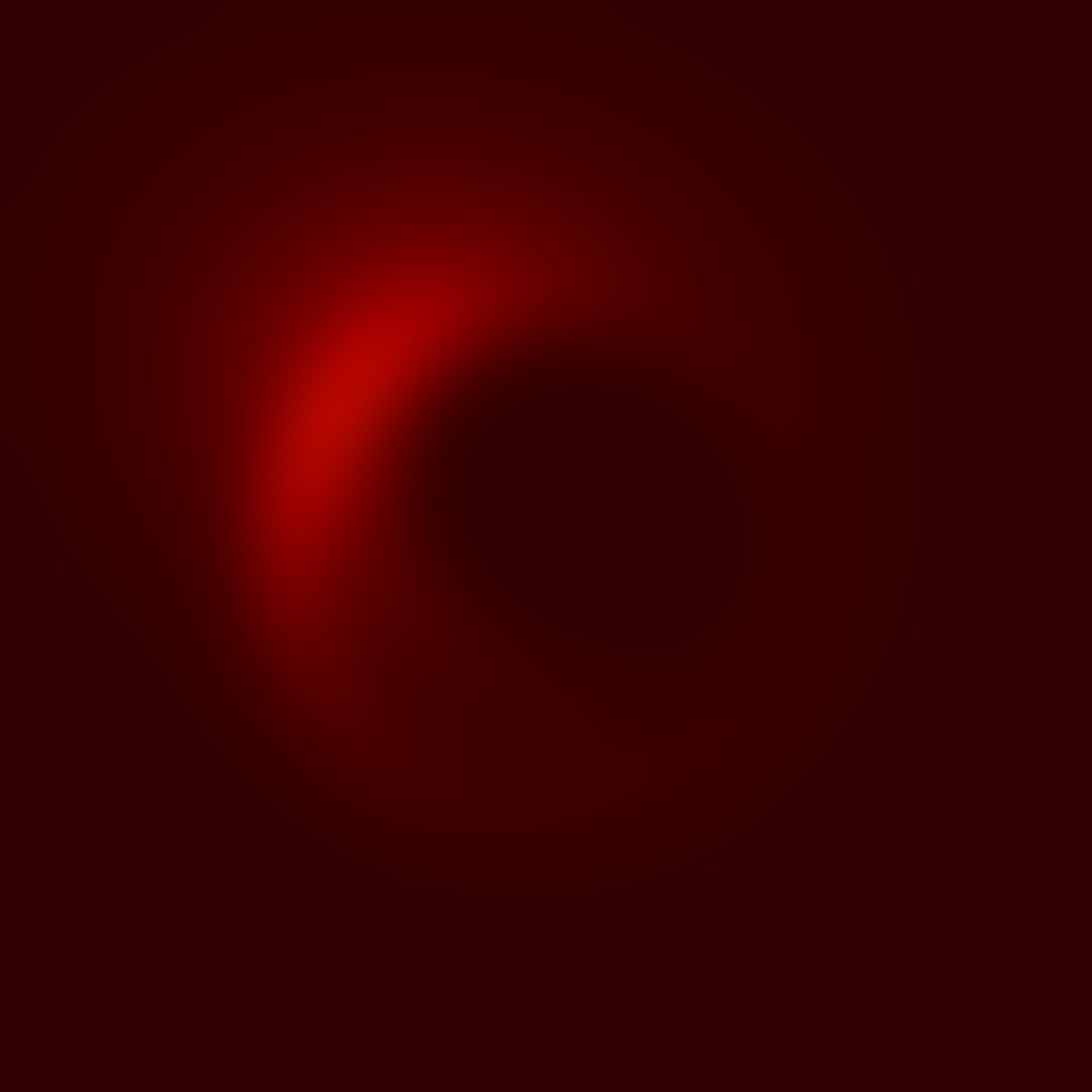}
\includegraphics[width=1.8cm]{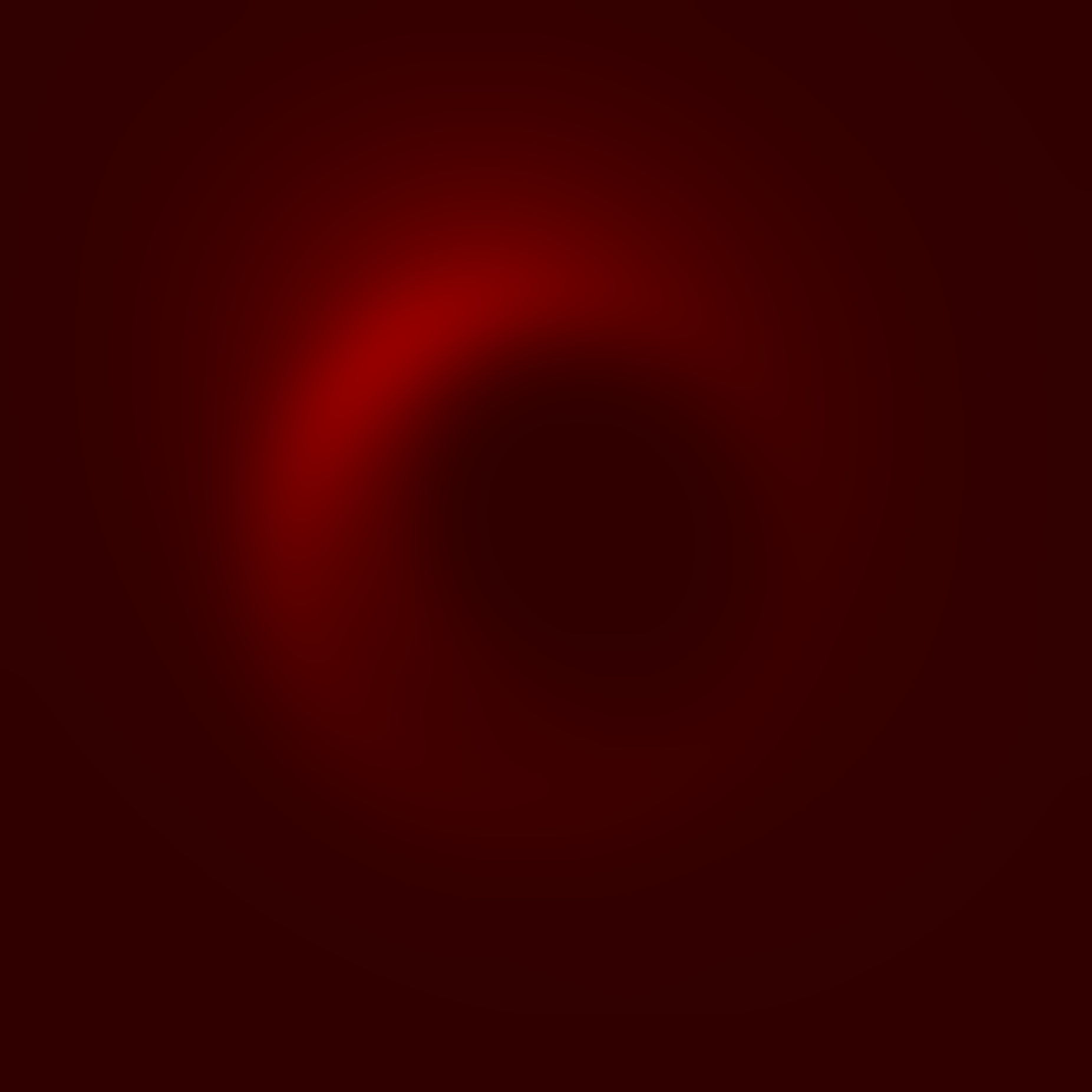}
\includegraphics[width=1.8cm]{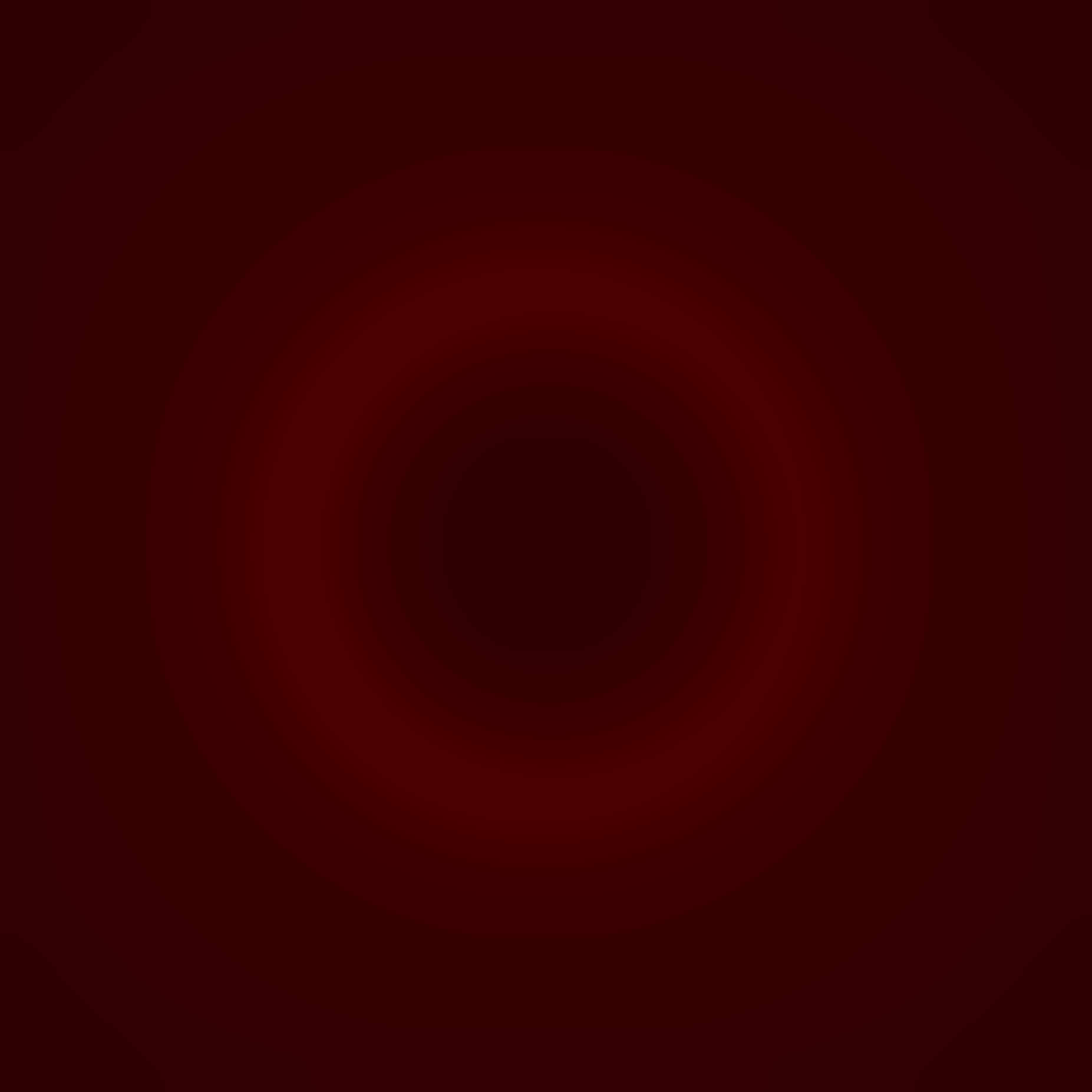}
\includegraphics[width=1.8cm]{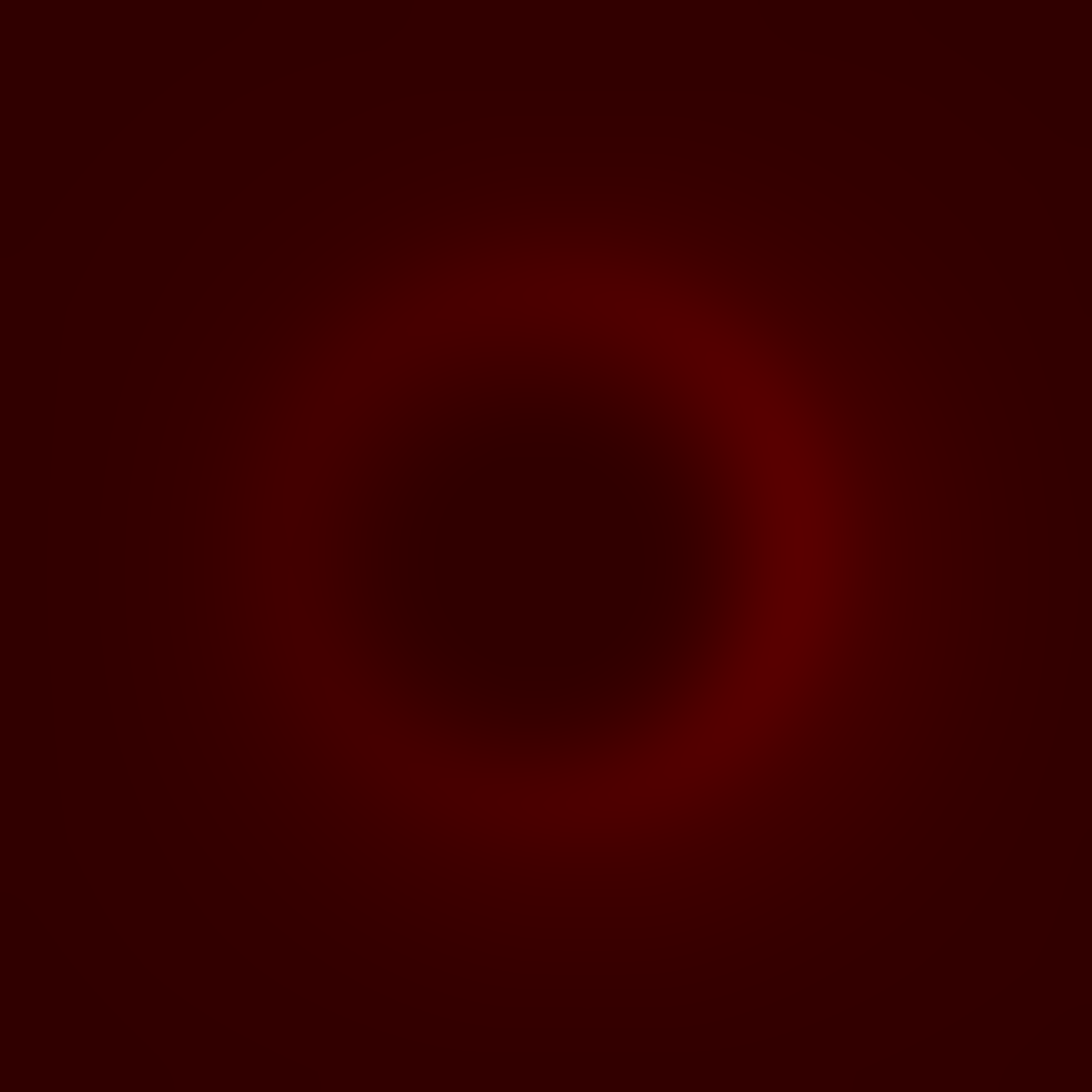}
\includegraphics[width=1.8cm]{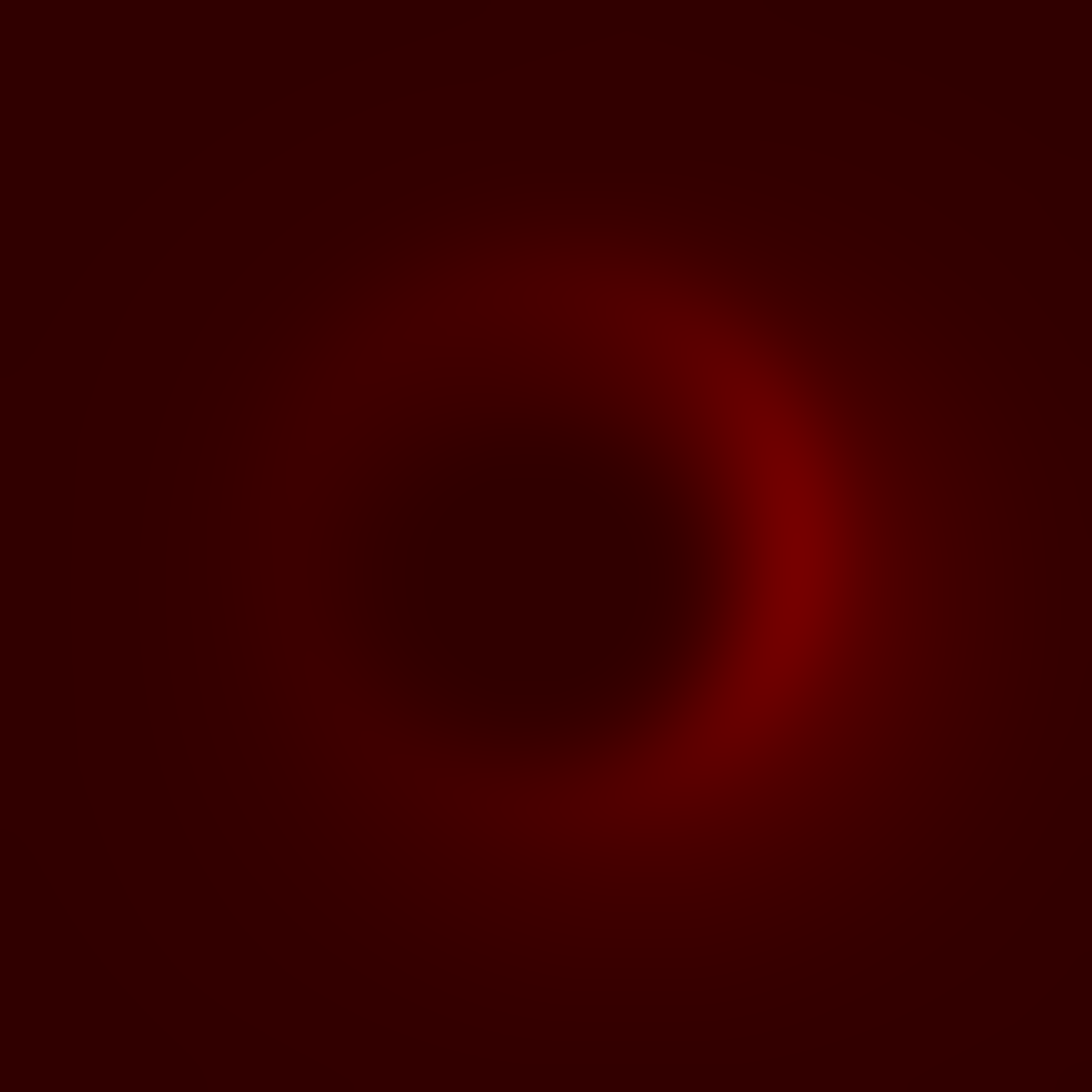}
\includegraphics[width=1.8cm]{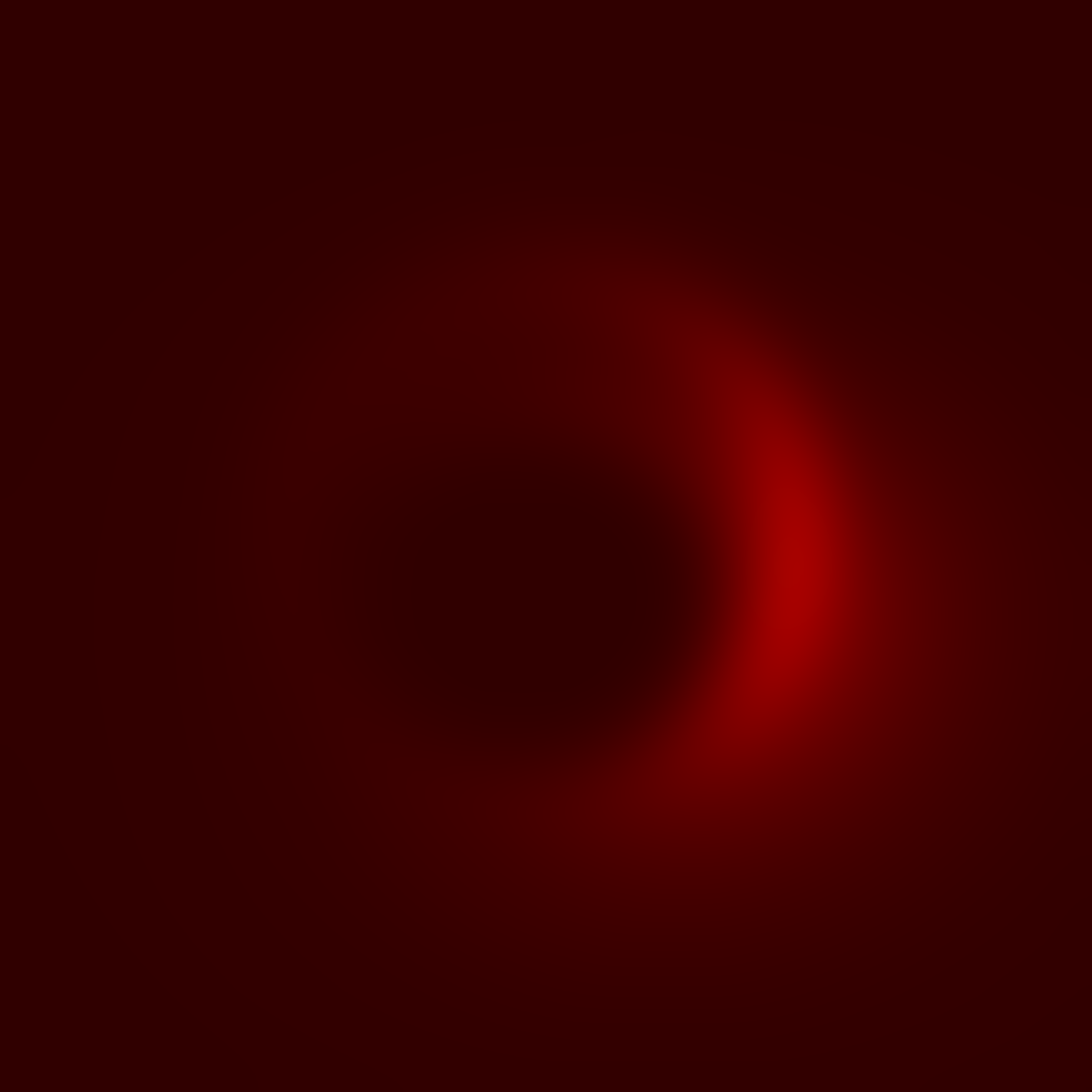}
\includegraphics[width=1.8cm]{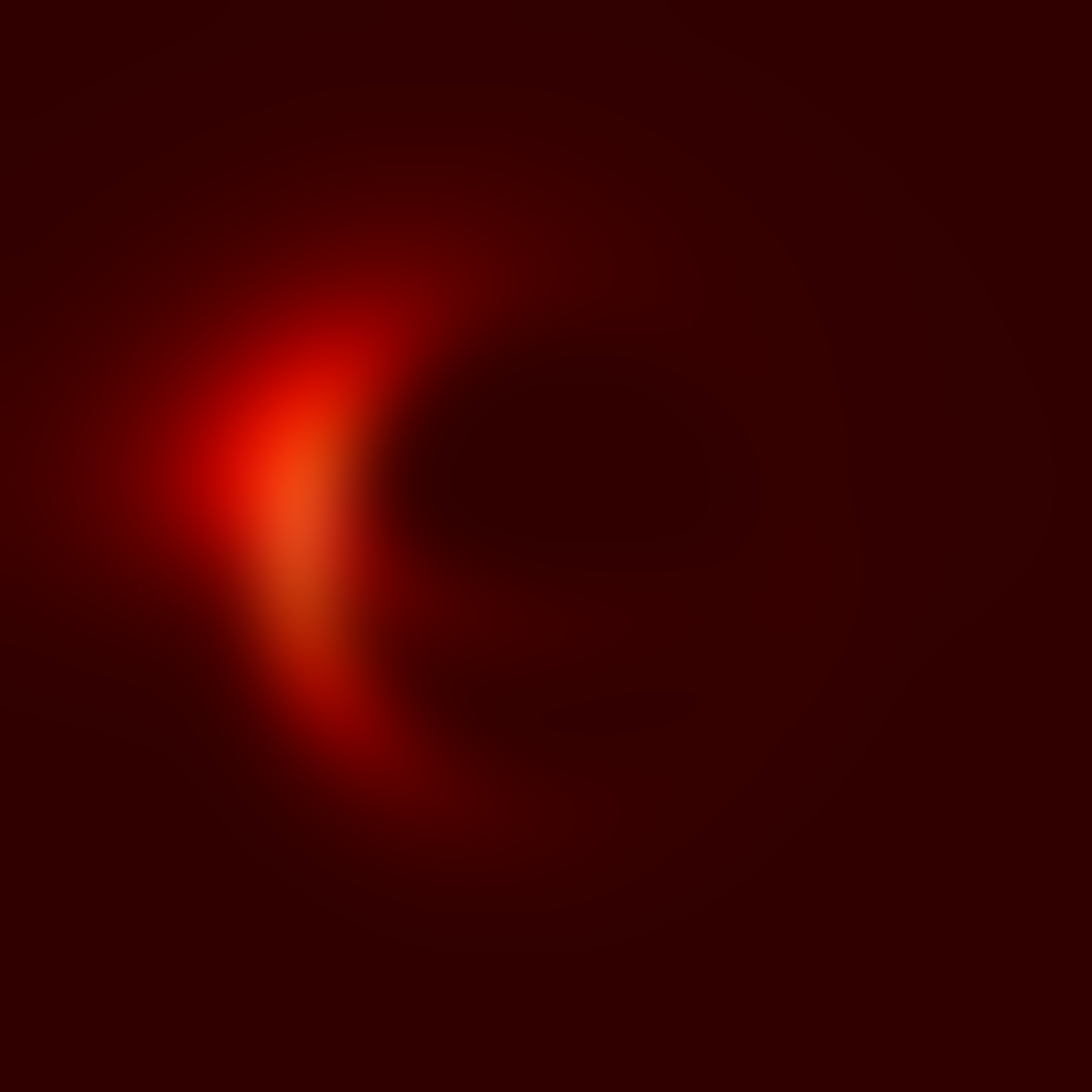}
\includegraphics[width=1.8cm]{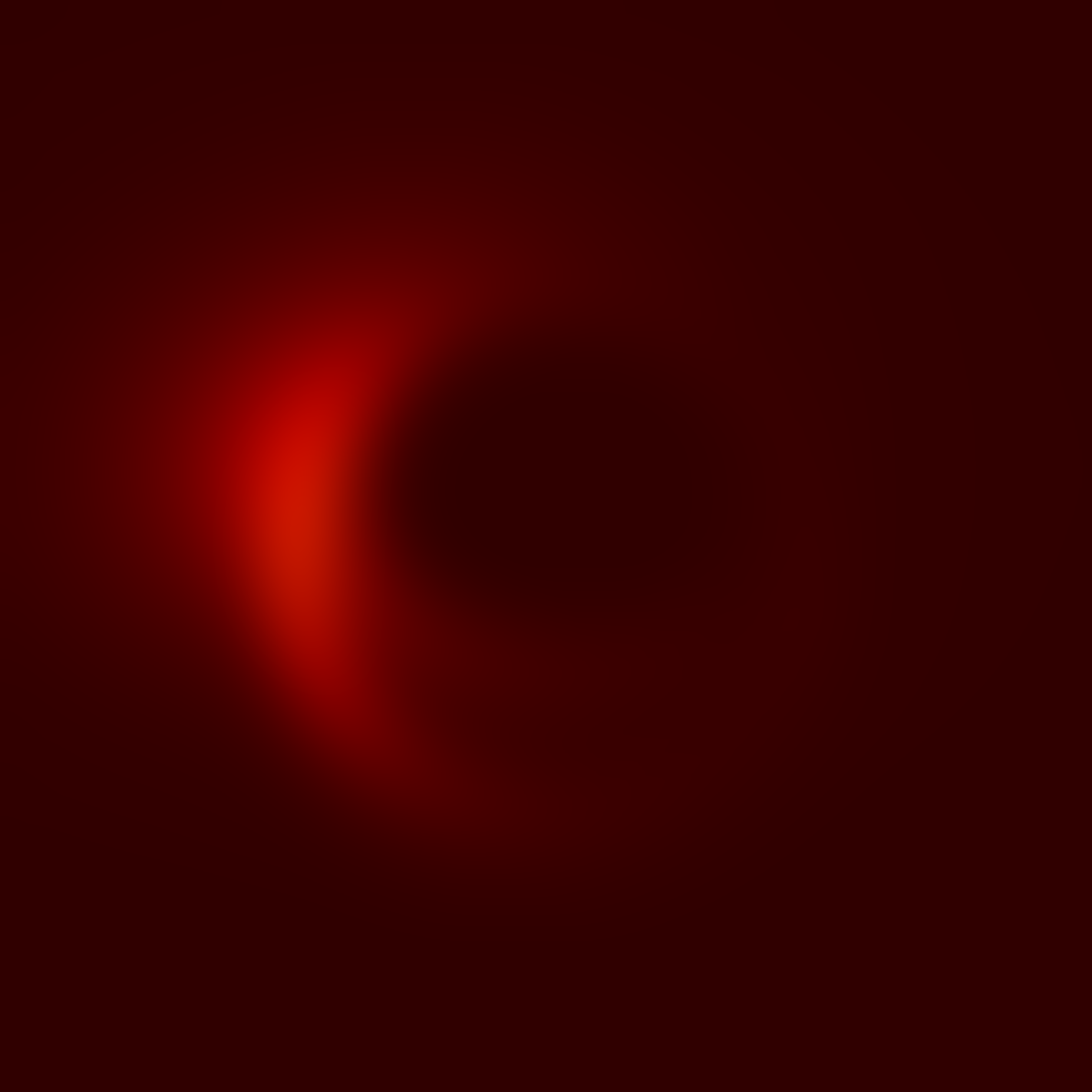}
\includegraphics[width=1.8cm]{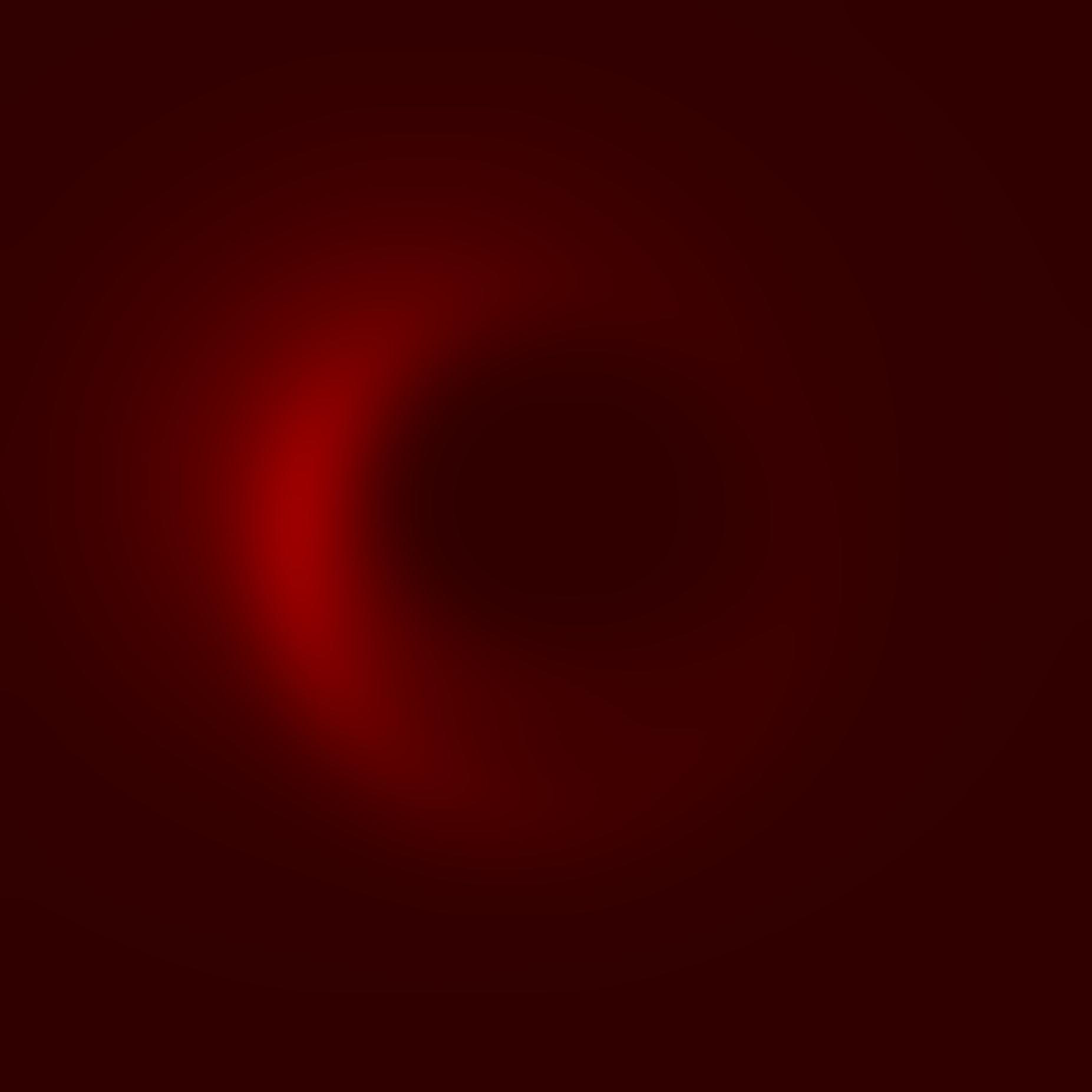}
\includegraphics[width=1.8cm]{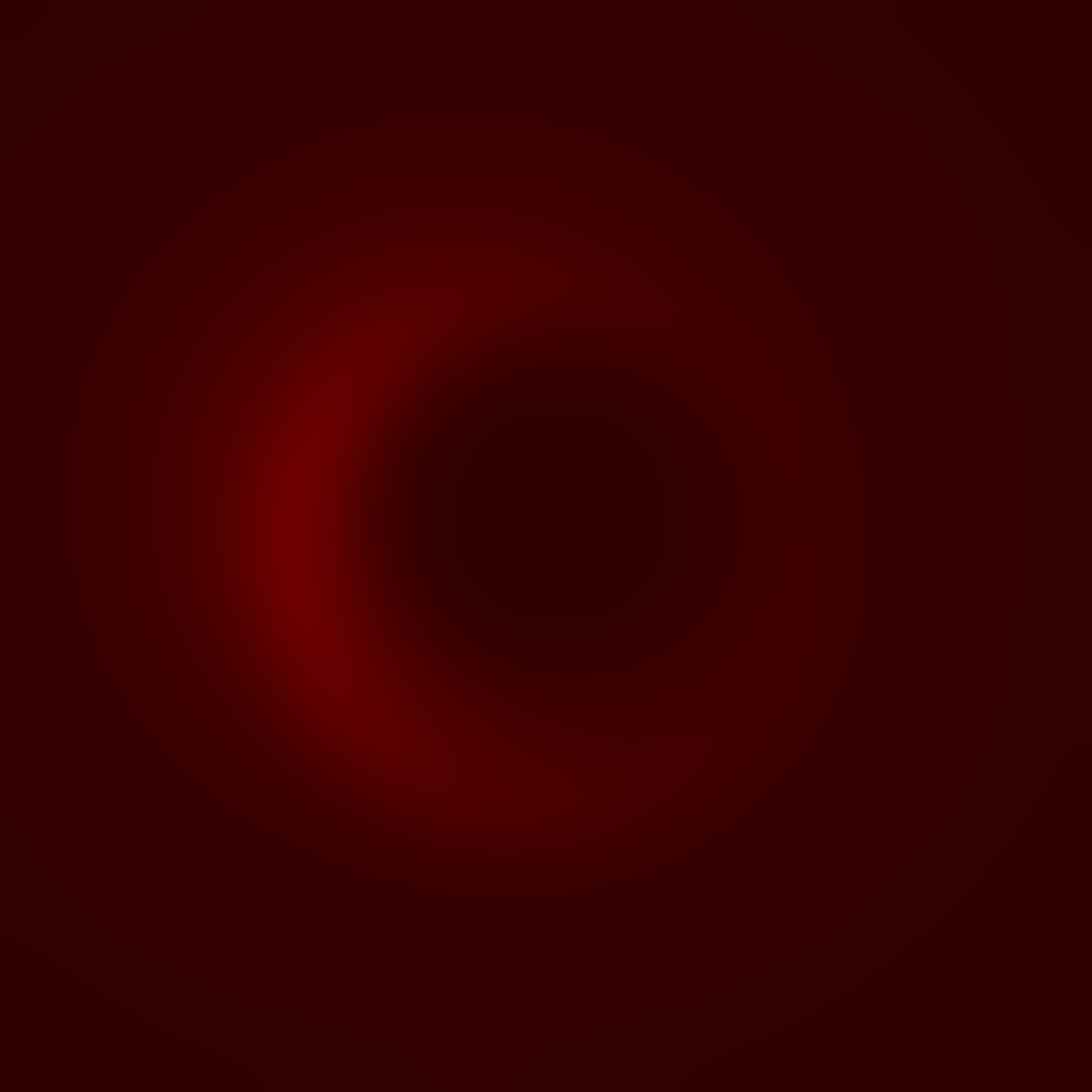}
\caption{Images of the hairy BH after being blurred with a Gaussian filter using a $20$ $\upmu$as kernel. The observation azimuths range from $30 - 180^{\circ}$ in steps of $30^{\circ}$, progressing from the top to the bottom, and remain consistent in each row. The observation angle is $17^{\circ}$ for the left four columns and $85^{\circ}$ for the right four columns. In blocks with the same observation angle, the disk tilts are $15^{\circ}$, $30^{\circ}$, $45^{\circ}$, and $60^{\circ}$ from left to right, respectively, and remain consistent in each column. Here, we use a scalar hair parameter of $h = -1$.}}\label{fig15}
\end{figure*}

Figure 16 shows the observational appearance of BHs surrounded by both an equatorial accretion disk and a tilted accretion disk with various inclinations. The first two rows correspond to the Schwarzschild case ($h = 0$), while the last two rows correspond to the hairy BH with the scalar hair parameter of $h = -1$. We find that the images of the BH are brighter in the multiple accretion disk case compared to the single accretion disk case. At the same time, the extra accretion disk does not change the profile of the critical curve. However, it is interesting to note that the two accretion disks significantly modify the morphology of the inner shadow of the BH. For instance, in figure 9 (a), we recognize a circle-like inner shadow of the BH, but with the inclusion of a tilted accretion disk, the inner shadow undergoes compression in the vertical direction, transforming into an ellipsoid-like shape, as depicted in figure 16 (a). This alteration occurs due to the additional accretion disk obstructing the light rays that would have been absorbed by the BH, resulting in the emergence of bright streaks in the image and the erosion of the inner shadow. Moreover, when the observer's line of sight is positioned between two accretion disks, as in plots (d)-(f) and (j)-(l), the size of the inner shadow undergoes a substantial reduction when compared to the case of a single accretion disk. Specifically, the area of the inner shadow expands as the angle between the two accretion disks increases.
\begin{figure*}
\center{
\includegraphics[width=4.5cm]{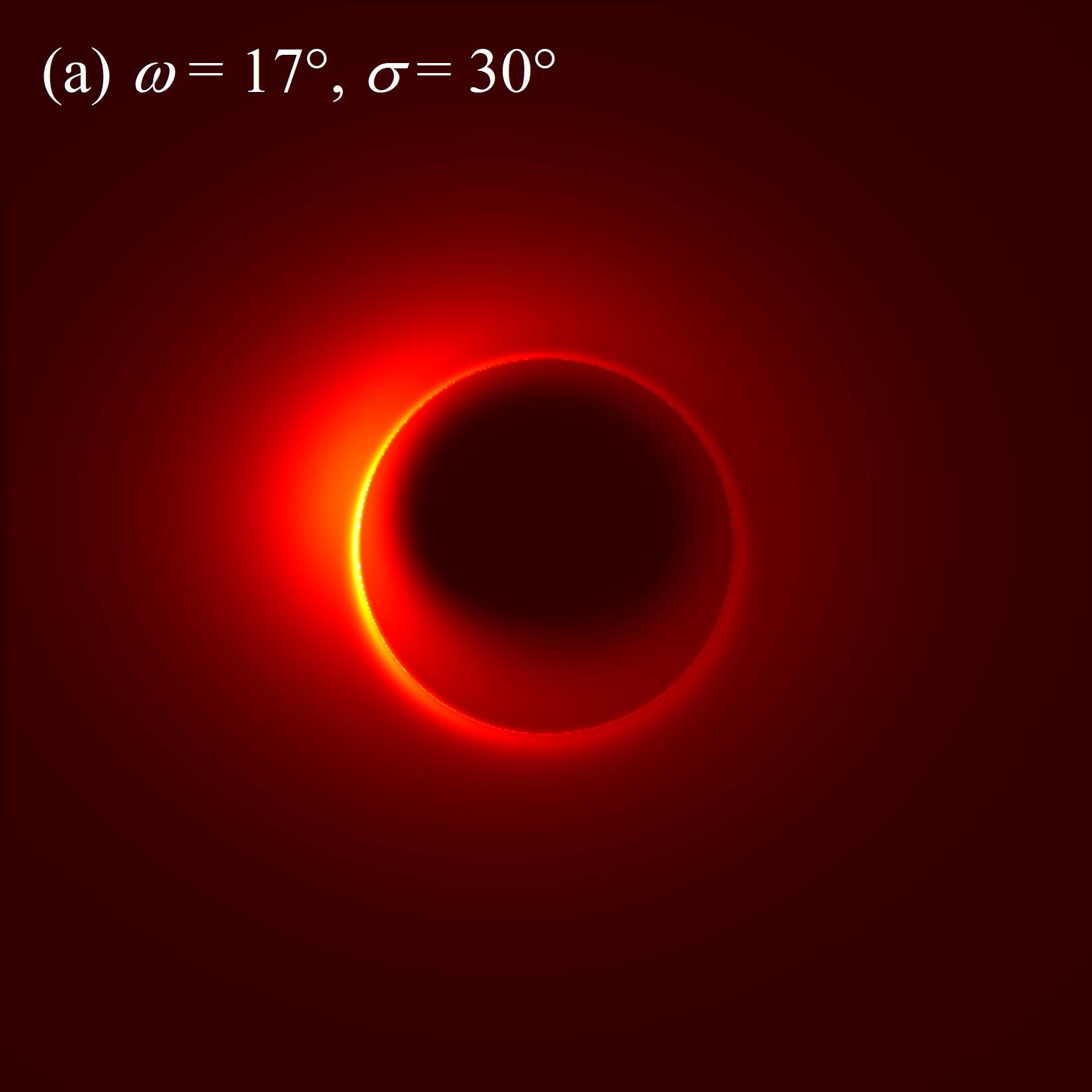}
\includegraphics[width=4.5cm]{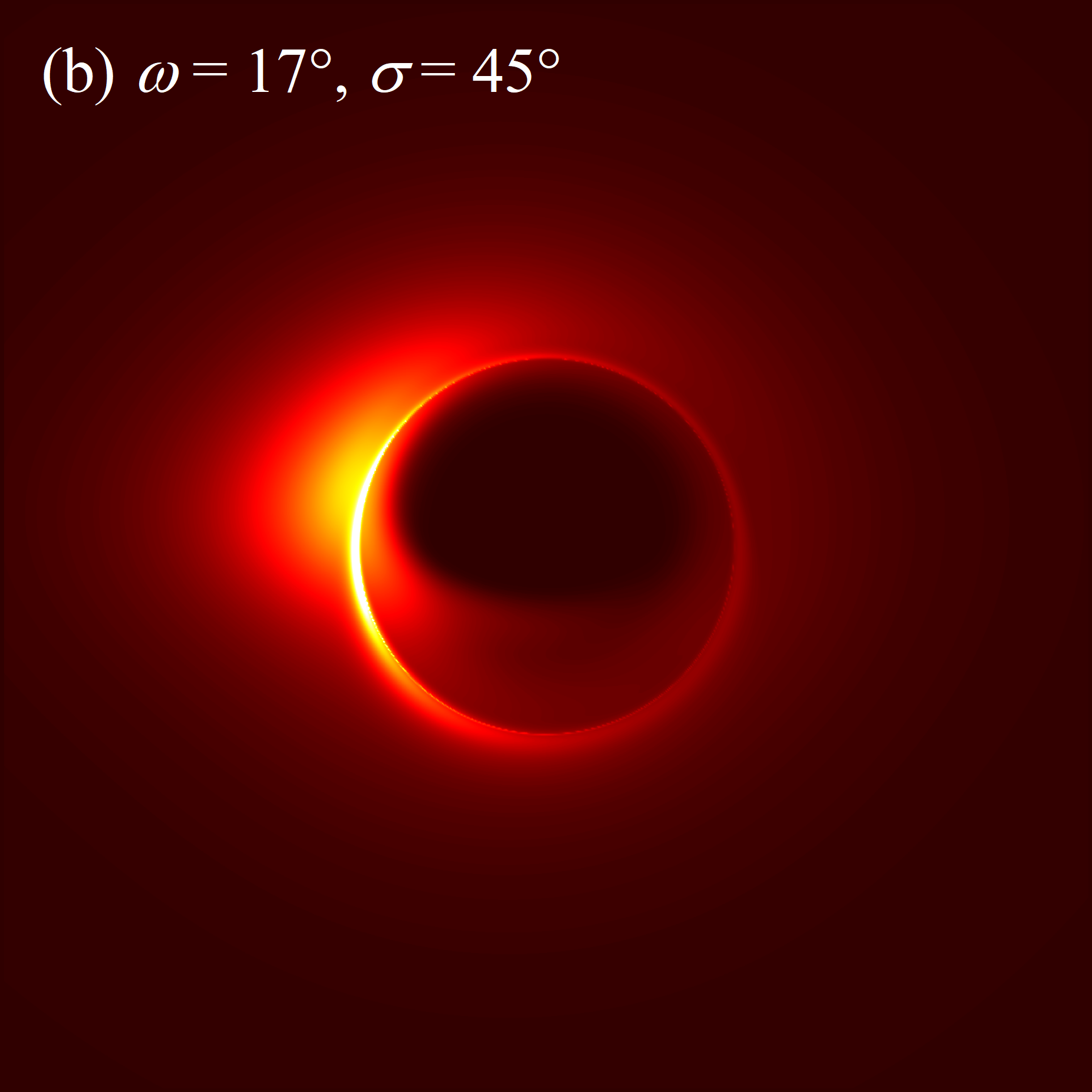}
\includegraphics[width=4.5cm]{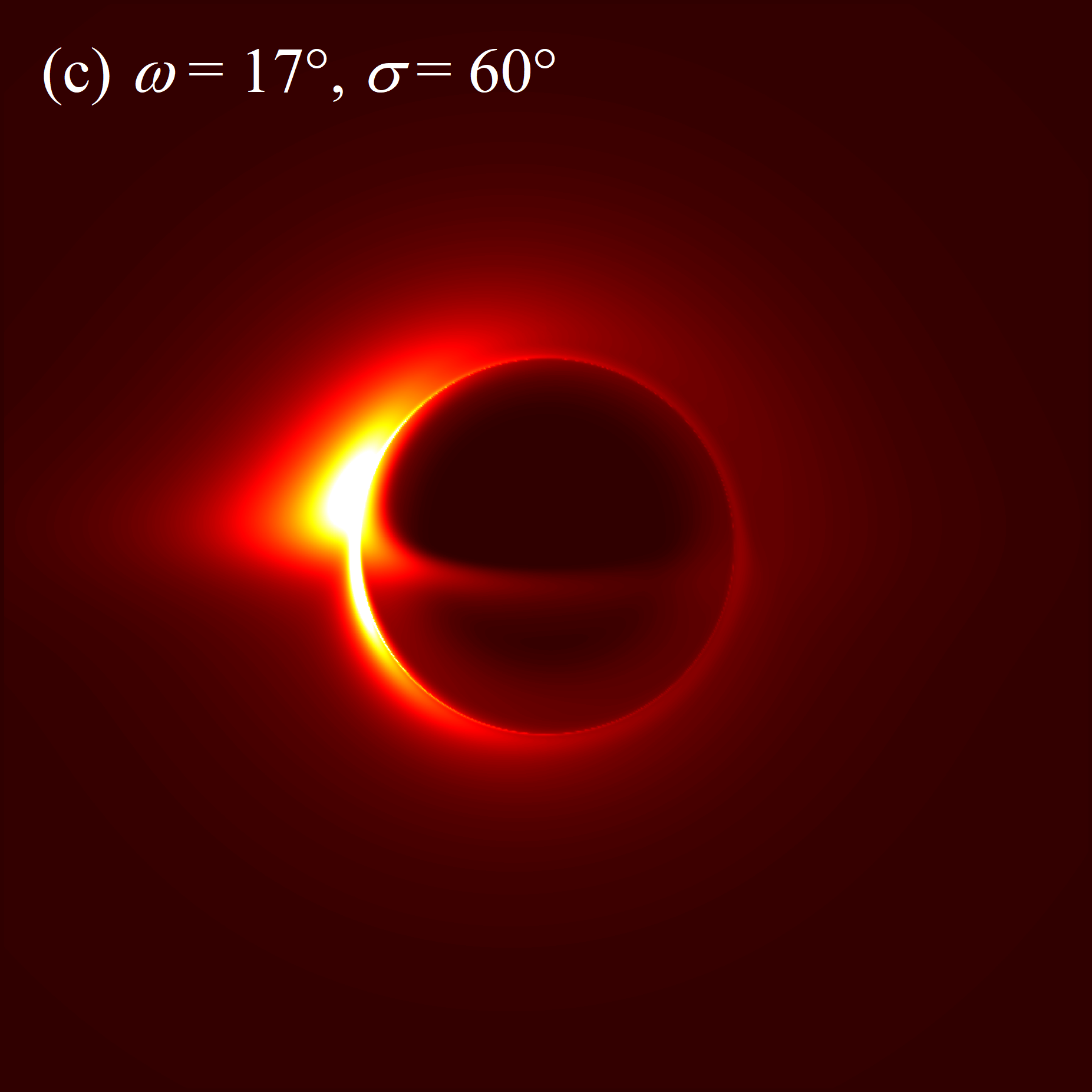}
\includegraphics[width=4.5cm]{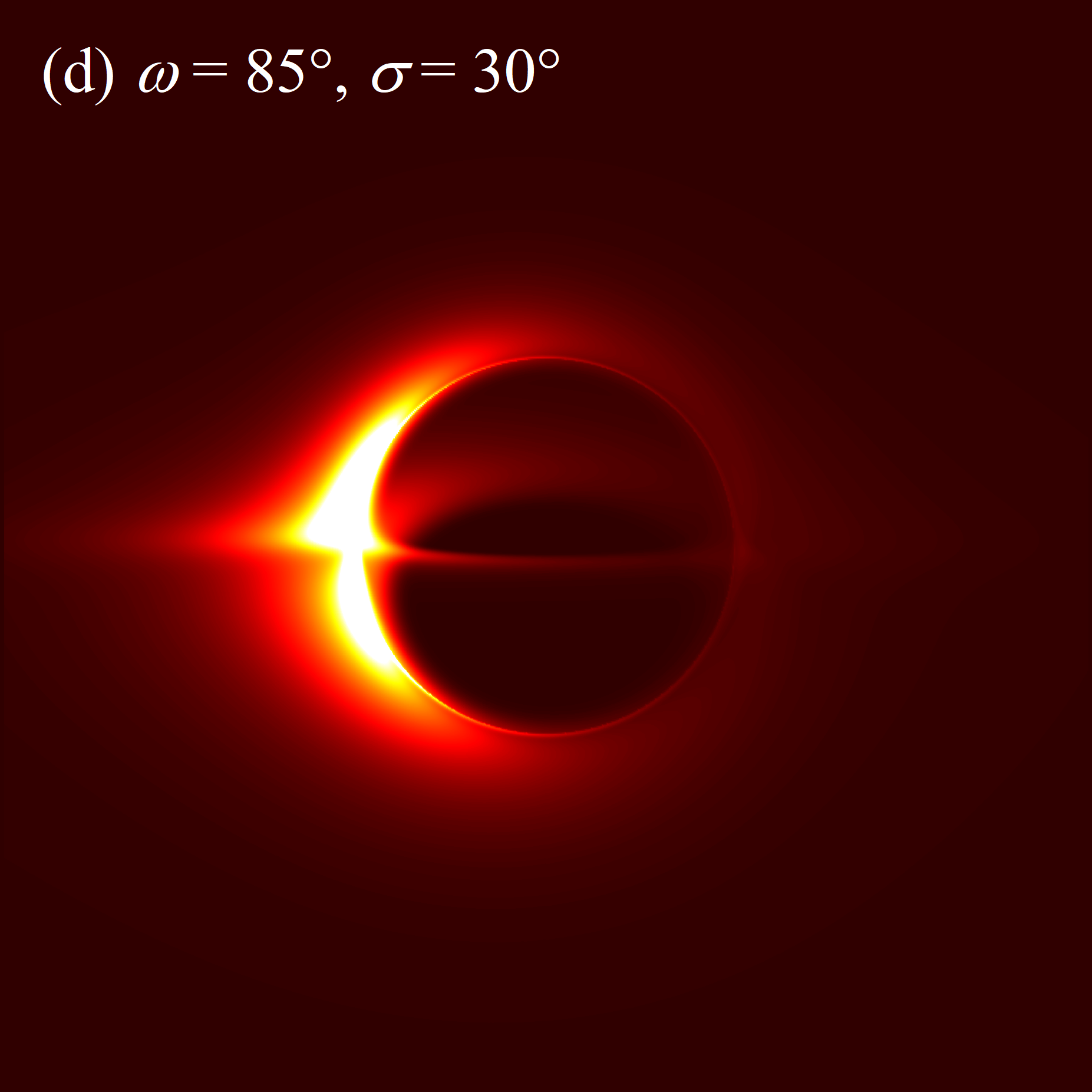}
\includegraphics[width=4.5cm]{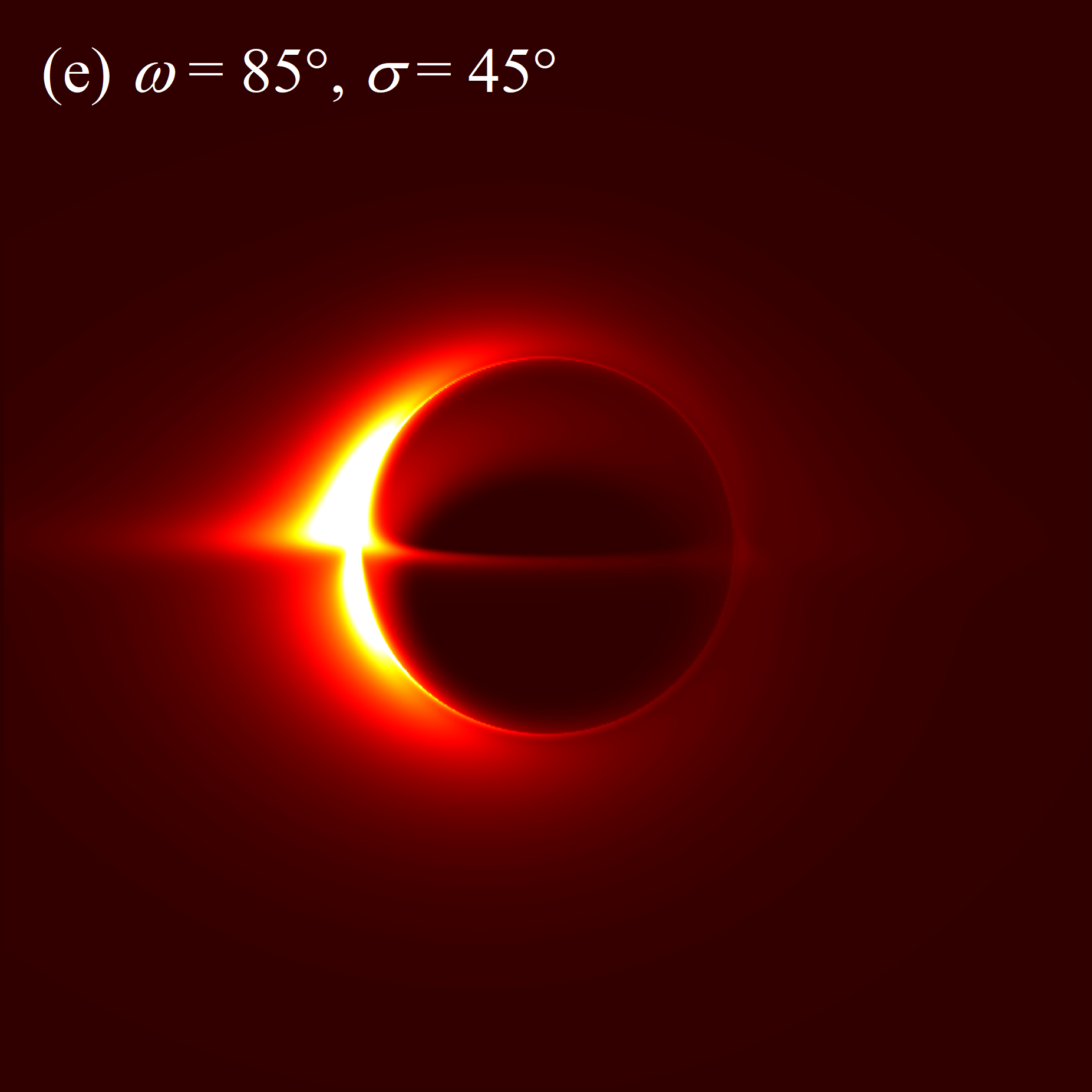}
\includegraphics[width=4.5cm]{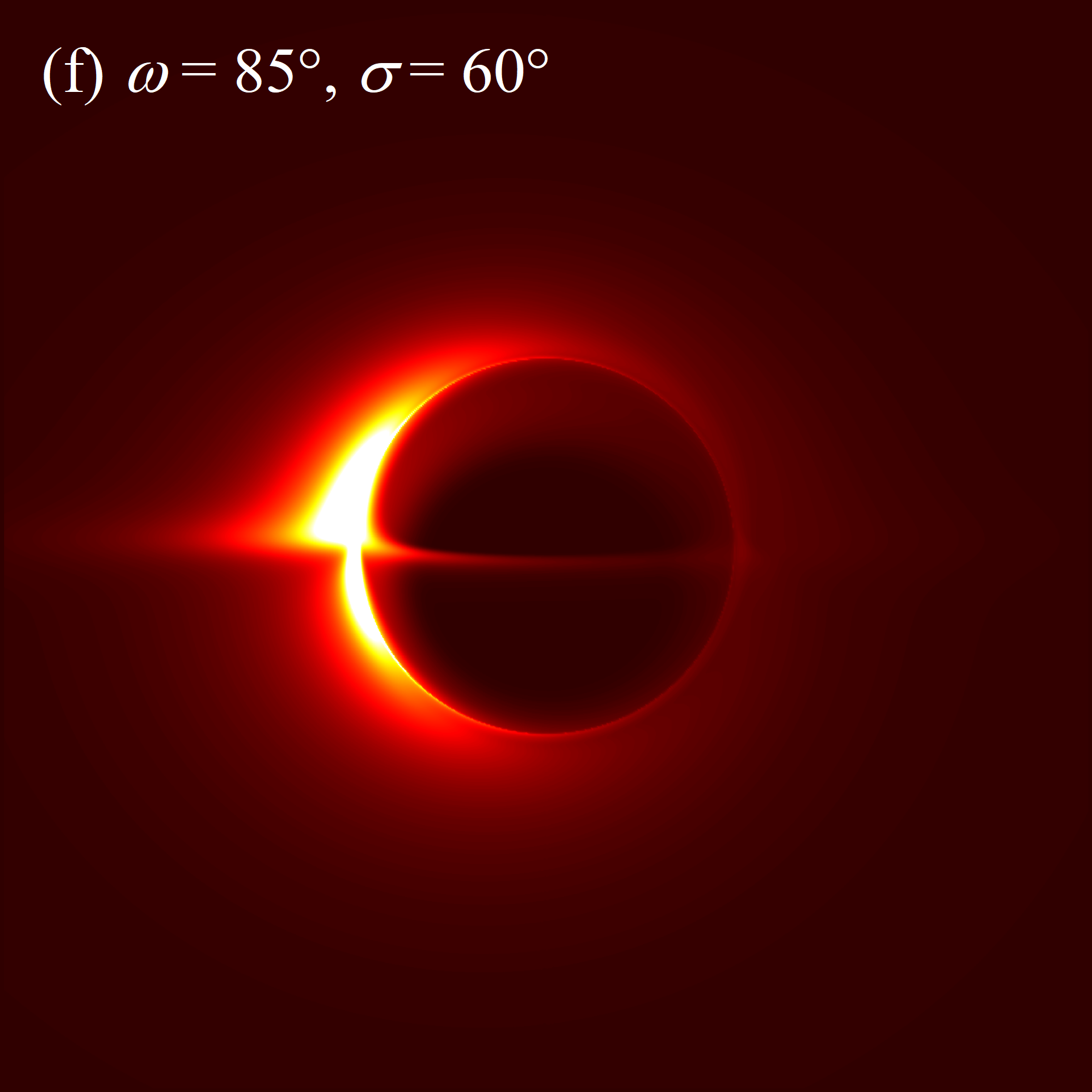}
\includegraphics[width=4.5cm]{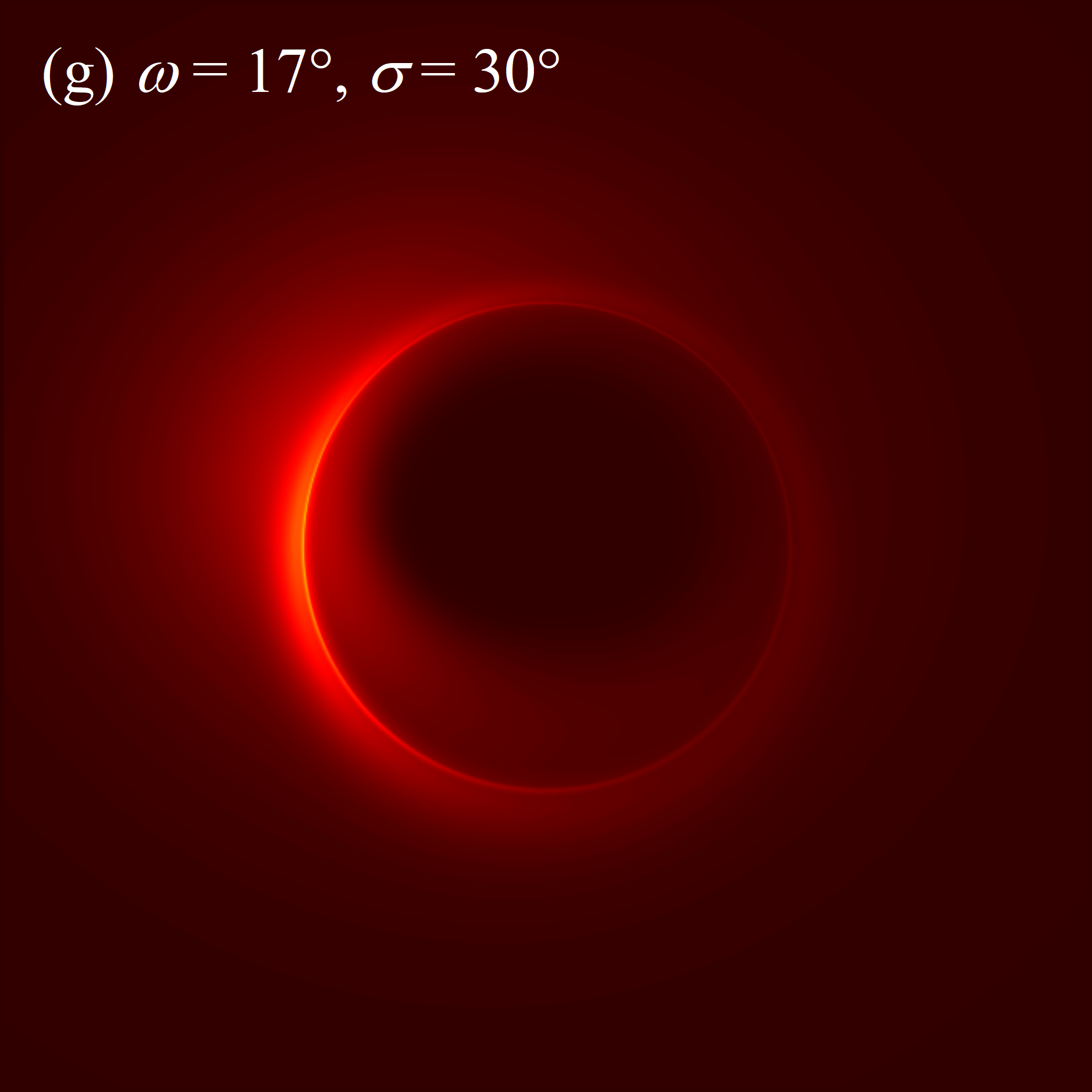}
\includegraphics[width=4.5cm]{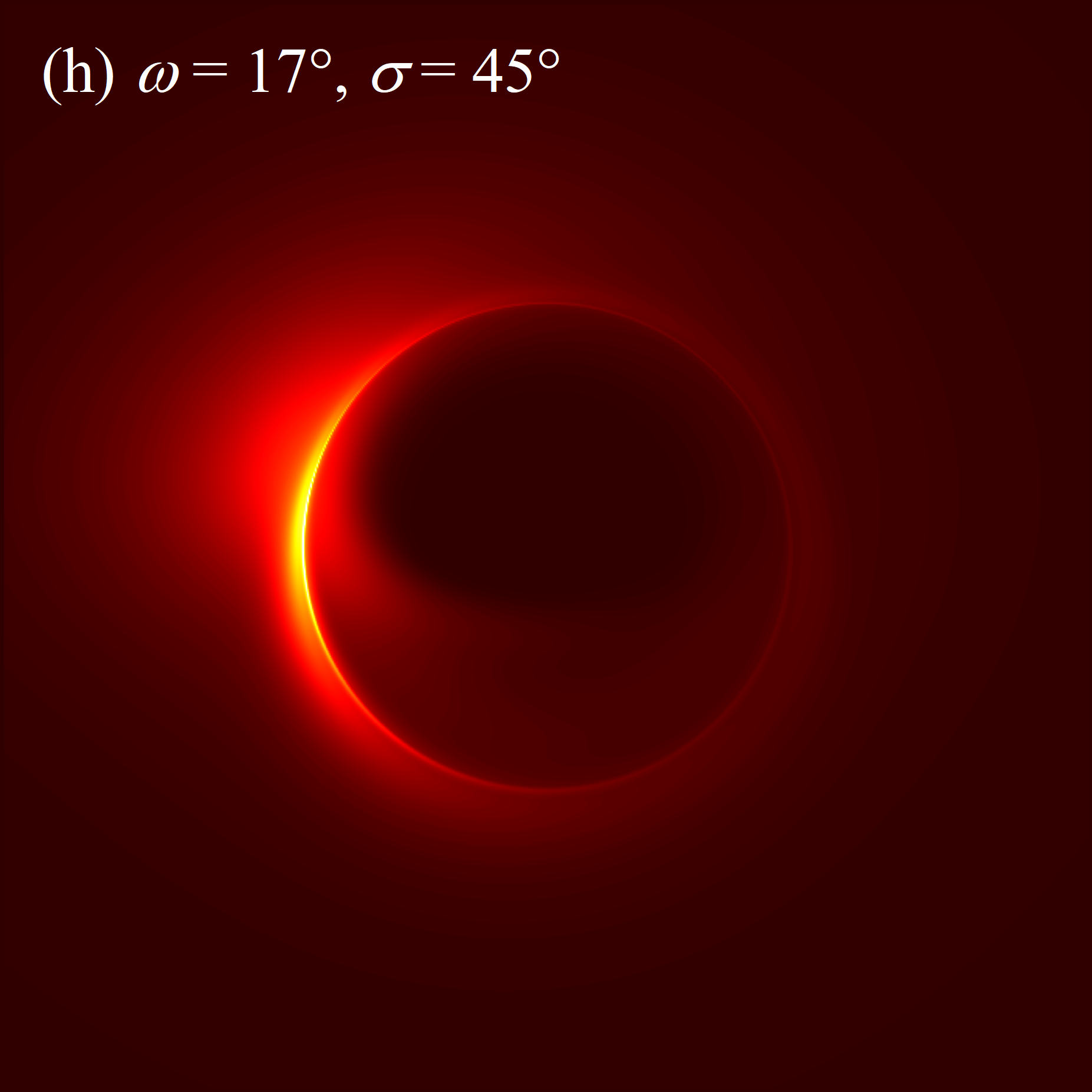}
\includegraphics[width=4.5cm]{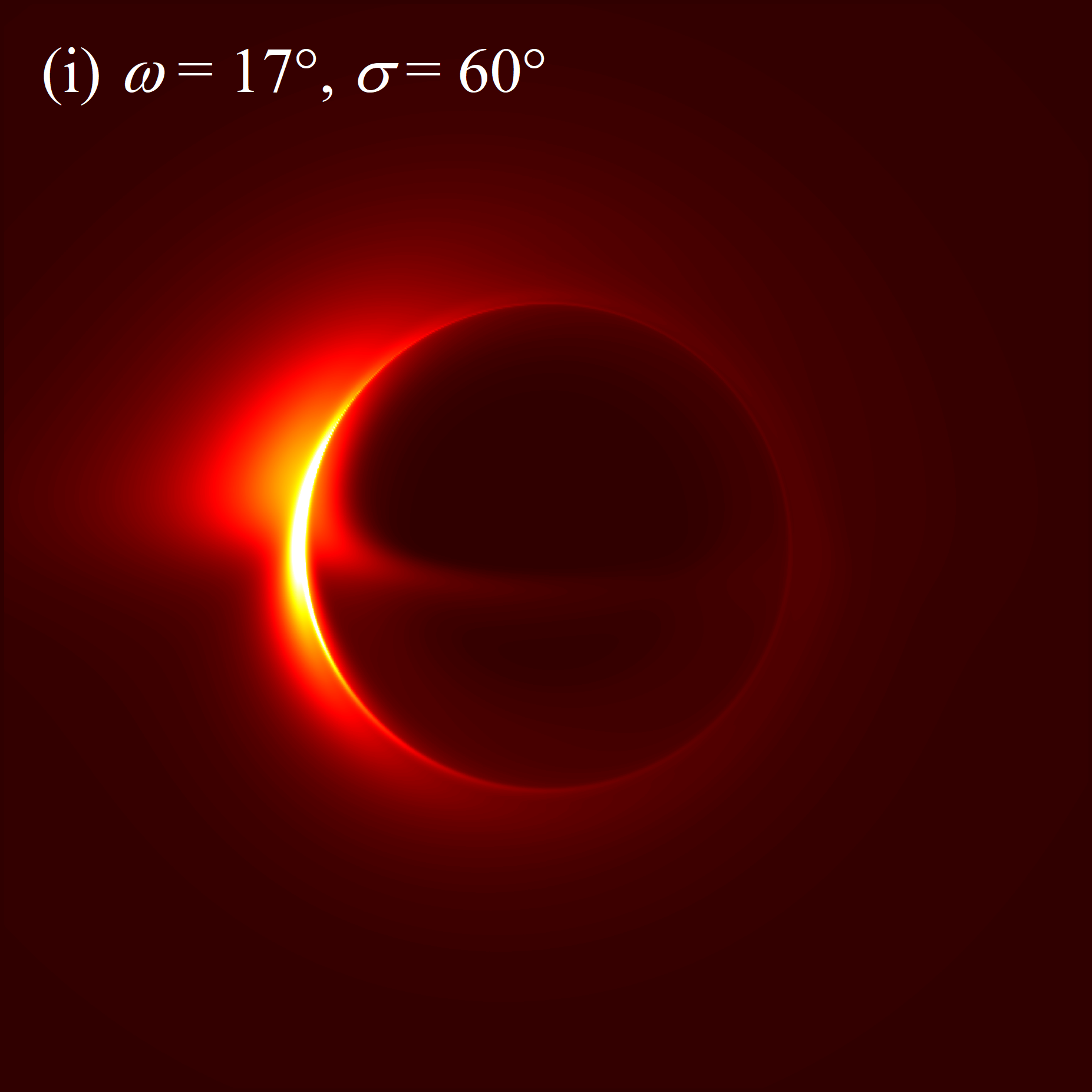}
\includegraphics[width=4.5cm]{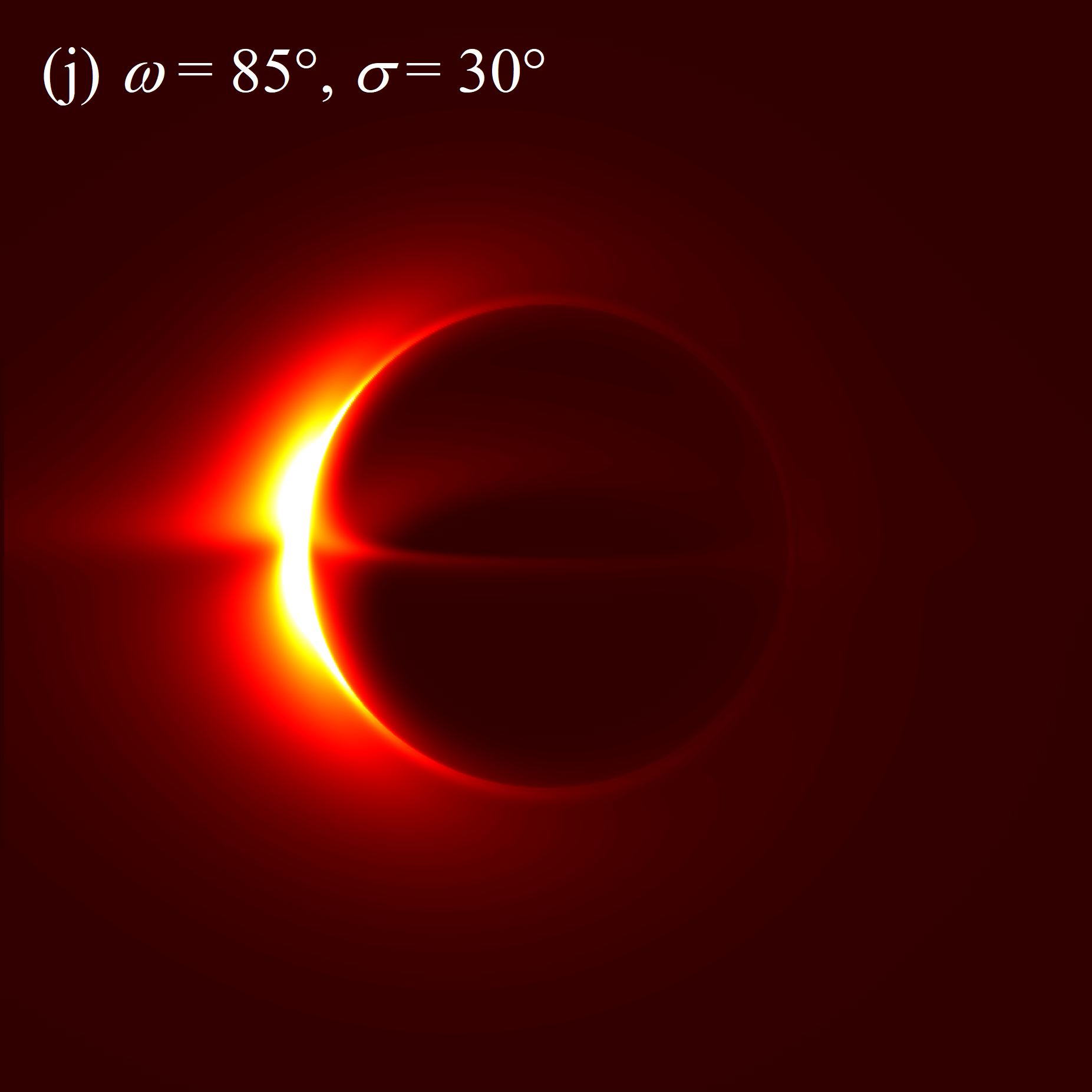}
\includegraphics[width=4.5cm]{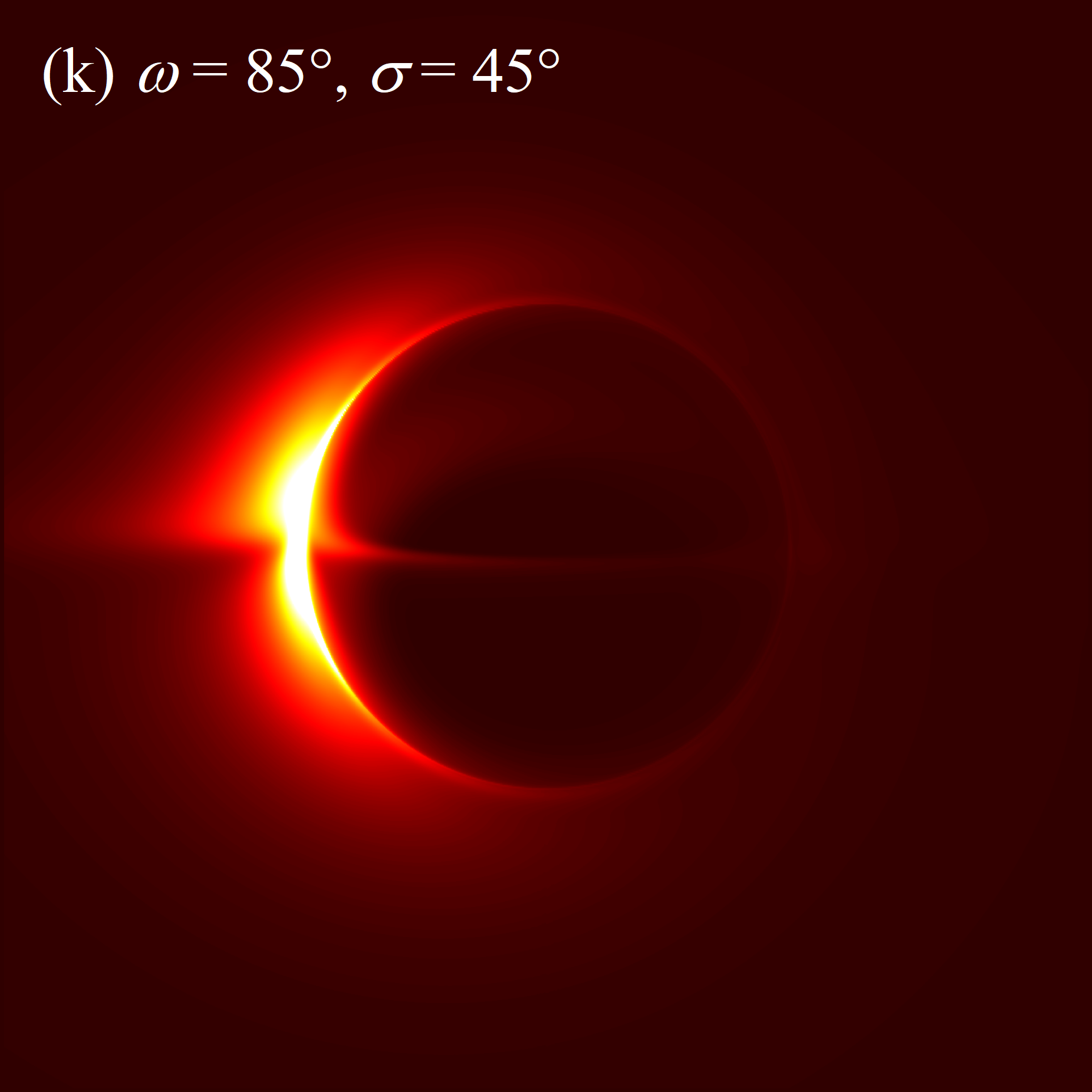}
\includegraphics[width=4.5cm]{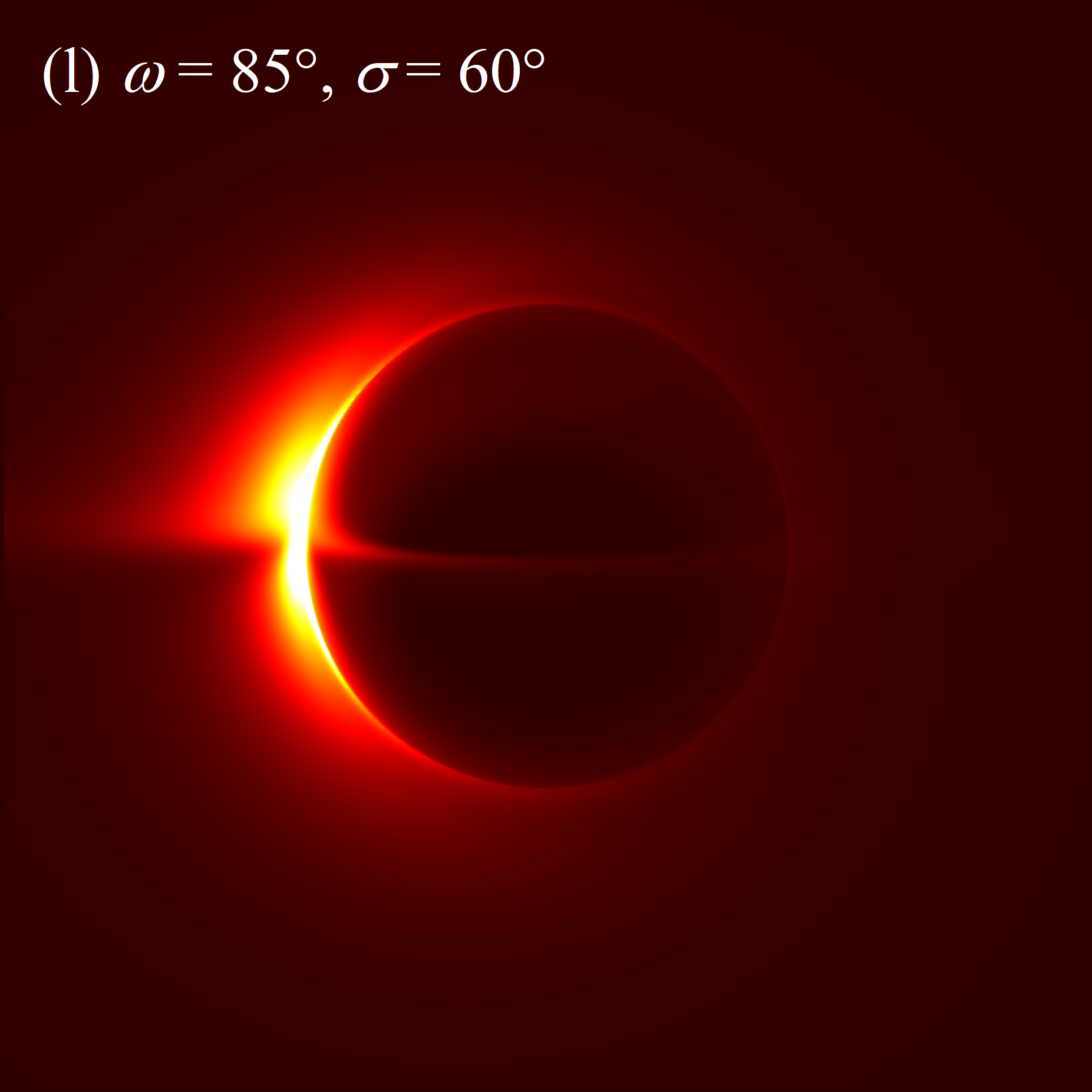}
\caption{Images of the Schwarzschild BH (first two rows) and the hairy BH (last two rows, $h = -1$) illuminated by two accretion disks, as viewed from different $\omega$. The emission regions of two accretion disks extend down to the event horizon. It is observed that the two accretion disks tend to obscure the inner shadow, with this effect becoming more pronounced when the observer is situated between the two accretion disks.}}\label{fig16}
\end{figure*}

In order to demonstrate the impact of the accretion environment on the inner shadow more effectively, we record the inner shadow profiles in different parameter spaces, as depicted in figure 17. The blue dashed line represents the boundary of the inner shadow cast by the BH with a single, equatorial accretion disk. The green and black solid lines represent the silhouettes of the inner shadow of the BH in the scenario involving multiple accretion disks. The upper halves of the blue, green, and black lines overlap when the observation angle is $17^{\circ}$, and the lower halves of these lines align when $\omega$ satisfies $\omega=85^{\circ}$. It is found that when the two accretion disks are located on the same side of the observer's line of sight, the observed inner shadow shrinks as the inclination of the tilted accretion disk increases. Conversely, when the observer is situated between the two accretion disks, an increase in $\sigma$ results in an enlarged inner shadow. This suggests that the obscuring effect of multiple accretion disks on the inner shadow of a BH is contingent not only upon the structure of the accretion environment but also on the relative position between the observer and the accretion disks. In light of these complexities, it becomes apparent that estimating BH masses and testing general relativity through the observed inner shadow is a challenging endeavor, as achieving a comprehensive and precise understanding of the accretion environment around BHs remains elusive.
\begin{figure*}
\center{
\includegraphics[width=3.7cm]{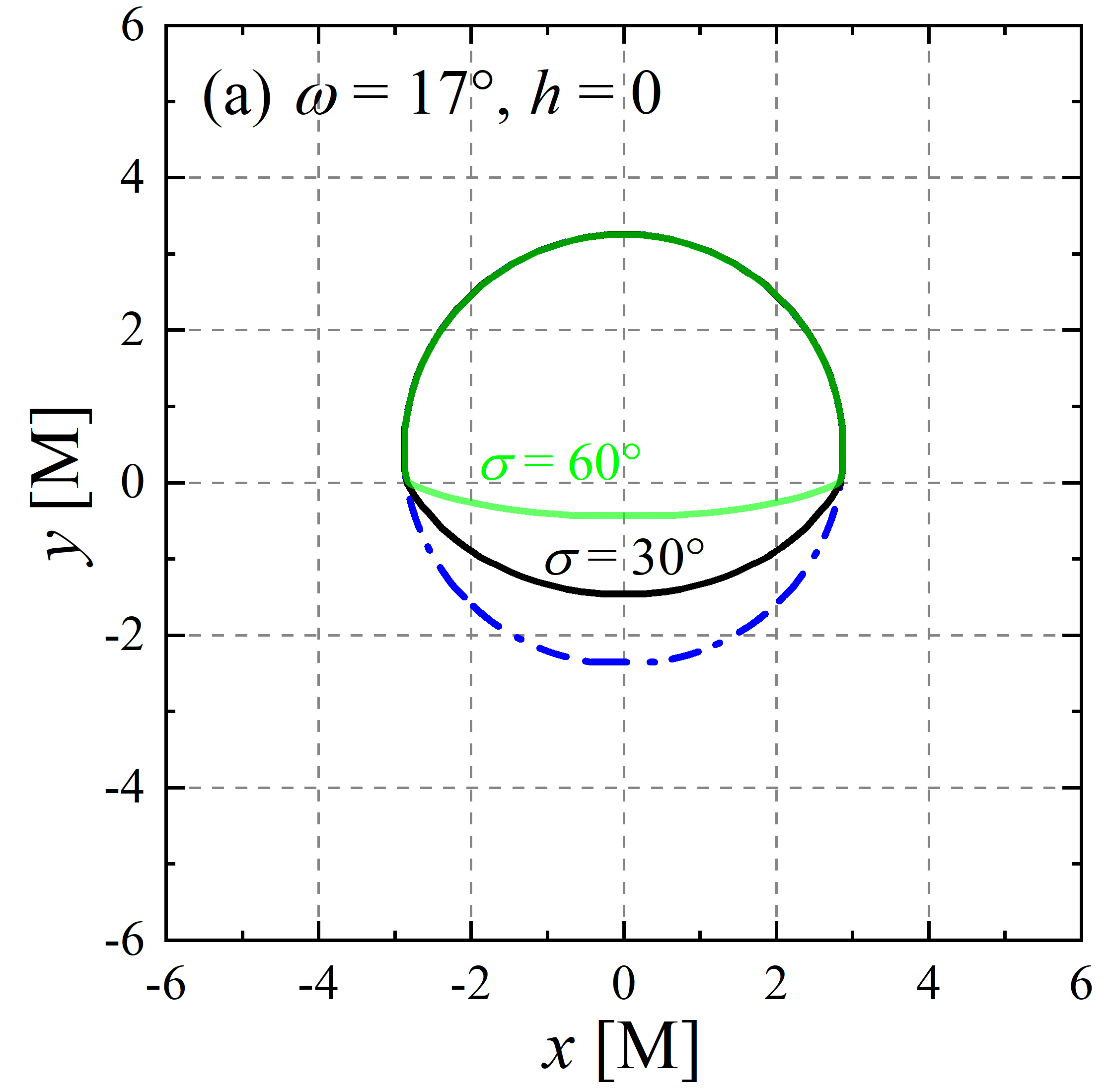}
\includegraphics[width=3.7cm]{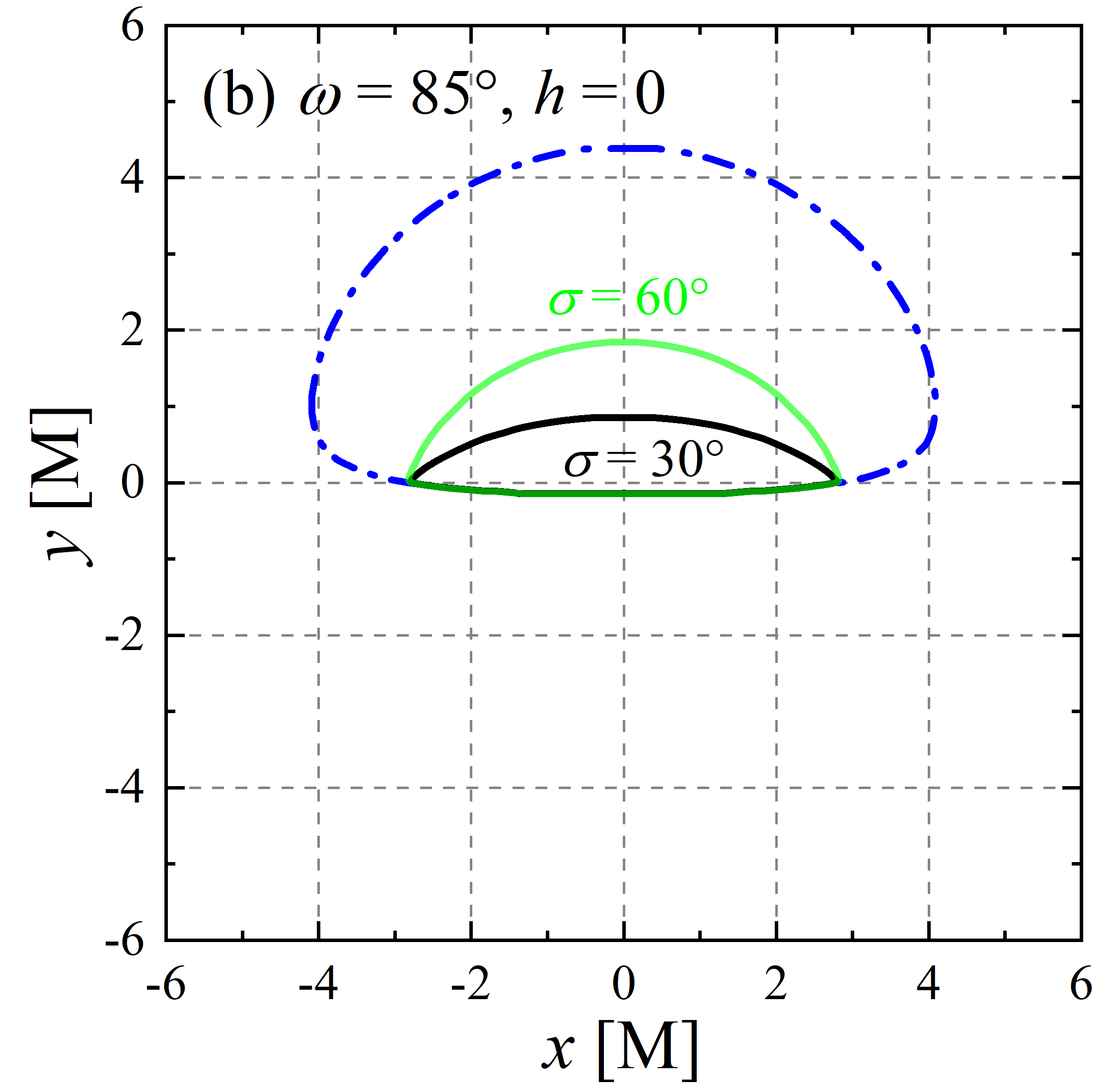}
\includegraphics[width=3.7cm]{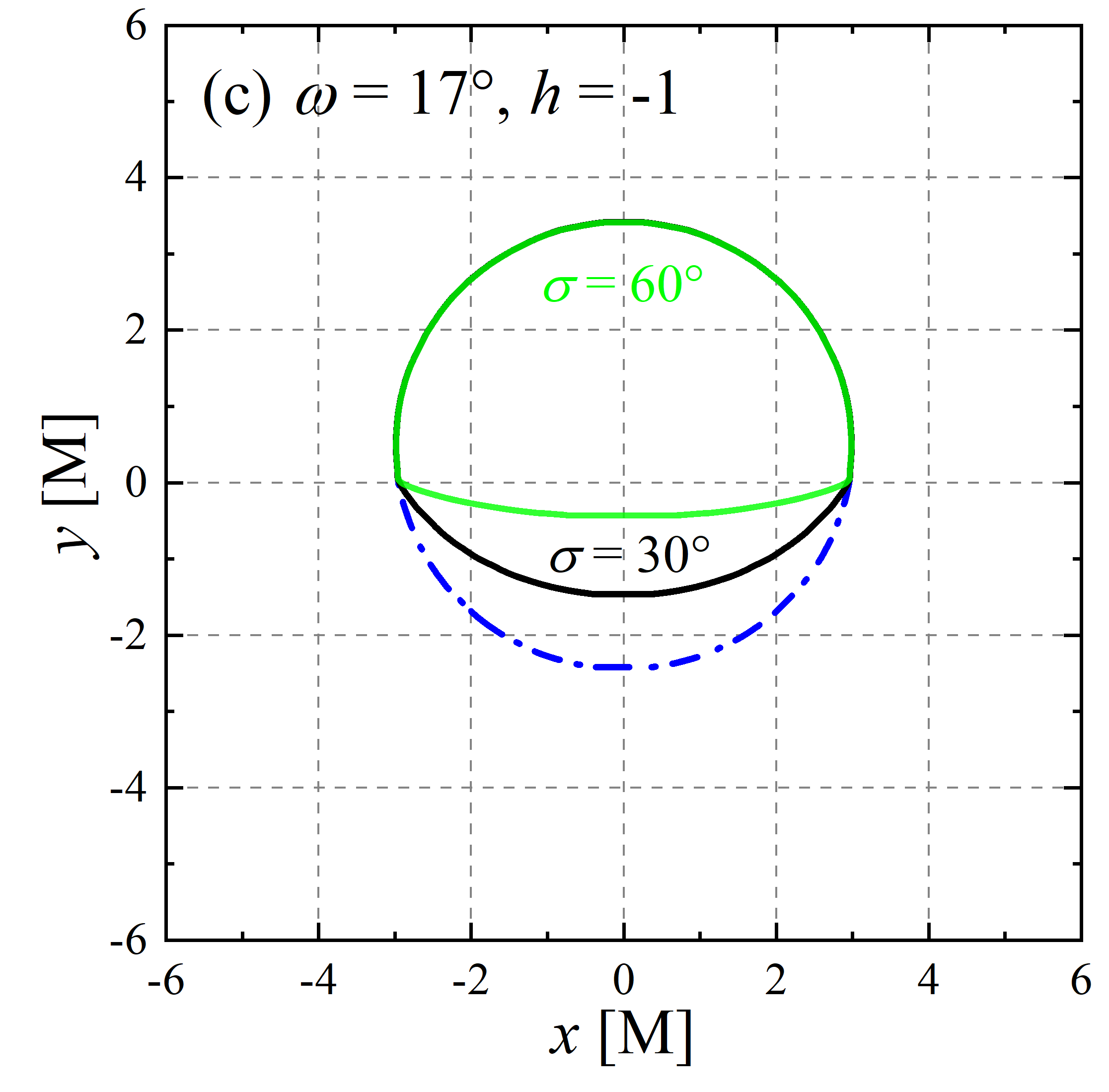}
\includegraphics[width=3.7cm]{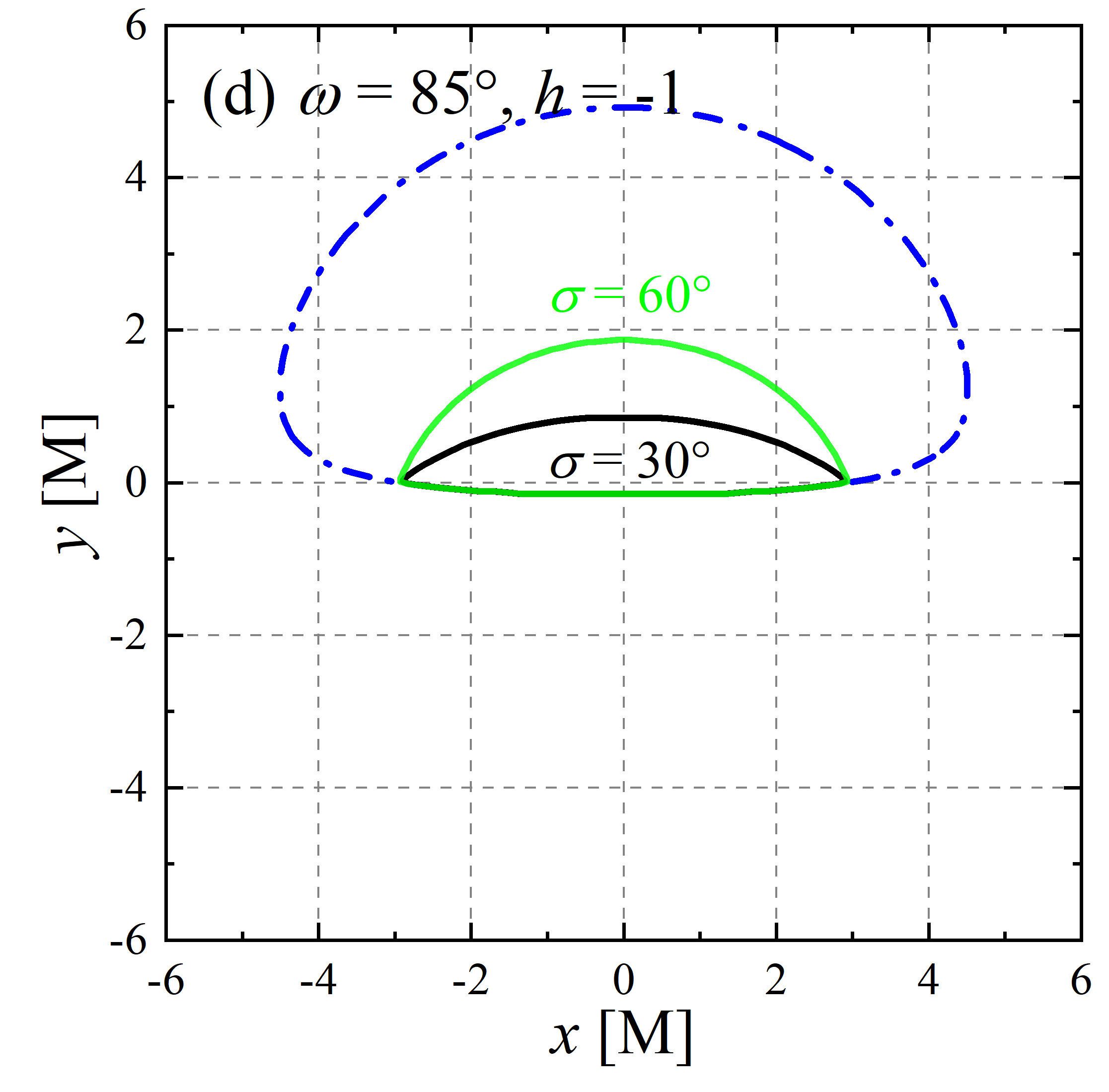}
\caption{Profiles of the inner shadows for the Schwarzschild BH (left two panels) and the hairy BH (right two panels) in the context of different accretion models. The blue dashed lines correspond to the single equatorial accretion disk case, while the green and black solid lines correspond to the multiple accretion disk scenario. It is shown that the obscuring effect of inner shadows is closely related to the inclination of the accretion disk.}}\label{fig17}
\end{figure*}
\section{Conclusions}
In this paper, we simulate $230$ GHz images of hairy BHs in Horndeski gravity with the help of a ray tracing method and discuss how tilted accretion disks affect the observational signatures of hairy BHs. It is found that a reduction in the scalar hair parameter leads to an expansion of both the inner shadow and the critical curve of the BH images. In contrast, the scalar hair parameter has a positive effect on the brightness of the images. As a result, the image brightness of hairy BHs is lower than that of the Schwarzschild BH. However, the inner shadow and critical curve are larger in the former case compared to the latter. We demonstrate that, in specific parameter spaces, these distinctions can be detected by the current EHT array. These findings furnish valuable theoretical and practical insights into the differentiation of hairy BHs from Schwarzschild BHs and facilitate the testing of the no-hair theorem.

We propose that there exists a degeneracy between the inclination of the tilted accretion disk and the observation angle based on simulated images of hairy BHs across various parameter spaces. Consequently, we introduce an effective observation angle (the angle between the line of sight and the tilted accretion disk) to elucidate the combined effects of the tilted disk inclination and the observation angle on the observational appearance of hairy BHs. That is, when the absolute value of the effective observation angle decreases, a notable enhancement in the brightness of the light spot in the image can be observed. This is accompanied by the presence of an inner shadow that extends horizontally and squeezes vertically. When the effective observation angle is negative, an inverted image appears. In addition, we would like to emphasize that the position of the bright spot in the image remains unaffected by the equatorial accretion disk. Instead, it is strongly correlated with the position of the tilted accretion disk. Specifically, the precession of the tilted accretion disk can cause the light spot to drift within the image. In other words, tracking the shift of the light spot can roughly infer the evolution of the accretion disk.

In the context of multiple thin disk accretion scenario, it is interesting to note that the inner shadow of the hairy BH can be obscured by the accretion flow, resulting in a significantly smaller inner shadow compared to the case of a single accretion disk. Therefore, we suggest that great caution is needed when diagnosing the observed inner shadow to identify BHs and test general relativity.

\acknowledgments
The authors are very grateful to the referee for insightful comments and valuable suggestions. This research has been supported by the National Natural Science Foundation of China [Grant Nos. 11973020 and 12133003], and the National Natural Science Foundation of Guangxi (No. 2019GXNSFDA245019).


\end{document}